\documentclass[a4paper,twoside,ngerman,british]{scrbook}
\usepackage{pibonn-thesis}
\usepackage{thesis_defs}
\usepackage[backend=bibtex8,hyperref=true,bibencoding=latin1,
    style=numeric-comp,sorting=none,block=ragged,firstinits=true]{biblatex}
\usepackage{biblatex-num-v2009}
\bibliography{thesis_refs.bib}

\begin{document}
{\addtolength{\oddsidemargin}{1.0cm}\addtolength{\topmargin}{1.0cm}
\thispagestyle{empty}
\begin{titlepage}
\begin{center}
\vspace{2.5cm}
\vspace{1.5cm}
\rule{\linewidth}{1mm} \\
\vspace{0.3cm}
{\Huge \textbf{Effective Models for Dark Matter}} \\
\vspace{0.2cm}
{\Huge \textbf{at the International Linear Collider}} \\
\vspace{0.3cm}
\rule{\linewidth}{1mm} \\
\vspace{3.5cm}
{\Large \textbf{Masterarbeit in Physik}}

 \vspace{0.1cm}

\textbf{von }

\vspace{0.1cm}

{\Large \textbf{Daniel Schmeier}}

\vspace{3.5cm}

\textbf{angefertigt am}

 \vspace{0.1cm}

{\Large \textbf{Physikalischen Institut der Universität Bonn}}

\vspace{1.5cm}
\textbf{vorgelegt der }

\vspace{0.1cm}

{\Large \textbf{Mathematisch--Naturwissenschaftlichen Fakultät} }

\vspace{0.1cm}

\textbf{der }

\vspace{0.1cm}

{\Large \textbf{Rheinischen Friedrich--Wilhelms--Universität Bonn}} \vspace{1cm}

\vspace{1.5cm}
\textbf{eingereicht}

\vspace{0.1cm}
{\Large \textbf{Oktober 2012}} 

\vspace{1cm}

\textbf{überarbeitet}
\vspace{0.1cm}

{\Large \textbf{ August 2013}}
\end{center}
\end{titlepage}

\pagebreak

\pagebreak
\thispagestyle{empty}
\null
\pagebreak
\hfill\begin{minipage}[t]{0.5\textwidth}
\begin{flushright}
\emph{“What happens if a big asteroid hits Earth? Judging from realistic
simulations involving a sledge hammer and a common laboratory frog, we
can assume it will be pretty bad.”} \\
 -- Dave Barry, 1947
\end{flushright}
\end{minipage}
\vfill
\begin{minipage}[b]{0.5 \textwidth}
1. Gutachter: Prof. Herbert K. Dreiner, PhD \\
2. Gutachter: Prof. Dr. Michael Krämer
\end{minipage}

\pagebreak
\null
\thispagestyle{empty}
\pagebreak
\thispagestyle{empty}
\pagebreak
\thispagestyle{empty}

\chapter*{Acknowledgements}
\label{sec:ack}
I first want to thank Prof. Herbi Dreiner for giving me the opportunity to work on this thesis in his research group. I thank all its members for the great athmosphere I was allowed to experience each day. In particular, I owe Jamie a lot of respect and gratitude for advising me over the whole time and proof reading this thesis without ever losing temper. Also, I have to thank Nicolás with regard to his assistance for both the original and the revised version of this work. I also want to thank Prof. Michael Krämer for being the second assessor of this thesis and Moritz Huck for his contributions to this project.

Furthermore I want to mention my family and friends for all they did to me my whole life. A special word of thanks goes to Larissa, Leandra, Michael and Monika for all kind of support over the last years. I want to close with a big thank you to Elena, who deserves my full gratefulness for all her support and love without which I would never have been able to complete this work.



\tableofcontents

\mainmatter
\pagestyle{scrheadings}

\chapter{Introduction}
\label{sec:intro}

The history of dark matter dates back to the early 1930s, with the first experimental evidence found by J. Oort \cite{Oort}. By measuring the velocity of stars within the Milky Way by looking at their Doppler shifted radiation, he could derive a lower bound on the total galactic mass. This bound is formed by arguing that for lower masses the fastest measured stars should have escaped the gravitational potential well. It turned out to be twice as large as the amount of visible, luminous mass. This discrepancy was confirmed by F. Zwicky one year later \cite{Zwicky}. Observing around 1000 galaxies within the Coma cluster, he tried to deduce its total mass by measuring the velocity dispersion of the individual galaxies and using the virial theorem
\begin{align}
\langle T \rangle = - \frac{\langle V \rangle}{2}.
\end{align}
This equation states that for sufficiently stable systems (like galaxy clusters), the time average of the kinetic and the potential energies have a simple relation. After estimating the kinetic energy using the galaxies' velocity measured through radioscopy, Zwicky deduced a total mass within the cluster of about \num{4.5e13} solar masses. This was also puzzling, since the (at that time) standard mass measurement procedure claimed a value of about a factor 100 smaller by measuring the total luminosity $L$ of the cluster and comparing to similar objects with well--known $M/L$--ratios, so called \emph{standard candles}.

The breakthrough in the experimental claim for dark matter was made by V. Rubin and W. Ford Jr.\ around forty years later \cite{Rubin}. By measuring spectral lines of the outer stars at the edge of the Andromeda Nebula --- a rotating spiral galaxy ---, they compared the relation between velocity and distance to the galactic center. According to Newton's laws of gravity there is a simple relation between those two observables by identifying the gravitational with the centripetal force:
\begin{align}
v(r) = \sqrt{G_\text{N} \frac{M(r)}{r}}.
\end{align}
With an approximate model for the mass $M(r)$ included within the radius $r$ of such a spiral galaxy, one can compare the theoretical expectation with the measured values. In particular one expects that outside the ``visible radius'' $R$ the total mass $M_\text{tot}$ of the galaxy should be included. In that case, $M(r > R) = M_\text{tot}$ should stay constant and velocity should fall with $r^{-1/2}$. But instead they observed an approximative constant behaviour for large radii, hinting at an invisible dark halo with a mass contribution that grows linearly with distance (results from a similar analysis are shown in figure \ref{img:rotcurve}). One of the most promising explanations is a halo of \emph{dark matter}, possibly made out of \emph{weakly interacting massive particles} (\textsc{Wimp}s), i.e.\ new particles beyond the Standard Model that interact weakly enough to be considered non--luminous and whose number and mass densities are high enough to explain the velocity curves. 
\begin{figure}[bt]
\centering
\includegraphics[width=0.4\textwidth]{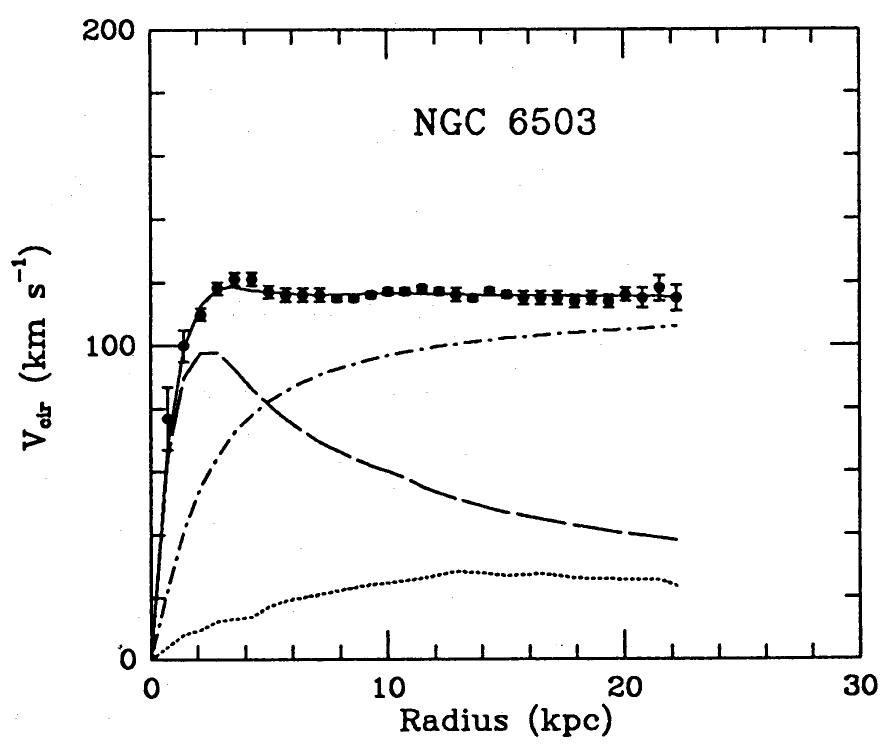}
\caption{Measured rotation curves $v(r)$, taken from \cite{rotationcurve}. The dashed curve describes the visible component, the dotted curve a (negligibly small) gas contribution and the dash--dotted curve gives a fitted estimate on the dark matter halo curve.}
\label{img:rotcurve}
\end{figure}
Although the Standard Model of particle physics is very successful in explaining many different experimental observations, it is not able to explain this large additional mass component by its particle content. In fact neutrinos are the only long--lived particles that interact purely via the weak force, but they are nearly massless within the Standard Model. Current experimental bounds derived from tritium $\beta$--decay show that they are most probably not heavier than \SI{2}{\eV} \cite{PDG}, which is too light to explain the total missing mass component \cite{KolbTurner}. Furthermore, it is hard to explain the structure formation of galaxies inside our universe using a model with only relativistic dark matter \cite{caldwell2001current}. 

Different experiments try to explore the parameter space of dark matter mass and interaction strength to Standard Model particles by different kinds of processes (see figure \ref{img:darkmatterdetection}). Direct detection experiments (e.g.\ \textsc{Cdms} \cite{CDMS}, \textsc{Xenon} \cite{Xenon100}, $\ldots$) try to measure recoil energies from elastic scattering of heavy nuclei with \textsc{Wimp}s, whereas indirect detection experiments (e.g.\ Fermi \textsc{Lat} \cite{Fermilat}, \textsc{Pamela} \cite{Pamela}, $\ldots$) look for gamma--rays or leptons as remnants of dark matter annihilation that took place in various places (e.g.\ white dwarfs or galactic centers). These methods are usually less sensitive to light dark matter masses below \SI{10}{\GeV} because of intrinsic threshold effects, such that their exclusion limits become weak in that area. This is where collider searches come into play; dark matter particles are produced in pairs and their properties are measured through additional final state objects, usually photons or jets. Due to phase space arguments, this method works particularly well in case of low mass \textsc{Wimp}s and gives weaker bounds, the closer the \textsc{Wimp} mass comes to the collider energy threshold $\sqrt{s}/2$. 

All three types of measurements have been used in different studies to analyse a variety of models. Some fundamental extensions of the Standard Model, like Supersymmetry \cite{Wess197439,Martin:1997ns,Drees:2004jm,Nilles:1983ge}, Universal Extra Dimensions \cite{Appelquist:2000nn} or Little Higgs Models \cite{ArkaniHamed:2001nc,Cheng:2003ju}, naturally lead to good candidates for \textsc{Wimp}s and the cosmological requirements for the \textsc{Wimp} abundance in the universe can be used to set constraints on the parameter space within that theory. Model independent searches have also become quite prominent (e.g.\ \cite{DMClass, TevatronDarkMatter, WWapproach, RelicDensity, DMChina1, DMChina2, effmodels2, effmodels3, effmodels4, BartelsThesis, BartelsList}). Often this is performed by parametrising the \textsc{Wimp}--Standard Model interaction through effective four--particle vertices. Thanks to their simple vertex structure and the relatively small number of free parameters, these models can easily be analysed in all the different types of experiments described above in order to receive complementary information about allowed parameter values for the \textsc{Wimp} mass and coupling. 
\begin{figure}
\begin{center}
\includegraphics[width=0.32\textwidth]{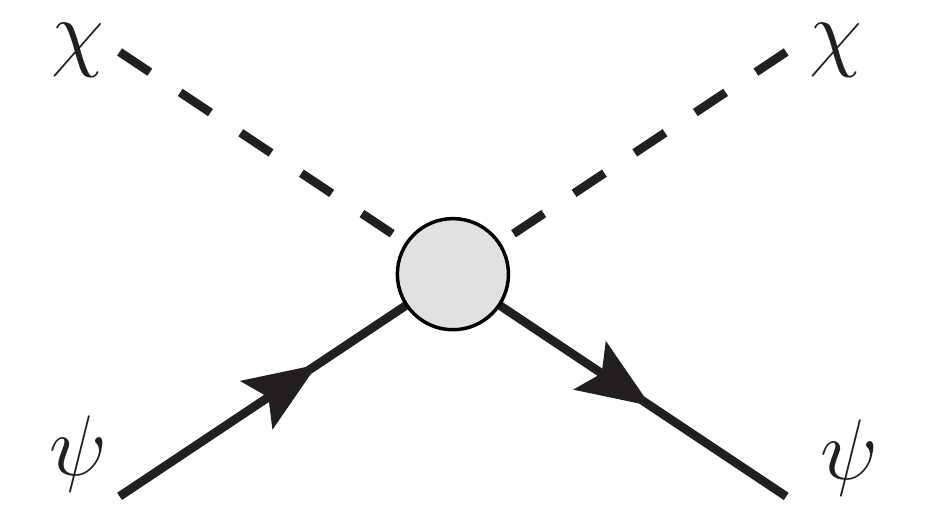}
\includegraphics[width=0.32\textwidth]{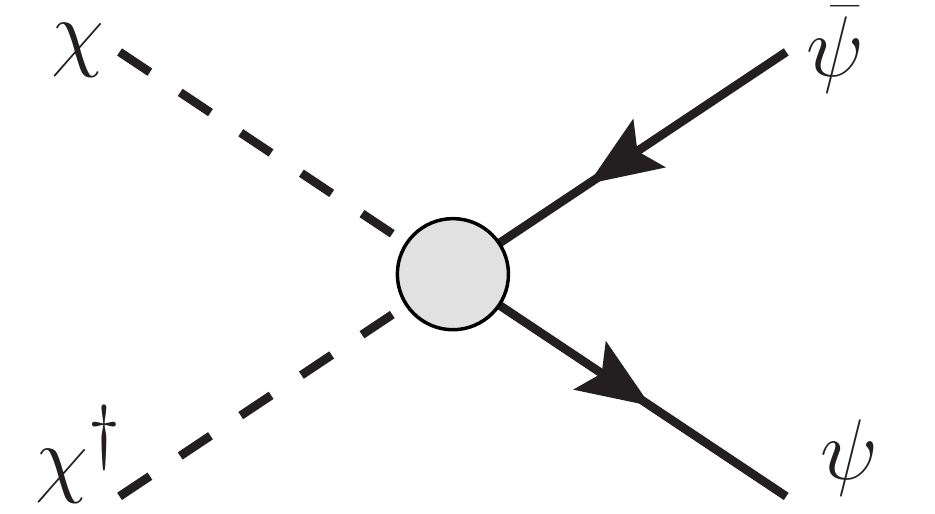}
\includegraphics[width=0.32\textwidth]{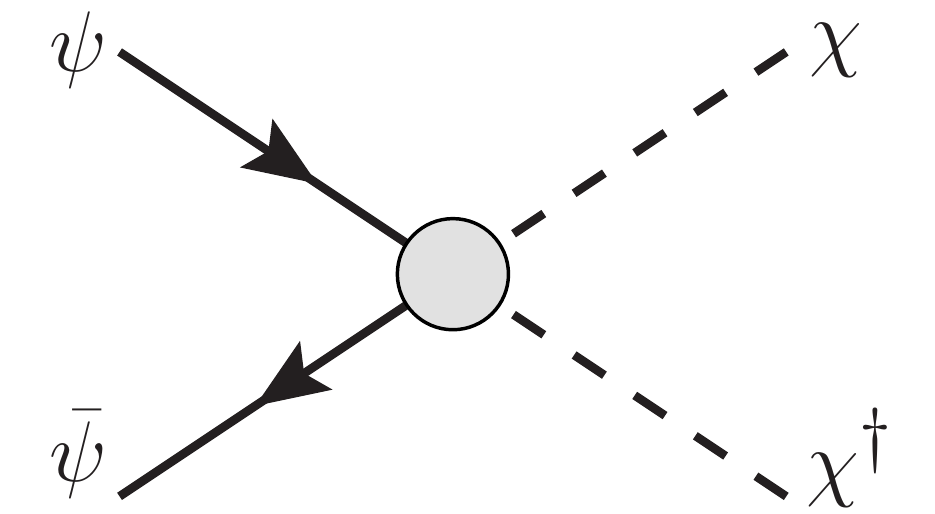}
\end{center}
\begin{flushleft}
\hspace{0.02\textwidth} a) \hspace{0.29\textwidth} b) \hspace{0.29\textwidth} c)
\end{flushleft}
\centering
\caption{Different methods of \textsc{Wimp} detection. a) Direct detection through elastic scattering. b) Indirect detection through measurement of final state particles from annihilation. c) Direct production at a collider.}
\label{img:darkmatterdetection}
\end{figure}

In this thesis, we analyse a list of such effective models, with a particular focus on the International Linear Collider (\textsc{Ilc}, \cite{ILCReport1, ILCReport2, ILCReport3, ILCReport4}) as a next--generation electron--positron--collider. It is expected that the center of mass energy will be \SI{500}{\GeV} with a potential upgrade to \SI{1}{\TeV}. In addition the tunable spin polarisation of the incoming leptons will lead to a strong exclusion potential for dark matter couplings that could improve the current leading collider bounds from the Large Electron Positron Collider (\textsc{Lep}, \cite{LEPShinesLight}) and the Large Hadron Collider (\textsc{Lhc}, \cite{CMSDM, ATLASDM}). 

This thesis is structured as follows: In chapter \ref{chap:darkmatter} we discuss standard \textsc{Wimp} cosmology and which constraints it sets on the interaction strength of a potential dark matter candidate. We formulate our effective models in chapter \ref{chap:appeffop} by writing down fundamental Lagrangians with a Standard Model--\textsc{Wimp} interaction mediated by a single heavy field. We assume different spin and mass for both the mediator and the dark matter particle as well as different interaction mechanisms. Due to its virtuality, the heavy mediator can be integrated out from the path integral, which leads to effective four particle vertices with a pair of Standard Model fermions and a pair of dark matter particles. These operators are analysed with respect to the determined cosmological bounds in chapter \ref{chap:wmap} by evaluating the respective coupling strength for \textsc{Wimp}--masses in the range of \SI{1}{\GeV} to \SI{500}{\GeV} which fits the currently well--measured dark matter relic abundance $\Omega^0_\text{DM} h^2 \approx 0.11$, given by the \textsc{Wmap} experiment \cite{WMAPPara, WMAPSkymap, WMAPCosmo}. We continue with the analysis of radiative pair production of dark matter with an additional final state photon, with general considerations on the cross section evaluation in chapter \ref{chap:chichigamma} and the analysis of the expected \textsc{Ilc} exclusion potential in chapter \ref{chap:ilcanalysis}. The size of Standard Model background contributions is discussed together with beam and detector resolution effects. In addition, the impact of the incoming lepton polarisation on both the signal to background ratio and expected systematic uncertainties is studied. Finally we translate our expected \textsc{Ilc} exclusion limits into bounds on the dark matter proton cross section in chapter \ref{chap:xenon} in order to compare with the leading exclusion bounds from the \textsc{Xenon}--experiment for \textsc{Wimp} masses above \SI{10}{\GeV}. We will finish with our conclusions in chapter \ref{chap:conclusions}.

\chapter{Dark Matter Cosmology}
\label{chap:darkmatter}
There is a considerable amount of experimental evidence for the presence of dark matter in our universe. Some examples are mass measurements from weak gravitational lensing \cite{gravitationallensing, gravitationallensing2}, the matter distribution of colliding galaxies \cite{bulletcluster} or the still existing rotation curve problem \cite{Persic:1995ru, Persic:1995td, Borriello:2000rv}  explained in the introduction. 

In this chapter we want to concentrate on cosmological hints for dark matter, namely the implications from measuring the global structure of the universe. In section \ref{sec:matterdensity} we start with the description of the latter by giving a short introduction to the Friedmann--equation, which relates the curvature of the universe to its energy density. The experimentally measured global flatness today points to a particular critical density that must be present and which normal, baryonic matter is not able to account for alone. Significant contributions from both radiation and dark matter are needed to fill the remaining gap. In section \ref{sec:energydensity} we describe how the total energy density of the universe as a thermal bath is composed of its constituents. The conclusion will be that massive objects must depart from equilibirum to give sizable energy contributions today. If we assume thermal dark matter that once was in equilibrium with the thermal bath of the universe, its departure from equilibrium can be quantitatively described by the freeze--out effect introduced in section \ref{sec:relicdensity}. This relates the measured dark matter abundance today to the underlying theoretical interaction model and allows us to analyse the parameter space of particular dark matter models later.

  For the calculations within this introductory chapter, we follow the derivations in \cite{KolbTurner, Dodelson}. 
\section{Energy Density in a Flat Universe}
\label{sec:matterdensity}
Einstein's theory of general relativity, formulated in 1916, relates the metric of the universe to its matter and energy constituents through the following tensor equation:
\begin{align}
R_{\mu \nu} - \frac{1}{2} \mathcal{R} g_{\mu \nu} = 8 \pi G_\text{N} T_{\mu \nu} \label{eqn:einstein},
\end{align}
where $R_{\mu \nu}$ and $\mathcal{R}$ are the so called \emph{Ricci tensor} and \emph{Ricci scalar}, which are related to the global metric tensor $g_{\mu \nu}$ as follows:
\begin{align}
\Gamma^\mu_{\alpha \beta} &\equiv \frac{1}{2} g^{\mu \sigma} \left( \partial_\beta g_{\sigma \alpha} + \partial_\alpha g_{\sigma \beta} - \partial_\sigma g_{\alpha \beta} \right), \\
R_{\mu \nu} &\equiv \partial_\sigma \Gamma^\sigma_{\mu \nu} - \partial_\nu \Gamma^\sigma_{\mu \sigma} + \Gamma^\sigma_{\mu \nu} \Gamma^{\lambda}_{\sigma \lambda} - \Gamma^\lambda_{\mu \sigma} \Gamma^\sigma_{\nu \lambda}, \\
\mathcal{R} &\equiv g^{\mu \nu} R_{\mu \nu}.
\end{align}
$G_\text{N}$ denotes Newton's gravity constant and $T_{\mu \nu}$ the \emph{energy--momentum--tensor} of all fields besides gravity. If we assume a homogenous and isotropic universe, we can formulate the latter as an ideal fluid with an energy density $\rho$ and a pressure $p$ through $T^{\mu}_{\nu} = \text{diag}(\rho, -p, -p, -p)$. In such a universe, the metric can be described by the \emph{Friedmann--Lemaître--Robertson--Walker--metric}\footnote{Throughout this thesis, we will work with the metric signature $(+, -, -, -)$}:
\begin{align}
x_\mu &= (t, r, \theta, \phi), \\
g_{\mu \nu} &= \text{diag} \left(1, -\frac{a^2(t)}{1-kr^2}, -a^2(t) r^2, -a^2(t)r^2 \sin^2 \theta \right). \label{eqn:metric}
\end{align}
For $k = 0$, this is the metric of flat 4--dimensional Minkowksi--space in spherical coordinates. For $k = 1$, $g_{\mu \nu}$ describes a closed 3--dimensional sphere with a radius proportional to $a(t)$. The time dependence of this so called \emph{scale factor} describes a potential expansion or contraction of the finite volume of such a sphere over time. For negative $k$ we receive a space of negative curvature and infinite volume.

This metric leads to the following nonvanishing elements of $R_{\mu \nu}$ and $\mathcal{R}$ which  we need for the Einstein equations:
\begin{align}
R_{00} &= -3 \frac{\ddot{a}}{a}, \\
R_{ij} &= -g_{ij} \frac{\ddot{a} a + 2 \dot{a}^2 + 2 k}{a^2}, \\
\mathcal{R} &= -6 \frac{\ddot{a} a + \dot{a}^2 + k}{a^2}.
\end{align}
Inserting the building blocks of our universe model into (\ref{eqn:einstein}) leads to two independent equations: 
\begin{align}
\frac{\dot{a}^2}{a^2} + \frac{k}{a^2} &= \frac{8 \pi G_\text{N}}{3} \rho, \\
2 \frac{\ddot{a}}{a} + \frac{\dot{a}^2}{a^2} + \frac{k}{a^2} &= - 8 \pi G_\text{N} p.
\end{align}
The first of these equations is the so called \emph{Friedmann--equation} and describes a relation between energy density $\rho$, curvature $k$ and expansion rate $a$ of the universe. Defining the \emph{Hubble--parameter} $H(t) \equiv \dot{a}/a$, we can reformulate it as follows:
\begin{align}
\frac{k}{H^2 a^2} = \frac{8 \pi G_\text{N}}{3 H(t)^2} \rho - 1 \equiv \Omega - 1. \label{eqn:finalFriedmann}
\end{align}
For a nearly flat universe with $k \approx 0$, $\Omega$ must have a value close to one. Smaller or larger values lead to an open or a closed universe, respectively. Different astrophysical observations indicate that the current universe is almost or exactly flat and therefore the present value $\Omega_0$ is close to 1. From redshifted light emitted by objects from different luminosity and distance, we can also derive the present value of the Hubble parameter $H_0 \equiv h\  \unit{100}{\kilo\meter\per\second\per\mega\parsec}$ with $h \approx \num{0.743(21)}$ \cite{Spitzer} and $\unit{1}{\mega\parsec} = \unit{3e19}{\kilo\meter}$ . Using this information we expect a matter density of the present universe with a value of $\rho \approx \unit{1e-29}{\gram\per\cm}$.

Big Bang Nucleosynthesis (BBN) can be used to get an estimate for the amount of baryonic matter in our universe. Since it has proven very successful in predicting the amount and relative abundance of light atoms in stars, the method is a reasonable tool to calculate the total amount of baryonic matter. This leads to an abundance of $\Omega_{b} \approx 0.05$, which is only about \unit{5}{\%} of the matter we need for agreement with the measured flatness of the universe. Since this number is in agreement with different experiments measuring the baryonic abundance, we think that a component other than baryons must be responsible for the rest. 

\textsc{Wmap} measures fluctuations in the cosmic microwave background radiation (see chapter \ref{chap:wmap}), and deduces a composition of about \unit{73}{\%} energy, whose source is yet unknown and therefore called \emph{dark energy},  and \unit{22}{\%} non--baryonic and non--relativistic matter, which is denoted as \emph{cold dark matter}. Any object that is supposed to contribute to cold dark matter must fulfill the following criteria:
\begin{itemize}
\item It must be long--lived enough to survive the history of the universe until today.
\item It must interact very weakly with our ordinary matter to explain why it has not been detected yet.
\item It needs a sufficiently large mass density today to give the correct present total energy density.
\item It must be non--relativistic at the time of freeze--out in order to explain the formation of clumped objects like galaxies from the uniformly distributed thermal soup in the early universe.
\end{itemize}
In this thesis we are looking at a \emph{weakly interacting massive particle} (\textsc{Wimp}) as the dark matter dandidate: We assume that there exists a new particle beyond the Standard Model, which has a mass in the \GeV{}--\TeV{} range and whose interaction with the Standard Model is such that the aforementioned conditions are fulfilled. Even though the coupling must be weak, it should still be strong enough to form a \emph{thermal} dark matter relic, i.\/e.\ the \textsc{Wimp} should be in thermal equilibrium with the Standard Model particles during the early universe. This assumption allows us to calculate the present energy density $\Omega_0^\text{\textsc{Wimp}}$ according to the procedure explained in the following sections.
However, we note that it is not mandatory for dark matter to be thermally produced and in fact there exist various \emph{non--thermal} dark matter models (i.\/e.\/ primordial black holes \cite{Hawkins:2011qz, Frampton:2010sw}, axions \cite{ Steffen:2008qp,Rosenberg:2000wb}, \textsc{Wimpzillas} \cite{Kolb:1998ki}, \textsc{Fimps} \cite{Hall:2009bx}, $\ldots$), which use different mechanisms to explain the large dark matter abundance.

\section{Energy Density in Thermal Equilibrium}
\label{sec:energydensity}
We are interested in how the present dark matter energy abundance $\Omega_\text{DM}^0$ depends on the properties of a \textsc{Wimp} candidate. Knowing this relation will allow us to set limits on parameters of the underlying model by using the experimentally measured value of \unit{22}{\%}.

We start by thinking of the universe as a uniform and isotropic thermal bath of Standard Model and dark matter particles. In that case, principles of statistical physics can be applied to derive the energy density as a macroscopic observable. We begin at very early times when the universe was hot enough that all particles were in thermal equilibrium. The energy density for a specific particle type within such a thermal bath is then given as follows:
\begin{align}
\rho(T) &= \frac{g}{(2 \pi)^3} \int \de^3 p \ E(\vec{p}) f(\vec{p}, T). \label{eqn:defrho}
\end{align}
Here, $g$ counts the total degrees of freedom for a particular particle species, for example taking into account spin and color. $E(\vec{p})$ denotes the general energy momentum relation $E = \sqrt{|\vec{p}|^2 + M^2}$ and $f(\vec{p})$ describes the statistical weight of the momentum $\vec{p}$, which follows usual Bose--Einstein or Fermi--Dirac statistics in thermal equilibrium\footnote{We omit any chemical potential $\mu$ during this and the following calculation, since we do not take into account matter--antimatter--asymmetries. The chemical potentials are therefore equal for particles and antiparticles and will drop out in any physical quantity.}:
\begin{align}
f_{\text{B/F}}(\vec{p}, T) &= \frac{1}{e^{E/T} \mp 1}.
\end{align}
The total matter density is the sum of all particles $i$, each contributing with different degrees of freedom and indiviual weights of the occupancy function $f$, caused by different masses $M_i$ and possibly different temperatures $T_i$. Inserting everything into $\rho$ and changing the integration variable to $E$, we find:
\begin{align}
\rho_\text{tot} &= \frac{\pi^2}{30} g_* T^4, \label{eqn:rhoT}\\
g_* &\equiv \sum_i \left(\frac{T_i}{T} \right)^4 g_i \frac{15}{\pi^4} \int_{M_i}^\infty \frac{\sqrt{E^2-M_i^2}}{e^{E/T} \mp 1} E^2 \de E.
\end{align}
We call $g_*$ the \emph{effective degrees of freedom}. Its value can be evaluated numerically after choosing the corresponding set of contributing particles, since particles with a mass much larger than the temperature get exponentially supressed and are therefore negligible. An analysis of the relation $g_*(T)$ between \unit{10}{\keV} and \unit{100}{\GeV} has been performed in \cite{effectivedegrees}. The results are approximated by the fit shown in figure \ref{img:relicdensity:effdeg}. For small temperatures, only massless particles contribute, leading to a value of about \num{3.4} with contributions from photons and neutrinos only. As the temperature increases, leptons as well as light hadrons contribute. At the so called \emph{critical temperature} of about \unit{235}{\MeV}, QCD deconfinement takes place and free quarks and gluons contribute instead of hadronic degrees of freedom. For temperatures above the \GeV--scale, successively all Standard Model particles including the Higgs and the heavy gauge bosons contribute and increase $g_*$ to above \num{100} at \unit{200}{\GeV}.
\begin{figure}
\centering
\includegraphics[width=0.45\textwidth]{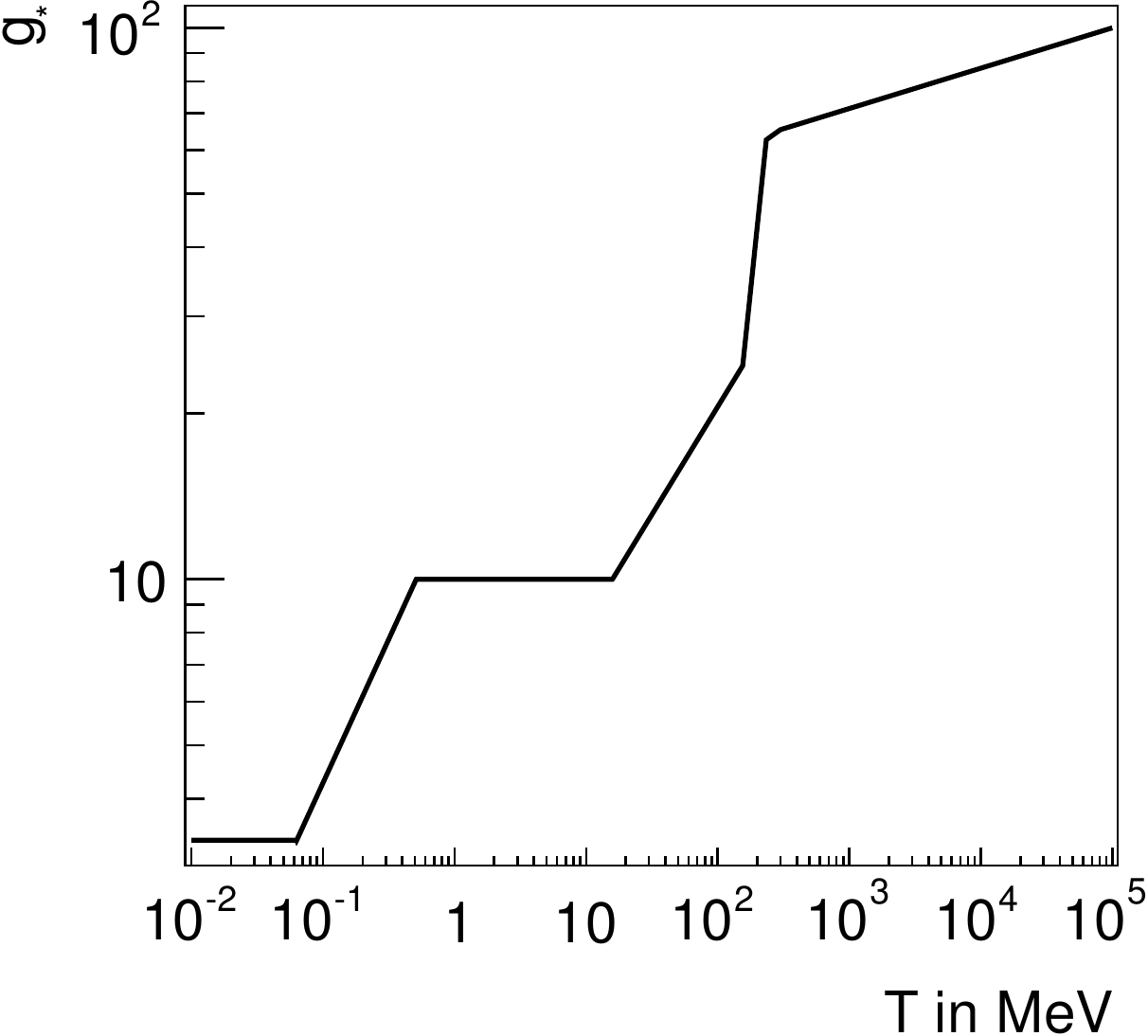}
\caption{Relation between effective degrees of freedom and temperature, given as an approximative fit to results in \cite{effectivedegrees}.}
\label{img:relicdensity:effdeg}
\end{figure}

The temperature evolution of the universe is strongly linked to its expansion. To evaluate this relation, we will use the generalised law of energy momentum conservation to relate energy density to the scale factor $a(t)$ and link it to the overall temperature $T$ via (\ref{eqn:rhoT}):  In an arbitrary metric $g_{\mu \nu}$, conservation of energy is given by the vanishing \emph{covariant derivative} of the energy--momentum--tensor:
\begin{align}
D_\mu T^\mu_\nu \equiv \partial_\mu T^\mu_\nu + \Gamma^\mu_{\alpha \mu} T^\alpha_\nu - \Gamma^\alpha_{\nu \mu} T^\mu_\alpha &\stackrel{!}{=} 0. \\
\Rightarrow \pabl{\rho}{t} + 3 \frac{\dot{a}(t)}{a(t)} \left( \rho + p \right) &= 0. \label{eqn:energyconservation}
\end{align}
In the temperature region we are interested in, energy and pressure are mostly dominated by relativistic particles and we call it the \emph{radiation dominated era}. Then, total pressure and energy density are approximately related through $p =  \rho/3$ and we can solve (\ref{eqn:energyconservation}) for $\rho(a)$:
\begin{align}
\frac{\dot{\rho}}{\rho} &= - 4 \frac{\dot{a}}{a}. \\
\Rightarrow\ \rho &\propto a^{-4}. \\
\Rightarrow\ T &\propto a^{-1}. \label{eqn:Tproptoa}
\end{align}
The last relation is derived using $\rho \propto T^4$ according to (\ref{eqn:rhoT}) and holds only if $g^*$, which itself depends implicitly on $T$, stays approximately constant.
\section{Cold Dark Matter Today}
\label{sec:relicdensity}
In thermal equilibrium, the energy density of particles with masses large compared to the temperature is exponentially suppressed. Due to the expansion and resulting cooling of the universe over time, most particle species will encounter an individual so called \emph{freeze out moment}, during and after which the particles are no longer able to remain in equilibrium with the rest of the thermal bath. Any potentially new weakly interacting massive state therefore is not subject to the exponential thermal damping in $\rho_\text{eq}$ and may give significant energy contributions today.

 We will now calculate the size of this contribution by starting with the general evolution of the number density $n_\chi$ given by the Boltzmann equation:
\begin{align}
\abl{n_\chi}{t} + 3 H n_\chi = - \left< \sigma v \right> \left(n_\chi^2 - n_{\chi, \text{eq}}^2\right). \label{eqn:boltzmann}
\end{align}
It states that any change in the number density is given by two contributions: On the one hand there is the time dependent Hubble parameter $H$ which describes the expansion rate of the universe. The volume taken by a fixed number of particles will increase over time and therefore the number density dilutes. On the other hand $\left< \sigma v \right>$ --- denoting the thermally averaged annihilation cross section multiplied by velocity --- is a measure of the annihilation and creation rates of dark matter particles through interactions within the thermal bath. The larger the annihilation cross section is, the faster the number density decreases. The annihilation rate is proportional to the density of two dark matter particles $n_\chi^2$, whereas the creation rate is given by the equilibrium density within the thermal bath $n_{eq, \chi}^{2}$. Their difference measures the net effect of both processes. 

We now try to find a semi--analytical solution for (\ref{eqn:boltzmann}). Let us first absorb the expansion factor $H = \dot{a}/a$ by rewriting its LHS:
\begin{align}
\dot{n} + 3 n \frac{\dot{a}}{a} = a^{-3} \  \abl{\left(n a^3\right)}{t}. \label{eqn:narelation}
\end{align}
Using (\ref{eqn:Tproptoa}) we conclude that $a T = \text{const}$, so it is convenient to rewrite  (\ref{eqn:boltzmann}) as an equation for the variable $Y \equiv n/T^3$. This leads to:
\begin{align}
\abl{Y}{t} = T^3 \left<\sigma v \right> \left(Y_\text{eq}^2 - Y^2 \right).
\end{align}
It will also proove to be useful to use the dimensionless variable $x \equiv M_\chi/T$ instead of $t$. With (\ref{eqn:Tproptoa}) we find $\mathrm{d}x =  H x \ \mathrm{d}t$ and we derive:
\begin{align}
\abl{Y}{x} = \frac{M_\chi^3 \left<\sigma v \right>}{H x^4} \left(Y_\text{eq}^2 - Y^2 \right).
\end{align}
Finally we use the Friedmann--equation of a flat universe (\ref{eqn:finalFriedmann}) and the total energy density (\ref{eqn:rhoT}) to extract the $x$--dependent factor of $H = H(M_\chi)/x^2$ with $H(M_\chi) = \sqrt{4/45\  \pi^3\ G_\text{N}\ g_*} M_\chi^2$ and arrive at the final equation for $Y(x)$ to solve:
\begin{align}
\abl{Y}{x} = \frac{\lambda(x)}{x^2} \left(Y_\text{eq}^2 - Y^2 \right). \label{eqn:finalBoltzmann}
\end{align}
Here we defined $\lambda(x) \equiv M_\chi^3 \left<\sigma v\right> / H(M_\chi)$. The velocity of astrophysical particles is small enough to allow the expansion $\left< \sigma v \right> \approx a + b v^2 = a + 6 b/x$. $\lambda$ is therefore not constant and we know its leading order dependence on $x$. We solve this differential equation (\ref{eqn:finalBoltzmann}) approximately at both early and late times and try to match the individual solutions at the intermediate freeze out temperature. 
\newcommand{\Yeq}{Y_\text{eq}}
\paragraph{Early times:} Far before freeze out occurs, all dark matter particles are close to equilibrium. Rewriting (\ref{eqn:finalBoltzmann}) in terms of the difference--to--equilibrium $\Delta \equiv Y - \Yeq $ leads to
\begin{align}
\Delta^\prime = - \frac{\lambda(x)}{x^2} \Delta (\Delta + 2 \Yeq) - \Yeq^\prime,
\end{align}
where ${}^\prime$ denotes the derivative  with respect to $x$. Since $Y$ is close to equilibrium and $x$ is large, $\Delta$ is relatively small and does not change much. We also know the analytical expression for $\Yeq$ in case of nonrelativistic particles by solving (\ref{eqn:defrho}) for $M_\chi \gg T$ or equivalently $x \gg 1$:
\begin{align}
\Yeq = \frac{g}{(2 \pi)^{\nicefrac{3}{2}}} x^{\nicefrac{3}{2}} e^{-x}.
\end{align}
Using this information and the approximation $\Delta^\prime \approx 0$, we can solve for $\Delta$:
\begin{align}
\Delta = - \frac{x^2}{\lambda(x)} \frac{\Yeq^\prime}{2 \Yeq + \Delta} \approx \frac{x^2}{2 \lambda(x)}. \label{eqn:delta}
\end{align}
With the intermediate step $\Yeq^\prime/\Yeq = 3/2x - 1 \approx -1$ for large $x$.
\paragraph{Late times:}
After freeze out occured, the number density is stabilised and is not suppressed exponentially like it is in equilibrium. We can therefore solve (\ref{eqn:finalBoltzmann}) for $Y \gg \Yeq$:
\begin{align}
Y^\prime &= - \frac{\lambda(x)}{x^2} Y^2.
\end{align}
Using the explicit $x$--dependence in $\lambda(x) \equiv \lambda_0 + \lambda_1 x^{-1}$, we can separate variables, integrate from freeze out $x_f$ to a time $x_\infty$ sufficiently after freeze out and solve for $Y_\infty \gg Y_\text{eq}$:
\begin{align}
Y_\infty \approx \frac{x_f}{\lambda_0 + \frac{\lambda_1 x_f}{2}}. \label{eqn:yinfty}
\end{align}
We see that $Y$ stays constant after freeze out in this approximation.
\paragraph{Matching the solutions:}
We define the freeze out temperature $x_f$ as the scale at which $Y$ deviates significantly from $\Yeq$, quantitatively formulated as follows:
\begin{align}
\Delta(x_f) = c \Yeq(x_f) \label{eqn:matchcondition}
\end{align}
Here we use $c$ as a matching parameter of order unity. We use the formula for $\Delta$ from (\ref{eqn:delta}) but we do not approximate $\Delta \ll \Yeq$ but instead we replace $\Delta$ with $c \Yeq$ according to (\ref{eqn:matchcondition}):
\begin{align}
\frac{x_f^2}{(2+c) \lambda} = c \Yeq(x_f).
\end{align}
With $H(M_\chi) = \sqrt{4/45\  \pi^3\ G_\text{N}\ g_*} M_\chi^2$ we can insert $\lambda$ and formulate a recursive equation for $x_f$ in the canonical form using the Planck mass $m_\text{pl} \equiv 1/\sqrt{G_\text{N}}$ instead of Newton's constant:
\begin{align}
x_f = \text{ln} \left[ c (c+2) \sqrt{\frac{45}{8}} \frac{1}{2 \pi^3} \frac{g\ m_\text{pl} M_\chi (a + 6b/x_f)}{\sqrt{x_f} \sqrt{g_*(x_f)}} \right]. \label{eqn:x0}
\end{align}

This relation is sufficient to perform numerical studies: Using explicit values for the parameters, constants and functions, the recursive relation can be iterated to get a numerical solution for $x_f$.

\paragraph{Relic density today}
With our approximative solutions at hand, we can now derive the value of $\Omega_0$ today after dark matter freeze out by straightforwardly using the above results to determine $\rho_\chi = M_\chi \cdot n_\chi$ and using the definition of $\Omega$ in (\ref{eqn:finalFriedmann}). In order to to this, both the number density $n_{\chi,\infty}$ and the temperature $T_\infty$ after freeze out, which appear as $Y_\infty$ in (\ref{eqn:yinfty}), have to be translated into today's values first: 

We use (\ref{eqn:boltzmann}) to determine the number density today: Since no annihilation takes place after freeze--out,
\begin{align}
\abl{n_\chi}{t} = - 3 \frac{\dot{a}}{a}n_\chi,
\end{align} 
with the solution $n_{\chi, 0} = n_{\chi, \infty} \cdot (a_0/a_\infty)^3$.

For the temperature evolution, we argued before that (\ref{eqn:Tproptoa}) holds for constant $g_*$. However, in our case the effective degrees of freedom\footnote{To be exact, not the value of $g_*$ but of $g_*^s$ is important, with $g_*^s \equiv \sum_i \left(\frac{T_i}{T} \right)^3 g_i \frac{15}{\pi^4} \int_{M_i}^\infty \frac{\sqrt{E^2-M_i^2}}{e^{E/T} \mp 1} E^2 \de E$. However, since they only differ by the exponent of the $(T_i/T)$ factor, $g_*^s$ can be approximated by $g_*$ as long as the particles' individual temperatures $T_i$ do not differ too much from the overall temperature $T$.} change from the freeze out scale until now. It follows that $(a_0 T_0^3) =  (a_\infty T_\infty^3) \cdot g_*(x_f)/g_0$ with $g_0 = 3.36$. 

Combining the above relations leads to the following final result for the dark matter abundance, typically given as $\Omega_0 h^2$ (where $h$ was defined with the numerical value of the hubble parameter $H_0$):
\begin{align}
\Omega^\text{DM}_0 h^2 \approx \unit{1.04e9}{\per\GeV} \frac{x_f}{m_\text{Pl} \sqrt{g_*(x_f)} (a + 3b/x_f)}. \label{eqn:Omega0}
\end{align}
$\Omega_0$ does not depend explicitely on the dark matter mass $M_\chi$ but still implicitly through the cross section and $x_f$. As expected we see that the larger the annihilation cross section, the smaller the relic density is today. For a given \textsc{Wimp} model, the relic density today can be evaluated by calculating the total annihilation cross section, reading off the expansion coefficients $a$ and $b$ and then finally solving (\ref{eqn:x0}) and (\ref{eqn:Omega0}) with the corresponding value for $g$ depending on the type of \textsc{Wimp} particle.

\chapter{Effective Models for Dark Matter Interaction}
\label{chap:appeffop}
\newcommand{\Geff}{$G_\text{eff}$\ } 
\newcommand{\Geffb}{$\mathbf{G_\textbf{eff}}$\ }
In this work we use effective vertices to describe the interaction of the Standard Model with the dark matter sector. Advantages of this approach are described in section \ref{sec:generaleffectivewords}, along with a motivation for the particular effective ansatz we use in comparison to previous work in this field. In section \ref{sec:derivingeffectiveLagrangians} we explicitly calculate effective vertices out of fundamental Lagrangians for theories with different spin for the \textsc{Wimp} and different interaction modes. In section \ref{sec:benchmarks} we define specific benchmark models for the general Lagrangians that lead to theories with only one coupling constant \Geff.
\section{Effective Theories for \textsc{Wimp}s}
\label{sec:generaleffectivewords}
\textsc{Wimp}s are of particular interest in modern particle physics since they naturally arise in some currently probed extensions of the Standard Model. A famous example is \emph{Supersymmetry} \cite{Martin:1997ns,Drees:2004jm,Nilles:1983ge,Wess197439}, where every Standard Model particle gets a partner with all the same quantum numbers except with the spin differing by $\nicefrac{1}{2}$. If one assumes an additional discrete symmetry that only allows an even number of supersymmetric particles in any interaction, the lightest supersymmetric particle becomes automatically stable, leading to a potential dark matter candidate \cite{Ellis1984453}. However, even the smallest supersymmetric extension of the Standard Model with one supersymmetric partner for each Standard Model field, called the \emph{Minimal Supersymmetric Standard Model} \cite{Fayet:1976et, Fayet:1977yc}, introduces about 120 new mass--, interaction-- and mixing--parameters \cite{Dimopoulos:1995ju}. This huge number of degrees of freedom and the large number of production mechanisms makes it extremely difficult to derive general statements about individual parameters within the theory from experimental dark matter constraints.

Effective theories avoid this problem by employing valid approximations and staying in a general framework to describe interactions with as few parameters and vertices as possible. In our case, we will use a theory with only one effective vertex and two unknown parameters, the mass of the dark matter candidate $M_\chi$ and the effective vertex coupling \Geff.

The idea of parametrising the interaction of a dark matter particle with
Standard Model particles by using effective operators has been used in previous analyses (see for example
\cite{DMClass, RelicDensity, DMChina1, DMChina2, LEPShinesLight, TevatronDarkMatter}). Many authors
construct a list of 4--particle-interactions with Lorentz--invariant
combinations of $\gamma^\mu$, $\partial_\mu$ and
spinor--/vector--indices up to mass dimension 5 or 6. In many cases there is
no explanation of how these operators may arise in an underlying
fundamental theory. That makes it difficult to judge how exhaustive the lists
of operators are, whether interferences between different operators should
be taken into account and how the effective model is connected to realistic fundamental theories and their couplings.

We therefore follow a more sophisticated approach which was introduced in
\cite{DMClass}. We start with different fundamental theories with given
renormalisable interactions which are mediated by a very massive particle from the
Standard Model fermions to the dark matter particles. By assuming energies $\sqrt{s}$ much smaller than the mass $M_\Omega$ of the heavy
mediator, we receive effective
4--particle--vertices. Using these operators, one has a simple framework at hand with only one new vertex and two free parameters. These can be easily  analysed and probed by different experiments. However, we still have the possibility to propagate any information back onto the parameters of the corresponding underlying
fundamental theory. 

\section{Deriving Effective Lagrangians}
\label{sec:derivingeffectiveLagrangians}
We start with a list of fundamental Lagrangians motivated in \cite{DMClass}. We are interested in the phenomenology of a
high energy experiment in chapter \ref{chap:ilcanalysis}, so we do not perform a non--relativistic approximation as it is done in other studies. Therefore the final results for the effective operators may differ. We
also do not apply equations of motion to derive the effective vertices, but we make use of the path integral formalism as motivated in
\cite{EffectiveOperators}, which is argued to be more stable against quantum effects that can make the use of equations of motion invalid. 

Since models with the same mediator involve similar calculations, the following subsections are ordered according to the mediator's spin. 
%
\subsection*{Scalar Mediator}
Let $\psi$ denote the
Standard Model fermion spinor and $\phi$ correspond to a real\footnote{The
  calculation for a complex mediator field is similar. Factors of 2 will
  appear at different steps but will at the end lead to the same result.} scalar field for the
mediating particle. Regardless of the mediator's spin, we will use $M_\Omega$ to denote its mass throughout the whole chapter. We now derive the effective operators for different spins
 for the dark matter candidate $\chi$:
\paragraph{Complex Scalar Dark Matter}
\begin{center}
\includegraphics[width=0.3\textwidth]{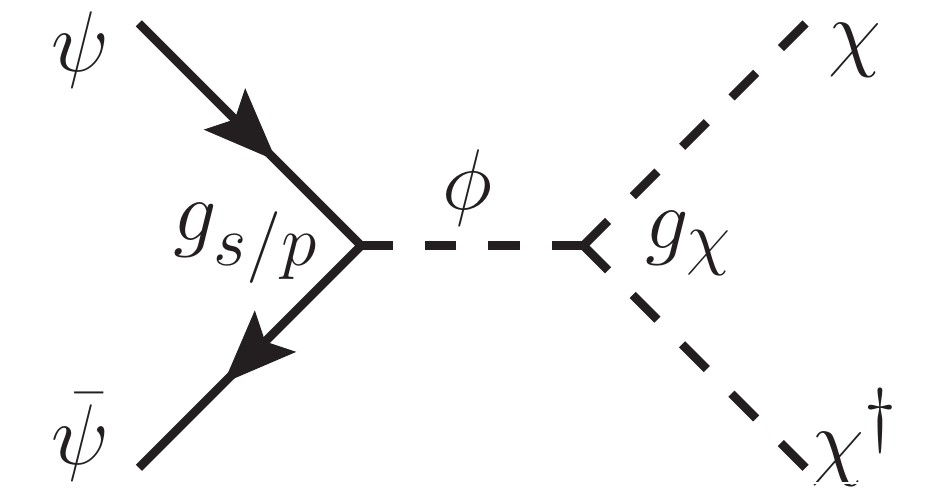}
\end{center}
A simple general Lagrangian with both scalar and pseudoscalar couplings to the
fermion can be formulated as follows:
\begin{align}
\mathscr{L} &= \mathscr{L}_\text{free} + \half \left(\partial_\mu \phi \right)^2 - \half M_\Omega^2 \phi^2 - g_\chi \chi^\dagger \chi \phi - \bar{\psi} \left( g_s +
 i g_p \gamma^5 \right) \psi \phi \\
&= \mathscr{L}_\text{free} - \half \phi \square \phi - \half M_\Omega^2 \phi^2 - F \phi, \label{app:effectivemodels:eqn4}\\
F &\equiv  g_\chi \chi^\dagger \chi + \bar{\psi} \left( g_s + i
  g_p \gamma^5 \right) \psi. 
\end{align}
$\mathscr{L}_\text{free}$ describes the kinetic parts for
  $\psi$ and $\phi$, which are of no interest for this analysis. We now use the Green
  function $D_S(x-y)$ for the scalar propagator to rewrite the linear term in $\phi$. Using its definition\footnote{Note that (\ref{app:effectivemodels:eqn1}) and (\ref{app:effectivemodels:eqn2}) denote operator equations, and all derivatives (here as well as in the following sections) act to the right. In order to perform calculations for explicit representations of the Green function, i.\/e.\/ (\ref{eqn:greends}), partial integrations have to be applied first to make the derivatives act to the left. This is particularly important in the later discussion of fermionic mediators.}
\begin{align}
(- \square_x - M_\Omega^2 ) D_S(x-y) &= i \delta^4(x-y), \label{app:effectivemodels:eqn1} \\
 D_S(y-x) (-\square_x - M_\Omega^2) &= i \delta^4(x-y), \label{app:effectivemodels:eqn2}
\end{align}
we can complete the square in $\mathscr{L}$:
\begin{align}
\mathscr{L}_{\text{int}} = &\half \left[ \phi(x) + i \int \de^4y \ F(y) D_S(y-x) \right]
\left[ - \square_x - M_\Omega^2 \right]\left[ \phi(x) + i \int \de^4z \  F(z) D_S(z-x)
\right] \nonumber \\
&+ \frac{i}{2} \int \de^4y \ F(y) D_S(y-x) F(x). \label{app:effectivemodels:eqn3}
\end{align}
Using relations (\ref{app:effectivemodels:eqn1}), (\ref{app:effectivemodels:eqn2}) and
performing the integrals over the delta functions, one can easily verify that
(\ref{app:effectivemodels:eqn3}) indeed is identical to the interaction part in 
(\ref{app:effectivemodels:eqn4}).

The path integral sums up all possible configurations for $\phi(x)$. Thus we may
shift the function $\phi$ arbitrarily without changing any physical
quantity. We apply the following substitution rule:
\begin{align}
\phi(x) \rightarrow \phi(x) - i \int \de^4y \ F(y) D_S(y-x).
\end{align}
With the abbreviation $\mathcal{D} \mathcal{F}  \equiv  \mathcal{D}
\psi \mathcal{D} \bar{\psi} \mathcal{D} \chi \mathcal{D} \chi^\dagger$ to sum
up the functional integration measures of the remaining fields, we then arrive at the
following form of the path integral:
\begin{align}
\int \mathcal{D} \phi \ \mathcal{D}  \mathcal{F} \ \exp \left( i \int \de^4 x\  \mathscr{L}_{\text{int}} \right) = \int
\mathcal{D} \phi \ \mathcal{D} \mathcal{F} \exp & \left( \frac{i}{2} \int \de^4 x \ \phi(x)
   \left[ - \square - M_\Omega^2 \right] \phi(x) \right. \nonumber \\ & \left. - \half \int
   \de^4 x \ \de^4 y \ F(x)
  D_S(x-y) F(y) \right).
\end{align}
 We factorised the $\phi$--dependence and now use the path integral version of the Gaussian integral:
\begin{equation}
\int \ \mathcal{D} \phi \ \exp \left(- \frac{i}{2} \int \de x \  \phi(x) M(x, y) \phi(y)
\right) \propto \left( \det M \right)^{-\nicefrac{1}{2}}.
\end{equation}
Field--independent factors like the proportionality constant as well as the determinant itself will cancel in any calculation of a physical
correlation function. They are therefore irrelevant for the remaining
calculation. The path integral has now been brought to the following form:
\begin{align}
\int \mathcal{D} \phi \ \mathcal{D}  \mathcal{F} \ \exp \left( i \int \de^4 x\  \mathscr{L}_{\text{int}} \right) &= \int
\mathcal{D} \mathcal{F} \exp \left( - \half \int \de^4 x \ \de^4 y \ F(x) D_S(x-y)
  F(y) \right)
\end{align}
Next we apply the effective approximation that the energy within the system is signficantly smaller than the mass of the mediator. Thus it is never produced on--shell. We do so by expanding the path integral in $\nicefrac{p}{M_\Omega}$. After transforming all functions into momentum space
\allowdisplaybreaks
\begin{align}
F(x) &= \int \frac{\de^4 k}{\left(2 \pi\right)^4} \ \tilde{F}(k)  e^{- i k x}, \\
D_S(x-y) &= \int \frac{\de^4 q}{\left(2 \pi\right)^4} \ \frac{i}{q^2-M_\Omega^2} e^{-
  i q (x - y)} \label{eqn:greends}
\end{align}
 we can insert these into the action $S_\text{int} = \displaystyle \int \de^4x \ \mathscr{L}_\text{int}$ and perform the space integral over the complex
exponentials:
\begin{align}
\int \de^4x \ e^{i p x} = (2 \pi)^4 \delta^{(4)}(p).
\end{align} Using the delta functions to reduce the number of momentum
integrals, we find:
\begin{align}
i S_\text{int}  &= - \half \int \frac{\de^4 k}{\left(2 \pi \right)^4} \ 
\tilde{F}(-k) \frac{i}{k^2-M_\Omega^2} \tilde{F}(k).
\end{align}
The scale of $k$ is now restricted to physical values through the field--dependence of $\tilde{F}(k)$. It is
therefore reasonable to use the effective approximation $k^2 - M_\Omega^2 \approx
-M_\Omega^2$. Transforming back into real space, one gets the effective action and can read off
the corresponding effective Lagrangian:
\begin{align}
i S_\text{int} &= + \half \frac{i}{M_\Omega^2} \int \de^4 x \  F^2(x), \\
\mathscr{L}_\text{eff} &= \frac{1}{2 M_\Omega^2} F^2 \\
&\supset \frac{g_\chi}{M_\Omega^2} \chi^\dagger
 \chi \bar{\psi} \left(g_s + i g_p \gamma^5 \right) \psi.
\end{align}
%
\paragraph{Dirac Fermion Dark Matter}
\begin{center}
\includegraphics[width=0.3\textwidth]{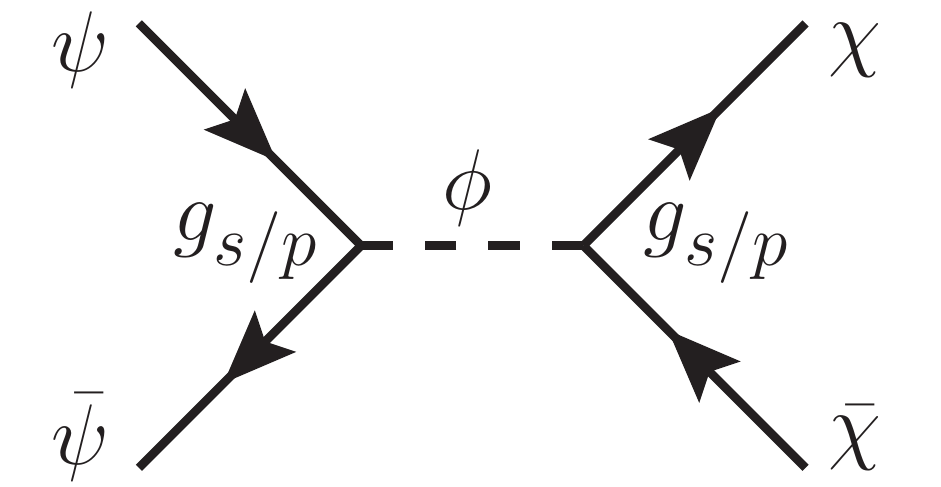}\qquad \includegraphics[width=0.3\textwidth]{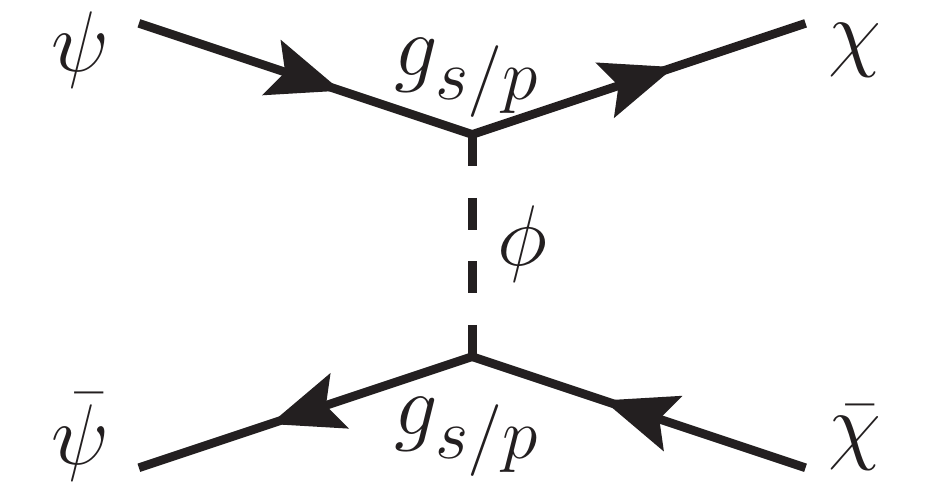}
\end{center}
Fermionic dark matter can interact via s-- or t--channel interactions:
\begin{align}
\mathscr{L}^s &= \mathscr{L}_{\text{free}} +\half \left(\partial_\mu \phi
\right)^2 - \half M_\Omega^2 \phi^2 - \left[\bar{\chi} \left(g_s + i g_p
  \gamma^5 \right) \chi + \bar{\psi} \left( g_s +
  i g_p \gamma^5 \right) \psi \right] \phi, \\
\mathscr{L}^t &=\mathscr{L}_{\text{free}} + \half \left(\partial_\mu \phi
\right)^2 - \half M_\Omega^2 \phi^2 - \left[\bar{\chi} \left(g_s + i g_p
  \gamma^5 \right) \psi  + \bar{\psi} \left( g_s +i 
  g_p \gamma^5 \right) \chi \right]\phi. \\
\intertext{Both Lagrangians can be summarised as follows:}
\mathscr{L}^i &= \mathscr{L}_{\text{free}} + -\half \phi \square \phi - \half M_\Omega^2 \phi^2 - F^i \phi, \\
\intertext{with two different auxiliary fields $F^i$:}
F^s &\equiv  \bar{\chi} \left(g_s + i g_p
  \gamma^5 \right) \chi + \bar{\psi} \left( g_s +
  i g_p \gamma^5 \right) \psi, \\
F^t &\equiv  \bar{\chi} \left(g_s + i g_p
  \gamma^5 \right) \psi  + \bar{\psi} \left( g_s + i
  g_p \gamma^5 \right) \chi.
\end{align}
For the s--channel operator, one could even assume individual couplings $g_s^\chi$ and $g_s^\psi$ for the respective $\chi \chi$ and $\psi \psi$ bilinears. However, since we won't analyse the full parameter space of all $g_i$ but only particular benchmark scenarios later, we can already apply the simplifying assumption here that $g_i^\chi = g_i^\psi$. For the t--channel operator, the two couplings have to be equal to keep the Lagrangian real.

The calculation of the effective Lagrangian is identical to the previous case with a different auxiliary
field $F$. The result reads as follows:
\begin{align}
\mathscr{L}^i_\text{eff} &= \frac{1}{2 M_\Omega^2} {F^{i}}^2, \\
\mathscr{L}_\text{eff}^s &\supset \frac{1}{M_\Omega^2} \bar{\chi}
\left(g_s + i g_p \gamma^5 \right) \chi
\bar{\psi} \left( g_s + i g_p \gamma^5 \right)\psi, \\
\mathscr{L}_\text{eff}^t &\supset \frac{1}{M_\Omega^2} \bar{\psi}
\left(g_s + i g_p \gamma^5 \right) \chi
\bar{\chi} \left( g_s + i g_p \gamma^5 \right)\psi.
\end{align}

\paragraph{Complex Vector Dark Matter}
\begin{center}
\includegraphics[width=0.3\textwidth]{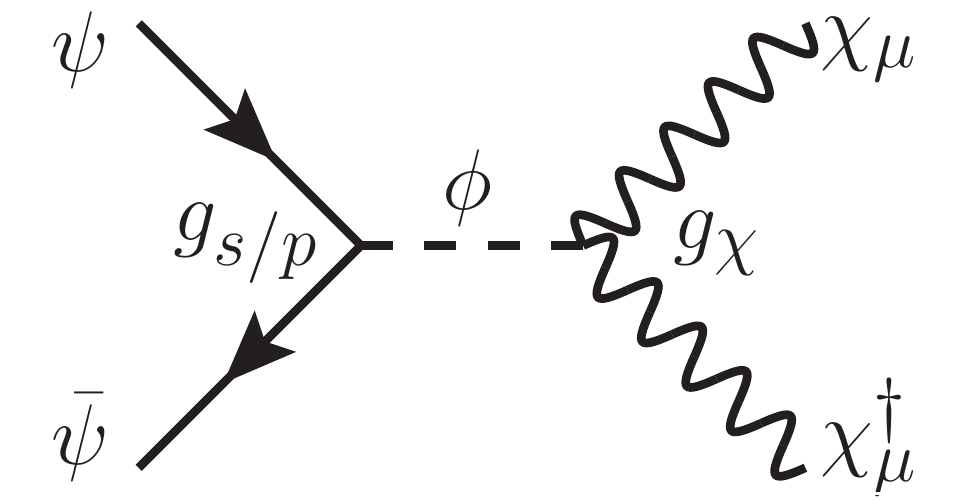}
\end{center}
This case is evaluated analogously:
\begin{align}
\mathscr{L} &= \mathscr{L}_\text{free} + \half \left(\partial_\mu \phi
\right)^2 - \half M_\Omega^2 \phi^2 + g_\chi \chi^\mu \chi_\mu \phi -
\bar{\psi} \left( g_s + i
  g_p \gamma^5 \right) \psi \phi, \\
&= \mathscr{L}_\text{free} - \phi \square \phi - \half M_\Omega^2 \phi^2 - F \phi, \\
F &\equiv - g_\chi \chi^\mu \chi_\mu + \bar{\psi} \left( g_s + i
  g_p \gamma^5 \right) \psi, \\
\mathscr{L}_\text{eff} &= \frac{1}{2 M_\Omega^2} F^2 \supset - \frac{1}{M_\Omega^2} \chi^\mu \chi_\mu 
\bar{\psi} \left( g_s + i g_p \gamma^5 \right)\psi.
\end{align}
\clearpage
\subsection*{Fermionic Mediator}
We now look at models with a Dirac fermion mediator, $\eta$. The difference in
the Green function will significantly change the result.
\paragraph{Complex Scalar Dark Matter}
\begin{center}
\includegraphics[width=0.3\textwidth]{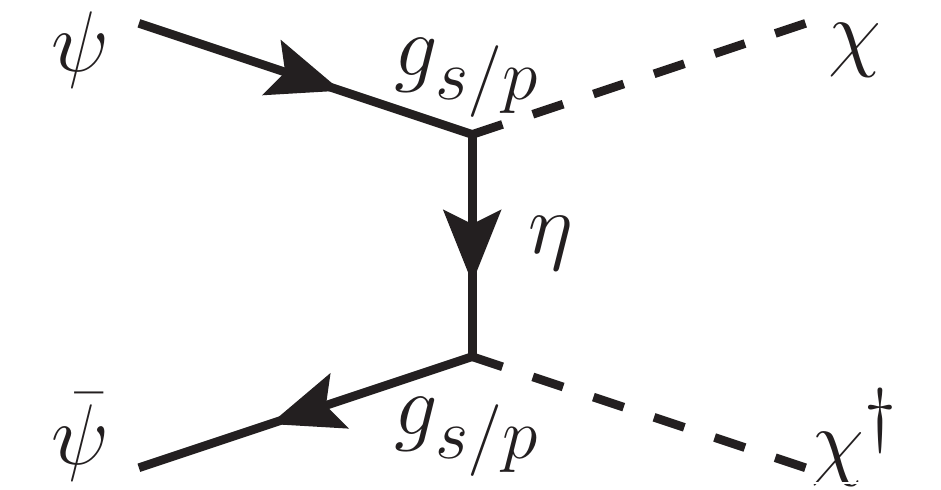}
\end{center}
A possible Lagrangian for complex 
scalar dark matter looks like this:
\begin{align}
\mathscr{L} &= \mathscr{L}_\text{free} + \bar{\eta} \left( i \slashed{\partial} -
  M_\Omega \right) \eta - \bar{\eta} \left( g_s  + i g_p \gamma^5 \right) \psi \chi
   - \bar{\psi} \left( g_s + i g_p \gamma^5 \right) \eta \chi^\dagger \label{eqn:fermmedlag}\\
&= \mathscr{L}_\text{free} + \bar{\eta} \left( i \slashed{\partial} -
  M_\Omega \right) \eta - \bar{\eta} F - \bar{F}  \eta,
 \label{app:effectivemodels:eqn5}\\
F &\equiv \left( g_s + i g_p \gamma^5 \right) \psi \chi
\end{align}
Note that $F$ is now a Dirac spinor and therefore $\bar{F} \equiv F^\dagger \gamma^0$. We follow the same calculation as in the
case of a scalar mediator but with different Green functions: 
\begin{align}
 (i \slashed{\partial}_x - M_\Omega ) D_F(x-y) &= i \delta^4(x-y), \\
 D_F(y-x) ( i \slashed{\partial}_x - M_\Omega) &= i \delta^4(y-x).
\end{align}
We again rewrite (\ref{app:effectivemodels:eqn5}) to
reformulate the linear term: 
\begin{align}
\mathscr{L}_{\text{int}} = &\left[ \bar{\eta}(x) + i \int \de^4y \ \bar{F}(y) D_F(y-x) \right]
\left[ i \slashed{\partial}_x - M_\Omega \right]\left[ \eta(x) + i \int \de^4z \ D_F(x-z) F(z)
\right] \nonumber \\
&+ i \int \de^4y \ \bar{F}(y) D_F(y-x) F(x).
\end{align}
We may perform shifts for $\eta$ and $\bar{\eta}$ in the path integral similarly to the scalar
case: 
\begin{align}
\eta(x) &\rightarrow \eta(x) - i \int \de^4z \ D_F(x-z) F(z), \\
\bar{\eta}(x) &\rightarrow \bar{\eta}(x) - i \int \de^4y \ \bar{F}(y)  \ D_F(y-x).
\end{align}
However, the Gaussian path integral formulation for spinor fields has a different form:
\begin{equation}
\int \ \mathcal{D} \psi \mathcal{D} \bar{\psi} \ \exp \left(- \int \de x \ \de
  y \   \bar{\psi}(x) M(x, y) \psi(y)
\right) \propto \left( \det M \right).
\end{equation}
The determinant appears with a different power but is again of no interest since it does not contain any fields and will drop out in any correlator. Similar to the scalar case we end up with
\begin{align}
\int \mathcal{D}  \mathcal{F} \ \exp \left( i \int \de^4 x\  \mathscr{L}_{\text{int}} \right) &= \int
\mathcal{D} \mathcal{F} \exp \left( - \int \de^4 x \ \de^4 y \ \bar{F}(x) D_F(x-y)  F(y) \right).
\end{align}
We again switch to momentum space\footnote{Note that the complex conjugate spinor field $\bar{F}$ has a Fourier transform with a positive exponent $e^{+ i k x}$, which leads to an opposite sign for the momentum in comparision to the real scalar of the previous calculation.} using the momentum representation of the Dirac propagator $S$:
\begin{align}
D_F(x-y) &= \int \frac{\de^4 q}{\left(2 \pi\right)^4} \ \frac{i \left( \slashed{q} + M_\Omega \right)}{q^2-M_\Omega^2} e^{-i q (x - y)},\\
i S_\text{int}  &= - \int \frac{\de^4 k}{\left(2 \pi \right)^4}
 \tilde{\bar{F}}(k) \frac{i\left( \slashed{k} + M_\Omega \right)}{k^2-M_\Omega^2} \tilde{F}(k).
\end{align}
Expanding the denominator in $\nicefrac{k}{M_\Omega}$ up to order $\nicefrac{k^2}{M_\Omega^2}$, we now find two terms:
\begin{align}
i S_\text{int}&\approx i  \int \frac{\de^4 k}{\left(2 \pi \right)^4}
\tilde{\bar{F}}(k) \left( \frac{\slashed{k}}{M_\Omega^2} + \frac{1}{M_\Omega} \right)  \tilde{F}(k).
\end{align}
If we now go back to real space, $\slashed{k}$ becomes $i
\slashed{\partial}_x$. Acting on $F$, it acts on both fields $\phi$ and
$\psi$. \footnote{Contrary to \cite{DMClass}, derivatives on the Standard Model fermion
fields are not neglected. They only vanish if the Dirac equation $i
\slashed{\partial} \psi = m \psi$ is applicable (i.e. not for off-shell
particles) and if the mass is negligibly small}.
We can now read off the effective Lagrangian:
\begin{align}
\mathscr{L}_\text{eff} &= \frac{1}{M_\Omega^2} \bar{F} \left( i \slashed{\partial} + M_\Omega \right) F  \\
&\supset \frac{1}{M_\Omega^2} \bar{\psi}\left(g_s + i g_p \gamma^5 \right) i \gamma^\mu
\left(g_s + i g_p \gamma^5 \right) \psi \chi^\dagger \partial_\mu \chi \nonumber \\ 
&+ \frac{1}{M_\Omega^2} \bar{\psi}\left(g_s + i g_p \gamma^5 \right) i \gamma^\mu
\left(g_s + i g_p \gamma^5 \right) (\partial_\mu \psi) \chi^\dagger \chi \nonumber  \\ &+
\frac{1}{M_\Omega} \bar{\psi}\left(g_s + i g_p \gamma^5 \right) \left(g_s + i g_p \gamma^5 \right) \psi \chi^\dagger \chi \\
&= \frac{g_{s}^2 - g_{p}^2}{M_\Omega} \bar{\psi} \psi \chi^\dagger \chi + \frac{i}{M_\Omega^2} \bar{\psi} \left(g_{s}^2 + g_{p}^2 - 2 g_s g_p \gamma^5 \right)
\gamma^\mu \psi \chi^\dagger \partial_\mu \chi \nonumber \\ 
&+\frac{i}{M_\Omega^2} \bar{\psi} \left(g_{s}^2 + g_{p}^2 - 2 g_s g_p \gamma^5 \right)
\gamma^\mu (\partial_\mu \psi) \chi^\dagger \chi.
\end{align}

\paragraph{Vector Dark Matter}
\begin{center}
\includegraphics[width=0.3\textwidth]{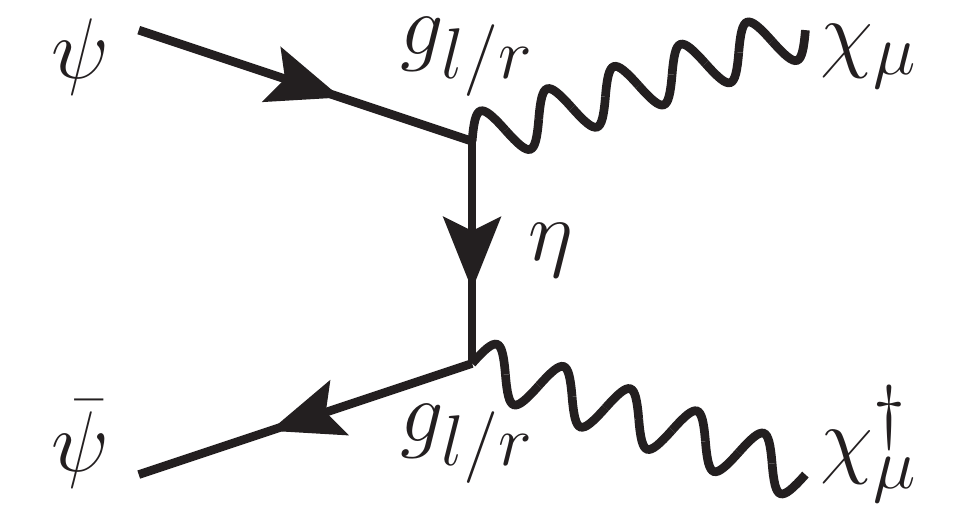}
\end{center}
The Lagrangian changes to
\begin{align}
\mathscr{L} &= \mathscr{L}_\text{free} + \bar{\eta} \left( i \slashed{\partial} -
  M_\Omega \right) \eta + \bar{\eta} \gamma^\mu  \left( g_l P_L  + g_r P_R \right) \psi \chi_\mu
   + \bar{\psi}  \left( g_l P_R + g_r P_L \right) \gamma^\mu  \eta \chi^\dagger_\mu \label{eqn:fermmedlag}\\
&= \mathscr{L}_\text{free} + \bar{\eta} \left( i \slashed{\partial} -
  M_\Omega \right) \eta - \bar{\eta} F - \bar{F}  \eta,
 \\
F &\equiv \gamma^\mu \left( g_l P_L + g_r P_R  \right)  \psi \chi_\mu.
\end{align}
With the above $F$, we can directly use and evaluate the result from the previous
calculation:
\begin{align}
\mathscr{L}_\text{eff} &= \frac{1}{M_\Omega^2} \bar{F} \left( i \slashed{\partial} + M_\Omega \right) F  \\
&\supset \frac{g_l g_r}{M_\Omega} \bar{\psi} \gamma^\nu \gamma^\rho \psi \ \chi^\dagger_\nu
\chi_\rho + \frac{i }{M_\Omega^2} \bar{\psi}  \gamma^\nu
\gamma^\mu \gamma^\rho \left(g_l^2P_L + g_r^2P_R \right) \psi \ \chi^\dagger_\nu \partial_\mu \chi_\rho  \nonumber \\
&+\frac{i }{M_\Omega^2} \bar{\psi}  \gamma^\nu
\gamma^\mu \gamma^\rho \left(g_l^2 P_L + g_r^2 P_R \right) (\partial_\mu
\psi) \ \chi^\dagger_\nu  \chi_\rho 
\end{align}
\pagebreak
\subsection*{Vector Mediator}
\paragraph{Scalar Dark Matter}
\begin{center}
\includegraphics[width=0.3\textwidth]{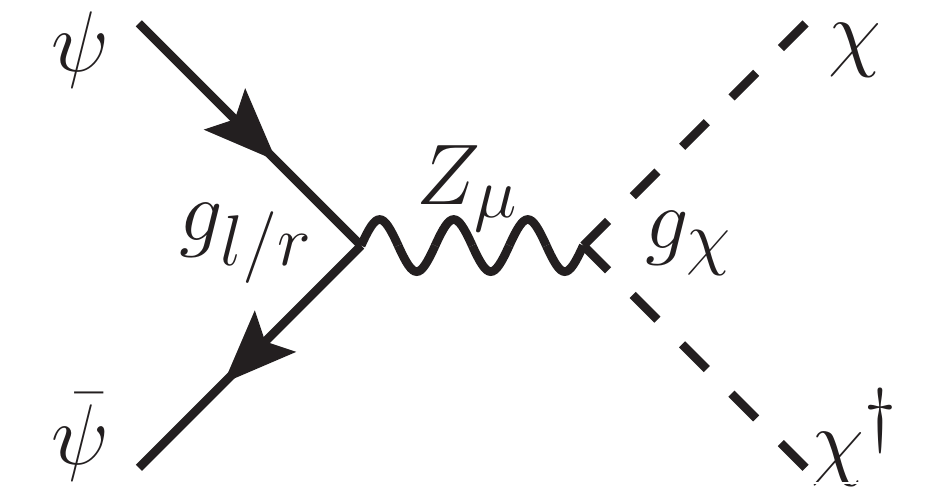}
\end{center}
$Z_\mu$ denotes a single mediating real vector field with field strength
$F_{\mu \nu} = \partial_\mu Z_\nu - \partial_\nu Z_\mu$ . A possible Lagrangian for
scalar dark matter then is as follows:
\begin{align}
\mathscr{L}_\text{int} &=  - \frac{1}{4} F^{\mu \nu} F_{\mu \nu} +
\half 
M_\Omega^2 Z^\mu Z_\mu  + g_\chi \left(\chi^\dagger \partial_\mu \chi -
  \chi \partial_\mu \chi^\dagger \right) Z^\mu + \bar{\psi} \gamma^\mu  \left( g_l P_L  +
  g_r P_R \right) \psi Z_\mu \\
&= \half Z^\mu \left[ \left(\square +
    M_\Omega^2\right)g_{\mu \nu} - \partial_\mu \partial_\nu \right]Z^\nu + F_\mu  Z^\mu,
 \\
F_\mu &\equiv g_\chi \left(\chi^\dagger \partial_\mu \chi -
  \chi \partial_\mu \chi^\dagger \right) + \left( g_s + g_p \gamma^5 \right) \psi \chi.
\end{align}
The sign structure appearing in the $Z \chi \chi^\dagger$ term can be
understood by taking a gauge theory as an example and expanding the covariant derivative $D_\mu \equiv \partial_\mu + i g_\chi Z_\mu$ in the 
kinetic term  $\left(D_\mu \chi\right)^\dagger \left( D^\mu \chi\right)$. 

The
corresponding Green functions are then defined as
\begin{align}
 \left[(M_\Omega^2 + \square)g_{\alpha \beta} - \partial_\alpha \partial_\beta\right]
 D_V^{\beta \gamma}(x-y) &= i \delta^4(x-y) \delta_\alpha^\gamma, \\
 D_V^{\alpha \beta}(x-y) \left[(M_\Omega^2 + \square)g_{\beta \gamma}
   - \partial_\beta \partial_\gamma \right]
  &= i \delta^4(x-y) \delta^\alpha_\gamma. 
\end{align}
We now proceed as usual by completing the square:
\begin{align}
\mathscr{L}_{\text{int}} = & \half \left[ Z^\mu(x) + i \int \de^4y \
  \bar{F}_\rho(y) D_{V}^{\rho \mu}(y-x) \right]
\left[\left(\square +
    M_\Omega^2\right)g_{\mu \nu} - \partial_\mu \partial_\nu  \right] \nonumber \\ &  \ \ \left[ Z^\nu(x) + i
    \int \de^4z \ D_V^{\nu \lambda}(x-z) F_\lambda(z) \right] 
+ \frac{i}{2} \int \de^4y \ \bar{F}_\mu(y) D_V^{\mu \nu}(y-x) F_\nu(x).
\end{align}
Shifting the field and performing the Gaussian integral leads to the following action:
\begin{align}
\int \mathcal{D}  \mathcal{F} \ \exp \left( i \int \de^4 x\  \mathscr{L}_{\text{int}} \right) &= \int
\mathcal{D} \mathcal{F} \exp \left(\frac{1}{2} \int \de^4y \ F_\mu(y)
  D_V^{\mu \nu}(y-x) F_\nu(x) \right).
\end{align}
Now we need the momentum representation for the vector propagator\footnote{Our form of the propagator in (\ref{eqn:vectorprop}) assumes a particular gauge fixing condition ($\xi \rightarrow \infty$ in $R_\xi$ gauge). Only the term propotional to $q^\alpha q^\beta$ is affected by that choice. Since this term vanishes after performing the effective approximation in (\ref{eqn:vectoreff}), our results are manifestly gauge invariant.}:
\begin{align}
D_V^{\alpha \beta}(x-y) &= \int \frac{\de^4 q}{\left(2 \pi\right)^4} \
\frac{-i }{q^2-M_\Omega^2} \left(g^{\alpha \beta} - \frac{q^\alpha
    q^\beta}{M_\Omega^2}\right) e^{i q (x - y)}. \label{eqn:vectorprop}
\end{align}
Transforming into momentum space, expanding in $k/M_\Omega$ and only using terms up to order
$k^2/M_\Omega^2$ leads to the effective Lagrangian for this model:
\begin{align}
i S_\text{int}  &= - \half \int \frac{\de^4 k}{\left(2 \pi \right)^4}
 \tilde{F}_\alpha(k) \frac{ i }{k^2-M_\Omega^2} \left(g^{\alpha \beta} -
   \frac{k^\alpha k^\beta}{M_\Omega^2}\right)  \tilde{F}_\beta(k) \\
&\approx \frac{i}{2} \int \frac{\de^4 k}{\left(2 \pi \right)^4}
 \tilde{F}_\alpha(k) \frac{ 1 }{M_\Omega^2} g^{\alpha \beta} \tilde{F}_\beta(k), \label{eqn:vectoreff} \\
\mathscr{L}_\text{eff} &= \frac{1}{2 M_\Omega^2} F^\mu F_\mu \\
&\supset \frac{g_\chi}{M_\Omega^2} \bar{\psi} \gamma^\mu \left( g_l P_L + g_r P_R \right) \psi \
\left( \phi^\dagger \partial_\mu \phi - \phi \partial_\mu \phi^\dagger \right).
\end{align}

\paragraph{Fermion Dark Matter}
\begin{center}
\includegraphics[width=0.3\textwidth]{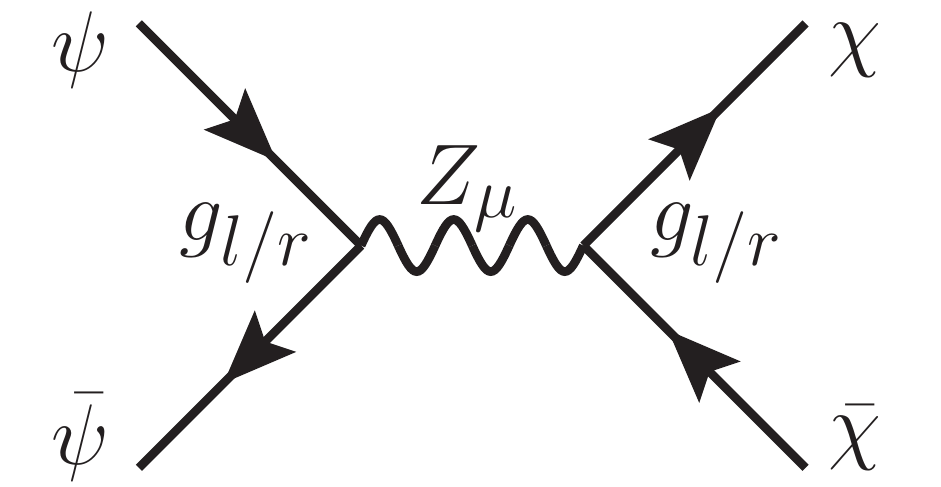} \qquad \includegraphics[width=0.3\textwidth]{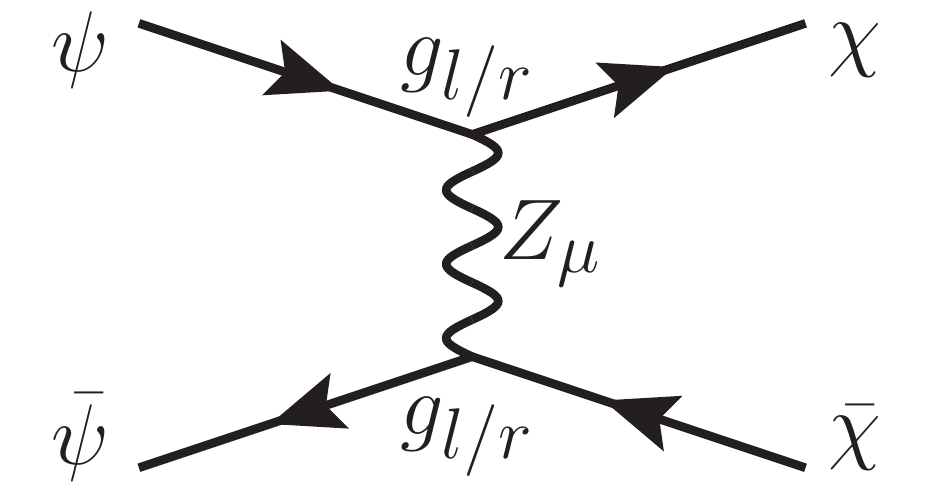}
\end{center}
In this case the Lagrangian can again either describe an s-channel\footnote{Again, we do not consider the possible generalisation of using individual couplings $g_{l/r}^\chi$ and $g_{l/r}^\psi$ for the s--channel case.} or a t-channel interaction
\begin{align}
\mathscr{L}^s_{\text{int}} &=  - \frac{1}{4} F^{\mu \nu} F_{\mu \nu} +
\half 
M_\Omega^2 Z^\mu Z_\mu  + \left[\bar{\psi} \gamma^\mu \left(g_l P_L + g_r P_R
    \right) \psi + \bar{\chi} \gamma^\mu \left(g_l P_L + g_r P_R
    \right) \chi \right] Z_\mu, \\
\mathscr{L}^t_{\text{int}} &=  - \frac{1}{4} F^{\mu \nu} F_{\mu \nu} +
\half 
M_\Omega^2 Z^\mu Z_\mu  + \left[\bar{\chi} \gamma^\mu \left(g_l P_L + g_r P_R
    \right) \psi + \bar{\psi} \left(g_l P_R + g_r P_L
    \right)  \gamma^\mu  \chi \right] Z_\mu \\
\intertext{We can sum up both cases as follows}
\mathscr{L}^i_{\text{int}} &= \half Z^\mu \left[ \left(\square +
    M_\Omega^2\right)g_{\mu \nu} - \partial_\mu \partial_\nu \right]Z^\nu + F_\mu^i  Z^\mu, \\
\intertext{where}
F^s_\mu &\equiv \bar{\psi} \gamma^\mu \left(g_l P_L + g_r P_R
    \right) \psi + \bar{\chi} \gamma^\mu \left(g_l P_L + g_r P_R
    \right) \chi,  \\
F^t_\mu &\equiv \bar{\chi} \gamma^\mu \left(g_l P_L + g_r P_R
    \right) \psi + \bar{\psi} \gamma^\mu \left(g_l P_L + g_r P_R
    \right)   \chi. 
\end{align}

The result then reads
\begin{align}
\mathscr{L}_\text{eff} &= \frac{1}{2 M_\Omega^2} F^\mu F_\mu, \\
\mathscr{L}^s_\text{eff} &\supset \frac{1}{M_\Omega^2} \bar{\psi} \gamma^\mu \left(
  g_l P_L + g_r P_R \right) \psi \ \bar{\chi} \gamma_\mu \left(g_l P_L + g_r P_R
    \right) \chi, \\
\mathscr{L}^t_\text{eff} &\supset  \frac{1}{M_\Omega^2} \bar{\psi}  \gamma^\mu \left(g_l P_L + g_r P_R
    \right)   \chi  \bar{\chi} \gamma_\mu \left(g_l P_L + g_r P_R
    \right) \psi.  
\end{align}

\paragraph{Vector Dark Matter}
\begin{center}
\includegraphics[width=0.3\textwidth]{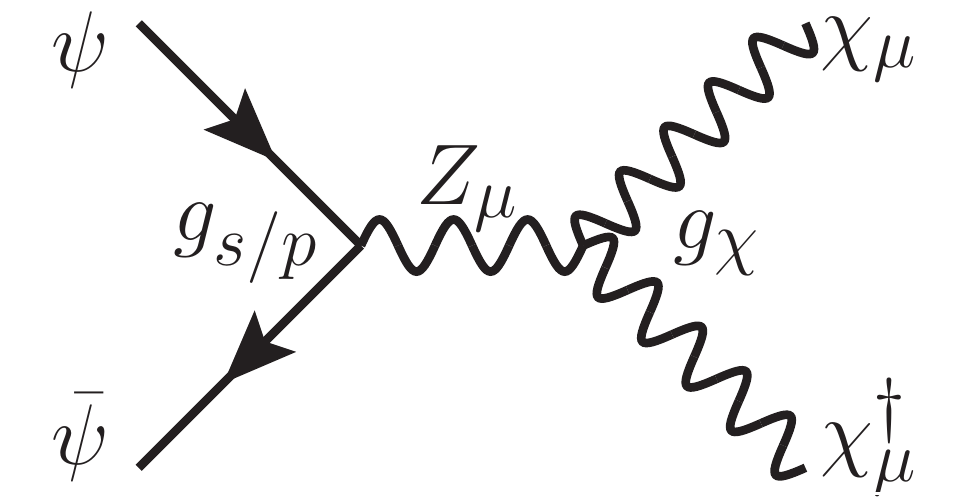}
\end{center}
We follow the idea of the scalar case and introduce the coupling between the
 dark matter field and the mediator using an analogy to gauge couplings: We combine the two
real degrees of freedom of the complex dark matter field with the single real
degree of freedom of the mediator to define a vector triplet as follows:
\begin{align}
\chi_\mu & \equiv  \frac{1}{\sqrt{2}} \left( B_1^\mu + i B_2^\mu \right), \\
\chi^\dagger_\mu & \equiv  \frac{1}{\sqrt{2}} \left( B_1^\mu - i B_2^\mu \right), \\
A_\mu^a & \equiv  \left( Z^\mu, B_1^\mu, B_2^\mu \right)  = \left( Z^\mu,
  \frac{1}{\sqrt{2}} \left[ \chi^\mu + \chi^{\dagger \mu} \right],
  \frac{i}{\sqrt{2}} \left[ \chi^{\dagger \mu} - \chi^{\mu} \right] \right).
\end{align}
The field--strength--tensor and the gauge kinetic term are constructed similar to
an SU(2) gauge theory with the structure constants $f_{abc} \equiv g \epsilon_{abc}$.  We continue by writing these tensors out explicitely in terms of $\chi_\mu$,
$\chi_\mu^\dagger$ and $Z^\mu$ to read off the triple vector boson vertex
 we need for our model:
\begin{align}
F_{\mu \nu}^a &\equiv \partial_\mu A_\nu^a - \partial_\nu A_\mu^a + f^{abc} A_\mu^b A_\nu^c  \\
F_{\mu \nu}^1 &= \partial_\mu Z_\nu - \partial_\nu Z_\mu + g \left( B_\mu^1
  B_\nu ^2 - B_\mu^2 B_\nu^1 \right) \\
&= \partial_\mu Z_\nu - \partial_\nu Z_\mu + i g \left( \chi_\mu
  \chi^\dagger_\nu - \chi^\dagger_\mu \chi_\nu \right), \\
F_{\mu \nu}^2 &= \frac{1}{\sqrt{2}} \left[ \partial_\mu \chi_\nu
  - \partial_\nu \chi_\mu + \partial_\mu \chi^\dagger_\nu - \partial_\nu
  \chi^\dagger_\mu + ig \left(\chi^\dagger_\mu Z_\nu - Z_\mu \chi^\dagger_\nu
    - \chi_\mu Z_\nu + Z_\mu \chi_\nu \right) \right], \\
F_{\mu \nu}^3 &= \frac{1}{\sqrt{2}} \left[ 
  \partial_\nu \chi_\mu - \partial_\mu \chi_\nu + \partial_\mu \chi^\dagger_\nu - \partial_\nu
  \chi^\dagger_\mu + ig \left(- \chi^\dagger_\mu Z_\nu + Z_\mu \chi^\dagger_\nu
    - \chi_\mu Z_\nu + Z_\mu \chi_\nu \right) \right], \\
F_{\mu \nu}^a F^{a \mu \nu} &= \half \left(\partial_\mu \chi^\dagger_\nu
  - \partial_\nu \chi^\dagger_\mu \right) \left( \partial^\mu \chi^\nu
  - \partial^\nu \chi^\mu \right) + \frac{1}{4} \left(\partial_\mu Z_\nu
  - \partial_\nu Z_\mu \right)^2 \\ &+ i g \left( \partial_\mu \chi^\dagger_\nu
  - \partial_\nu \chi^\dagger_\mu \right) Z^\mu \chi^\nu + i g \chi_\mu
\chi^\dagger_\nu \partial^\mu Z^\nu + \text{h.c.} + \text{quartic couplings}.
\end{align}
We will neglect any quartic couplings since they won't give any contribution
to our effective operators for pairwise coupling. This leads to the following Lagrangian:
\begin{align}
\mathscr{L}_\text{int} &= -\frac{1}{4} F_{\mu \nu} F^{\mu \nu} + \half M_\Omega^2
Z_\mu Z^\mu  -i g_\chi Z_\mu \chi^\dagger_\nu \left( \partial^\mu
  \chi^\nu - \partial^\nu \chi^\mu \right) + i g_\chi Z^\mu \chi^\nu \left( \partial_\mu
  \chi^\dagger_\nu - \partial_\nu \chi^\dagger_\mu \right) \nonumber \\
 & \quad - i g_\chi
\left(\partial^\mu Z^\nu - \partial^\nu Z^\mu \right) \chi^\dagger_\mu
\chi_\nu  + \bar{\psi} \gamma^\mu \left( g_l P_L + g_r P_R\right) \psi Z_\mu \\
&= + \half Z^\mu \left[ \left(\square +
    M_\Omega^2\right)g_{\mu \nu} - \partial_\mu \partial_\nu \right]Z^\nu + F_\mu  Z^\mu,
 \\
F_\mu &\equiv -i g_\chi \chi^\dagger_\nu \left( \partial_\mu
  \chi^\nu - \partial^\nu \chi_\mu \right) + i g_\chi \chi^\nu \left( \partial_\mu
  \chi^\dagger_\nu - \partial_\nu \chi^\dagger_\mu \right) + i g_\chi \partial_\nu \left(\chi^\dagger_\nu
\chi_\mu  - \chi^\dagger_\mu \chi_\nu \right)  \nonumber \\ & \qquad {}   + \bar{\psi} \gamma_\mu \left( g_l P_L + g_r P_R\right) \psi
\end{align}
One has to perform  partial integrations on the $\partial_\mu Z_\nu$ terms in
order to receive the full $F_\mu Z^\mu$ form. Using our previous result we are
able to read off the effective Lagrangian:
\begin{align}
\mathscr{L}_\text{eff} &= \frac{1}{2 M_\Omega^2} F^\mu F_\mu \\
&\supset \frac{g_\chi}{M_\Omega^2} \bar{\psi} \gamma^\mu \left( g_l P_L + g_r P_R
\right) \psi \left[i \chi^\nu \left( \partial_\mu
  \chi^\dagger_\nu - \partial_\nu \chi^\dagger_\mu \right) -i \chi^\dagger_\nu \left( \partial_\mu
  \chi^\nu - \partial^\nu \chi_\mu \right) \right. \nonumber \\
& \qquad  \left. {} + i \partial_\nu \left(\chi^\dagger_\nu \chi_\mu - \chi^\dagger_\mu \chi_\nu \right) \right].
\end{align}

\section{Benchmark Models}
\label{sec:benchmarks}
\begin{table}
\centering
\renewcommand{\arraystretch}{1.5}
\begin{tabular}{c c c l}
\toprule
\multirow{2}{*}{DM} & \multirow{2}{*}{Med.} & \multirow{2}{*}{Diagram} &
$-\mathscr{L}^{\text{int}}_{\text{UV}}$ \\
& & & $-\mathscr{L}^{\text{int}}_{\text{eff}}$ \\
\midrule \midrule
\multirow{2}{*}{S} & \multirow{2}{*}{S} & \multirow{2}{*}{\raisebox{-0.435\height}{\includegraphics[width=0.1\textwidth]{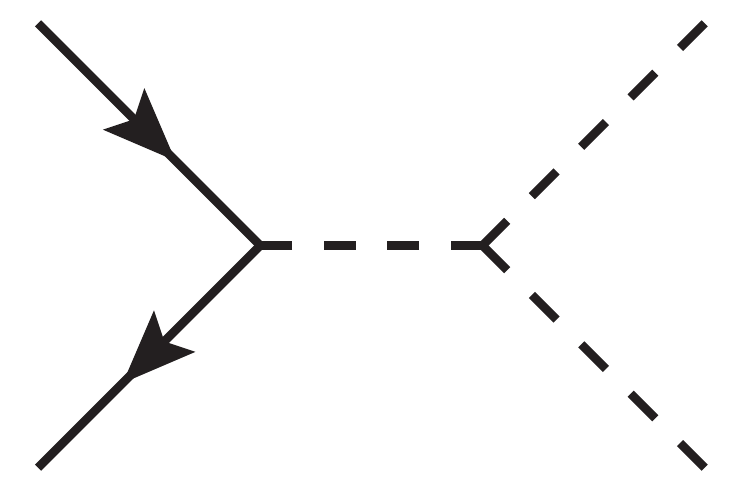}}} & $g_\chi \chi^\dagger \chi \phi + \bar{\psi} (g_s + i g_p \gamma^5) \psi \phi$ \\
& & &  $\displaystyle \frac{g_\chi}{M_\Omega^2} \chi^\dagger \chi \bar{\psi} (g_s + i g_p \gamma^5) \psi$ \\
\midrule
\multirow{2}{*}{S}       & \multirow{2}{*}{F} &
       \multirow{2}{*}{\raisebox{-0.435\height}{\includegraphics[width=0.1\textwidth]{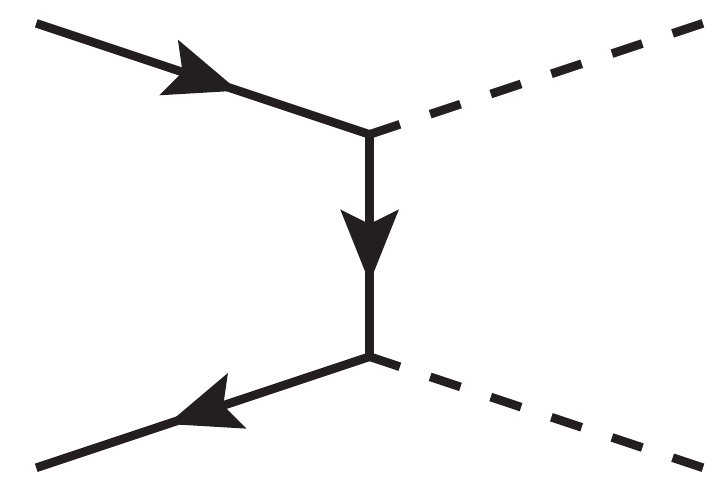}}}&
       $ \bar{\eta} (g_s + g_p \gamma^5 ) \psi \chi +   \bar{\psi} (g_s - g_p
       \gamma^5 ) \eta \chi^\dagger$ \\
& & &  $ \displaystyle   \frac{1}{M_\Omega} \left[ (g_s^2 - g_p^2) \bar{\psi} \psi \chi^\dagger \chi+ \frac{i}{M_\Omega} \chi^\dagger \bar{\psi} \left(g_s^2 + g_p^2 - 2     g_s g_p \gamma^5 \right)\gamma^\mu  \partial_\mu \left( \psi \chi \right) \right]$ \\
\midrule 
 \multirow{2}{*}{S}      & \multirow{2}{*}{V} &
       \multirow{2}{*}{\raisebox{-0.435\height}{\includegraphics[width=0.1\textwidth]{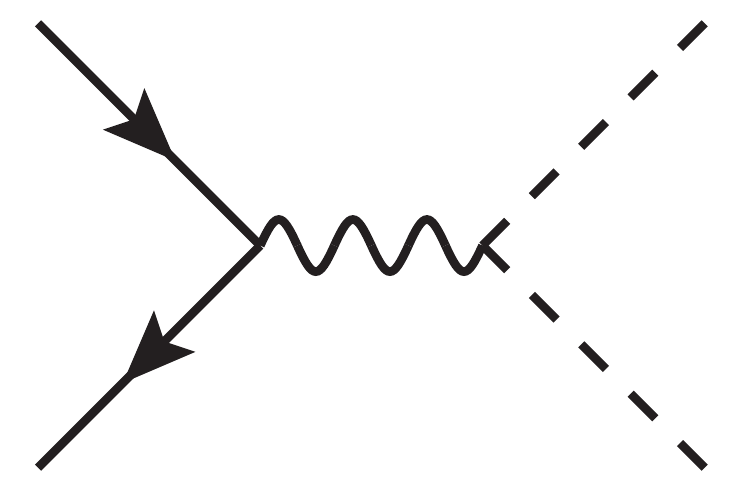}}}&$
       g_\chi (\chi^\dagger \partial_\mu \chi - \chi \partial_\mu
       \chi^\dagger) Z^\mu + \bar{\psi} \gamma^\mu (g_l P_L + g_r P_R) \psi
       Z_\mu$ \\
& & & $\displaystyle \frac{g_\chi}{M_\Omega^2} \bar{\psi} \gamma^\mu \left( g_l P_L + g_r P_R \right) \psi \left( \phi^\dagger \partial_\mu \phi - \phi \partial_\mu \phi^\dagger \right)$\\
\midrule \midrule
\multirow{2}{*}{F} & \multirow{2}{*}{S} &
\multirow{2}{*}{\raisebox{-0.435\height}{\includegraphics[width=0.1\textwidth]{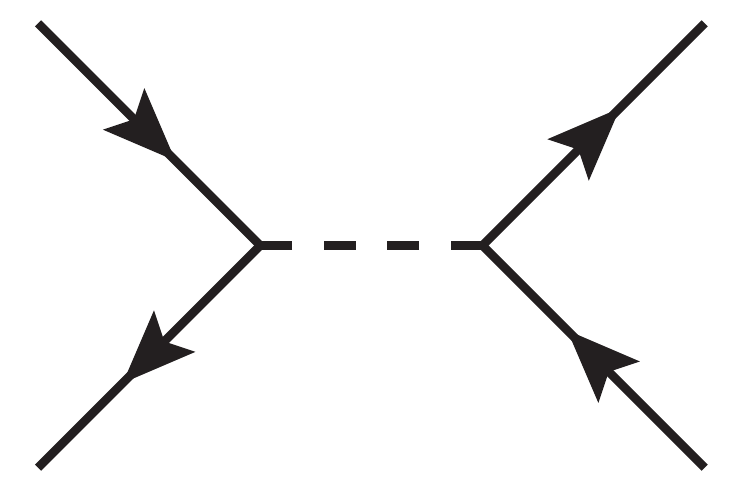}}}&
$\bar{\chi} \left(g_s + g_p \gamma^5 \right) \chi \phi + \bar{\psi} \left( g_s
  + g_p \gamma^5 \right) \psi \phi$ \\
& & & $\displaystyle \frac{1}{M_\Omega^2} \bar{\chi} \left(g_s + i g_p \gamma^5 \right) \chi \bar{\psi} \left( g_s + i g_p \gamma^5 \right)\psi$\\
\midrule
\multirow{2}{*}{F}        & \multirow{2}{*}{V} &
        \multirow{2}{*}{\raisebox{-0.435\height}{\includegraphics[width=0.1\textwidth]{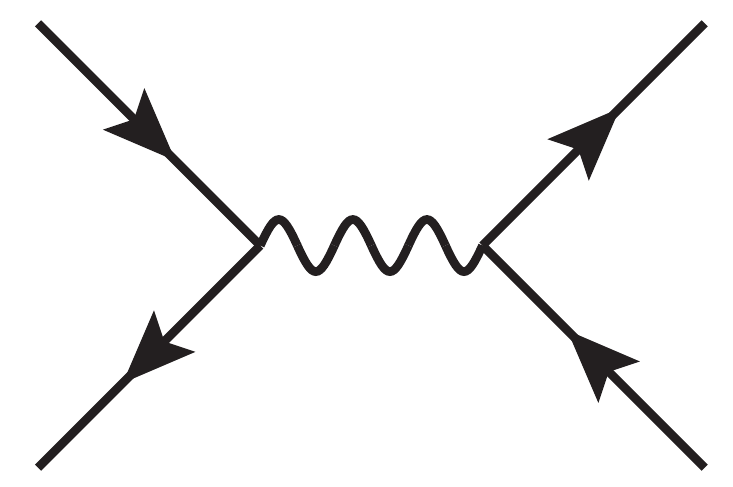}}}&
        $\bar{\psi} \gamma^\mu (g_l P_L + g_r P_R) \psi Z_\mu + \bar{\chi}
        \gamma^\mu (g_l P_L + g_r P_R) \chi Z_\mu $ \\
& & & $\displaystyle \frac{1}{M_\Omega^2} \bar{\psi} \gamma^\mu \left(  g_l P_L + g_r P_R \right) \psi \ \bar{\chi} \gamma_\mu \left(g_l P_L + g_r P_R    \right) \chi$\\
\midrule
\multirow{2}{*}{F}        & \multirow{2}{*}{tS} &
        \multirow{2}{*}{\raisebox{-0.435\height}{\includegraphics[width=0.1\textwidth]{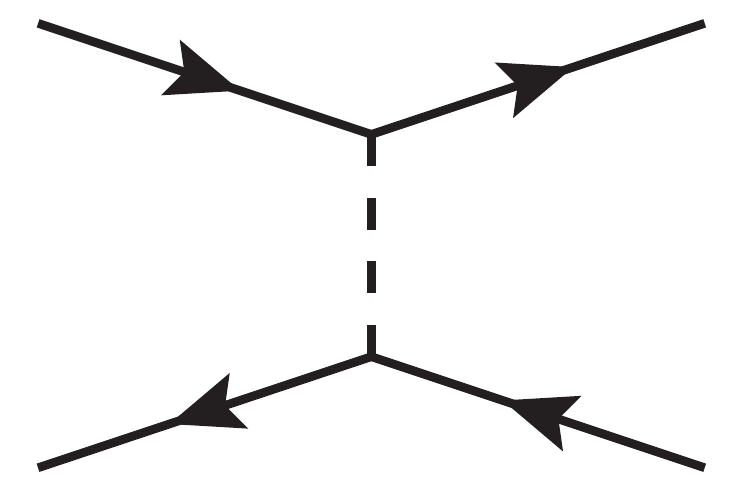}}}&$\bar{\chi}
        \left(g_s + g_p \gamma^5 \right) \psi \phi + \bar{\psi} \left( g_s +
          g_p \gamma^5 \right) \chi \phi$ \\
&&& $\displaystyle \frac{1}{M_\Omega^2} \bar{\psi} \left(g_s - g_p \gamma^5 \right) \chi \bar{\chi} \left( g_s + g_p \gamma^5 \right)\psi $ \\
\midrule
\multirow{2}{*}{F}        & \multirow{2}{*}{tV} &
        \multirow{2}{*}{\raisebox{-0.435\height}{\includegraphics[width=0.1\textwidth]{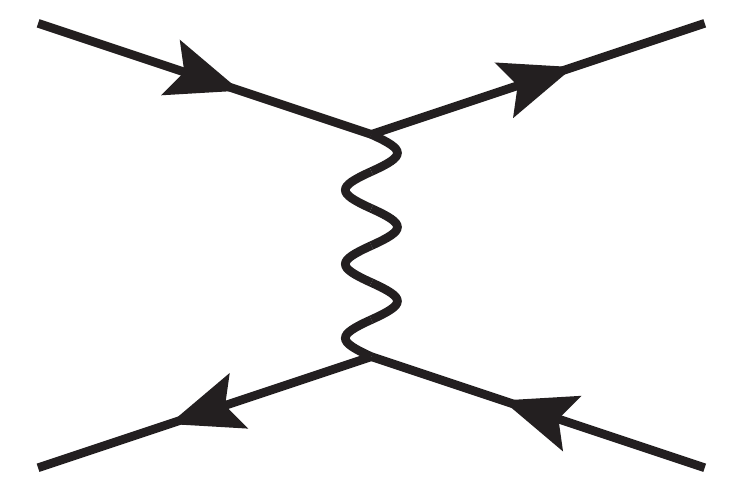}}}&$\bar{\psi}
        \gamma^\mu (g_l P_L + g_r P_R) \chi Z_\mu + \bar{\chi} \gamma^\mu (g_l
        P_L + g_r P_R) \psi Z_\mu $ \\
&&& $ \displaystyle \frac{1}{M_\Omega^2} \bar{\psi}
        \gamma^\mu \left(g_l P_L + g_r P_R \right)   \chi  \bar{\chi} \gamma_\mu \left(g_l P_L + g_r P_R  \right) \psi $ \\
\midrule \midrule
\multirow{2}{*}{V} & \multirow{2}{*}{S} &
\multirow{2}{*}{\raisebox{-0.435\height}{\includegraphics[width=0.1\textwidth]{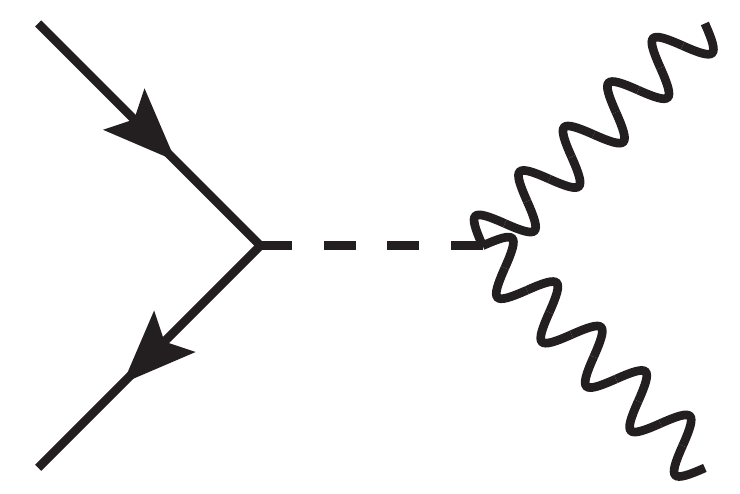}}}
& $-\chi^\mu \chi_\mu \phi + \bar{\psi} (g_s + i g_p \gamma^5)\psi \phi$ \\
&&& $\displaystyle - \frac{g_\chi}{M_\Omega^2} \chi^\mu \chi_\mu \bar{\psi} \left( g_s + i g_p \gamma^5 \right)\psi$\\
\midrule
\multirow{2}{*}{V}       & \multirow{2}{*}{F} &
       \multirow{2}{*}{\raisebox{-0.435\height}{\includegraphics[width=0.1\textwidth]{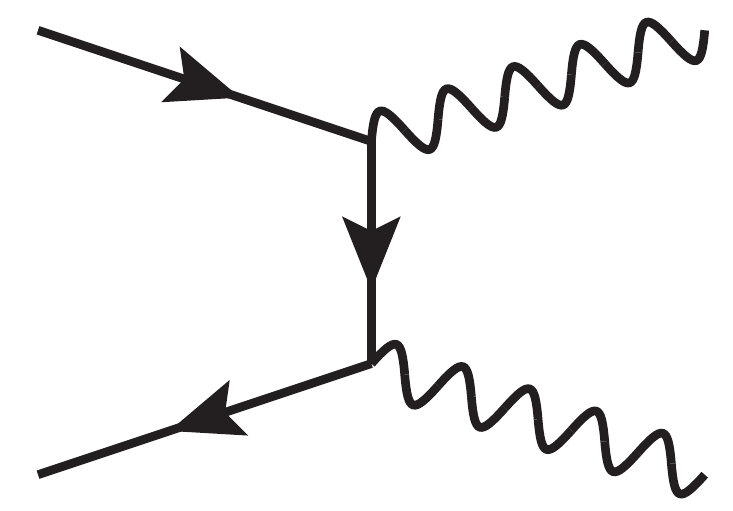}}}&
       $ - \bar{\eta} \gamma^\mu (g_l P_L + g_r P_R ) \chi_\mu
         + \bar{\psi} \gamma^\mu (g_l P_L + g_r P_R) \eta \chi_\mu^\dagger$      \\
&&& $\displaystyle \frac{1 }{M_\Omega} \left[g_l g_r \bar{\psi} \gamma^\nu \gamma^\rho \psi \ \chi^\dagger_\nu\chi_\rho + \frac{i}{M_\Omega} \chi^\dagger_\nu  \bar{\psi}
       \gamma^\nu\gamma^\mu \gamma^\rho \left(g_{L}^2 P_L + g_{R}^2P_R
       \right) \partial_\mu \left( \psi \chi_\rho \right) \right]$\\
\midrule
\multirow{2}{*}{V}       & \multirow{2}{*}{V} &
       \multirow{2}{*}{\raisebox{-0.435\height}{\includegraphics[width=0.1\textwidth]{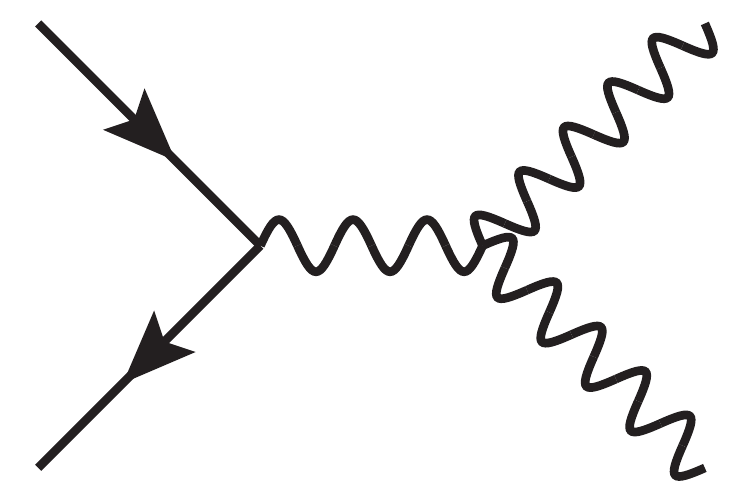}}}&
       $i g_\chi \left[ Z_\mu \chi^\dagger_\nu {\partial \chi}^{\mu \nu} + Z_\mu
         \chi_\nu  \partial \chi^{\mu \nu} + \chi^\dagger_\mu
         \chi_\nu \partial Z^{\mu \nu} \right]  + \bar{\psi} \gamma_\mu (g_l P_L + g_r P_R ) \psi$ \\
&&& $\displaystyle \frac{i g_\chi}{M_\Omega^2} \bar{\psi} \gamma^\mu \left( g_l P_L + g_r
  P_R \right) \psi \left[\chi^\nu \partial \chi^\dagger_{\mu \nu} - \chi^{\dagger, \nu} \partial \chi_{\mu \nu} + \partial^\nu \left(\chi^\dagger_\nu\chi_\mu - \chi^\dagger_\mu \chi_\nu \right) \right]$\\
\bottomrule
\end{tabular}
\caption{List of all fundamental and corresponding effective Lagrangians. $\partial X^{\mu \nu}  \equiv  \partial^\mu X^\nu
  - \partial^\nu X^\mu$.}
\label{tbl:allmodels}
\end{table}
Table \ref{tbl:allmodels} summarises the previously derived list of operators. We use the labelling scheme XY for a model with dark matter X and mediator Y which are either S(calars), F(ermions) or V(ectors). The mediator receives a prefix $t$  to denote t--channel interactions in fermion dark matter models in order to distinguish them from the corresponding s--channel interactions with the same mediator.  

These models still depend on multiple parameters $g_i$ appearing in linear combinations with different Dirac--structures. We restrict our analysis to specific benchmark models  (see table  \ref{tbl:constraints})  with
constraints on the individual couplings such that only one overall
multiplicative factor remains. The effective coupling constant $G_\text{eff}$ for each model is then defined as $g_1 g_2/ M_\Omega^2$. Models with fermionic mediators form an exception, because the leading order term depends only on $1/M_\Omega$. This is why in these cases we define  $G_\text{eff} \equiv g_1 g_2 / M_\Omega$. Since we have an additional $1/M_\Omega^2$--term in these models which we cannot factorise out, we choose two example values of $M_\Omega$ to model different supression strengths of those subleading terms which we call \emph{low} and \emph{high}. 

Note that SS and VS models have a triple scalar coupling with a dimensionful coupling constant $g_\chi$, which leads to a mass dimension of -1 for $G_\text{eff}$. This is different to other $1/M_\Omega^2$--models which have mass dimension -2.

In addition to the spin and interaction mode, we distinguish between complex and real dark matter fields (the latter denoted by the suffix r). However, models with real fields that are trivially connected to the corresponding complex cases by multiplicative prefactors are not taken into account separately. SS, FS, VS and FV axialvector models will always receive an additional factor of $2$ for all matrix elements, whereas SV, VV and FV vector interactions vanish\footnote{For scalar and vector dark matter, these statements are easily verified by constraining $\chi_{(\mu)} = \chi_{(\mu)}^\dagger$ and evaluating the operators in table \ref{tbl:allmodels}. For Majorana spinor fields, we follow the proper definitions and statements in \cite{MajoranaRules}. Note that for cross sections calculations, additional symmetry factors may appear for identical particles in the same final state}. The results therefore scale accordingly. We also omit models with left--chiral couplings, since they are identical to the respective right--chiral cases for polarisation independent analyses. In those cases, we call the interaction generally \emph{chiral}. For the \textsc{Ilc} study, which is polarisation dependent, we give the corresponding conversion factors and analyse the \emph{right}--chiral case only\footnote{Since FV is the only vector interaction model with different effects on the vector and axialvector parts for real particles, the right--chiral FV model as a superposition of both behaves non--trivially under the reality condition and is considered as a separate model.}.

We distinguish whether the coupling strength \Geff changes between different Standard Model particles:
\begin{description}
\item[Universal or Yukawa--like coupling strength:] In a universal model all particles couple with the same strength $G_\text{eff}$, whereas in a Yukawa--like model the coupling grows proportional to the respective Standard Model fermion mass. To keep the proper mass dimension of $G_\text{eff}$, we normalise the Yukawa--scaled coupling to the electron mass: 
\begin{align}
G_\text{eff}^X \equiv G_\text{eff}\ m_X / m_e.
\end{align}
In that case, \textsc{Ilc} results will be identical for both coupling schemes.
\item[Coupling to all fermions or leptons only:] The \emph{fermion}--case includes all Standard Model fermions as allowed interaction partners, whereas the \emph{leptons only}--scenario forbids couplings to quarks. The latter strongly reduces nuclear cross sections which are probed at direct detection experiments, which affects the relative exclusion potential of these experiments and the \textsc{Ilc}.
\end{description}

\begin{table}
\centering
\begin{tabular}{l@{\qquad}l@{\quad}l}
\toprule
Operators & Definition & Name \\
\midrule \midrule
SS, VS, FS, FtS, FtSr & $g_p = 0$ & Scalar \\
        & $g_s = 0$ & Pseudoscalar \\
\midrule
SF, SFr & $g_p = 0, M_\Omega = \unit{1}{\TeV}$ & Scalar, Low \\
        & $g_p = 0, M_\Omega = \unit{10}{\TeV}$ & Scalar, High \\
        & $g_s = 0, M_\Omega = \unit{1}{\TeV}$ & Pseudoscalar, Low \\
        & $g_s = 0, M_\Omega = \unit{10}{\TeV}$ & Pseudoscalar, High \\
\midrule
SV, FV, FtV, FtVr, VV & $g_l = g_r$ & Vector \\
        & $g_l = -g_r$ & Axialvector \\
        & $g_l = 0$ & Right (Chiral) \\
\midrule
VF, VFr &  $g_l = g_r, M_\Omega = \unit{1}{\TeV}$ & Vector, Low \\
        & $g_l = -g_r, M_\Omega = \unit{10}{\TeV}$ & Vector, High \\
        &  $g_l = g_r, M_\Omega = \unit{1}{\TeV}$ & Axialvector, Low \\
        & $g_l = -g_r, M_\Omega = \unit{10}{\TeV}$ & Axialvector, High \\
\midrule
FVr & $g_l = 0$ & Right (Chiral) \\
\bottomrule
\end{tabular}
\caption{Benchmark models with specific values for the coupling constants in
  table  \ref{tbl:allmodels}}
\label{tbl:constraints}
\end{table}
\chapter[Constraining \Geff with \textsc{Wmap}]{Constraining \Geffb with \textsc{Wmap} Data}
\label{chap:wmap}
With our effective interaction models at hand, we can now start to look at  the allowed parameter space $G_\text{eff}(M_\chi)$ that is in agreement with the expected dark matter relic density $\Omega^\text{DM}_0$, whose current best measured value is given by the \textsc{Wmap} experiment. In section \ref{sec:measureomega} we shortly describe how \textsc{Wmap} derived that value from a spectral analysis of the anisotropies within the cosmic microwave background radiation. Section \ref{sec:omegaresults} will then show how we can link that value to the effective vertices and give respective results for all benchmark models listed in the previous chapter.
\section[Measuring $\Omega^\text{DM}_0$ with \textsc{Wmap}]{Measuring $\mathbf{\Omega^\text{DM}_0}$ with \textsc{Wmap}}
\label{sec:measureomega}
The Wilkinson Microwave Anisotropy Probe--experiment (\textsc{Wmap}) measured the value and fluctuation of the cosmic microwave background radiation \cite{WMAPSkymap}. During the evolution of the universe, most particles once encountered the freeze--out phase described in chapter \ref{chap:darkmatter}. Among these, photons decouple last and set today's temperature of \SI{2.73}{\kelvin}.
\textsc{Wmap} measured this radiation over the whole sky from 2001 to 2010. In particular the telescope analysed polarisation and anisotropies in the spectrum which are caused by different effects during and after freeze out, with the aim to deduce cosmological parameters from these measurements. In figure \ref{img:wmapgeneral}a we show the temperature measurement of the full sky map. The deviations from the average temperature are below the per mille level and strongly support the theory of big bang cosmology that predicts a very homogenous distribution. In figure \ref{img:wmapgeneral}b we show the measured temperature spectrum in terms of multipole moments: Expanding the angle dependent spectrum in the base of spherical harmonics, one finds the given coefficients $C^{TT}_l$ as follows:
\begin{align}
T(\hat{n}) &= \sum_{l = 0}^\infty \sum_{m = -l}^l a^T_{lm} Y_{lm}(\hat{n}), \\
C_{l}^{TT} &= \frac{1}{2l+1}\sum_{m = -l}^l \left|a^T_{lm}\right|^2.
\end{align}
\begin{figure}
\centering
\includegraphics[height=0.21\textheight]{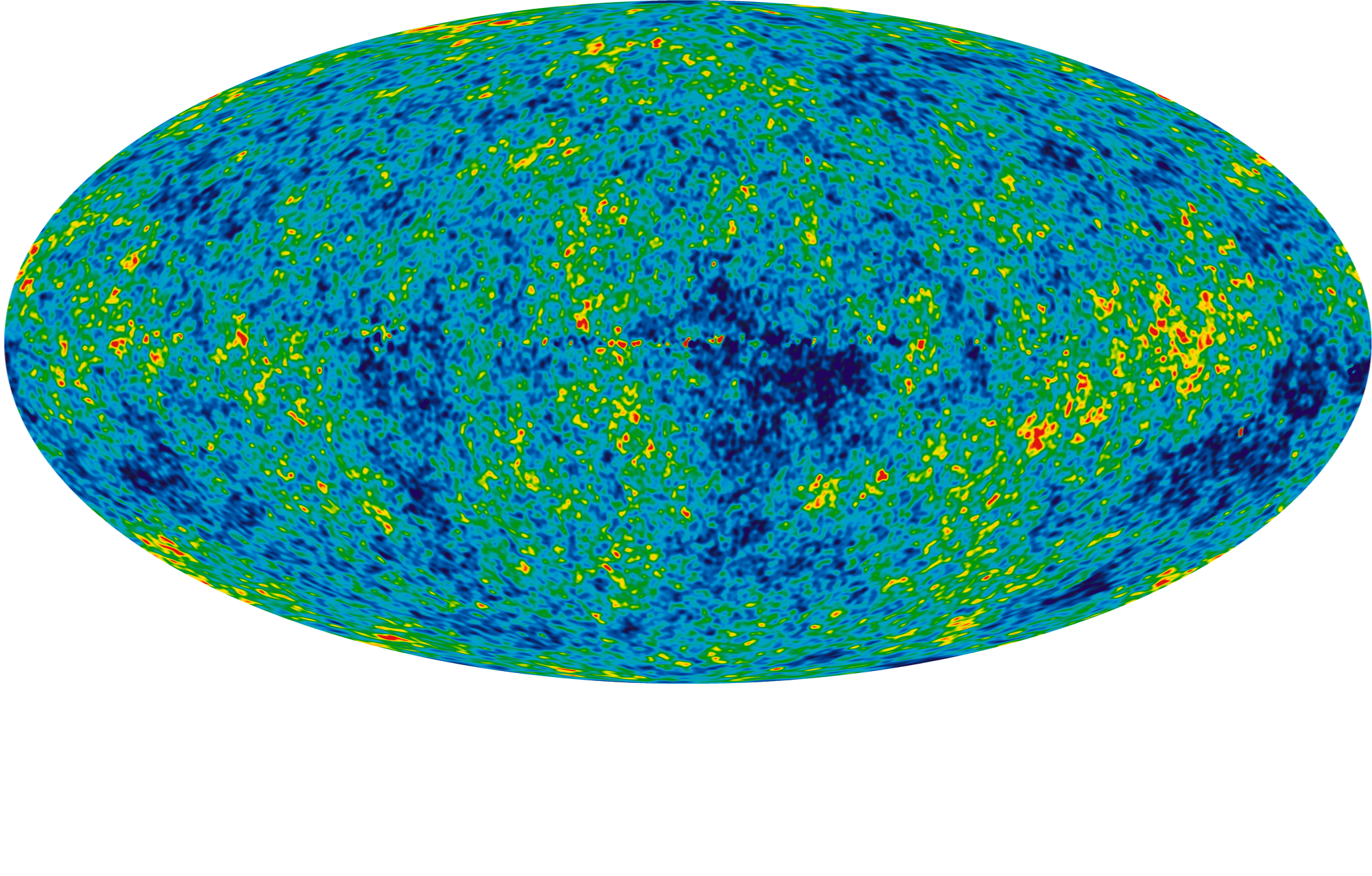}
\hspace{0.5cm}
\includegraphics[height=0.21\textheight]{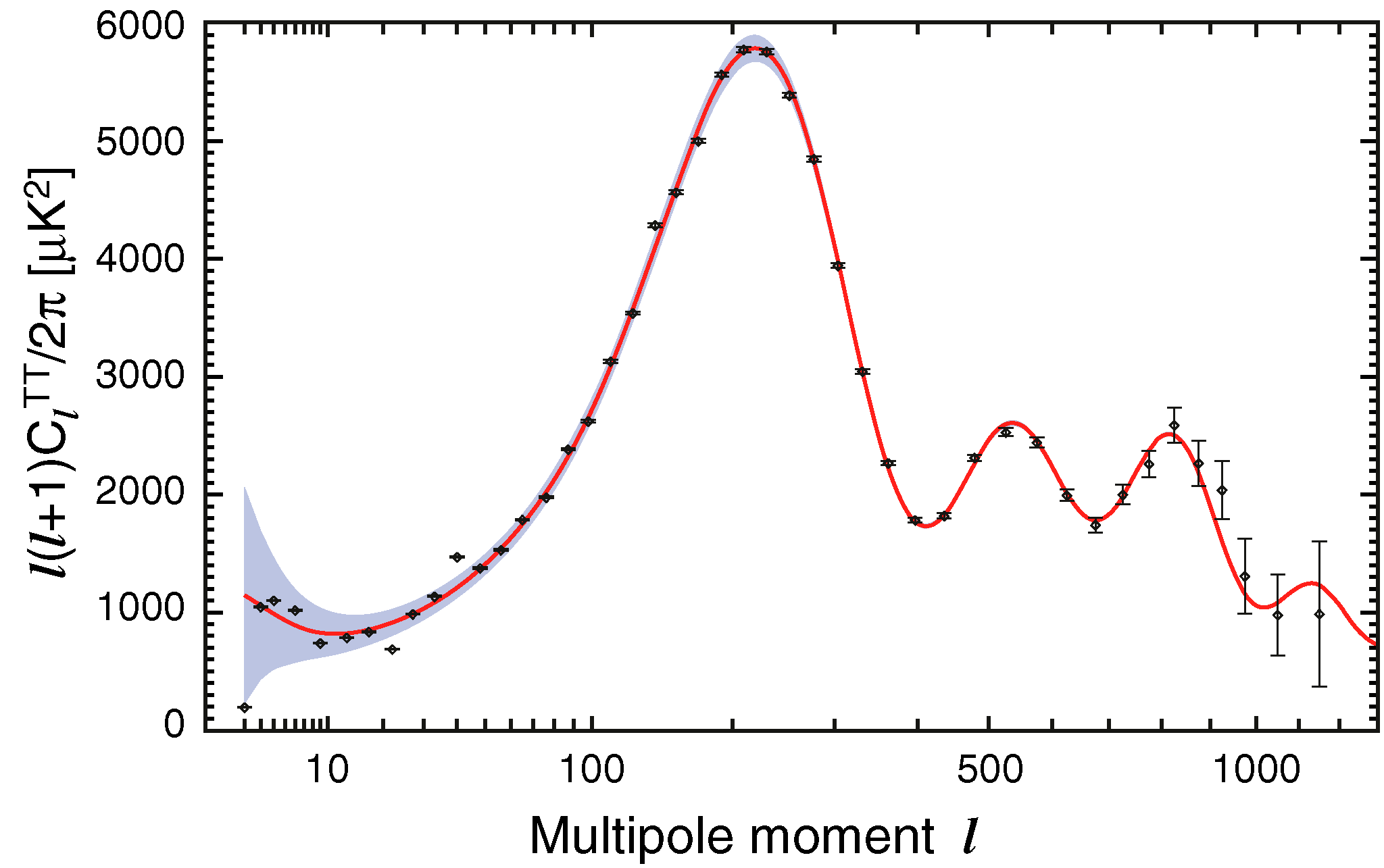} \\
a) \hspace{0.5\textwidth} b) \hspace{0.4\textwidth}
\caption{\textsc{Wmap} results after seven years of data taking. a) Galactic map and the measured temperature anisotropies from the median. The difference in color is linear proportional to the temperature anisotropy with a total range of $\pm$ \SI{200}{\micro \kelvin}. b) Temperature power spectrum in terms of multipole moments $l$. Both are taken from \cite{WMAPSkymap}.}
\label{img:wmapgeneral}
\end{figure}
The $C_l^{TT}$ spectrum can be analysed in the so called \emph{$\Lambda$\textsc{cdm}}--model, which assumes a universe with the Friedmann--Lemaître--Robertson--Walker--metric given in (\ref{eqn:metric}) and describes its evolution since the big bang with six degrees of freedom. At \textsc{Wmap}, the following parameter set is chosen:
\begin{itemize}
\item Relative abundance of dark energy $\Omega^0_\Lambda$.
\item Relative abundance of baryonic matter $\Omega^0_b h^2$.
\item Relative abundance of cold dark matter $\Omega^0_\text{DM}$.
\item Optical thickness $\tau$ for reionization\footnote{Structure formation has caused the emission of high energy photons that ionize atoms, leading to a reavailability of electrons for further scattering and changing the CMB spectrum.}.
\item Scalar spectral index\footnote{The early universe encountered a phase of rapid expansion, called inflation, to explain why the CMB is the same even in areas that have not been in causal contact. Macroscopic anisotropies are then caused by pre--inflationary primordial fluctuations, described by a power law with exponent $n_s - 1$ and amplitude $P_R(k_0) = (2 \pi^2) \Delta_R^2(k_0)/k_0^3$, $k_0 = \SI{0.002}{\mega\per\parsec}$. } $n_s$.
\item Curvature fluctuation amplitude\footnotemark[\value{footnote}] $\Delta_R^2(k_0)$.
\end{itemize}
Within this model a peak spectrum as in figure \ref{img:wmapgeneral}b is expected. Here, the size and position of the first three peaks depend on the relative values of the individual abundances for baryonic matter, cold dark matter and dark energy. The seven year measurement yielded the result
\begin{align}
\Omega_0^\text{DM} h^2 = \num{0.1109(56)}.
\end{align}

\section[Relating \Geff to $\Omega^\text{DM}_0$ and Results]{Relating \Geffb to $\mathbf{\Omega^\text{DM}_0}$ and Results}
\label{sec:omegaresults}
Using the list of effective operators from tables \ref{tbl:allmodels} and \ref{tbl:constraints}, we can calculate the total cross section for dark matter annihilation into standard model fermions given by the diagram in figure \ref{img:generalannihiloationdiagram}. The respective expansion terms $a$ and $b$ described in section \ref{sec:relicdensity} are given in appendix \ref{app:annixsects}. For the evaluation of the freeze out temperature and the dark matter abundance, we have to sum over the individual cross sections for each standard model fermion the dark matter particle can annihilate into.

We derive bounds on \Geff as follows: For each model, we numerically solve (\ref{eqn:x0}) for $x_f$ and calculate the corresponding value of $\Omega^\text{DM}_0 h^2$ using (\ref{eqn:Omega0}). We set $c = 1/2$ \cite{DMChina1}, but the actual value has no significant impact on the result as long as it is of order 1. For scalar (fermionic, vector) dark matter we set $g$ to 1 (2, 3) respectively. Also we use $\Omega^\text{DM}_\text{tot} = \Omega_\chi + \Omega_{\chi^\dagger}$, i.e. in the case of complex dark matter fields we double the final value of $\Omega_\text{DM}$ since both particle and antiparticle contribute individually with the same value \cite{DMChina1}. 

For dark matter masses $M_\chi$ between $\SI{1}{\GeV}$ and $\SI{500}{\GeV}$, which is the maximum mass accessible at ILC energies we want to compare with, we search for the value of \Geff that leads to 
$\Omega^\text{DM}_0 h^2 =  0.1181$, which is the upper one sided limit at the \unit{90}{\%} confidence level. Smaller values for \Geff are then excluded, since they lead to a too large abundances of dark matter today. This would overclose the universe. We allow stronger couplings which lower the abundances though, since there could be additional contributions to $\Omega_0^\text{DM}$ apart from the \textsc{Wimp} of our model. As an example, an additional neutrino--only coupling \textsc{Wimp} could evade any exclusion statements of both the \textsc{Ilc} and \textsc{Xenon} but affect $\Omega_0^\text{DM}$.

\begin{figure}
\centering
\includegraphics[width=0.35\textwidth]{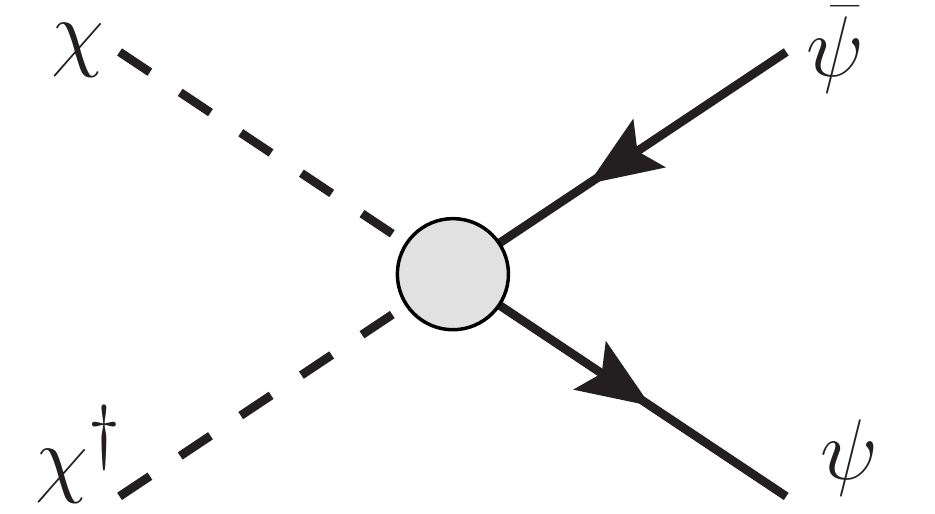}
\caption{Annihilation process through an effective four--particle vertex.}
\label{img:generalannihiloationdiagram}
\end{figure}
Figures \ref{img:wmapresults1} and \ref{img:wmapresults2} show the results for all benchmark models listed in table \ref{tbl:constraints} in the case of universal coupling to all standard model fermions. Models with an effective coupling of mass dimension -1 show only a small dependence on the mass of the dark matter particle, whereas the exclusion limits for other models decreasy by up to three orders of magnitude with increasing $M_\chi$. This is mostly caused by the additional $M_\chi^2$--dependence in the cross section of the latter.

In table \ref{tbl:constraints} we listed two different suppression scales $M_\Omega$ for models with fermionic t--channel operators. In this analysis, the size of that scale gives only a very small effect: The evaluation of the freeze--out temperature was performed in a small--velocity approximation, for which we expect a strong suppression of terms beyond the leading order within the effective theory. In particular we know for dark matter in the initial state that $s$ grows proportionally to $M^2_\chi$, such that the $s/M_\Omega^2$--suppression becomes weaker for larger \textsc{Wimp}--masses. This explains the small deviations between \emph{low} and \emph{high} XF models that arise at large values of $M_\chi$.

For \textsc{Wimp}--masses above \SI{1.2}{\GeV} (\SI{1.8}{\GeV}, \SI{4.2}{\GeV} or \SI{173.5}{\GeV}) the annihilation into charm--quarks (tau--leptons, bottom--quarks and top--quarks) become accessible and increase the total annihilation rate, leading to a smaller allowed value for $G_\text{eff}$. The size of these kinks depend strongly on how the cross section depends on the final state mass, i.\/e.\ if the coupling is universal or Yukawa--like. In figures \ref{img:wmapresults3} and \ref{img:wmapresults4} we give the results for the same set of models but with Yukawa--like couplings instead. In that case, the contribution from heavy particles is strongly enhanced through the larger coupling $G_\text{eff}$, such that the exclusion limits change by up to two orders of magnitude when the \textsc{Wimp} mass passes the top quark threshhold. This effect disappears for lepton--only couplings and the generally smaller annihilation rate leads to a weaker limit by up to a factor of 2.

A subset of the models we consider in this analysis has already been analysed before with respect to their compatibility with $\Omega^0_\text{DM}$ in \cite{DMChina1, DMChina2, RelicDensity}. We agree with their respective results, except for the slight deviations in \cite{RelicDensity} that are explained in \cite{DMChina1} through the missing definition of $\Omega_\text{complex} = 2 \Omega_\text{real}$.

\begin{figure}
\centering
\includegraphics[width=0.49\textwidth]{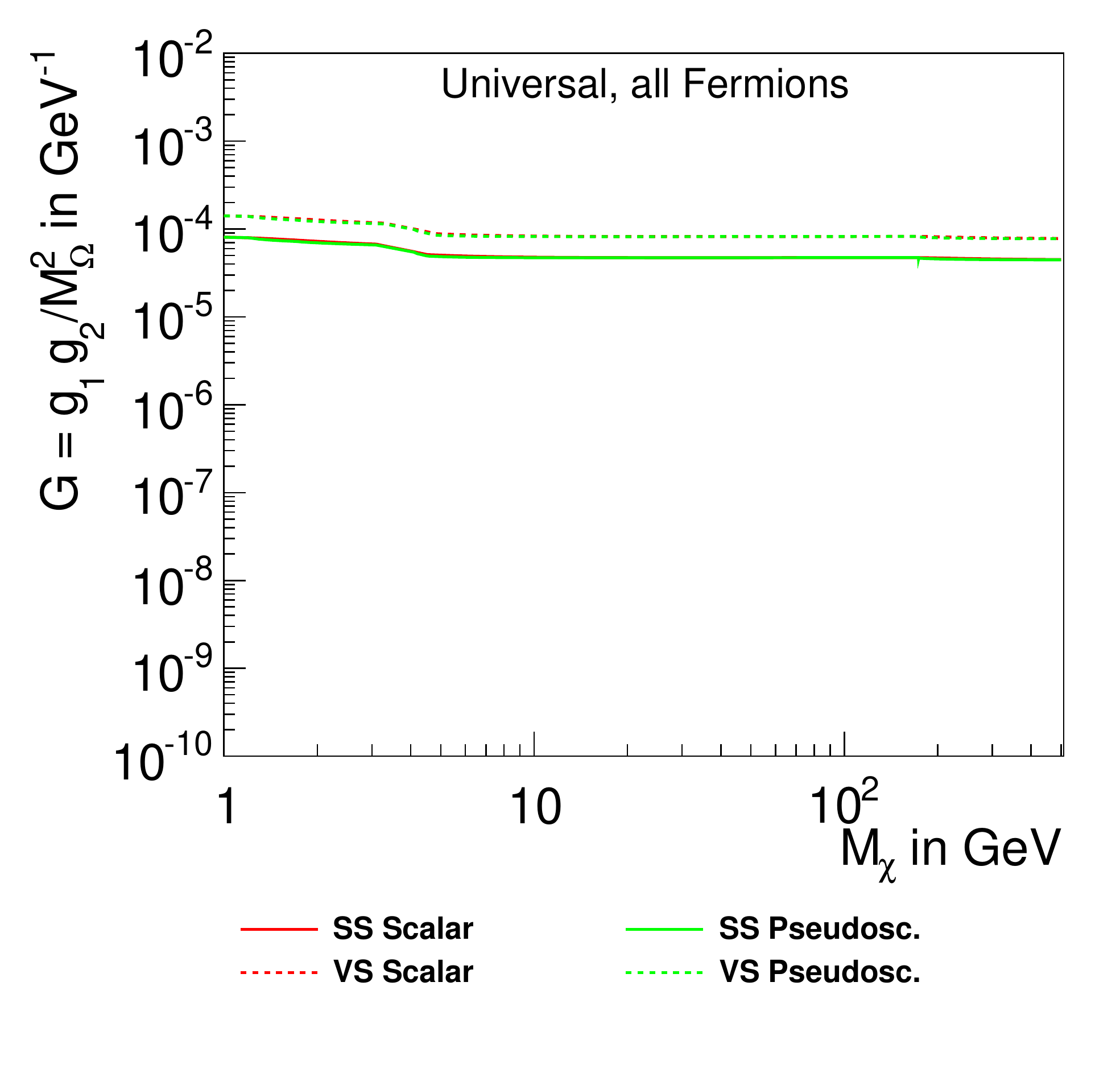}
\hfill
\includegraphics[width=0.49\textwidth]{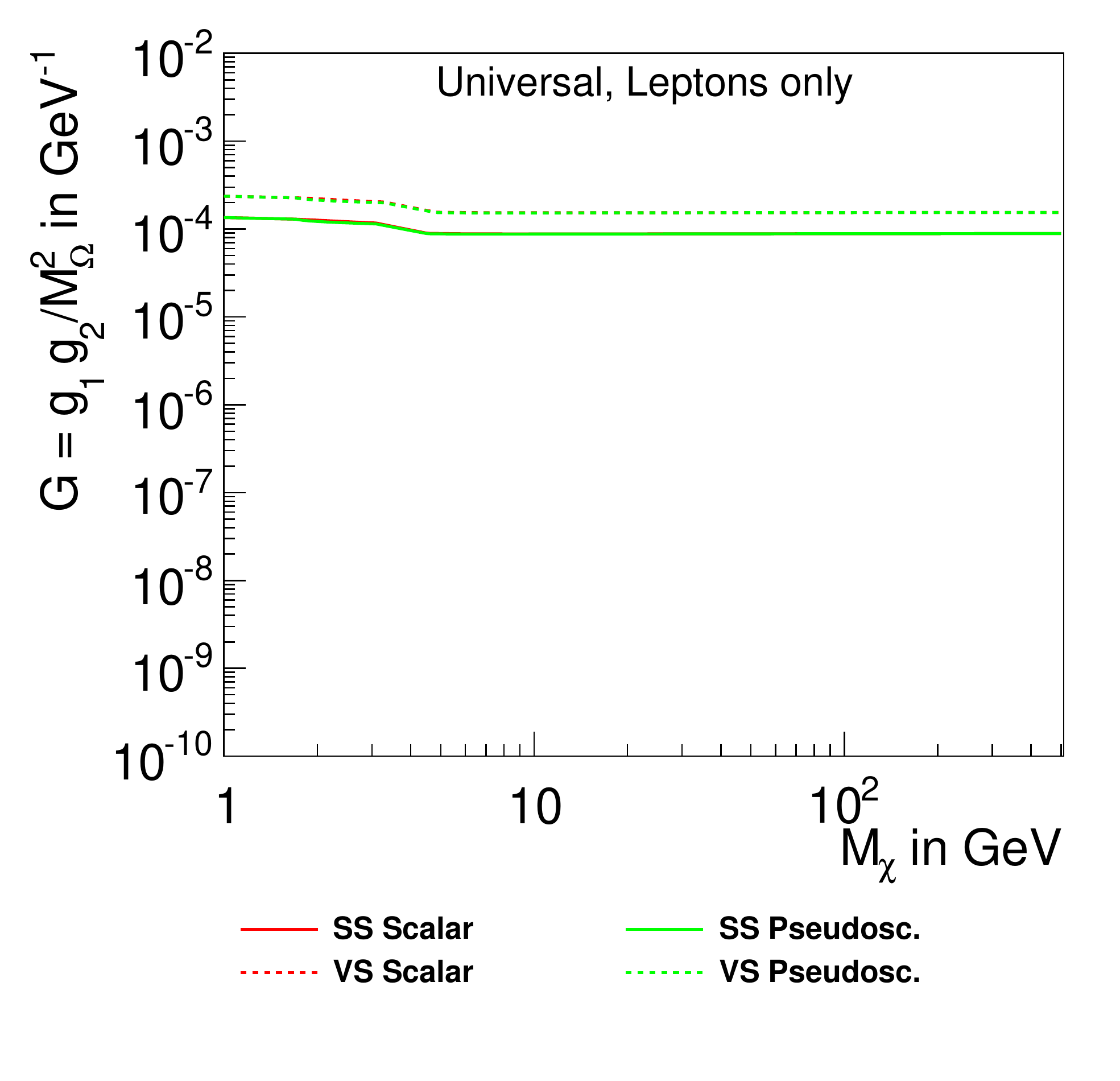} \\  \vspace{-0.25cm}
\includegraphics[width=0.49\textwidth]{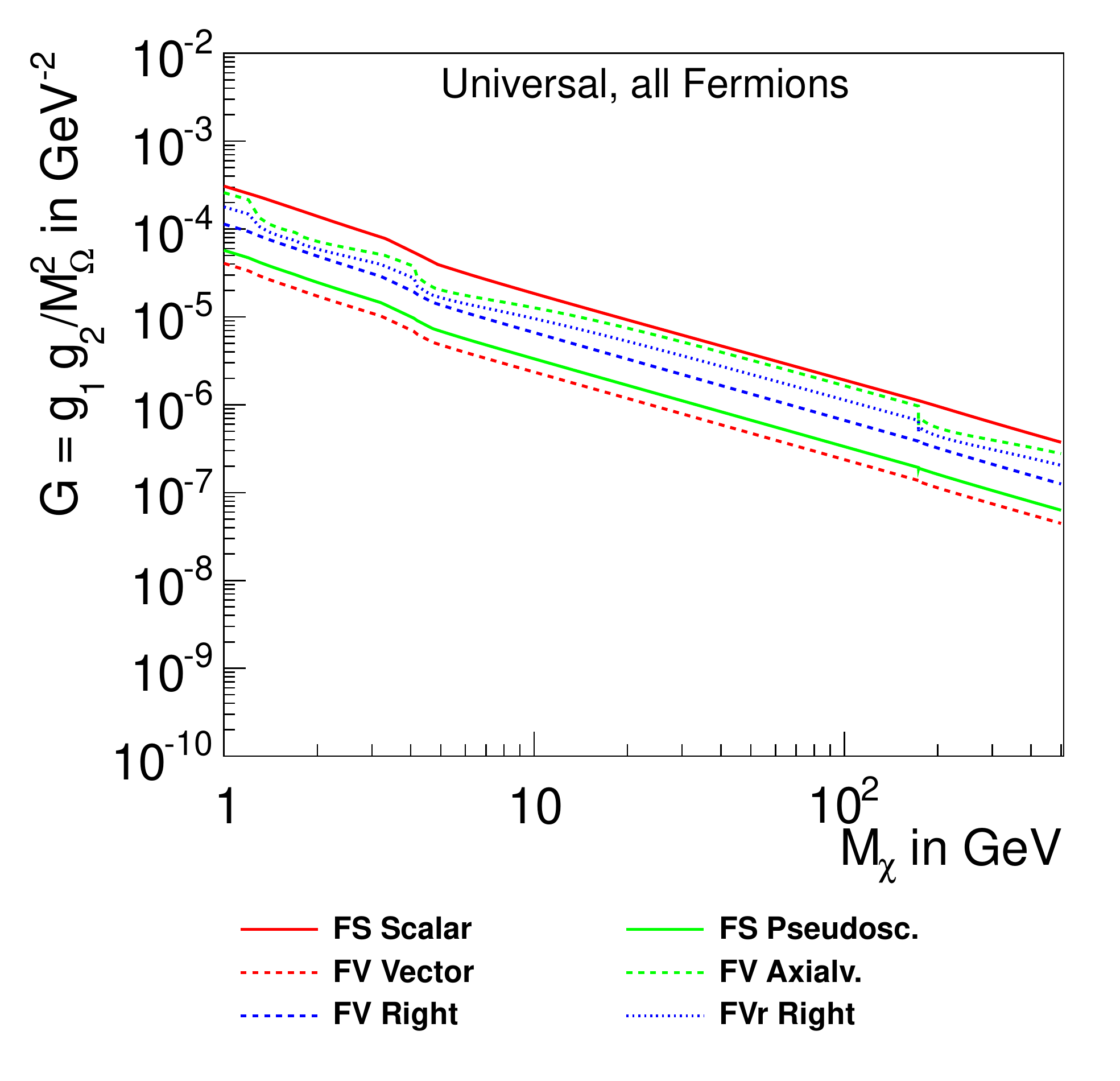}
\hfill
\includegraphics[width=0.49\textwidth]{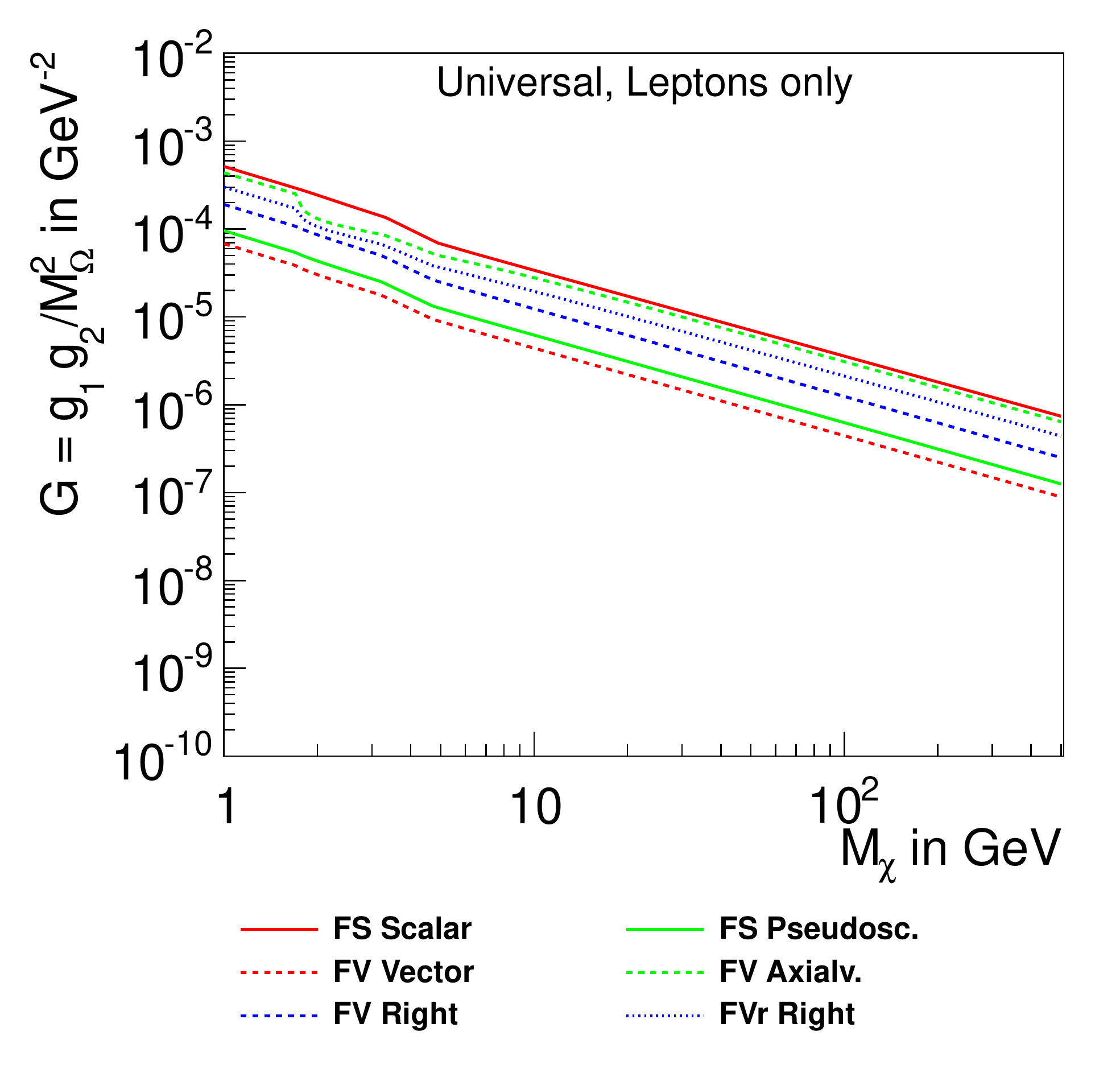} \\  \vspace{-0.25cm}
\includegraphics[width=0.49\textwidth]{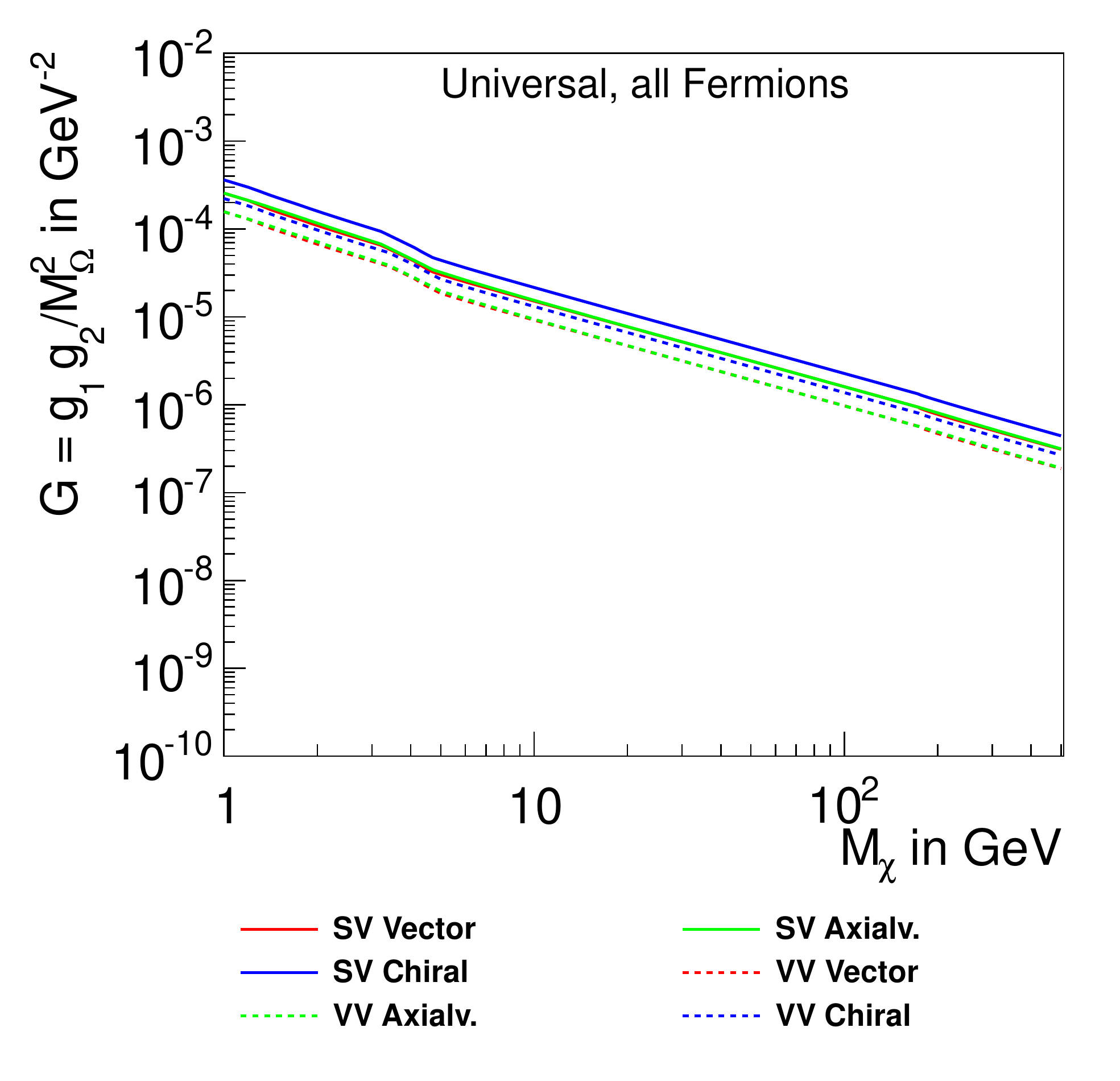}
\hfill
\includegraphics[width=0.49\textwidth]{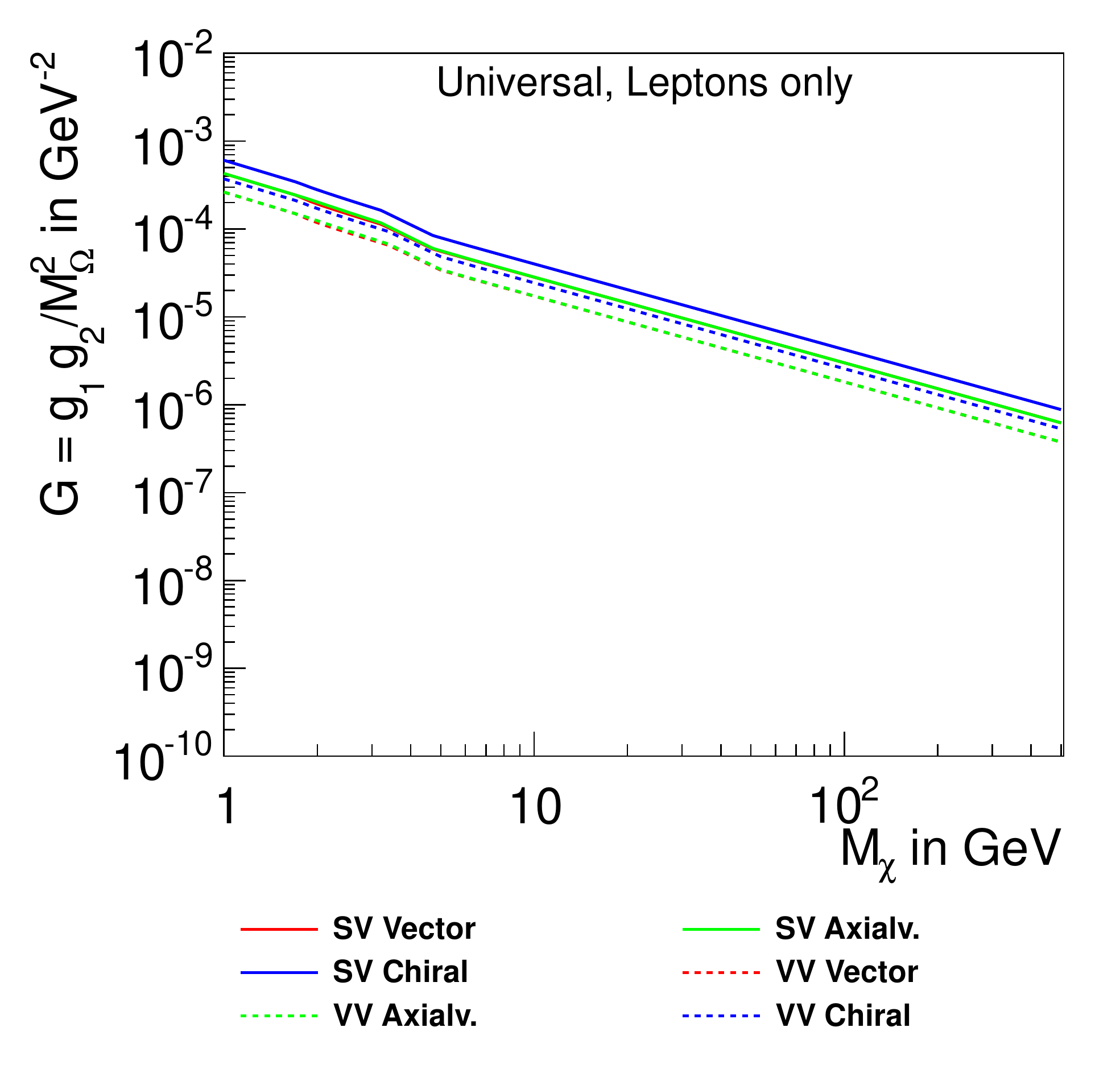} \\  \vspace{-0.25cm}
\caption{Minimum couplings in agreement with \textsc{Wmap}. Models with s--channel mediators and universal couplings.}
\label{img:wmapresults1}
\end{figure}
\begin{figure}
\centering
\includegraphics[width=0.49\textwidth]{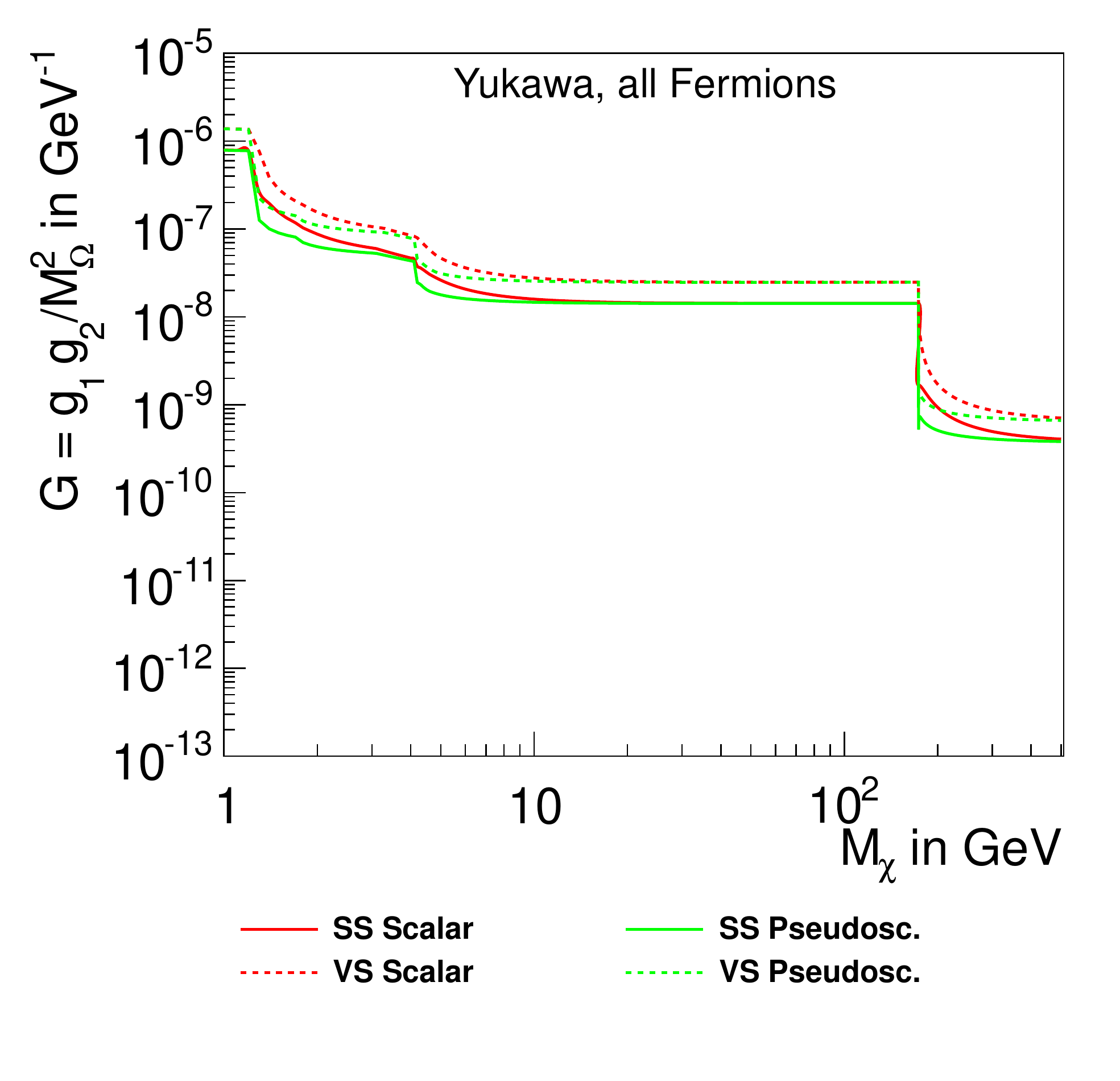}
\hfill
\includegraphics[width=0.49\textwidth]{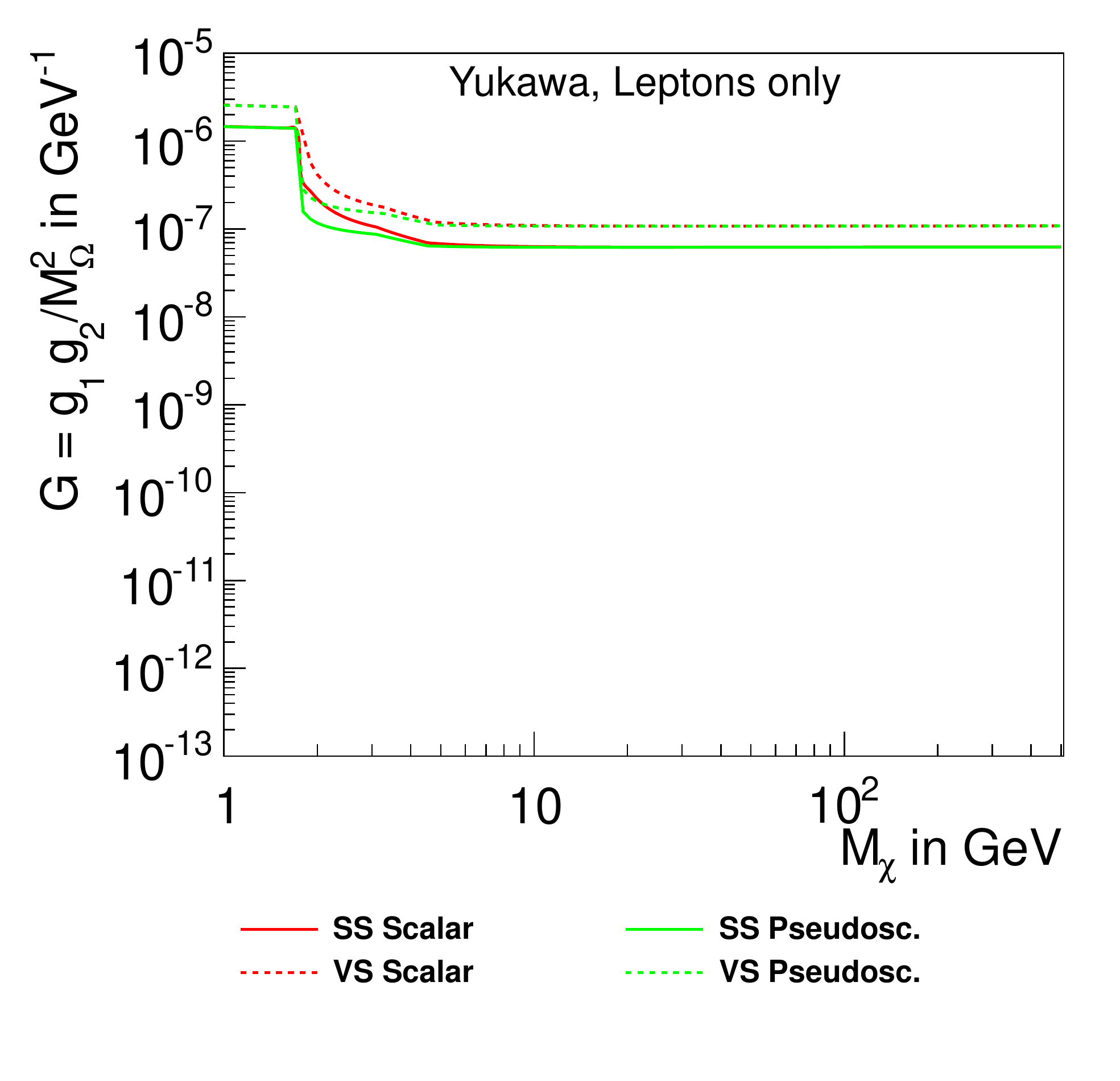} \\ \vspace{-0.25cm}
\includegraphics[width=0.49\textwidth]{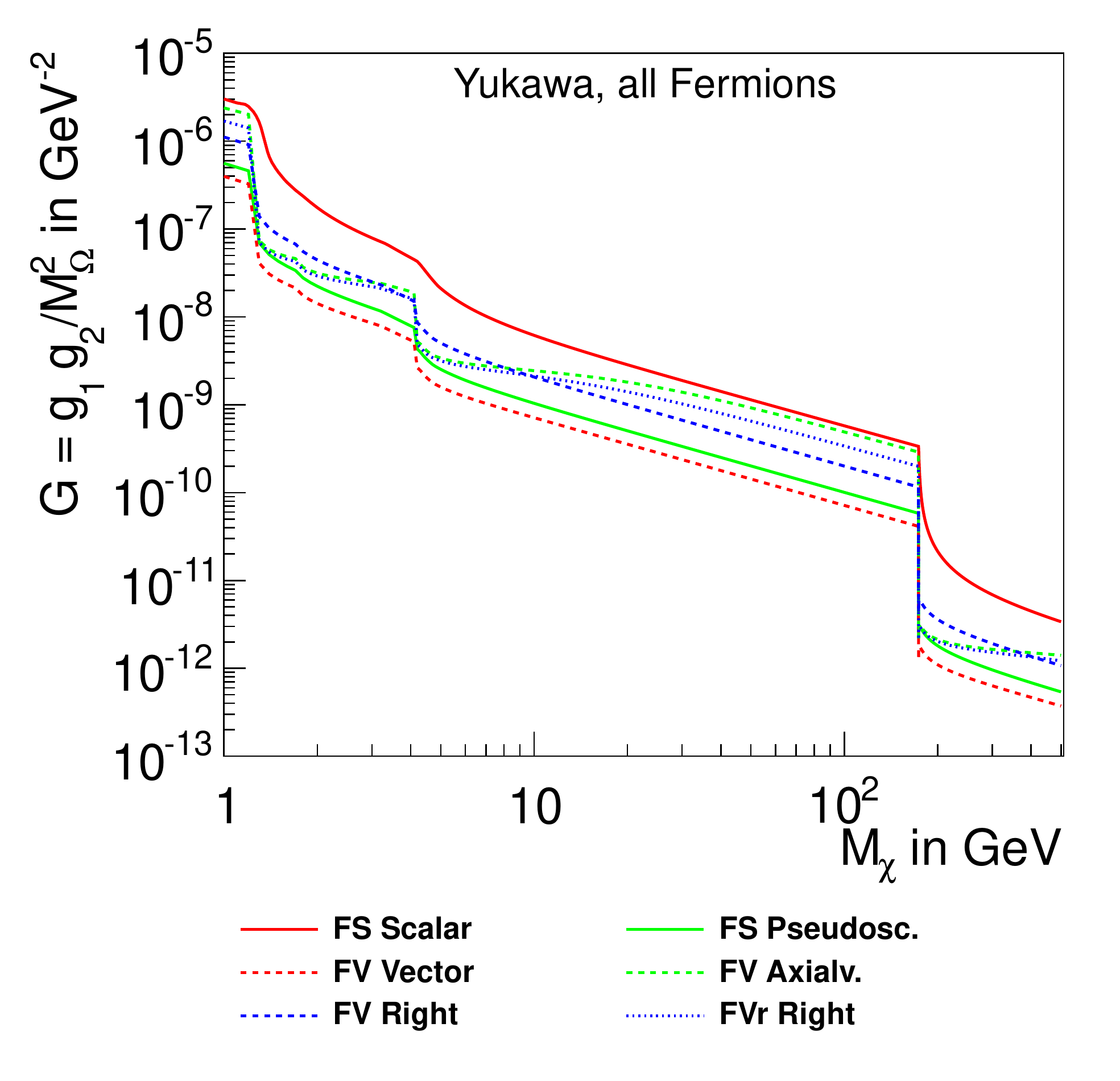}
\hfill
\includegraphics[width=0.49\textwidth]{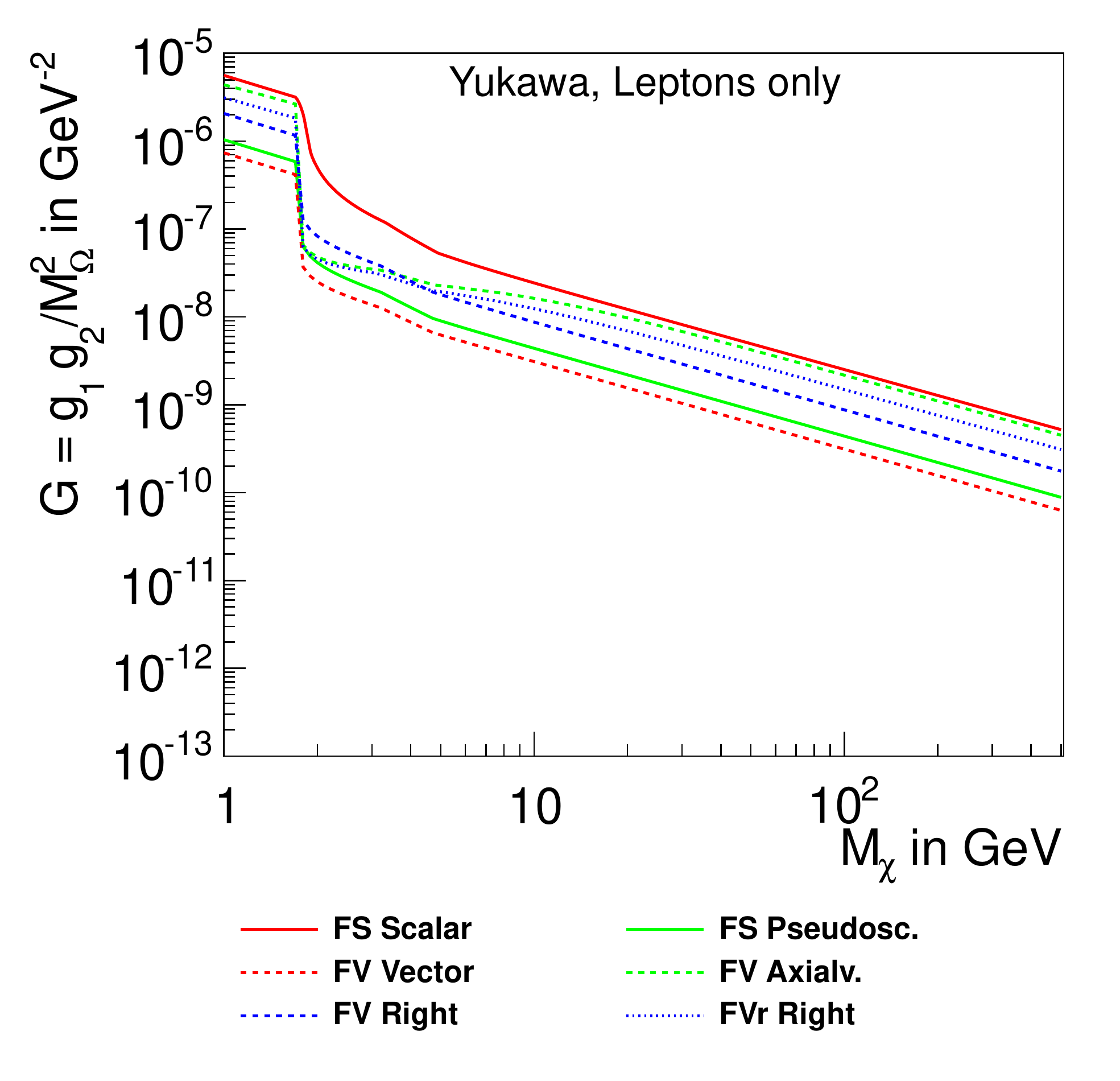} \\  \vspace{-0.25cm}
\includegraphics[width=0.49\textwidth]{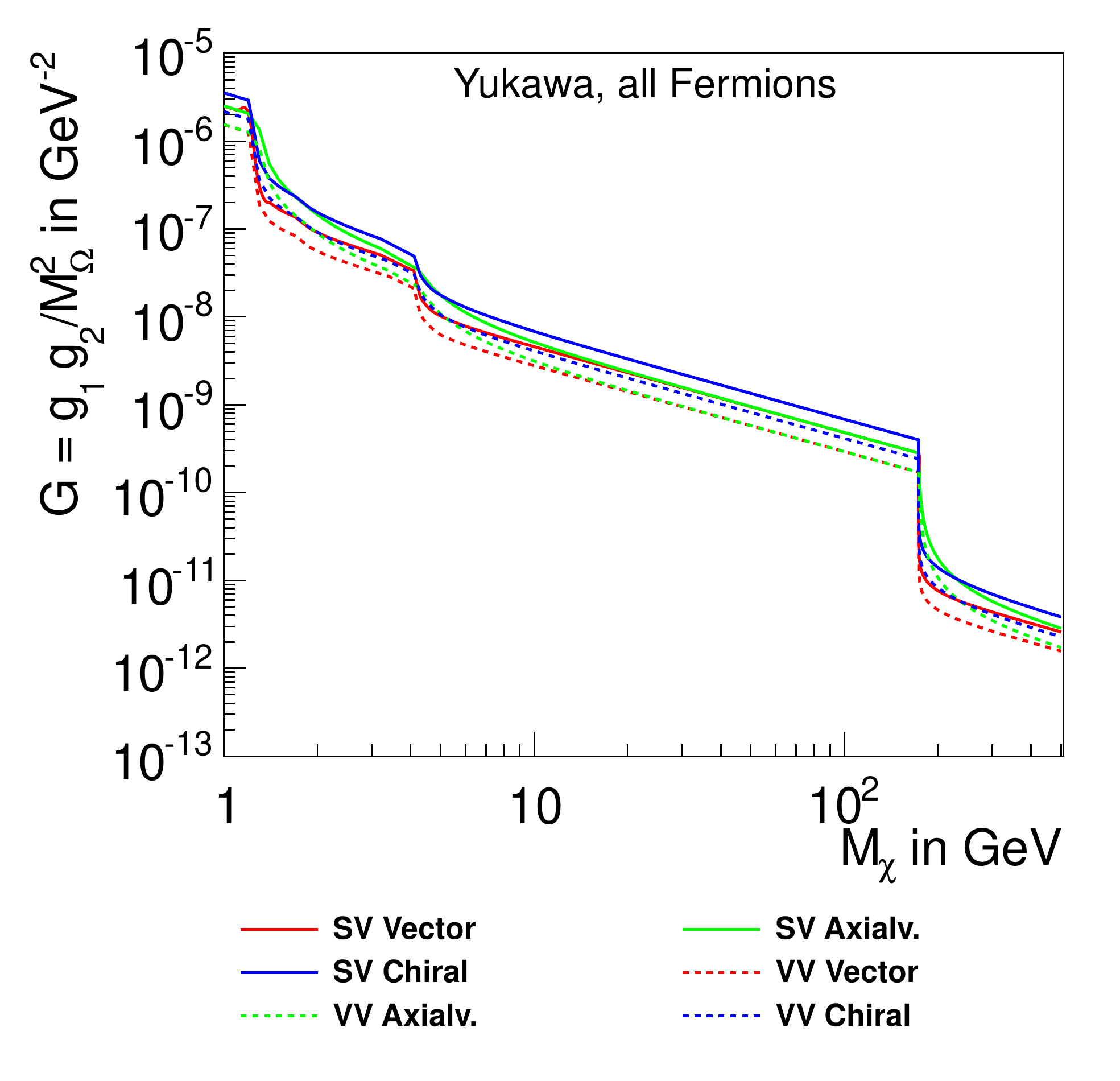}
\hfill
\includegraphics[width=0.49\textwidth]{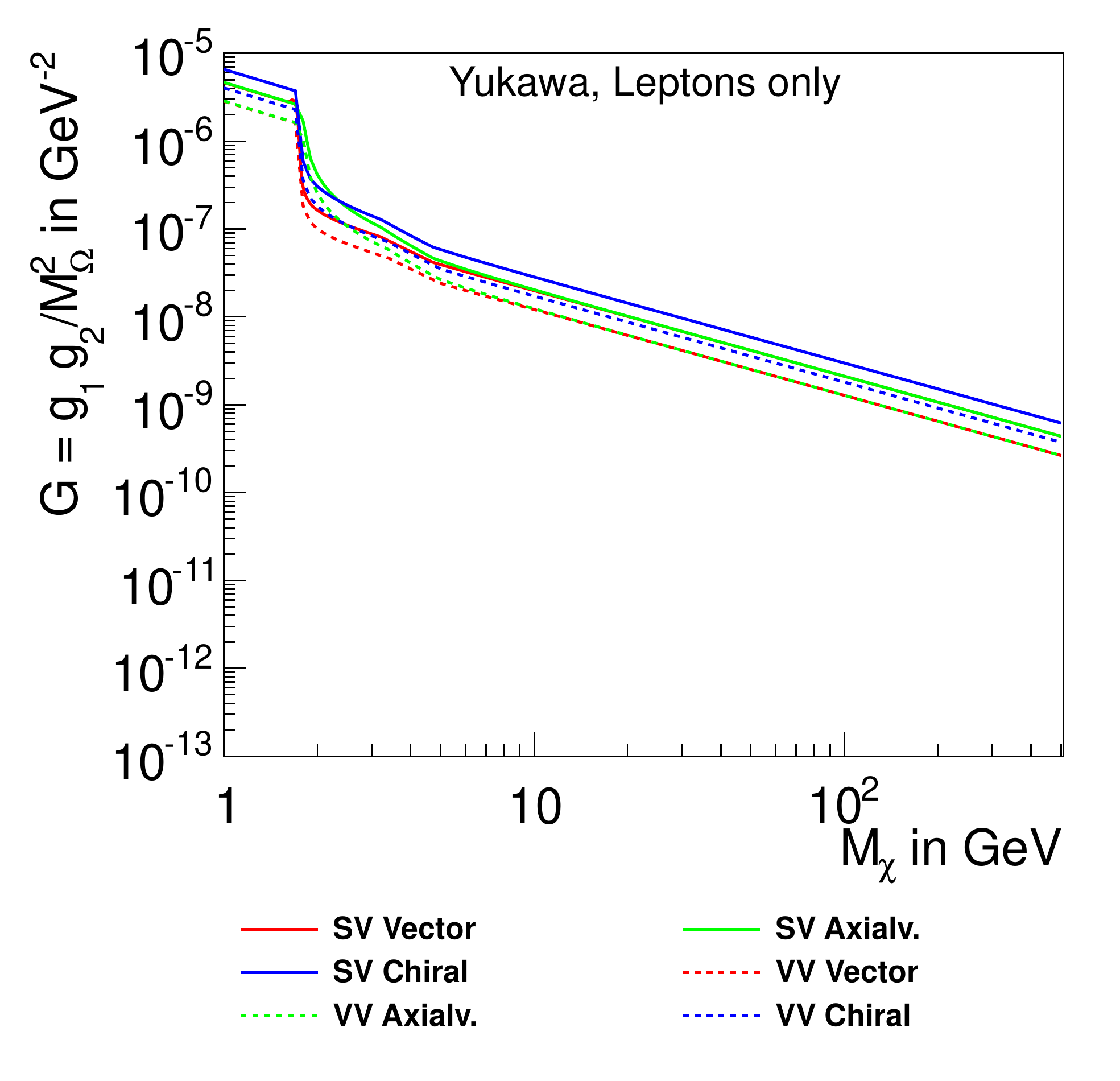} \\  \vspace{-0.25cm}
\caption{Minimum couplings in agreement with \textsc{Wmap}. Models with s--channel mediators and Yukawa--like couplings.}
\label{img:wmapresults2}
\end{figure}

\begin{figure}
\centering
\includegraphics[width=0.49\textwidth]{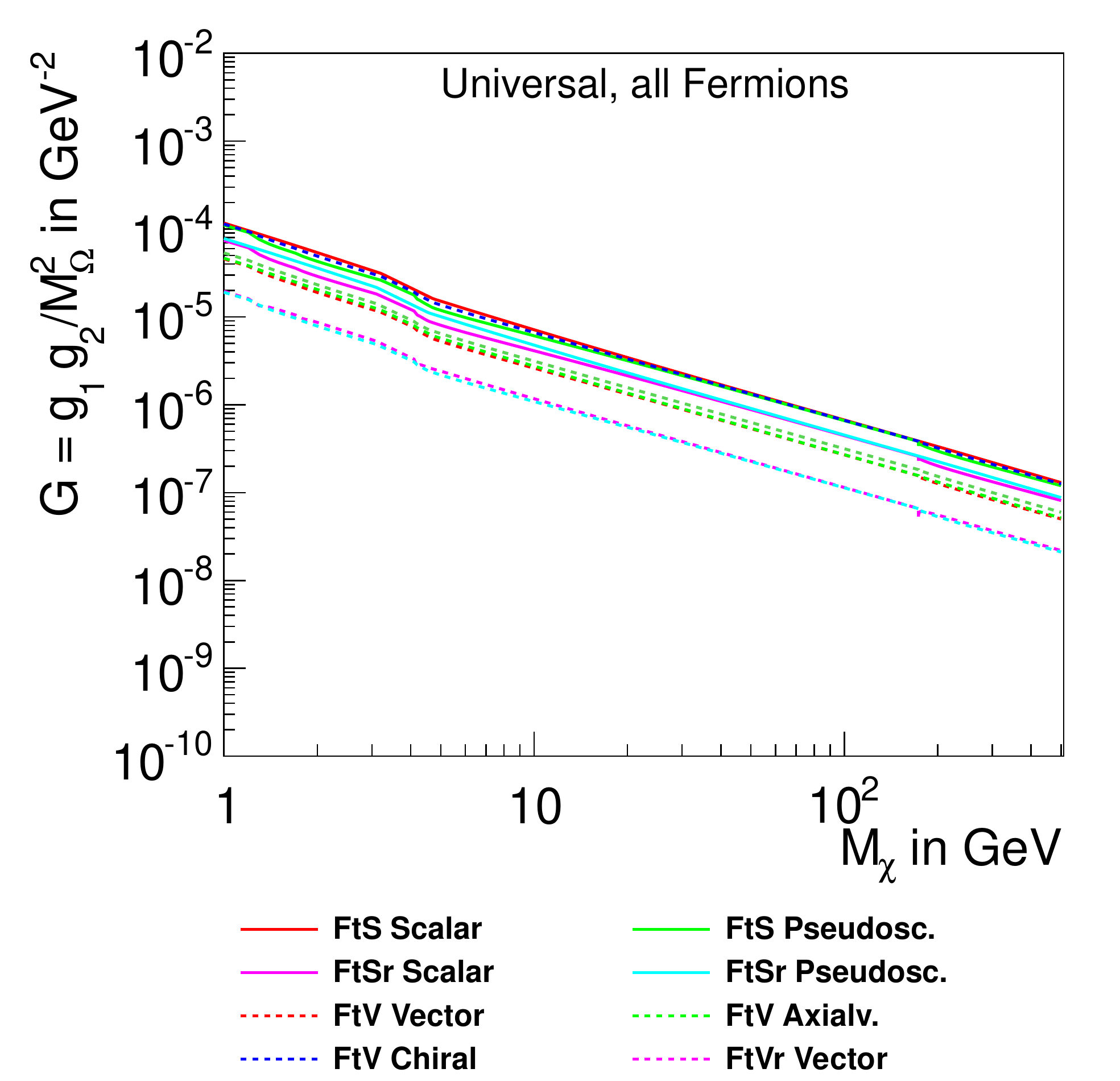}
\hfill
\includegraphics[width=0.49\textwidth]{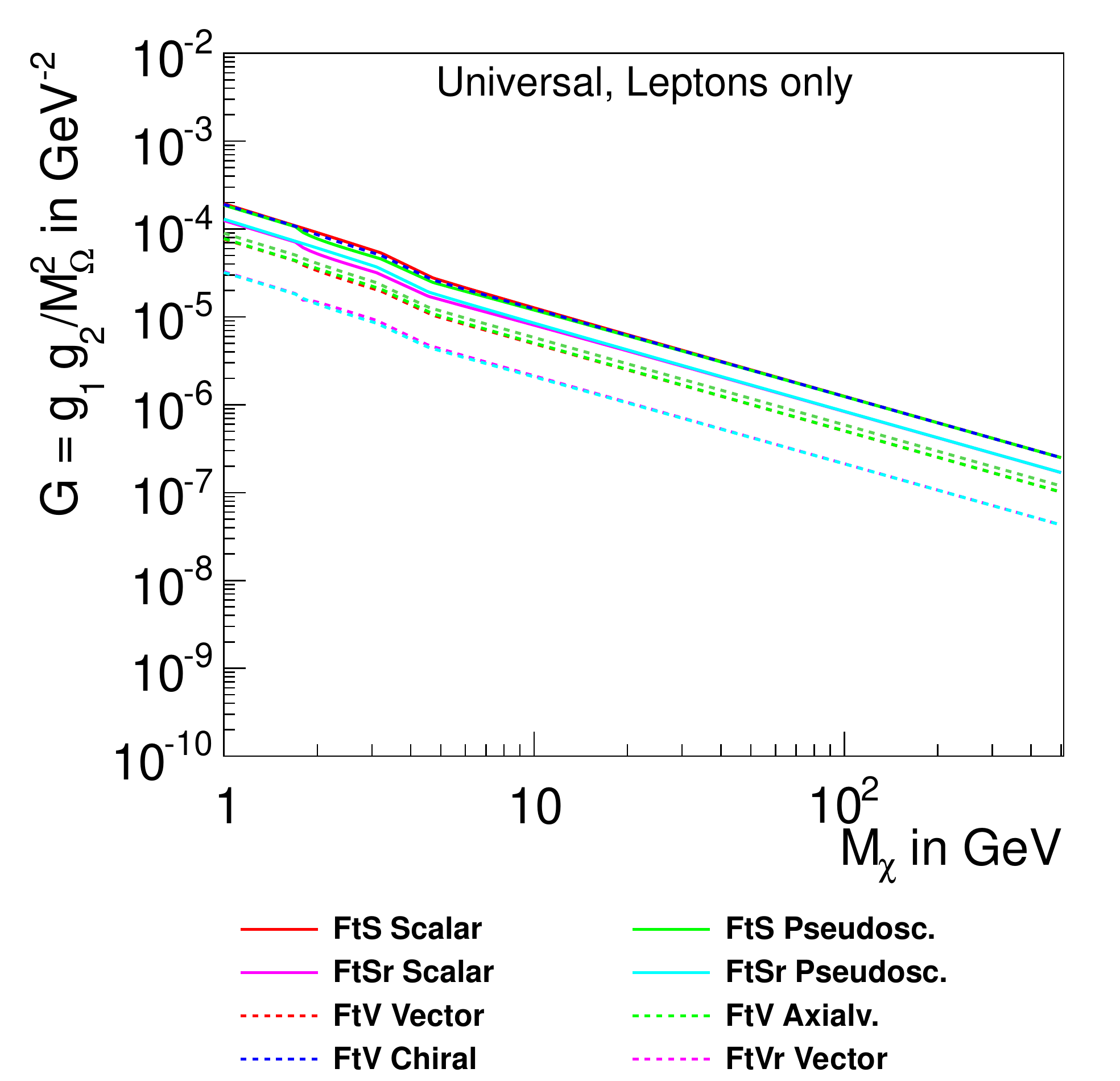} \\ \vspace{-0.25cm}
\includegraphics[width=0.49\textwidth]{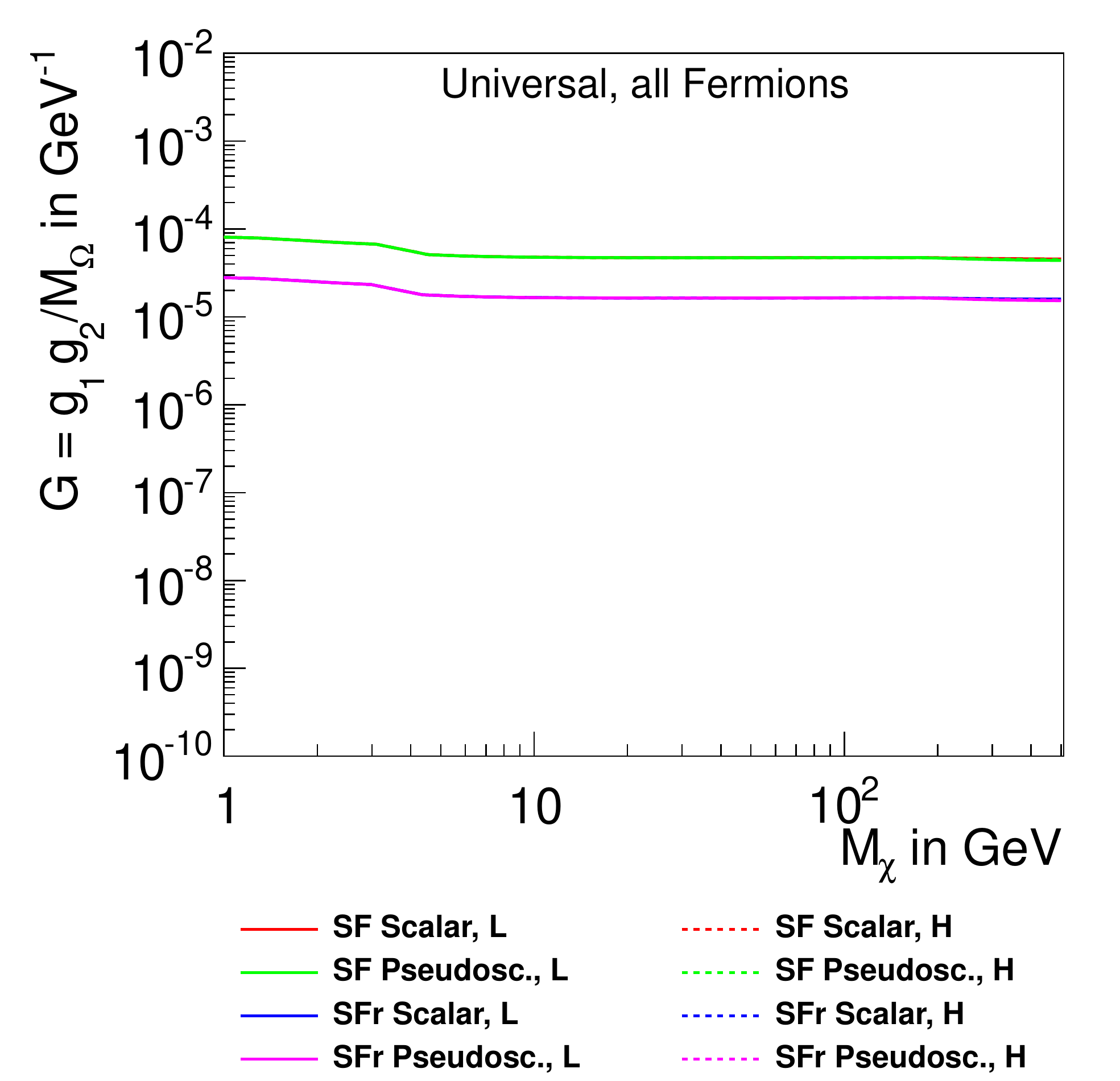}
\hfill
\includegraphics[width=0.49\textwidth]{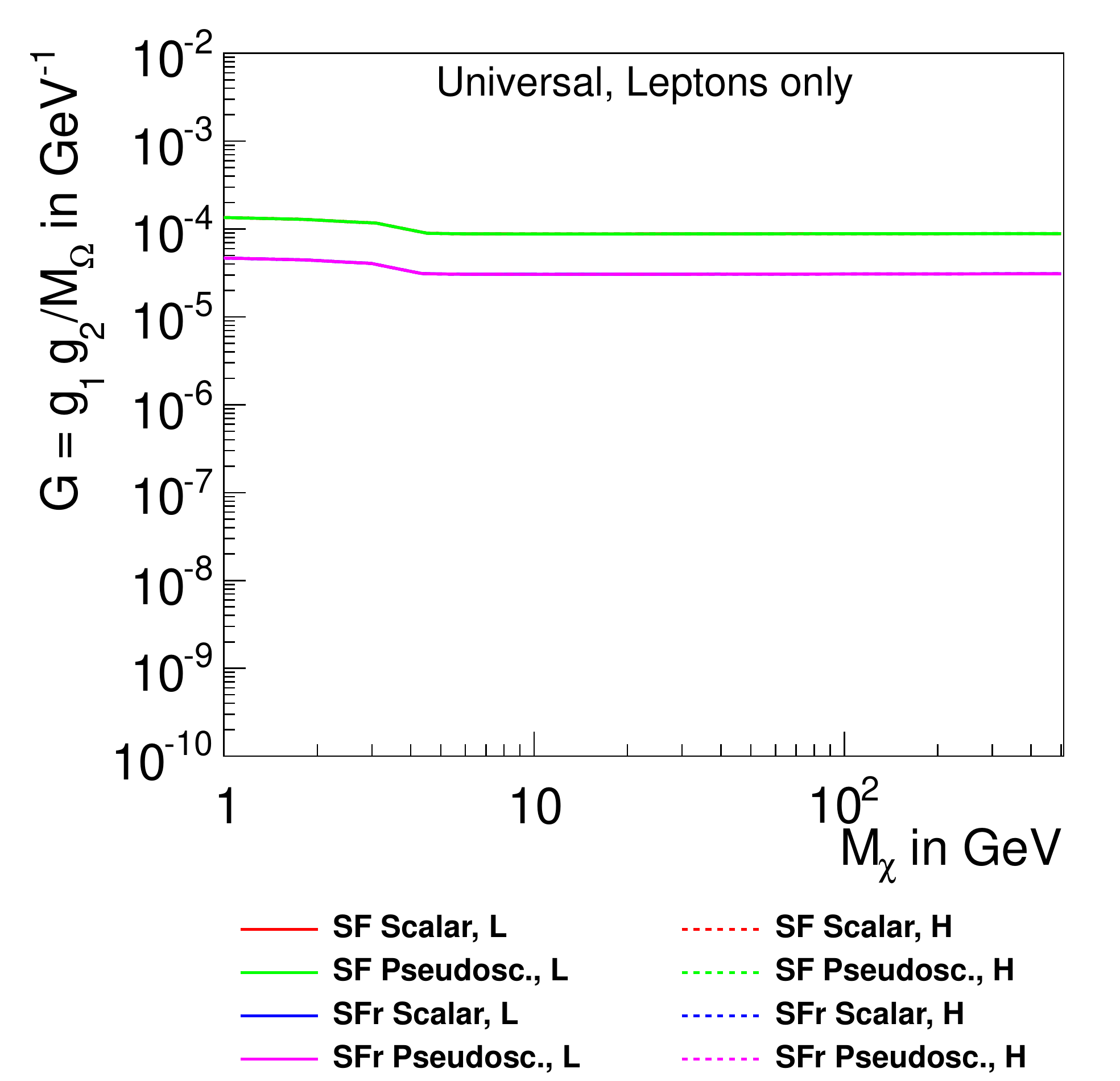} \\ \vspace{-0.25cm}
\includegraphics[width=0.49\textwidth]{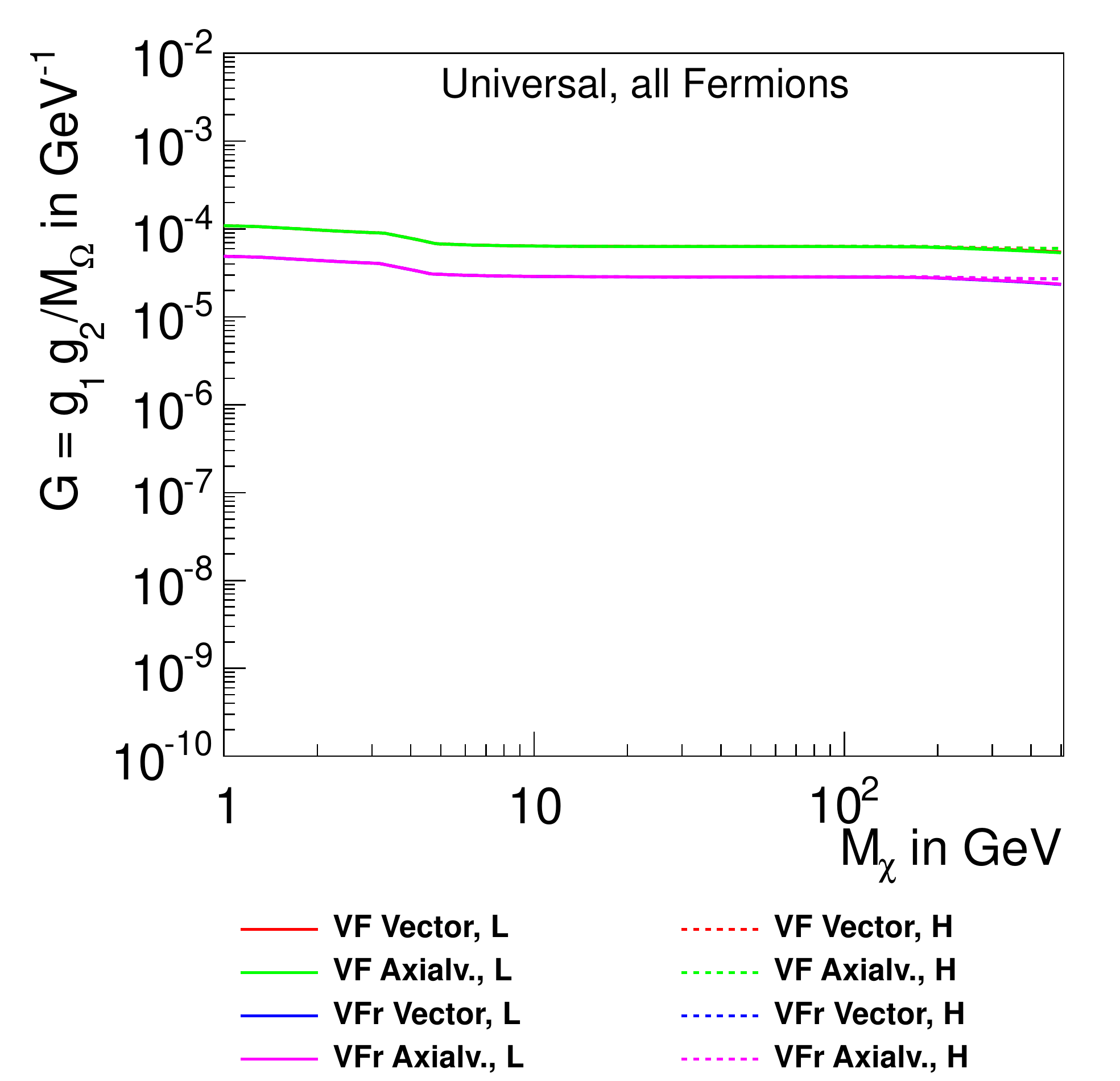}
\hfill
\includegraphics[width=0.49\textwidth]{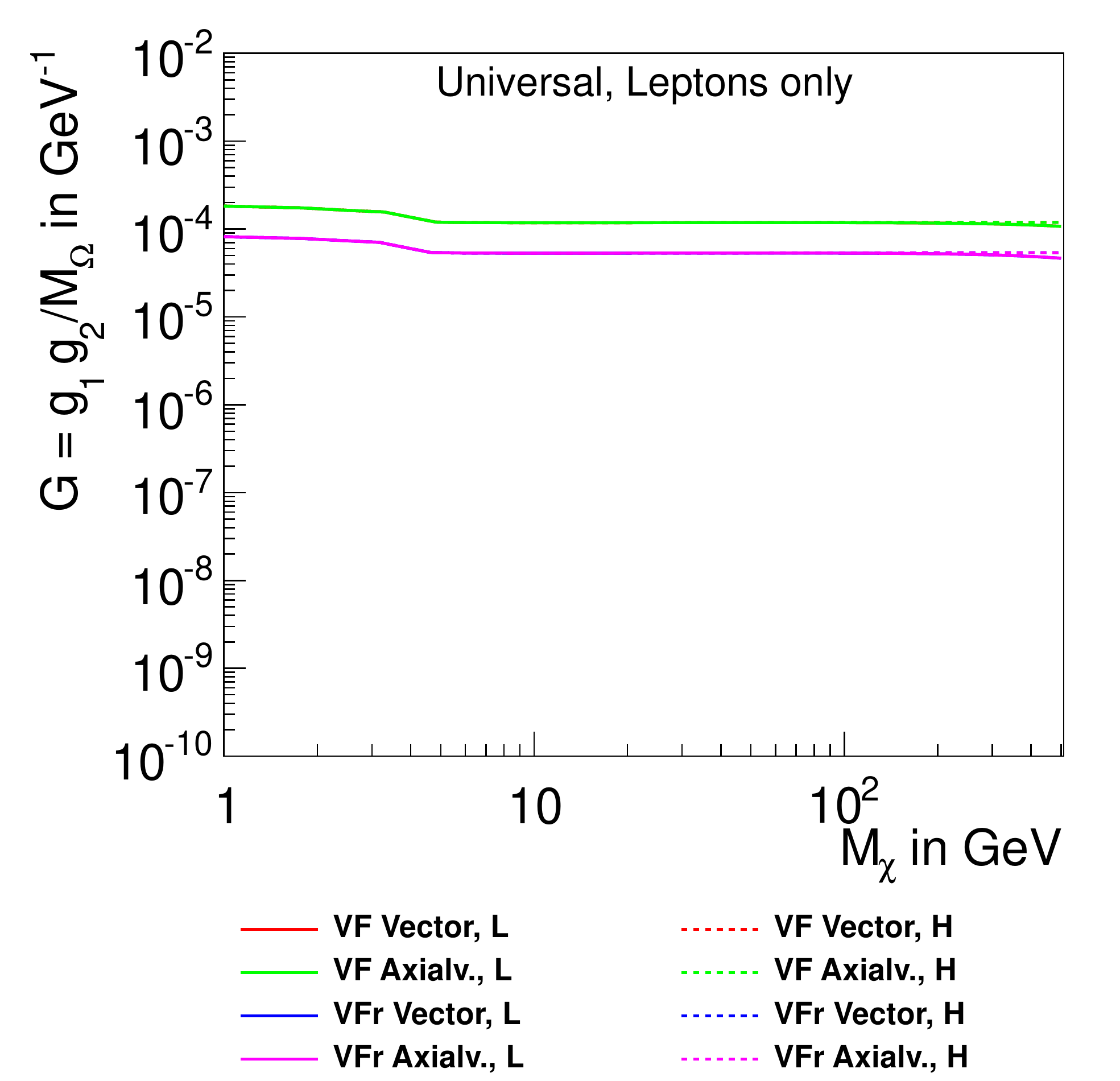} \\
 \vspace{-0.25cm}
\caption{Minimum couplings in agreement with \textsc{Wmap}. Models with t--channel mediators and universal couplings.}
\label{img:wmapresults3}
\end{figure}

\begin{figure}
\centering
\includegraphics[width=0.49\textwidth]{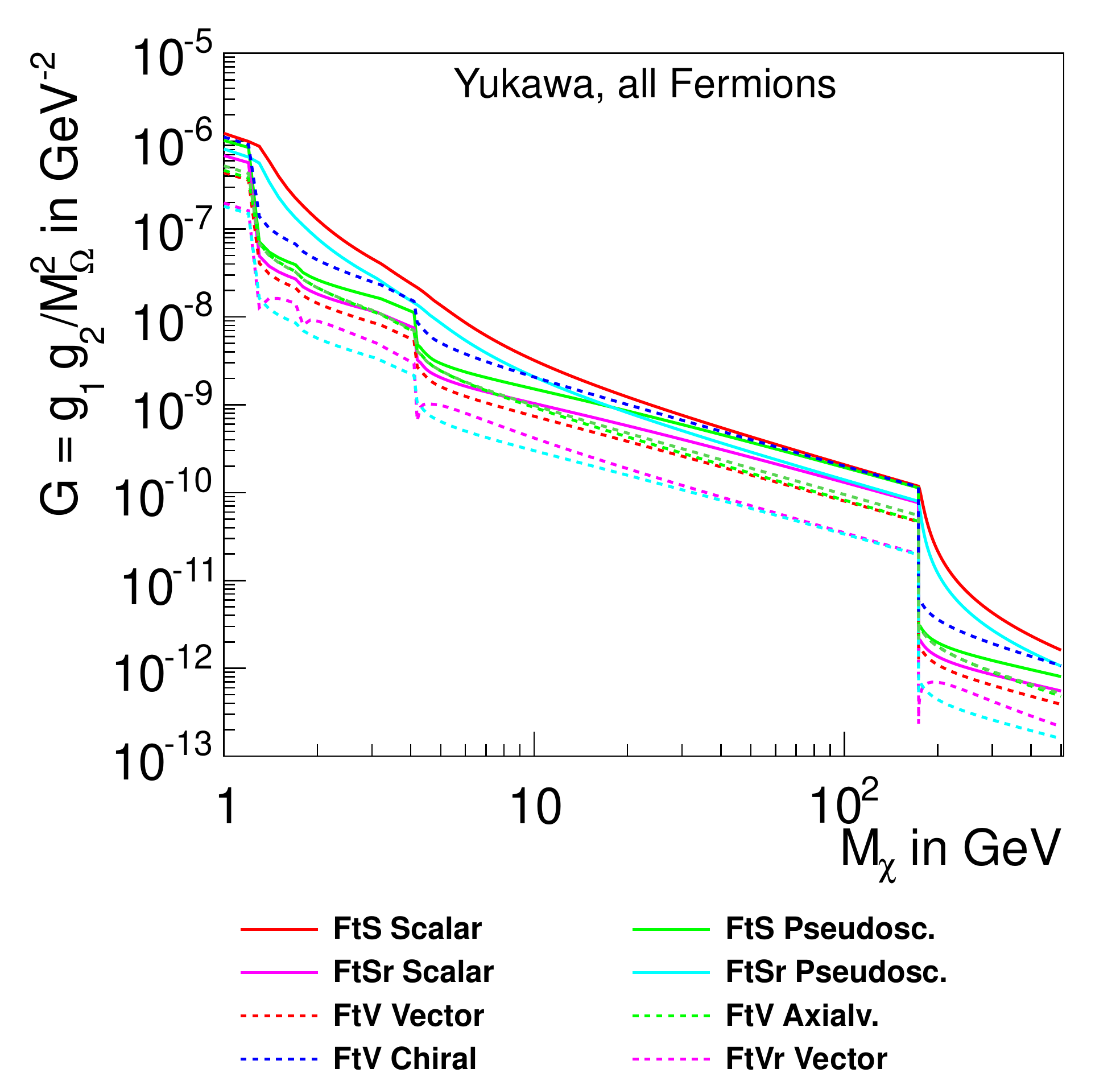}
\hfill
\includegraphics[width=0.49\textwidth]{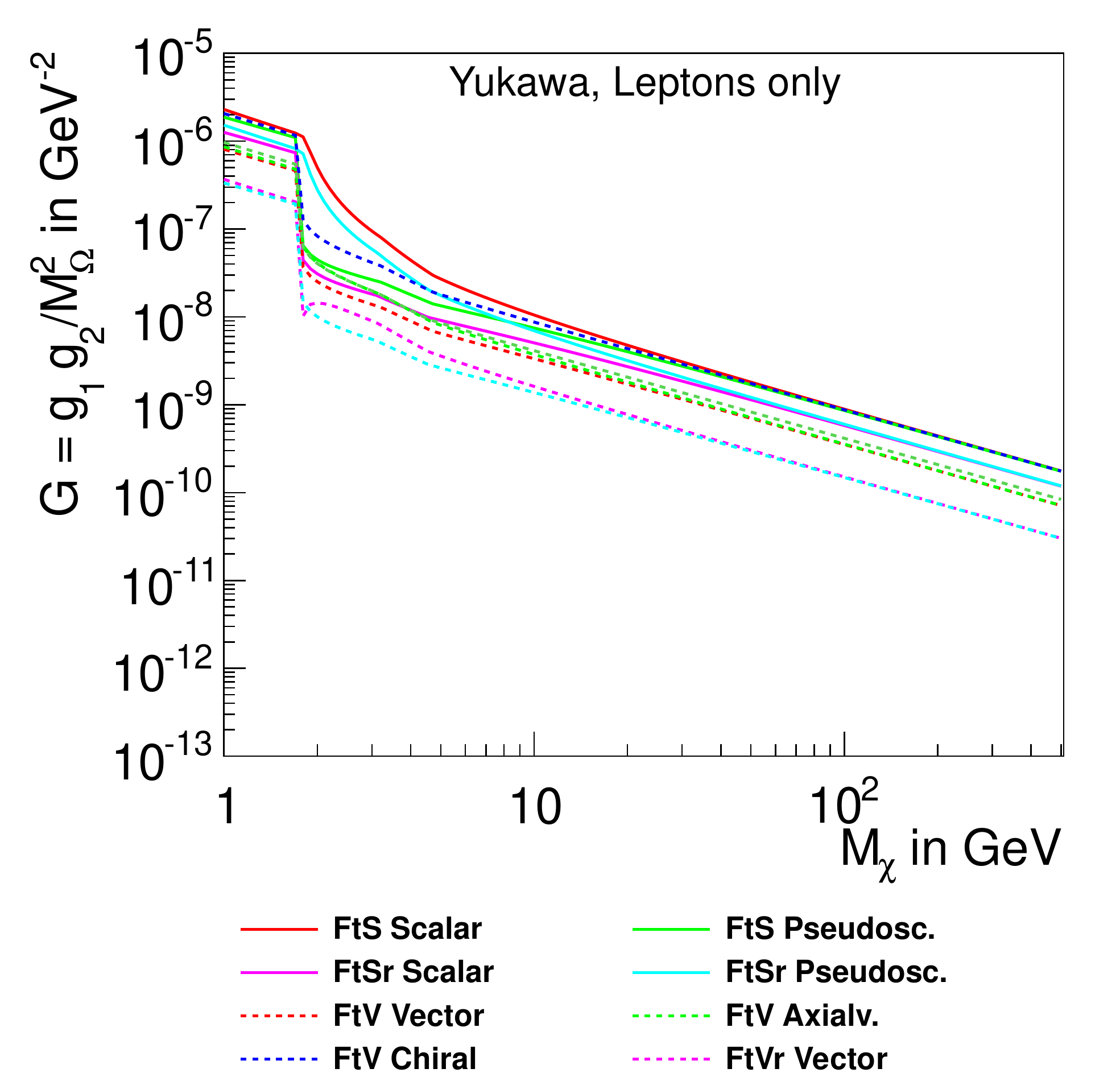} \\
\includegraphics[width=0.49\textwidth]{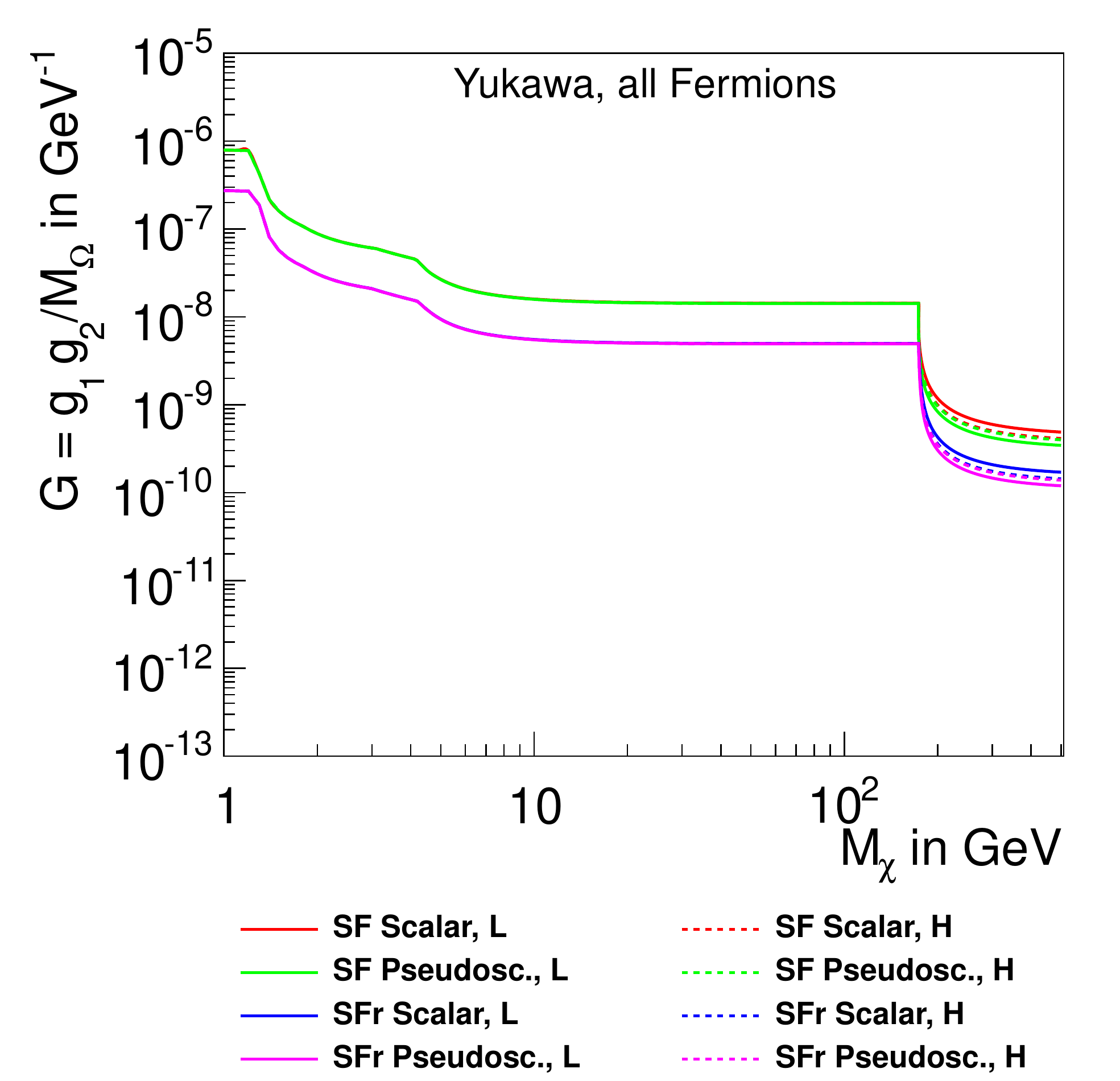}
\hfill
\includegraphics[width=0.49\textwidth]{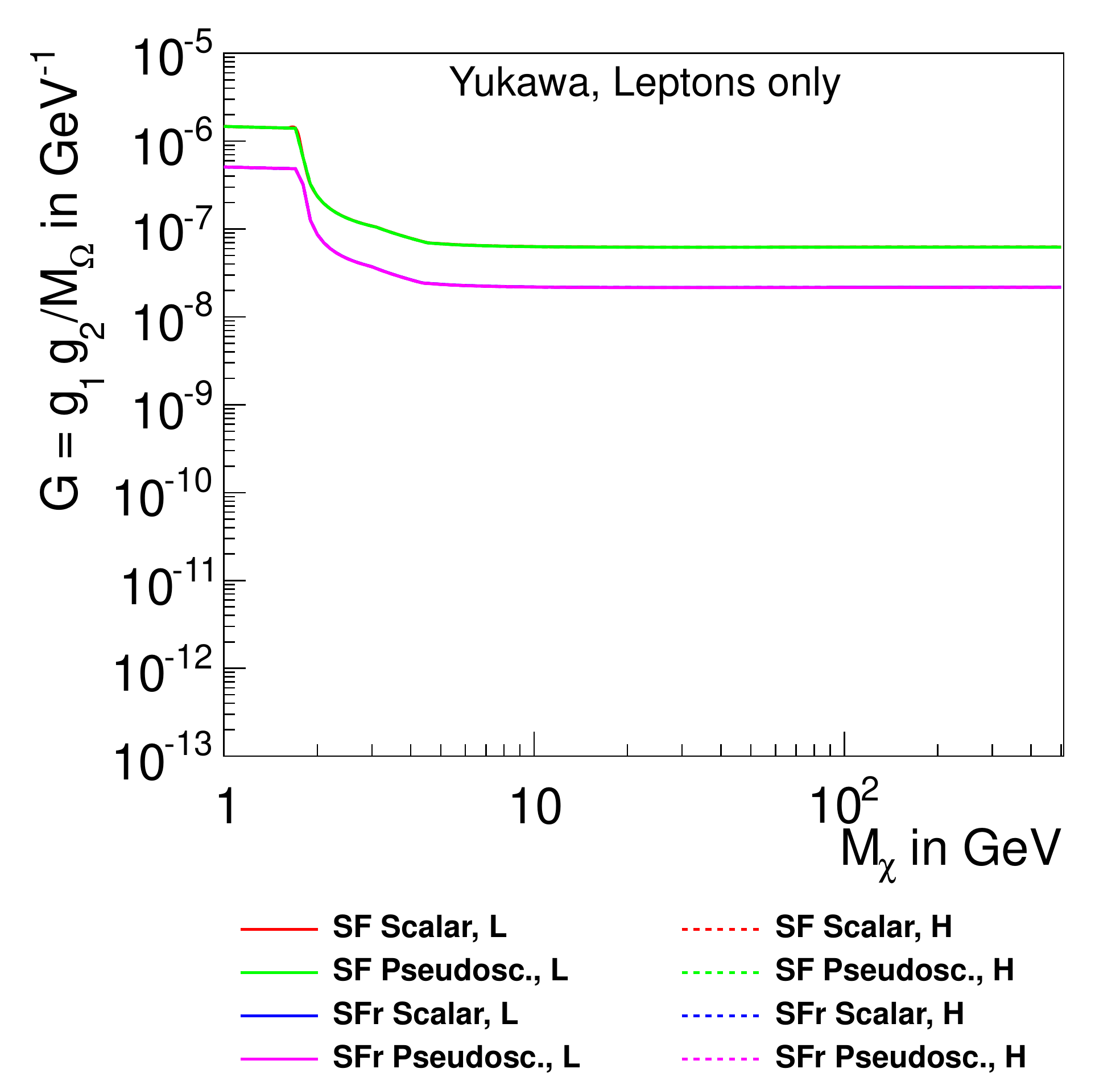} \\
\includegraphics[width=0.49\textwidth]{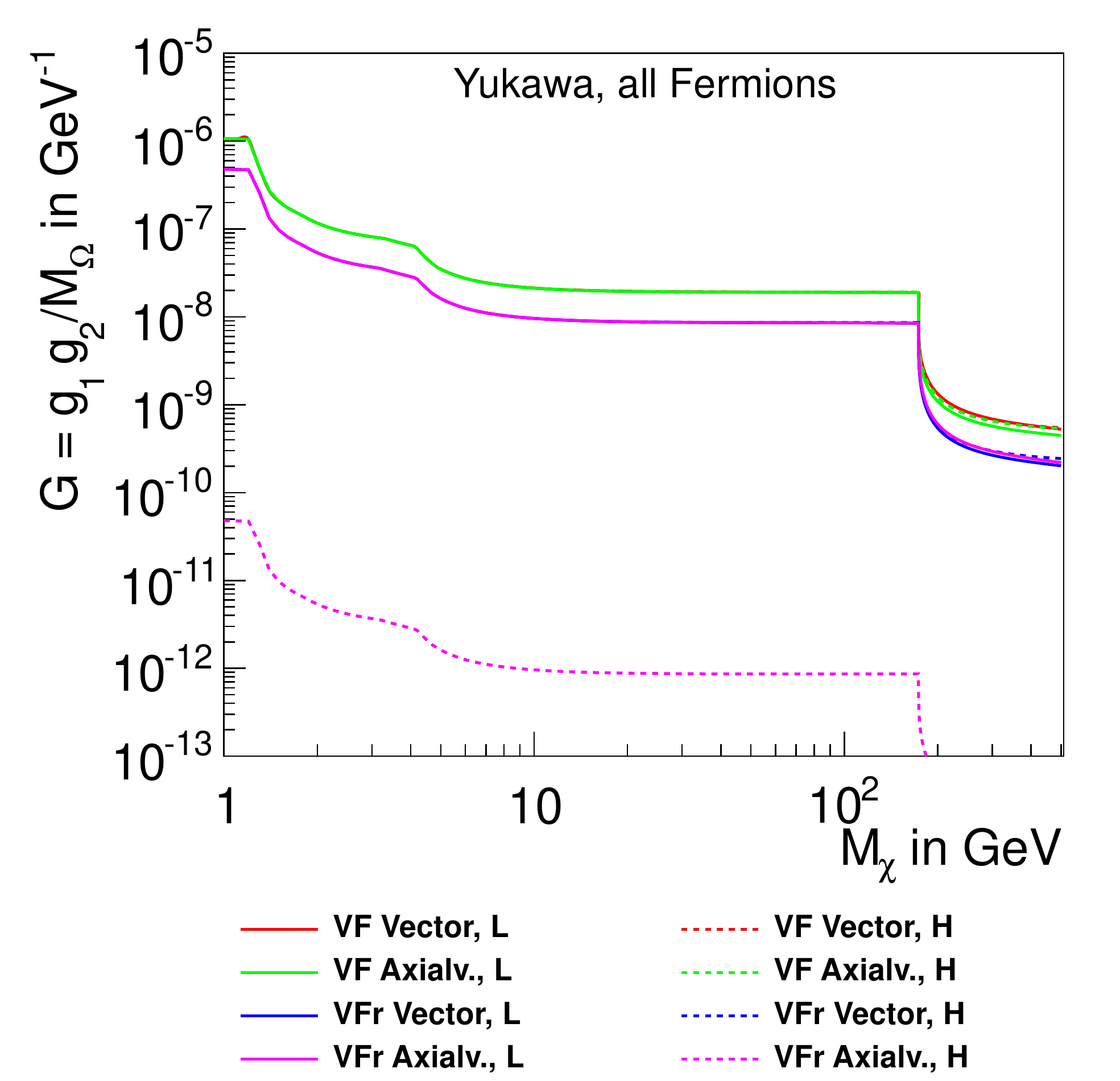}
\hfill
\includegraphics[width=0.49\textwidth]{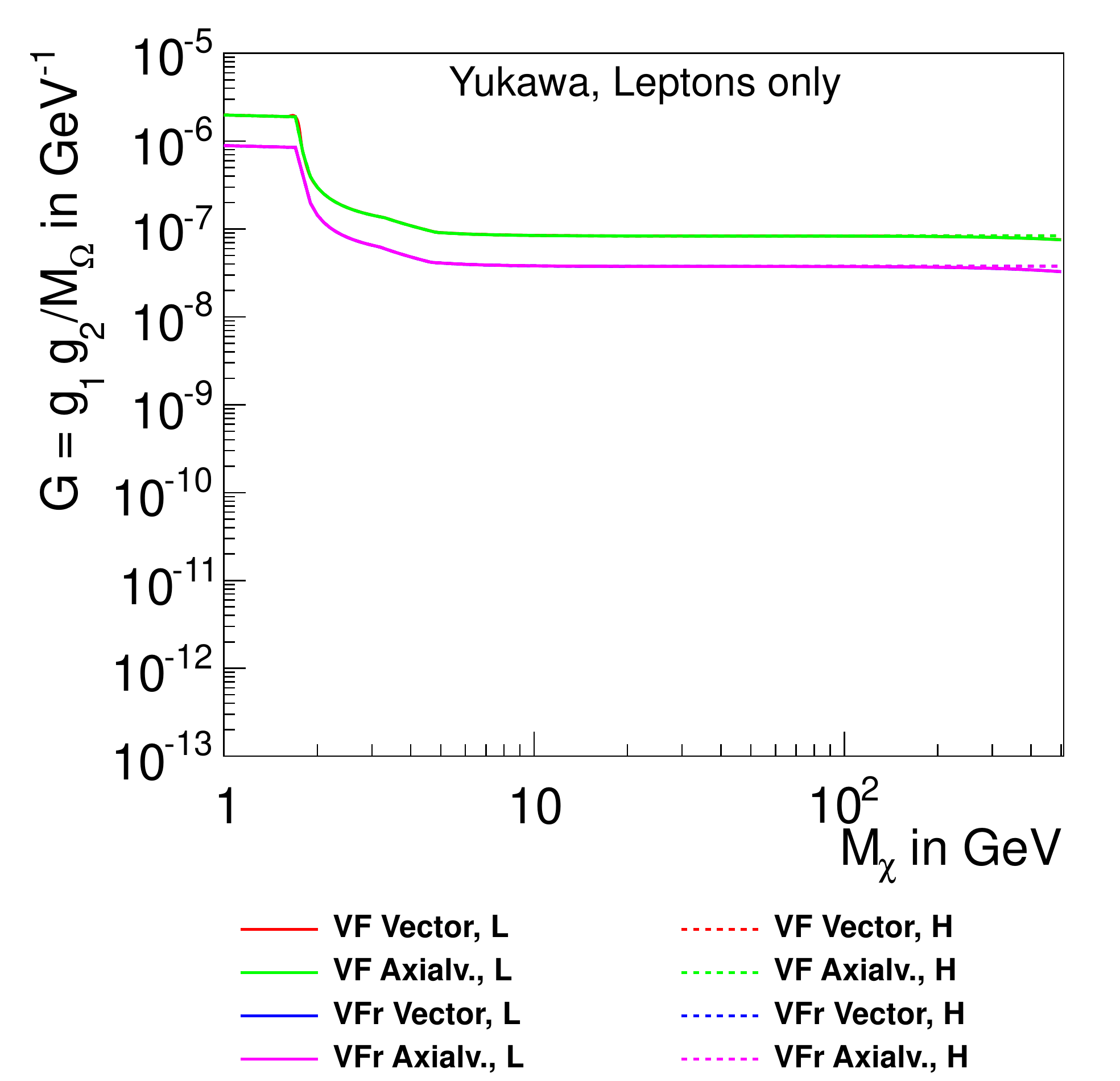} \\
\caption{Minimum couplings in agreement with \textsc{Wmap}. Models with t--channel mediators and Yukawa--like couplings.}
\label{img:wmapresults4}
\end{figure}

\chapter{Radiative Production of Dark Matter}
\label{chap:chichigamma}
\begin{figure}
\centering
\includegraphics[width=0.25\columnwidth]{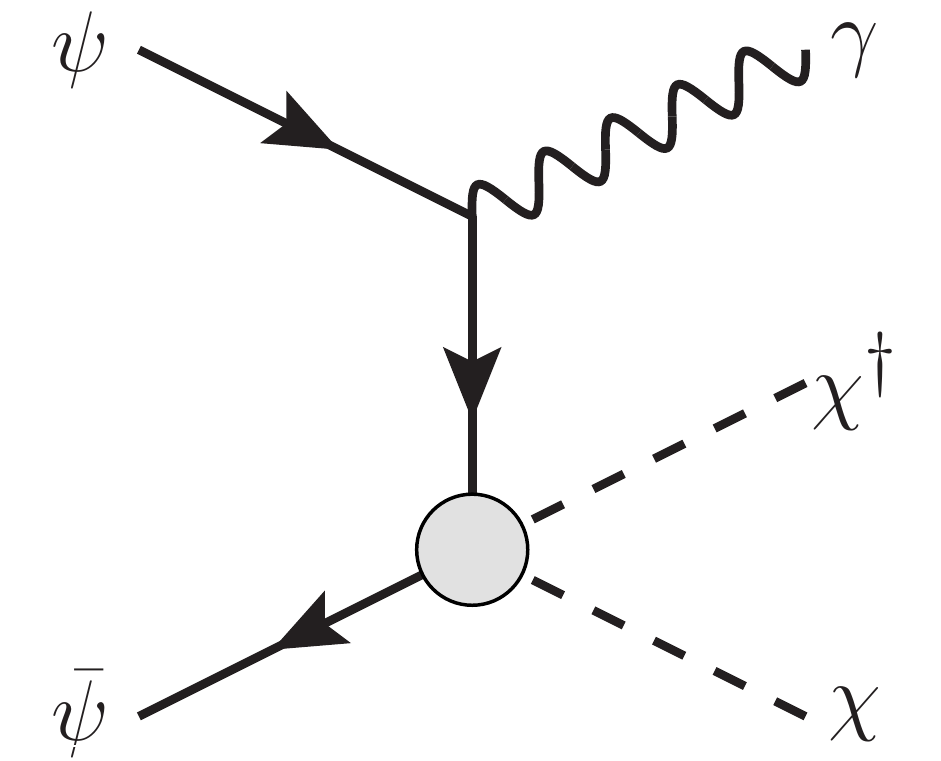}
\qquad \qquad
\includegraphics[width=0.25\columnwidth]{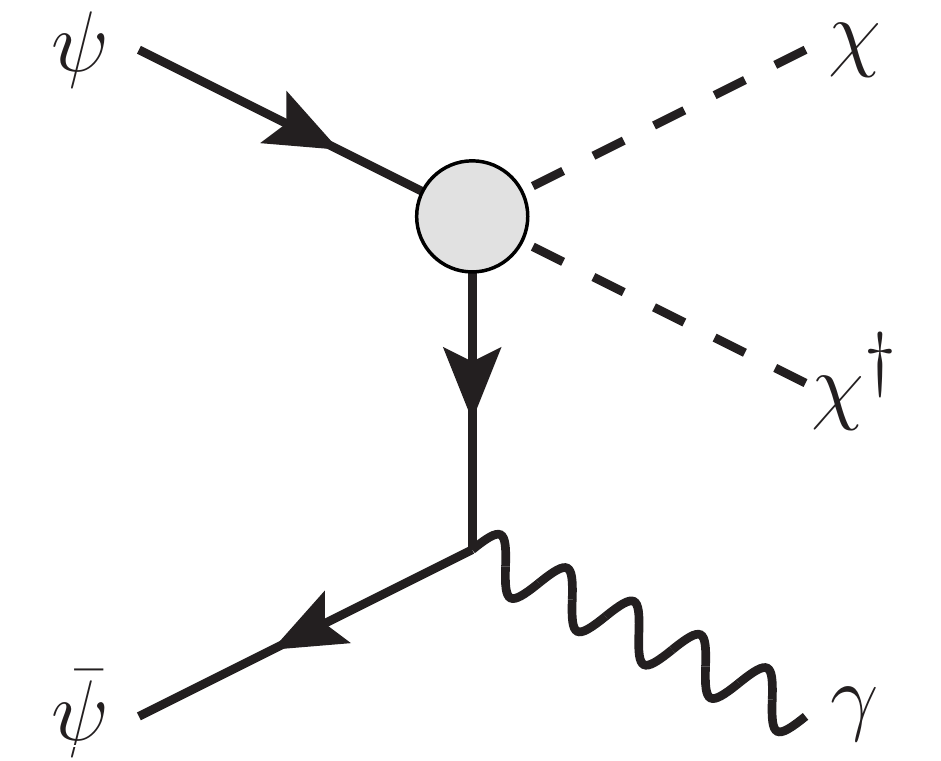}
\caption{Diagrams for radiative pair production of dark matter through effective four--particle--vertices.}
\label{fig:signalprocess_feynmandiagrams}
\end{figure}

Collider studies can analyse the pair production of dark matter particles. However, these particles only interact weakly and thus escape detection. At least one additional particle is needed to trigger the event selection and to give measurable observables. In this study we look at events with one additional hard photon in the final state radiated by one of the incoming leptons. 

For our effective 4--particle operators, this process is described by the two diagrams given in figure \ref{fig:signalprocess_feynmandiagrams}. In the following we discuss the calculation of the differential cross section for that process. In section \ref{sec:spinformalism} we give a short introduction to the spin formalism which is needed in this analysis because of the spin polarisation of the incoming particles at the \textsc{Ilc}. Section \ref{sec:dmphasespace} takes a short look at the parametrisation of the phase space of this particular $ 2 \rightarrow 3$ process. Besides the analytical calculation, we introduce a soft--photon--approximation in section \ref{sec:wwilliams} that is used in other analyses of effective \textsc{Wimp} theories. Finally we give the results of both the analytical and the approximative calculation and compare these in section \ref{sec:finalxsects}.


\section{Spin Formalism for Polarised Matrix Elements}
\label{sec:spinformalism}
We use spin density matrices to include the longitudinal spin polarisation $P^-$
and $P^+$ of both the
electron and positron in the initial state and follow the discussion in \cite{RadiativeNeutralinoILC}. In statistical quantum mechanics, the density matrix equals the sum of the projectors on all eigenspaces, weighted by the individual probability of the corresponding eigenstate. In our case, the states we project on are the unit spinors $ \begin{pmatrix} 1 \\ 0 \end{pmatrix}$ with probability $p_\text{up}$ and $\displaystyle \begin{pmatrix} 0 \\ 1 \end{pmatrix}$ with the respective complementary probability $1-p_\text{up}$:
\begin{align}
\rho = p_\text{up} \begin{pmatrix} 1 \\ 0 \end{pmatrix} \begin{pmatrix} 1 & 0 \end{pmatrix} + (1-p_\text{up}) \begin{pmatrix} 0 \\ 1 \end{pmatrix} \begin{pmatrix} 0 & 1 \end{pmatrix}.
\end{align}
The probability for the spin to point upwards\footnote{Note that the spin direction describes the \emph{helicity} state of a particle. This should not be confused with the \emph{chirality} of a particle, which is a statement about its behaviour under Poincaré--transformations. If the mass is negligible, chirality and helicity are identical for the electron but opposite for the positron. This is especially important to keep in mind for section 
\ref{sec:finalxsects} and after, where we classify interactions as \emph{left--/right-chiral}.} is assumed to be 0.5 in the case of no polarisation and shall vary between 0 and 1 for a polarisation between $-1$ and $+1$. We therefore replace $p_\text{up} = (1+P^\pm)/2$ with $P^\pm$ denoting the polarisation of the positron/electron and find
\begin{align}
\rho^\pm = \frac{1}{2} \begin{pmatrix} 1 & 0 \\ 0 & 1 \end{pmatrix} + \frac{P^\pm}{2} \begin{pmatrix} 1 & 0 \\ 0 & -1 \end{pmatrix}.
\end{align}
Replacing the two matrices by $\mathbb{1}$ and $\sigma_3$, we receive the following standard formulation for the polarisation density matrices of electrons and positrons in index formulation:
\begin{align}
\rho_{\lambda_-, \lambda_-^\prime} &= \frac{1}{2} \left( \delta_{\lambda_-,
    \lambda_-^\prime} + P^-   \sigma^3_{\lambda_-, \lambda_-^\prime}\right), \label{eqn:density1}\\
\rho_{\lambda_+, \lambda_+^\prime} &= \frac{1}{2} \left( \delta_{\lambda_+,
    \lambda_+^\prime} + P^+   \sigma^3_{\lambda_+, \lambda_+^\prime} \right). \label{eqn:density2}
\end{align}
After determining the helicity dependent matrix element, we calculate the full
averaged matrix element squared by contracting with the above density matrix:
\begin{align}
\left| \mathcal{M}\right|^2 = \sum_{\lambda_+, \lambda_+^\prime \lambda_-,
  \lambda_-^\prime}  \rho_{\lambda_+, \lambda_+^\prime}
\rho_{\lambda_-, \lambda_-^\prime} \sum_{i, j} \mathcal{M}_i^{\lambda_+  \lambda_-}
\mathcal{M}_j^{* \lambda_+^\prime \lambda_-^\prime}.
\end{align}
The helicity index will appear in the free spinor functions $u$ and $v$ of the
incoming leptons that come in combinations $u \bar{u}$ and $v
\bar{v}$. We also bring these into index form by using the
Bouchiat--Michel formulae for massless spinors \cite{SpinFormalism}
\begin{align}
u(p,\lambda_-) \bar{u}(p, \lambda_-^\prime) &= \frac{1}{2} \left(
  \delta_{\lambda_- \lambda_-^\prime} + \gamma^5 \sigma^3_{\lambda_-
    \lambda_-^\prime} \right) \slashed{p} + \ldots \label{eqn:uubar}\\
v(p,\lambda_+) \bar{v}(p, \lambda_+^\prime) &= \frac{1}{2} \left(
  \delta_{\lambda_+ \lambda_+^\prime} - \gamma^5 \sigma^3_{\lambda_+
    \lambda_+^\prime} \right) \slashed{p} +\ldots \label{eqn:vvbar}
\end{align}
Omitted additional terms have non--diagonal entries only and vanish when contracting with $\rho$. Contracting the helicity indices from the spinor bilinears (\ref{eqn:uubar}) + (\ref{eqn:vvbar}) with those from the density matrices (\ref{eqn:density1}) + (\ref{eqn:density2}) leads to the following substitution rule for polarised massless spinors:
\begin{align}
\sum_{\lambda_-, \lambda_-^\prime} \rho_{\lambda_-, \lambda_-^\prime} u(p,
\lambda_-) \bar{u}(p, \lambda_-^\prime) &= \frac{1}{2} \left( 1 - P^- \gamma^5
\right) \slashed{p}, \\
\sum_{\lambda_+, \lambda_+^\prime} \rho_{\lambda_+, \lambda_+^\prime} v(p,
\lambda_+) \bar{v}(p, \lambda_+^\prime) &= \frac{1}{2} \left( 1 + P^+ \gamma^5
\right) \slashed{p}.
\end{align}
These relations replace the usual rule $\sum_\text{pol} u(p) \bar{u}(p)  = \slashed{p}$ from unpolarised matrix elements and make it possible to evaluate the cross section with respect to the incoming polarisation.
\section{Dark Matter Phase Space Integration}
\label{sec:dmphasespace}
Since we do not measure the final state dark matter particles, we have to integrate over their total phase space in order to calculate the differential cross section for the photon with respect to its energy and angle. There are different possibilities to choose the coordinate system such that the phase space integral can be evaluated analytically. We use the center of mass system with total invariant mass $\sqrt{s}$ and choose the z--axis to coincide with the direction of the photon's momentum. Its angle $\theta_\gamma$ is measured with respect to the momentum of the incoming electron. In that case the 4--momenta in the process $\Pelectron(p) + \Ppositron(p^\prime) \rightarrow \chi(k) +
\bar{\chi}(k^\prime) + \Pphoton(q)$ are given as follows after applying energy-- and momentum conservation \cite{PhaseSpaceRef1, PhaseSpaceRef2, PhaseSpaceRef3, PhaseSpaceRef4}:

 \begin{align}
p &= \frac{\sqrt{s}}{2} \begin{pmatrix} 1 \\ \sin \theta_\gamma \\ 0 \\ \cos \theta_\gamma \end{pmatrix}, \quad p^\prime = \frac{\sqrt{s}}{2} \begin{pmatrix} 1 \\ - \sin \theta_\gamma \\ 0 \\ - \cos \theta_\gamma \end{pmatrix}, \quad q = E_\gamma \begin{pmatrix} 1 \\ 0 \\ 0 \\ 1 \end{pmatrix},\\ k &= \begin{pmatrix} E_\chi \\  p_\chi \sin \theta_\chi \cos \phi_\chi \\ p_\chi \sin   \theta_\chi \sin \phi_\chi \\ p_\chi \cos \theta_\chi \end{pmatrix}, \quad
k^\prime = \begin{pmatrix}\sqrt{s} - E_\gamma - E_\chi \\ -  p_\chi \sin \theta_\chi \cos
  \phi_\chi \\ - p_\chi \sin   \theta_\chi \sin \phi_\chi \\ -E_\gamma -p_\chi \cos
  \theta_\chi \end{pmatrix}, \\
p_\chi &\equiv \sqrt{E_\chi^2 - M_\chi^2}, \\
\theta_\chi &\equiv \frac{(\sqrt{s} - E_\gamma - E_\chi)^2 - E_\gamma^2 -
  E_\chi^2}{2 E_\gamma \sqrt{E_\chi^2-M_\chi^2}}.
\end{align}


Using the dimensionless quantity $x_\gamma \equiv 2 E_\gamma / \sqrt{s}$ for the
photon energy, we find the following phase space integral for the differential cross section:
\begin{align}
\frac{\mathrm{d} \sigma}{\mathrm{d} x_\gamma \ \mathrm{d} \cos \theta_\gamma} &= \frac{1}{512
  \pi^4 \sqrt{s}} \int_{E_-}^{E_+} \mathrm{d} E_\chi \ \int_0^{2 \pi}\mathrm{d}
\phi_\chi \ |\mathcal{M}|^2, \\
E_{\pm} &=  \frac{\sqrt{s}}{4}\left(2 -
x_\gamma \left[ 1 \pm \sqrt{1- \frac{4 M_\chi^2}{s(1-x_\gamma)}} \right]\right),
\end{align}
where $x_\gamma$ ranges from 0 to $1-4 M_\chi^2/s$. From now on we will omit the photon index and always refer to the photon's properties when using  $x$ and $\theta$. Note that an additional factor
of $1/2$ has to be added to the differential cross sections in the case of
real dark matter fields to take account of the indistinguishability of $k$ and $k^\prime$.

\section{Weizsäcker--Williams Approximation}
\label{sec:wwilliams}

Previous studies (e.g.\ \cite{BartelsList, WWapproach}) often worked in a model independent framework and therefore cannot evaluate an explicit matrix element to find the analytical differential cross section for the photon. They use the so called \emph{soft--photon--approximation} by Weizsäcker and Williams (WW) \cite{Peskin}, which factorises the properties of the photon in the process $\Ppositron \Pelectron \rightarrow \Pphoton \chi \chi$ into a kinematical function $F_{x \theta}$ and the total cross section $\hat{\sigma} \equiv \sigma(\hat{s})$ of the remaining interaction $\Ppositron \Pelectron \rightarrow \chi \chi$  with  reduced energy $\hat{s} \equiv s(1-x)$ according to
\allowdisplaybreaks
\begin{align}
\frac{\mathrm{d} \sigma \left[\Ppositron \Pelectron  \rightarrow \bar{\chi} \chi
    \gamma \right] }{\mathrm{d} x \ \mathrm{d}
   \cos \theta} &\approx F_{x \theta} \cdot  
 \hat{\sigma} \left[\Ppositron \Pelectron \rightarrow \bar{\chi} \chi \right], \label{eqn:wwapproximation}\\
F_{x \theta} &\equiv \frac{\alpha_\text{em}}{\pi} \frac{(x-1)^2+1}{x \sin^2 \theta}.
\end{align}
Here, $\alpha_\text{em}$ denotes the electromagnetic fine structure constant. The kinematics of $F_{x \theta}$ are universally predicted by the \textsc{Qed} structure of collinear photon radiation and are independent of the physics appearing in $\hat{\sigma}$. It is a valid approximation for small\footnote{Obvious divergences for $x, \theta \rightarrow 0$ arise from typical collinear divergences in \textsc{Qed}, which are only solved through resummation and consideration of next--to--leading order diagrams. They are avoided by experimentally required minimum values for those kinematical parameters for hard photons.} energies $x$ and angles $\theta$, but ignores any interference terms between different diagrams. 

\section{Cross Sections}
\label{sec:finalxsects}
We determine the polarised differential cross section both analytically and in the WW approximation. The results for this calculation are
given in table \ref{tbl:2t3crosssections}. For a compact representation, we use the following abbreviations:\\
\noindent
Polarisation Factors
\begin{align}
C_S \equiv 1+P^+ P^-&, \qquad C_L \equiv (1- P^-) ( 1+P^+), \\
C_V \equiv 1-P^+ P^- &, \qquad  C_R \equiv (1+ P^-) ( 1-P^+).
\intertext{Combined Coupling Constants}
G_{X \pm Y} \equiv g_{X}^2 \pm g_{Y}^2&,\qquad G_{XY} \equiv g_{X} g_{Y}.
\intertext{Velocity Functions}
\beta \equiv \sqrt{\displaystyle 1-\frac{4 M_\chi^2}{s}}&,\qquad \hat{\beta} \equiv \sqrt{\displaystyle 1-\frac{4 M_\chi^2}{\hat{s}}}. 
\intertext{Photon Kinematics:}
F_{x \theta} \equiv \frac{\alpha_\text{em}}{\pi} \frac{(x-1)^2+1}{x \sin^2 \theta}&, \qquad
V_{x \theta}  \equiv  \frac{x^2\cos(2 \theta) + (3x -8)x   + 8 }{4 \left((x-1)^2 +
    1\right)}.
\end{align}
\begin{table}
\centering
\renewcommand{\arraystretch}{1.5}
\begin{tabular}{r@{\quad}l}
\toprule
Model & $ \displaystyle \frac{ \mathrm{d}\sigma}{ \mathrm{d} x \ \mathrm{d}\cos \theta} $\\
\midrule
\midrule
SS & $ \displaystyle \frac{\hat{\beta}  F_{x \theta}}{32 \pi M_\Omega^4}
G_{s+p}  g_\chi^2 C_s $\\
SF & $\displaystyle \frac{  \hat{\beta}  F_{x \theta}}{32 \pi M_\Omega^2} 
\left[  G_{s-p}^2 C_s  + \frac{\hat{\beta}^2 \hat{s}}{12 M_\Omega^2} 
    \boldsymbol{ V_{x \theta}} \left[(g_s+g_a)^4 C_R + (g_s-g_a)^4 C_L \right]
    + \boldsymbol{A_{SF}} \right] $ \\
SFr & $\displaystyle \frac{  \hat{\beta}}{16 \pi M_\Omega^2} \left[ F_{x \theta}
G_{s-p}^2 C_s  + \boldsymbol{A_{SFr}}\right]$ \\
SV & $ \displaystyle \frac{\hat{s} \hat{\beta}^3 F_{x \theta}}{96 \pi M_\Omega^4}  \boldsymbol{V_{x \theta}} 
 \left[ g_l^2C_l + g_r^2C_r \right]  g_\chi^2 $ \\
\midrule
\midrule
FS &  $ \displaystyle \frac{\hat{s} \hat{\beta}  F_{x \theta}}{16 \pi M_\Omega^4} 
G_{s+p}  C_s \left[ g_{\chi s}^2 \hat{\beta^2} + g_{\chi p}^2 \right] $ \\
FV & $ \displaystyle \frac{ \hat{\beta} F_{x \theta} }{48 \pi
  M_\Omega^4}  \boldsymbol{V_{x \theta}} \left[
G_{l+r} \hat{s} \hat{\beta}^2 + 3 \left(g_l + g_r \right)^2 M_\chi^2 \right] \left[ g_l^2C_l + g_r^2C_r \right] 
$ \\
FVr & $ \displaystyle \frac{ \hat{s} \hat{\beta^3} F_{x \theta} }{48 \pi
  M_\Omega^4}  \boldsymbol{V_{x \theta}} \left(g_l - g_r \right)^2 \left[ g_l^2C_l + g_r^2C_r \right] $\\
FtS &  $\displaystyle \frac{  F_{x \theta} \hat{\beta}}{48 \pi M_\Omega^4} G_{s+p}^2
\left[\boldsymbol{V_{x \theta}}(\hat{s}-M_\chi^2) + \boldsymbol{A_{FtS}} \right]$ \\
FtSr & $\displaystyle \frac{ \hat{\beta} F_{x \theta}}{192 \pi M_\Omega^4}
G_{s+p}^2 \left[ 3 (\hat{s}-2 M_\chi^2) C_P + \boldsymbol{V_{x \theta}} 2 (\hat{s}
  - 4 M_\chi^2) C_V \right] $\\
FtV & $\displaystyle \frac{\hat{\beta} F_{x \theta}}{48 \pi M_\Omega^4} \left[ 
  6 G_{lr}^2 C_s
  (\hat{s} - 2 M_\chi^2 ) +  (\hat{s} - M_\chi^2) \boldsymbol{V_{x \theta}} (g_l^4 C_L + g_r^4 C_R) \right] $ \\
FtVr & $\displaystyle \frac{\hat{\beta} F_{x \theta}}{48 \pi M_\Omega^4} \left[ 
  12 G_{lr}^2 C_s   (\hat{s} - 2 M_\chi^2 ) + (\hat{s} - 4 M_\chi^2) \boldsymbol{V_{x \theta}} (g_l^4 C_L + g_r^4 C_R) \right] $\\
\midrule
\midrule
VS &  $ \displaystyle \frac{ \hat{\beta} F_{x \theta}} {128 \pi M_\chi^4 M_\Omega^4} 
G_{s+p}  g_\chi^2 C_s (12 M_\chi^4-4M_\chi^2\hat{s}+\hat{s}^2) $ \\
VF & $\displaystyle \frac{\hat{\beta}F_{x \theta}}{3840 \pi M_\chi^4 M_\Omega^2} 
\Big[\frac{1}{M_\Omega^2}\left(g_l^4C_l+g_r^4C_r\right)
  (40 M_\chi^6-22M_\chi^4\hat{s}+56 M_\chi^2 \hat{s}^2 + 3 \hat{s}^3) +  $\\

& $\displaystyle  40 G_{lr}^2 C_s (7 M_\chi^4 - 2 M_\chi^2 \hat{s} + \hat{s}^2)   + \boldsymbol{A_{VF}}\Big]$\\
VFr &  $\displaystyle \frac{\hat{\beta}F_{x \theta}}{3840 \pi M_\chi^4 M_\Omega^2} 
\Big[\frac{1}{M_\Omega^2}\left(g_l^4C_l+g_r^4C_r\right)
  (320 M_\chi^6-104^4\hat{s}+32 M_\chi^2 \hat{s}^2 + \hat{s}^3) +  $\\

& $\displaystyle  60 G_{lr}^2 C_s (12 M_\chi^4 - 4 M_\chi^2 \hat{s} + \hat{s}^2)   + \boldsymbol{A_{VFr}}\Big]$\\
VV & $ \displaystyle \frac{ \hat{s} \hat{\beta}^3 F_{x \theta} \boldsymbol{V_{x \theta}}}{384 \pi M_\chi^4 M_\Omega^4}\left[ g_l^2C_l + g_r^2C_r \right]  g_\chi^2 
( M_\chi^4 + 20 M_\chi^2 \hat{s} + \hat{s}^2)$ \\
\bottomrule
\end{tabular}
\caption{Analytical differential cross sections for the process $\Ppositron
  \Pelectron \rightarrow \chi \chi \gamma$ in the various effective
  models. Parts in bold do not appear in the Weizsaecker--Williams
  approach and are given in appendix \ref{app:2t3appendix}. Cross sections for SSr FSr and VSr are twice as large as in the complex case, SV and VV vanish completely for real particles.}.
\label{tbl:2t3crosssections}
\end{table}
\begin{figure}
\centering
\includegraphics[width=0.5\textwidth]{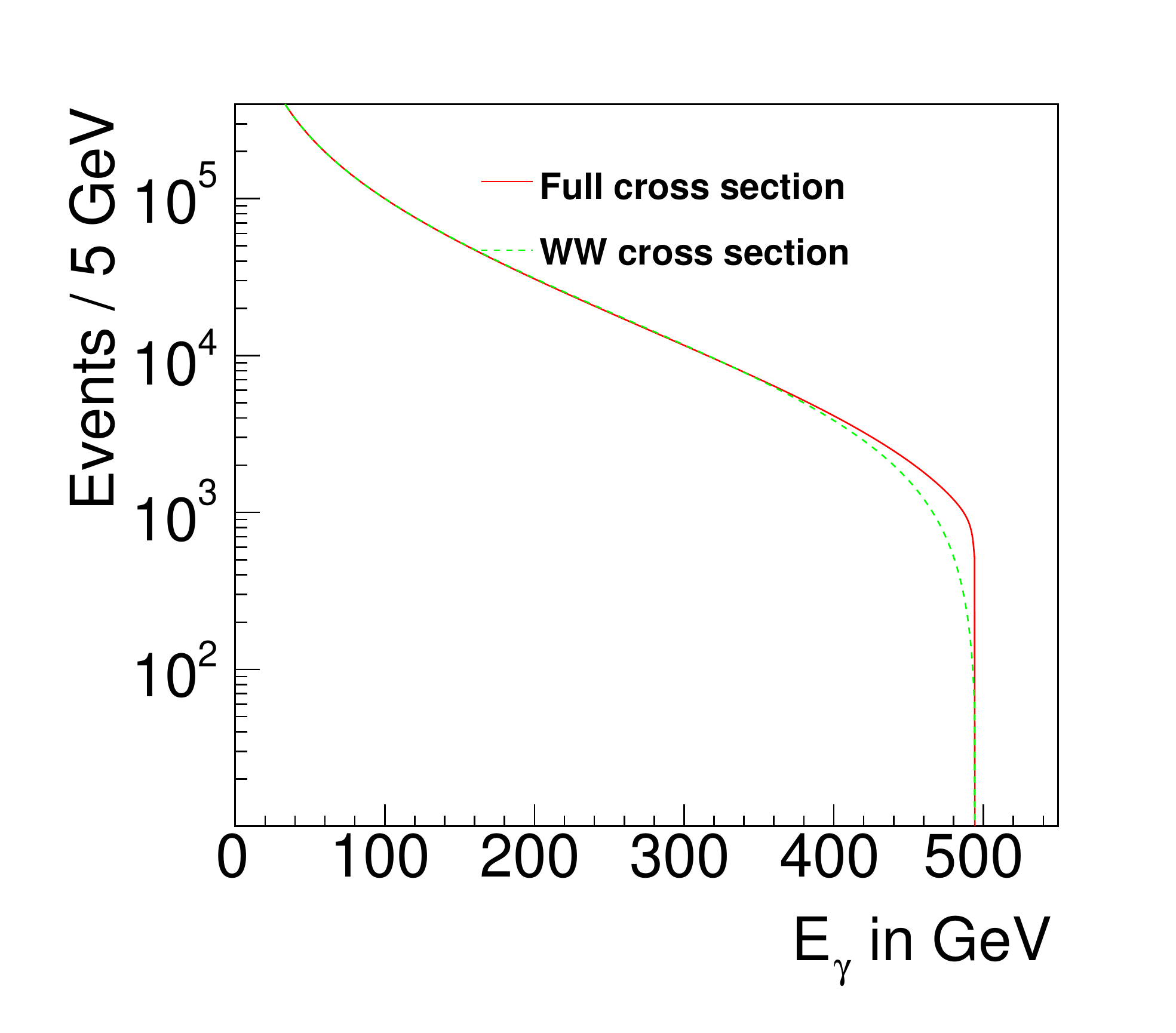} \hspace{-0.7cm}
\includegraphics[width=0.5\textwidth]{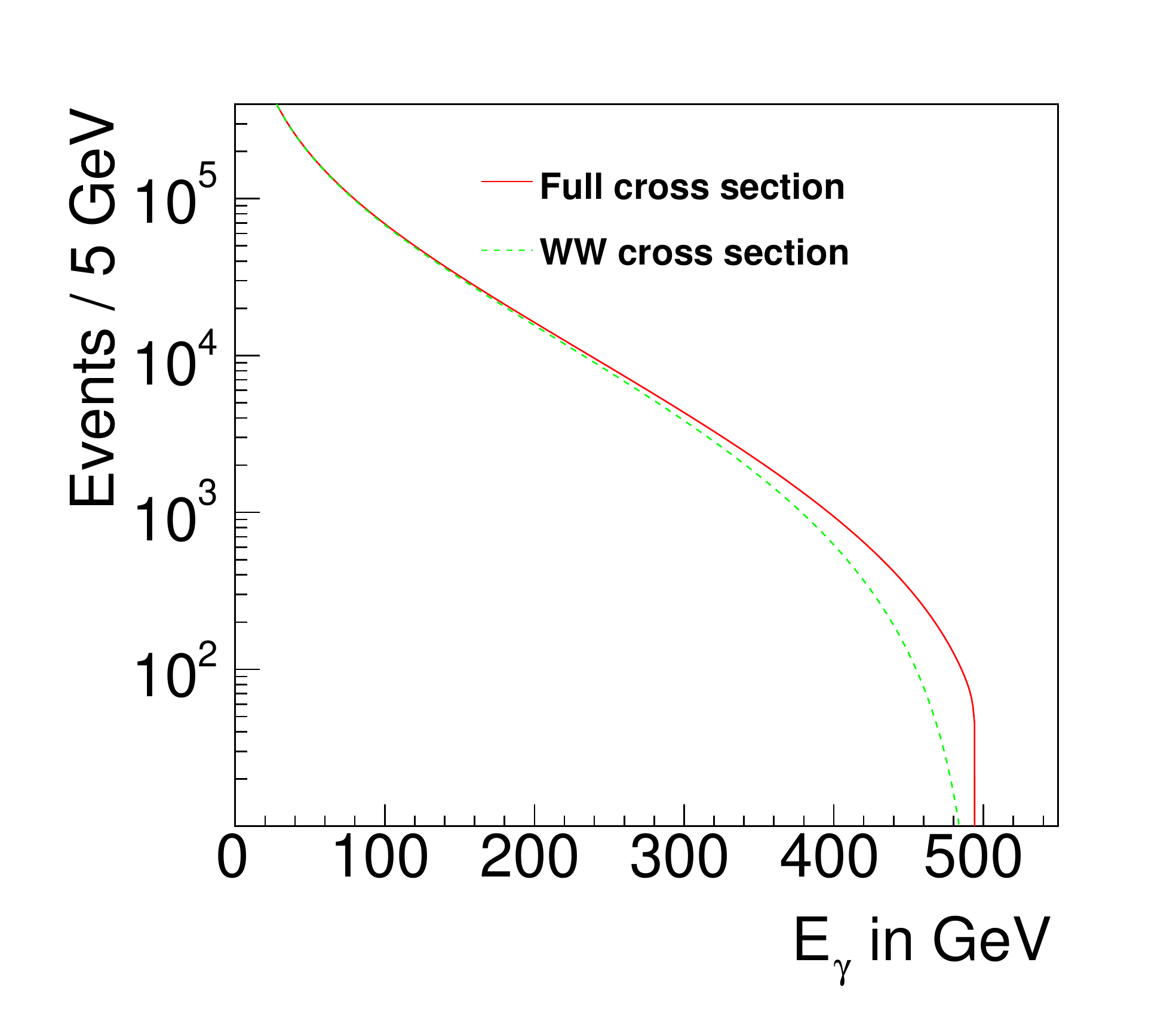}
\begin{flushleft}
\vspace{-0.85cm}
\hspace{0.1\textwidth}  (a) \hspace{0.44\textwidth} (b)
\end{flushleft}
\vspace{-0.35cm}
\caption{Comparision of tree level photon energy distributions in the
  WW--approximation and the analytical solution for $M_\chi = \unit{50}{\GeV}$, $|\cos \theta|_\text{max} = 0.98$ and $\sqrt{s} = \SI{1}{\TeV}$. (a) FtS, (b) VF
}
\label{img:comparison}
\end{figure}
We put parts in bold
if they are purely caused by the analytical calculation and do not appear in the soft photon approximation. Those corrections are either of the form of an additional
kinematical factor $V_{x \theta}$, mostly appearing in models with vector mediators, or completely new terms $A_i$ that typically appear in t--channel interactions. We list the latter in appendix \ref{app:2t3appendix}. They are generally expected in interactions with p--wave contributions, since then the spin configuration of all three final state particles is of importance to determine the orbital angular momentum. The impact of the spin needs a proper evaluation of the spinor structure and cannot be taken into account by the factorisation procedure. 

Since $\lim_{x \rightarrow 0} V_{x \theta} = 1$
and $\lim_{x \rightarrow 0} (A_i) = 0$, the
WW approximation is in agreement with our full result for small
energies, as predicted by the general derivation. In figure \ref{img:comparison} we compare the respective photon energy
distributions for two example models in both cases. The curves behave quite congruently with differences visible in the high energy sector. Since most of the signal events lie in the low energy region, using the approximation still gives accurate results for counting experiments. Shape dependent analyses would have to use the full analytical result to find the correct behaviour at the high energy threshold. Our further analysis is performed using the full analytical cross section.

In the general representation given in table \ref{tbl:2t3crosssections}, coupling constants appear with different polarisation dependent terms. For most of the benchmark models chosen in table \ref{tbl:constraints}, the
final cross section depends only on one factor $C_i$. To determine the polarisation for the best signal to background ratio,
we can do a combined analysis of all models with the same $C_i$, since they receive the same enhancement factor for equal polarisation settings.  We therefore
classify our models as follows:
\begin{align}
\text{scalar like}: \sigma_{\text{pol}} &= C_S \sigma_{\text{unpol}}, \label{eqn:polclasses} \\
 \text{vector like}: \sigma_{\text{pol}} &= C_V \sigma_{\text{unpol}}  \\
\text{right like}: \sigma_{\text{pol}} &= C_R \sigma_{\text{unpol}}, \\
 \text{left like}: \sigma_{\text{pol}} &= C_L \sigma_{\text{unpol}}. \label{eqn:polclassesleft}
\end{align}
Models with t--channel mediators usually have
multiple terms with different polarisation behaviour and do not fall into one
of the three basic polarisation classes given in (\ref{eqn:polclasses}). Models with fermionic mediators are classified according to their
  leading term, which is always scalar like. All other models have both scalar like and vector like parts of roughly
  the same size. We analyse them in a vector like scenario, since they naturally lead to a
  stronger background suppression, as will be shown in the following chapter.





\chapter{Dark Matter Search at the \textsc{Ilc}}
\label{chap:ilcanalysis}
Now that we know the analytical form of the photon energy distributions, we can continue with the analysis of the expected exclusion limits the \textsc{Ilc} can set. In Section \ref{sec:collider} we briefly summarise the history of collider searches for dark matter and their current state. We follow with a short introduction to the International Linear Collider and its expected advantages with respect to previous colliders in section \ref{sec:ilc}. 

Our simulation analysis is structured as follows: In section \ref{sec:smbackground} we look at the dominant Standard Model background contributions to single photon final states. We show in section \ref{sec:datamodelling} how we generate events and take into account beam and detector resolution effects. Our intention is to evaluate the total number of background events $N_\text{B}$ and the total error $\Delta N_\text{B}$ on that number. Under the assumption that we measure no signal events, we can exclude couplings that lead to a larger number of signal events $N_\text{S}$ than the total background uncertainty. On top of statistical fluctuations, sources of systematic errors are of great importance for this estimate. They are discussed in section \ref{sec:sysuncertainties} with their final impact on the result given in section \ref{sec:polarisation}. There we also discuss the optimum polarisation setting for the incoming leptons to maximise the exclusion power. After a short discussion about changes within this analysis for an increased center of mass energy of \SI{1}{\TeV} in section \ref{sec:1tev}, we close with our results in section \ref{sec:ilcresults}.

\section{Previous Collider Searches for \textsc{Wimp}s}
\label{sec:collider}
Analyses looking for dark matter through final state photons were performed first  in an \textsc{Ilc} scenario by using a model independent approach \cite{WWapproach}: The total \textsc{Wimp} production cross section can be thermally related to the annihilation cross section, which itself is estimated from the dark matter relic density (see section \ref{sec:relicdensity}). The WW approximation (\ref{eqn:wwapproximation}) can then be applied to estimate the kinematics of an additional final state photon. The total cross sections for signal and neutrino background are determined and used to derive statements on the discovery potential of the \textsc{Ilc}. A more elaborative consideration including Bhabha--background and resolution effects of both beam and the \textsc{Ild} detector was performed in \cite{BartelsList2, BartelsList3, BartelsList4, BartelsList}. We make use of these results concerning the detector effects, but with analytical effective models instead.

First collider studies with effective operators for \textsc{Wimp}--Standard Model interactions looked at the Large Hadron Collider (\textsc{Lhc}) and Tevatron sensitivities with jets instead of photons in the final state \cite{maverick}. Looking for jets is advantageous because of the hadronic initial state at these colliders. Further studies in the hadronic collider sector have been performed by searching for monojets, monophotons and monoleptons in the final state \cite{TevatronDarkMatter, LHCStudies, ColliderStudies, LHCStudies2, MonoLeptons, CMSDM, ATLASDM}. Different studies assume different effective interaction modes to set exclusion limits on the coupling strength of these models. Similar studies at \textsc{Lep} with monophotons have also been performed \cite{LEPShinesLight}. Results from collider studies are complementary to exclusion limits set from various direct or indirect detection measurements \cite{DMClass, effmodels3, effmodels4, DMChina1, DMChina2, RelicDensity}. This work will be the first to analyse the exclusion potential of the \textsc{Ilc} with respect to a large list of different effective operators beyond the analytical level under the consideration of the most important background-- and detector effects.

\section{The International Linear Collider}
\label{sec:ilc}
\begin{figure}
\centering
\includegraphics[width=0.9\textwidth]{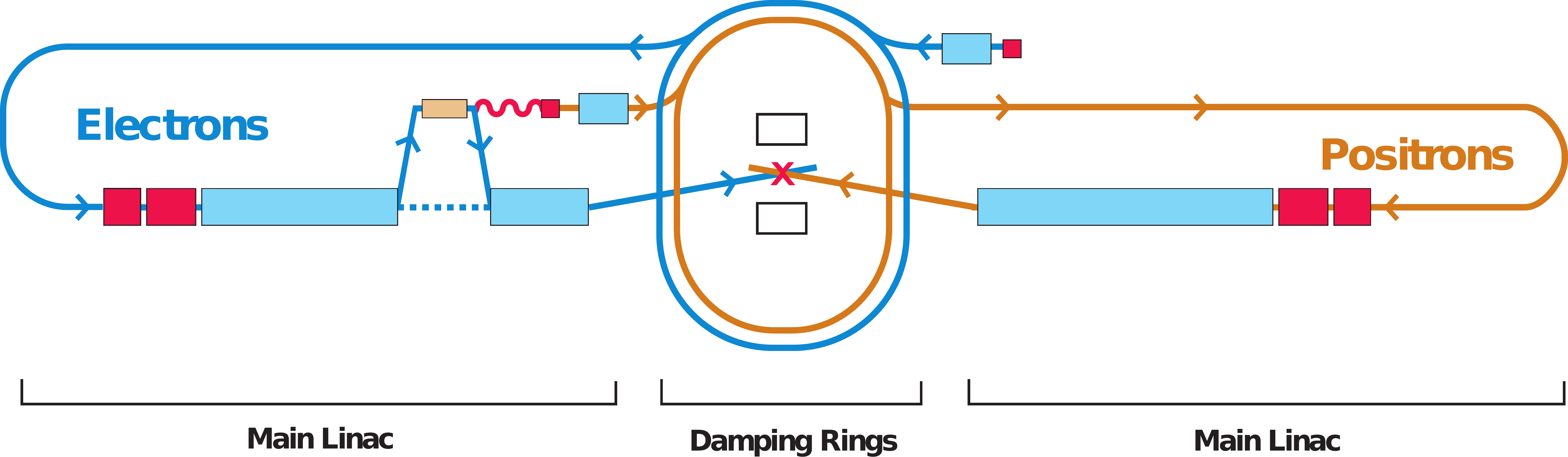}
\caption{Schematic Layout of the International Linear Collider \cite{ILCReport1}}
\label{fig:ilc}
\end{figure}
The International Linear Collider (\textsc{Ilc}, \cite{ILCReport1, ILCReport2, ILCReport3, ILCReport4}) is a proposed electron--positron--collider that is planned to operate at center of mass energies up to \SI{500}{\GeV} with a potential later upgrade to \SI{1}{\TeV}. This will be a significant improvement upon its lepton collider predeccessors, the \textsc{Slac} Linear Collider (\textsc{Slc}, \cite{SLCReport}) with $\SI{182}{\GeV}$ and the Large Electron Positron Collider (\textsc{Lep}, \cite{LEPReport}) with \SI{200}{\GeV}, leading to a much higher mass sensitivity. An additional promising feature is the possibility to polarise the incoming leptons, which can greatly improve the signal to background ratio for helicity--dependent processes. 

A schematic layout according to the current Reference Design report \cite{ILCReport1} is shown in figure \ref{fig:ilc}. After leaving their individual sources, electron-- and positron--bunches are squeezed by using two \SI{7}{\kilo\meter}  circumference damping rings. They are designed to produce beams with a high particle density for maximum interaction during the collision process. After compactification, the velocity of the bunches is then increased by two linear accelerators (\emph{linacs}) with a respective length of \SI{12}{\kilo\meter} to a total center of mass energy of \SI{500}{\GeV}. After a potential second phase upgrade uncluding an additional \SI{10}{\kilo\meter} per beam, the total energy could reach its designed maximum of \SI{1}{\TeV}. 

At the collision center, about \num{14000} interactions per second are expected with a peak luminosity of \SI{2e34}{\per\Square\centi\meter\per\second} or equivalently \SI{2e-5}{\per\femto\barn\per\second}. With an estimated running time of nine months per year and considering commissioning and ramp up time, it is expected to reach an integrated luminosity of \SI{500}{\per\femto\barn} within four years.

The collisions are recorded by two independent detectors, the \emph{International Large Detector} (\textsc{Ild}, \cite{ILD}) and the \emph{Silicon Detector} (\textsc{S}i\textsc{d}, \cite{SiD}), with different technologies to balance out individual advantages and disadvantages. They will most probably perform measurements one at a time with a \emph{push--pull} detector arrangement \cite{ILCReport3}. 

The currently running \textsc{Lhc} will hopefully give first indications on new physics beyond the Standard Model. However, due to hadronic collisions with theoretical uncertainties in both the parton distribution functions of the initial state and the associated multiple parton interactions of the final state will weaken the significance of the final results. The \textsc{Ilc} is being constructed as a high precision tool to accurately measure the masses and couplings of any new particles that may be detected at the terascale, with particular hope to find Supersymmetry, Large Extra Dimensions and/or dark matter \cite{ILCReport2}. 

\section{Standard Model Background for Monophotons}
\label{sec:smbackground}
\begin{figure}
\centering
\includegraphics[width=0.3\textwidth]{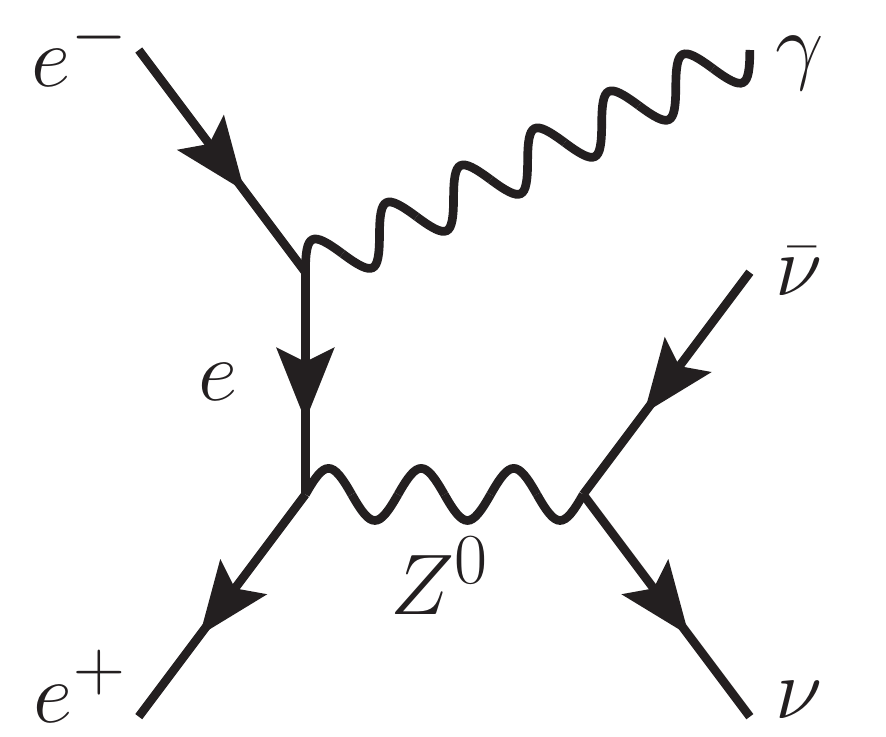} \quad
\includegraphics[width=0.3\textwidth]{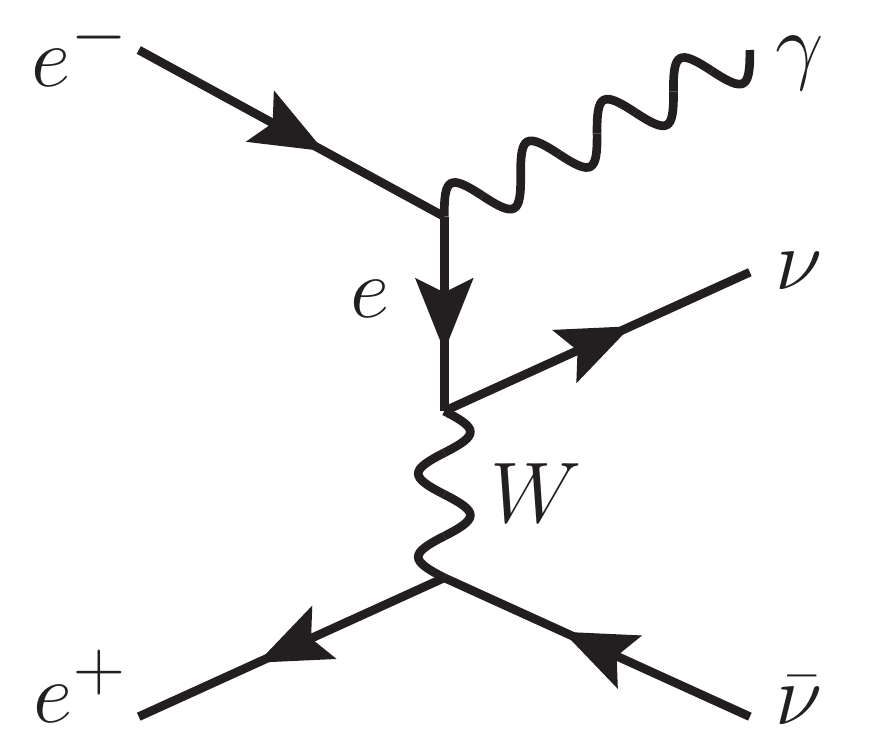} \quad
\includegraphics[width=0.3\textwidth]{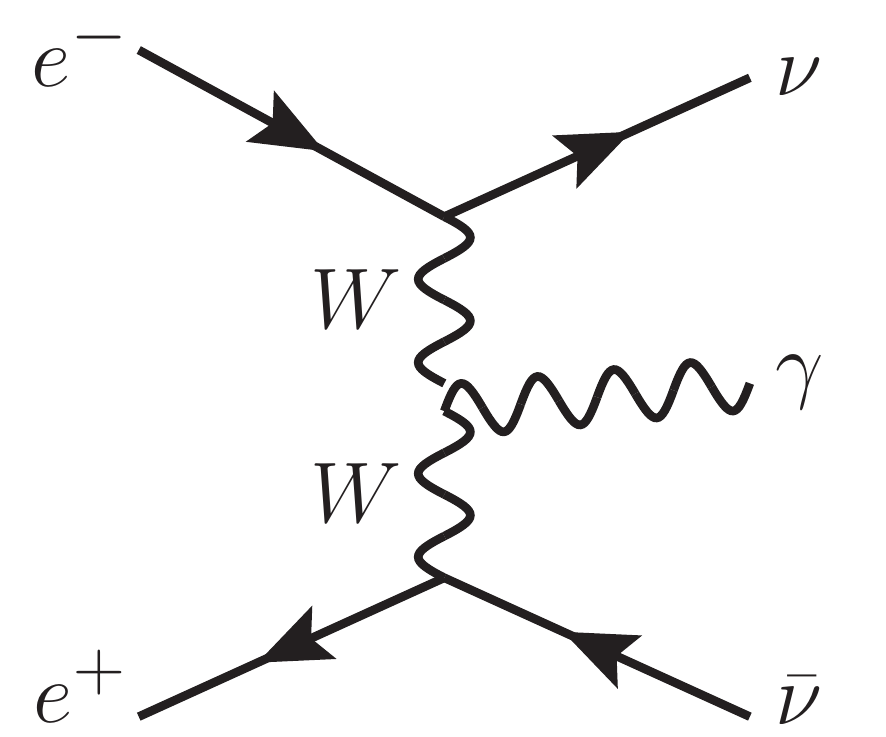} 
\caption{Example Feynman diagrams for radiative neutrino production as a Standard Model monophoton background. The first two diagrams have a second diagram of the same form but with the photon coupling to the positron instead.}
\label{img:nudiags}
\end{figure}
We consider the two dominant Standard Model background contributions
after selection cuts have been performed, determined with a full \textsc{Ild} simulation
\cite{BartelsThesis, BartelsList}.\footnote{All numbers here and in the following
paragraphs refer to the nominal center of mass energy of
\unit{500}{\GeV} \cite{ILCReport3}. Changes due to an increased energy to \unit{1}{\text{TeV}} are considered later in section \ref{sec:1tev}.}
\begin{itemize}
\item Neutrinos from $\Ppositron \Pelectron \rightarrow
  \Pnu \APnu \Pphoton (\Pphoton)$ form a polarisation dependent background and example diagrams are shown in figure \ref{img:nudiags}. The leading
  contribution is given by t--channel $\PW$--exchange, which couples only to
  left--chiral leptons. Additional smaller contributions come from s--channel
  $\PZ$--diagrams with both left-- and right--chiral couplings. We also
  consider the case of one additional undetected photon, which contributes
  with a total event number of about $\unit{10}{\%}$ compared to the single photon process.
\item Bhabha scattering of leptons with an additional hard photon, $\Ppositron \Pelectron \rightarrow
  \Ppositron \Pelectron \Pphoton$, has a large cross section but a small
  selection efficiency, since both final state leptons must be undetected or misidentified. It gives a final background contribution with the same order
  of magnitude as the neutrino background (after application of all selection
  criteria). It is mostly polarisation independent. 
\end{itemize}
 Other reducible background final states, like $\Pphoton \Pphoton + n \cdot \Pphoton$,  contribute with less
than \unit{1}{\%} to the number of events compared to the neutrino background and are therefore omitted.

\section{Data Modelling}
\label{sec:datamodelling}
To avoid the use of a full detector simulation, we build on the results presented in 
\cite{BartelsThesis, BartelsList}. For the signal and monophoton neutrino
background, we generate the events with identical phase space settings, apply the \textsc{Ild} estimates for the
energy resolution as well as the reconstruction and selection
efficiencies\footnote{From here on, the expression ``efficiency'' abbreviates
  ``reconstruction and selection efficiencies''.}  and compare the final energy
distributions. For the diphoton neutrino and Bhabha background, we model the
final distributions directly from the given results. 

\paragraph{Signal Generation:}
For the generation of signal and monophoton neutrino events we use
\textsc{C}alc\textsc{hep} \cite{CalcHEP}. This is a tool to generate event files evaluated from a model with given particle content and vertex structures. It is also able to calculate polarised cross sections which is required for this study. We produce signal events for all benchmark
scenarios with dark matter masses ranging from \unit{1}{\GeV} to
\unit{240}{\GeV}. To avoid collinear and infrared divergencies in the cross section for $x, \theta \rightarrow 0$ (see table \ref{tbl:2t3crosssections}),
we limit phase space in the event generation to \\ $E_\gamma \in \left[\unit{8}{\GeV}, \unit{250}{\GeV}
    \right]$ and $\cos \theta_\gamma \in \left[ -0.995, 0.995 \right]$. Initial State Radiation (\textsc{Isr})
    and beamstrahlung significantly lowers the position of the neutrino
    \PZzero--resonance into the signal region (see next paragraph) and is taken into account; we
    set the accessible parameters in \textsc{C}alc\textsc{hep} according to the \emph{\textsc{Ild}
    Letter of Intent} \cite{ILD} to \unit{645.7}{\nm} for the bunch size, \unit{0.3}{\mm}
    for the bunch length and a total number of particles per bunch of $2 \cdot
    10^{10}$. 

\begin{figure}
\centering
\includegraphics[width=0.5\columnwidth]{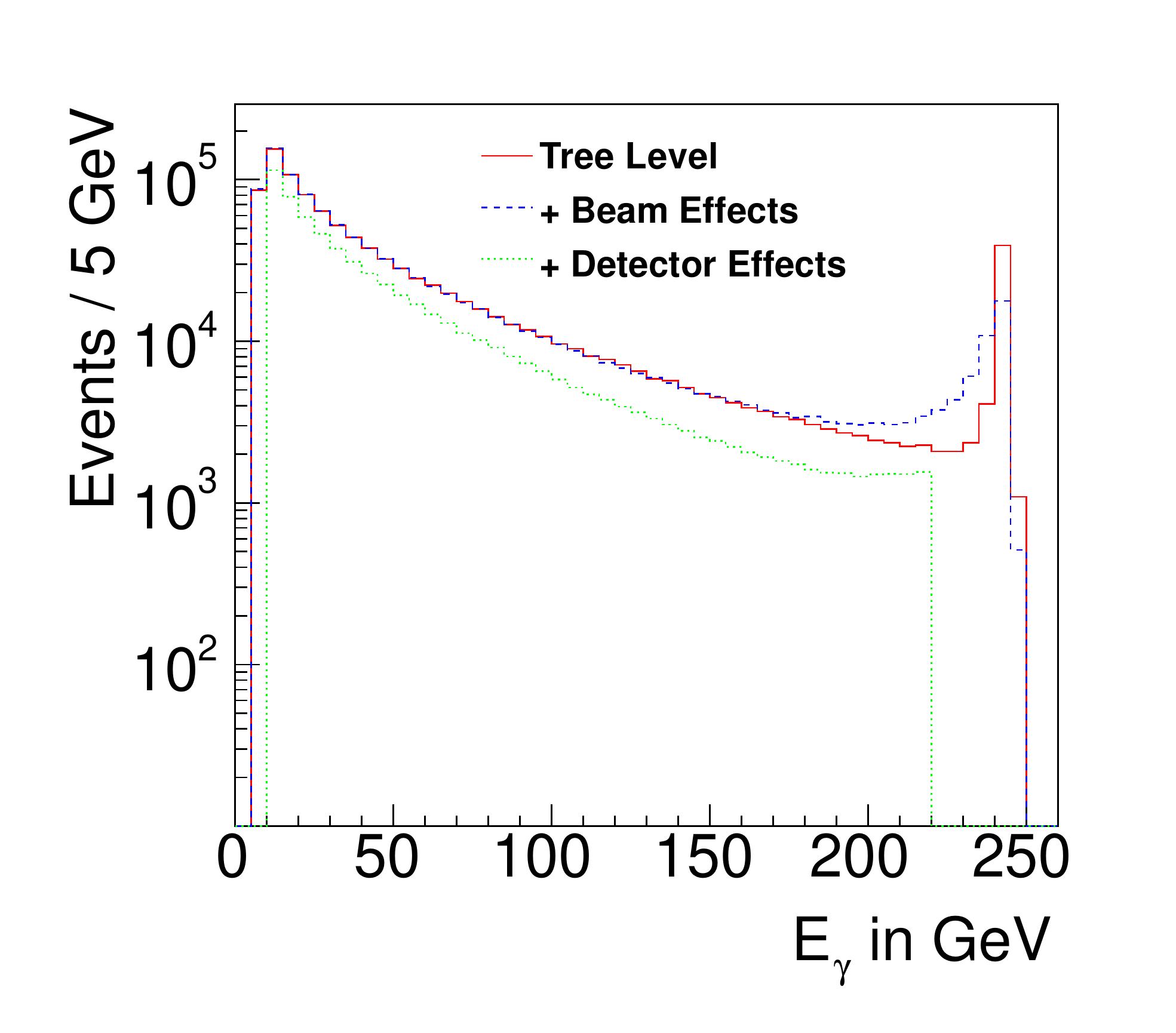} \hspace{-0.7cm}
\includegraphics[width=0.5\columnwidth]{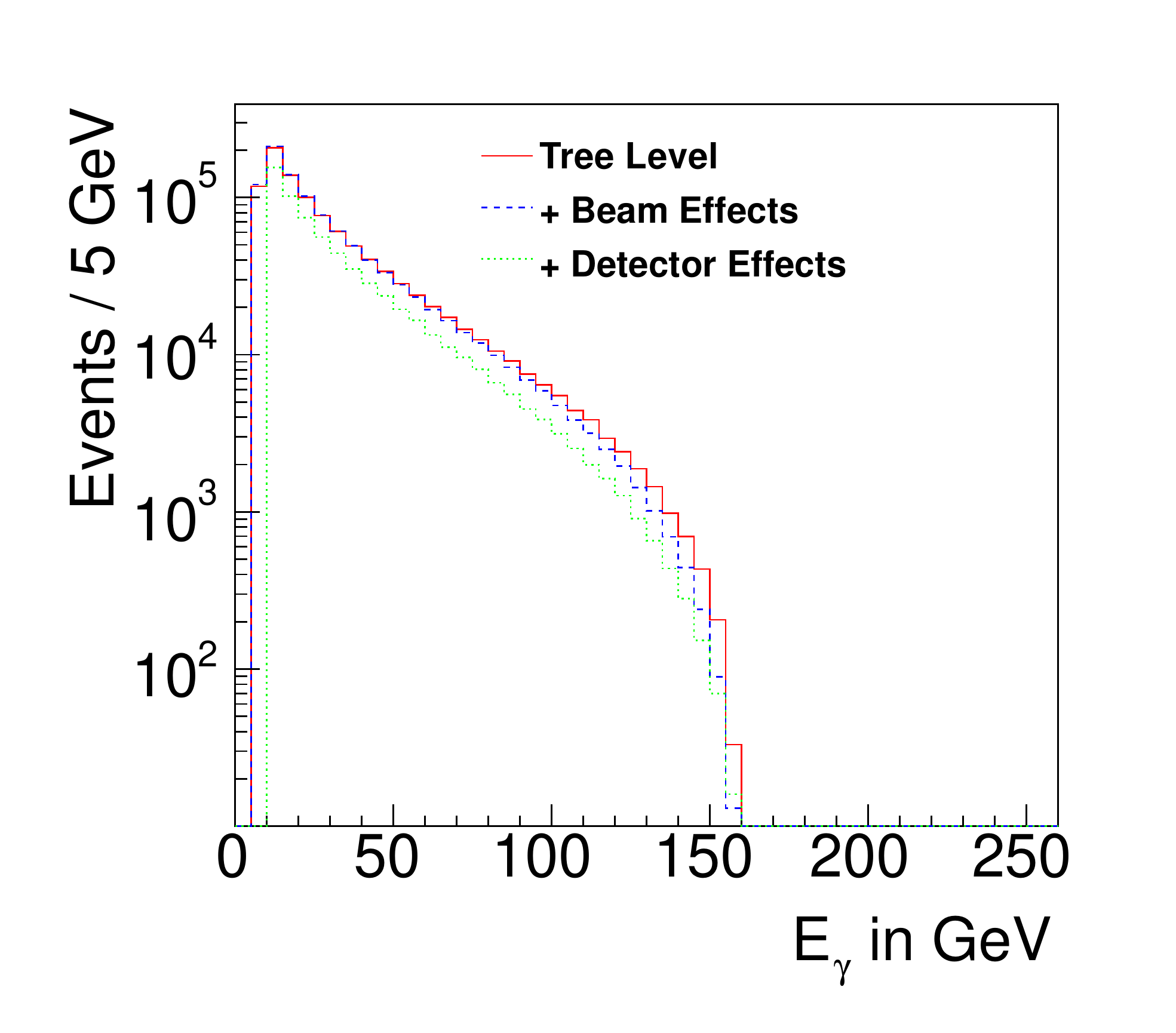}
\begin{flushleft}
\vspace{-0.045\textheight}
\hspace{0.1\textwidth}  (a) \hspace{0.44\textwidth} (b)
\end{flushleft}
\caption{Photon energy distribution before and after the application of beam effects
  (\textsc{Isr} + Beamstrahlung) and detector effects (resolution +
  efficiency) for (a) unpolarised neutrino background and (b) unpolarised FS
  scalar signal with $M_\chi = \unit{150}{\GeV}$. Distributions are normalised to
  \num{1e6} tree level events.}
\label{img:photonenergy}
\end{figure}
\paragraph{Detector Effects}
The finite resolution of the detector components and the use of 
selection criteria to reduce beam--induced background are taken into account by
applying the following steps to both signal and background data: We smear the photon
energy, given in \GeV, according to a Gaussian distribution by taking into account the estimated
  resolution of the \textsc{Ild} detector components given in \cite{ILD}:
\begin{align}
\frac{\Delta E}{E} =
\frac{\unit{16.6}{\%}}{\sqrt{E}} \oplus \unit{1.1}{\%}.
\end{align}
The neutrino background is partially produced through an s--channel $\PZzero$--exchange, which can be produced on resonance due to the large center of mass energy. This will produce a peak in the photon energy spectrum at $(\sqrt{s}-M_Z^2/\sqrt{s})/2 = \unit{242}{GeV}$, which is smeared out due to the intrinsic width of the $\PZzero$ as well as the beam energy spread and the finite detector resolution, as can be seen in figure \ref{img:photonenergy}a. Therefore, we further reduce the phase space to keep the background contribution from that resonance as small as possible:
\begin{align}
E &\in \left[ \unit{10}{\GeV}, \unit{220}{\GeV} \right], \\
\cos \theta &\in \left[ -0.98, 0.98 \right].
\end{align}
The additional angular restriction ensures a good photon reconstruction within the detector. In general, not all photons will be registred due to wrong tagging in the detector or misidentification during the reconstruction process. We succesively apply a random elimination of signal and background
  event records to simulate the following two independent efficiency factors:
\begin{align}
\epsilon_1 &= \unit{97.22}{\%} - (E \text{ in \GeV})
\cdot\unit{0.1336}{\%}, \label{eqn:efficiency}\\
\epsilon_2 &= \unit{96.8}{\%}.
\end{align}

\begin{figure}
\centering
\includegraphics[width=0.5\columnwidth]{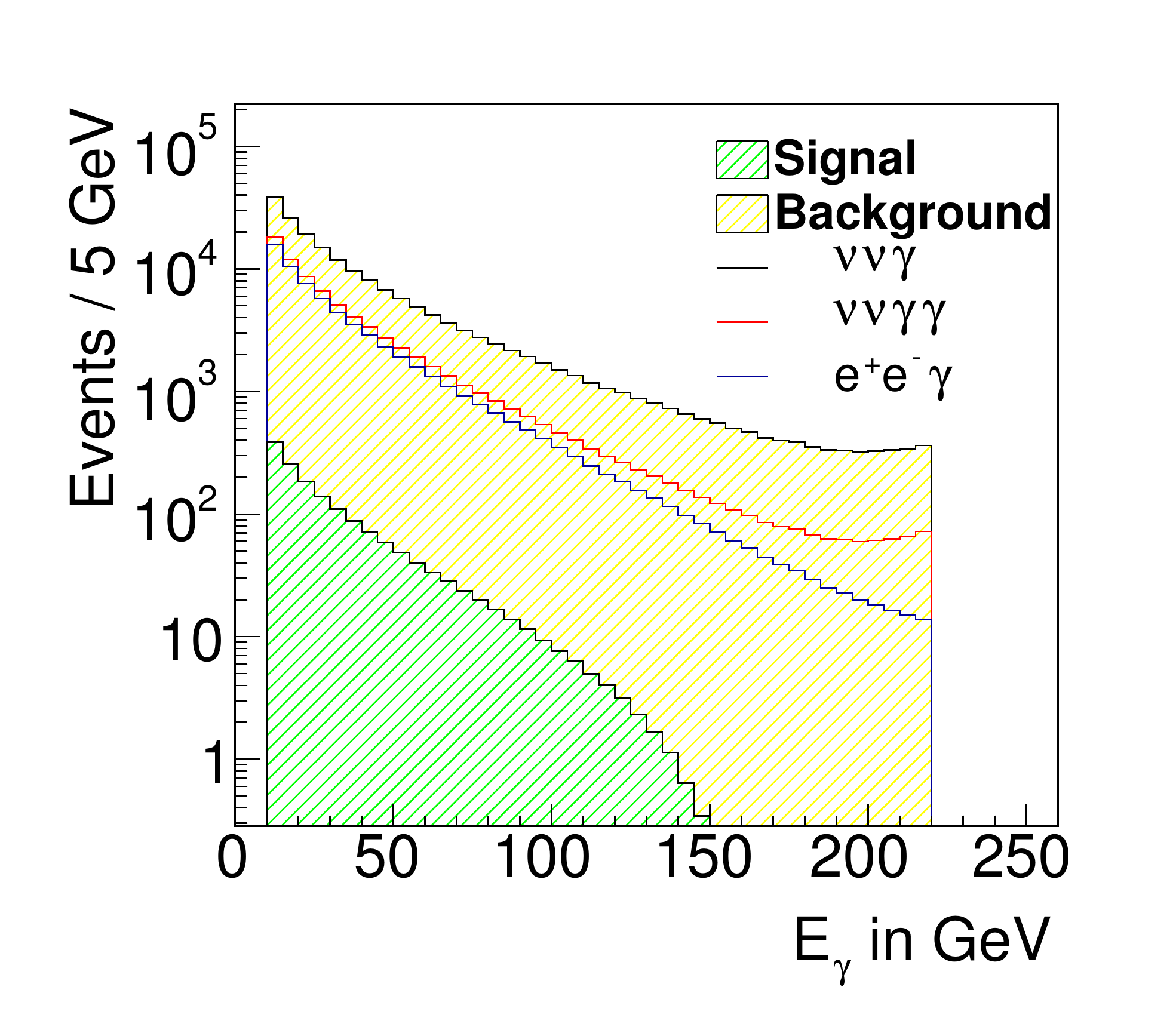}
\caption{Energy distribution of the most dominant background contributions
  after selection. Background histograms are stacked and compared to an example signal spectrum (FS scalar model with $M_\chi = \SI{150}{\GeV}$) with a tree level cross section of \SI{100}{\femto\barn}.}
\label{img:backgroundcontributions}
\end{figure}

\paragraph{Simulation of Reducible Backgrounds}
Diphoton neutrino and Bhabha processes can mimic monophoton signatures if additonal final state particles escape through the beam pipe or are not properly reconstructed. A full detector simulation is needed to evaluate the impact of the detector geometry and reconstruction algorithms on the total number of background events, which has been performed in \cite{BartelsThesis}. 
We estimate the diphoton background by
using the monophoton data rescaled by the corresponding global factor determined in the detector simulation. The Bhabha
background is found similarly by linearly deforming the monophoton neutrino distribution such that the shape and the
total number of events match. Table  \ref{tbl:neventsperscenario} gives the total number of events, after application of all the previously described effects, for
\SI{1}{\per\femto\barn} integrated luminosity in different polarisation settings. We compare these results to those given by the full detector simulation and see that they agree. Note that due to our discussion of the Bhabha background in section \ref{sec:smbackground}, we assume no polarisation dependence and therefore use a constant value for different settings of $P^\pm$.

The relative size of all different background contributions after
reconstruction and selection can be seen in
figure  \ref{img:backgroundcontributions}, compared to an example signal with \SI{100}{\femto\barn} total cross section.
Note that the full simulation \cite{BartelsThesis} encountered large statistical uncertainties in the simulation of the Bhabha background for large energies with only a small number of events, which makes it difficult to properly estimate the shape of the energy distribution in that region. This is why we do not perform a shape dependent analysis here. One would need more reliable information about the the energy dependence of all dominant background sources in order to analyse the \textsc{Ilc} sensitivity to separate them from a possible signal. 

\begin{table}[t]
\centering
\begin{tabular}{r@{\qquad}crr c@{\quad} rr c@{\quad}  rr }
\toprule
$P^-/P^+$ && \multicolumn{2}{c}{$\Pnu \Pnu \Pphoton$} && \multicolumn{2}{c}{$\Pnu \Pnu \Pphoton \Pphoton$} &&\multicolumn{2}{c}{$\Ppositron \Pelectron$} \\
\midrule
\midrule
$0/0$ && 2257 & (2240) && 226 & (228) && 1218 & (1229) \\
$+0.8/-0.3$ && 493 & (438) && 49 & (43) && 1218 & (1204) \\
$-0.8/+0.3$ && 5104 & (5116) && 510 & (523) && 1218 & (1227) \\
\bottomrule
\end{tabular}
\caption{Simulated and modelled number of events of the different background
  sources after application of all selection criteria. We show the numbers we determined for an integrated luminosity of \SI{1}{\per\femto\barn} in different
  polarisation settings, followed by numbers in parentheses, which are taken from \cite{BartelsThesis} and that
  show the numbers obtained when performing a proper detector simulation.}
\label{tbl:neventsperscenario}
\end{table}

\section{Systematic Uncertainties}
\label{sec:sysuncertainties}
Systematic errors play an important role in determining the uncertainty in the number of background events $\Delta N_\text{B}$
and for estimating exclusion bounds on the effective couplings. There are two dominant
contributions, motivated in \cite{BartelsThesis}.

In general, the efficiency given in (\ref{eqn:efficiency}) will be determined at
the real experiment by measuring the $\PZzero$--resonance peak, which is
theoretically known to a very good accurracy. Systematic uncertainties on that value are
given by the finite statistics of this measurement and further broadening of the
peak by unknown beam effects. These errors can be extrapolated down to the
dark matter signal region at small photon energies and, since the same
efficiency factor is used for signal and background, is highly correlated between
the two. This global uncertainty will therefore approximately cancel in the determination of the maximum coupling \Geff{}. 

However, cancellation will not take place for model--dependent effects. This is because the signal energy distribution depends on the
unknown mass of the dark matter particle and the underlying interaction model. Therefore the correct function $\epsilon(E)$
 for the signal will be different from the neutrino background efficiency given in (\ref{eqn:efficiency}). Since we do not know the model a priori, we use
the same value for both and therefore introduce an error on the determination of $N_\text{S}$. Compared to \cite{BartelsThesis}, we use a conservatively overestimated value for the efficiency uncertainty of $\Delta \epsilon = \unit{2}{\%}$. 

Since the neutrino spectrum depends on the leptons' polarisation $P^\pm$, any fluctuation $\Delta P$ within those
parameters will give additional systematic uncertainties in the number of expected
background events. One can not use the information from measuring the
$\PZzero$--resonance in this case to infer information in the low energy signal range because of
the polarisation dependence of the shape itself \cite{BartelsThesis}. Given the assumed accuracy of at least $\Delta P/P =
\unit{0.25}{\%}$ \cite{ILCReport3} with a possible improvement to \unit{0.1}{\%} at
the \textsc{Ilc} \cite{ILCPolarimetry}, we can derive
the corresponding error on the polarised number of background events. As an example, for ``left--like'' background (see (\ref{eqn:polclassesleft})):
\begin{align}
N_{\text{pol}} &= (1+P^+)(1-P^-)N_{\text{unpol}},  \\
\intertext{the corresponding error is given as}
\Delta N_{\text{pol}} &= \sqrt{\left[P^- (1+P^+) \right]^2 + \left[P^+ (1-P^-)
  \right]^2} \  \frac{\Delta P}{P} \ N_{\text{unpol}}. \label{eqn:polerr}
\end{align}
From the numbers in table \ref{tbl:neventsperscenario}, we assume an identical
polarisation dependence for $\Pnu \Pnu \Pphoton$ and $\Pnu \Pnu \Pphoton
\Pphoton$ events and no dependence for the Bhabha background.

\section{Optimum Polarisation Settings}
\label{sec:polarisation}
\begin{table}
\centering
\begin{tabular}{r@{\quad}r@{\quad}c rr@{\qquad}c rr@{\qquad}c rrrr}
\toprule 
$P^-/P^+$ & $N_{\text{B}}$ && $\Delta^\text{stat}_{50}$  & $\displaystyle\tilde{\Delta}^\text{stat}_{500}$ && $\Delta^\text{sys}_{P}$
& $\Delta^\text{sys}_{\tilde{P}}$  && $\Delta^\text{tot}_{50 P}$ & $\Delta^\text{tot}_{50 \tilde{P}}$ & $\displaystyle\tilde{\Delta}^\text{tot}_{500 P}$
& $\displaystyle \tilde{\Delta}^\text{tot}_{500 \tilde{P}}$ \\
\midrule \midrule
0/0 & \num{184998} && 430 & 136 && 0 & 0 && 430 & 430 & 136 & 136 \\
\midrule
$\num[retainplus]{+0.8}$/$\num[retainplus]{+0.3}$ & \num{97568} && 312 & 99 && 312 & 125 && 441 & 336 & 327 & 159 \\
$\num[retainplus]{+0.8}$/$\num[retainplus]{+0.6}$ & \num{102365} && 320 & 101 && 385 & 154 && 500 & 355 & 398 & 184 \\
\midrule
$\num[retainplus]{+0.8}$/$\num{-0.3}$ & \num{87974} && 297 & 94 && 169 & 68 && 341 & 304 & 193 & 116 \\
$\num[retainplus]{+0.8}$/$\num{-0.6}$ & \num{83177} && 288 & 91 && 104 & 42 && 307 & 291 & 138 & 100 \\
\midrule
$\num{-0.8}$/$\num[retainplus]{+0.3}$ & \num{341597} && 584 & 185 && 351 & 140 && 682 & 601 & 396 & 232 \\
$\num{-0.8}$/$\num[retainplus]{+0.6}$ & \num{404970} && 637 & 201 && 501 & 200 && 811 & 668 & 546 & 284 \\
\midrule
$\num{-0.8}$/$\num{-0.3}$ & \num{212851} && 461 & 156 && 233 & 93 && 517 & 471 & 275 & 173 \\
$\num{-0.8}$/$\num{-0.6}$ & \num{148478} && 385 & 122 && 337 & 135 && 512 & 408 & 359 & 182 \\
\bottomrule
\end{tabular}
\caption{Total number of background events ($N_{\text{B}}$) with statistical
  error $\Delta^\text{stat}$, the systematical error $\Delta^\text{sys}$ coming from the
  polarisation uncertainty and the total error combinations $\Delta^\text{tot}$. The subscripts 50 and 500 denote the integrated
  luminosity in inverse femtobarn. In case of a ten times larger luminosity,
  one will get ten times as many events in all channels; to better compare to the error
  of the low luminosity case,
  we show $\tilde{\Delta} \equiv \Delta/10$. The polarisation uncertainties are set to
  \unit{0.25}{\%} ($P$) and \unit{0.1}{\%} ($\tilde{P}$). }
\label{tbl:bkgevts}
\end{table}
Polarisation can be used to significantly increase the number of signal
events $N_\text{S}$ according to (\ref{eqn:polclasses}) but also increases the systematical contribution to $\Delta N_\text{B}$ with respect to (\ref{eqn:polerr}). We are interested in the setting for each individual
model that leads to the largest $N_S/\Delta N_B$ ratio, giving the strictest bounds on $G_\text{eff}$.

 We start with the impact on the background uncertainty: In table \ref{tbl:bkgevts} we
give the total number of background events in different polarisation settings
$P^-$ = $\pm 0.8$, $P^+$ = $\pm 0.3$/$\pm 0.6$ that are feasible at the \textsc{Ilc} \cite{ILCReport3}. We give the statistical
fluctuation for integrated luminosities of both \SI{50}{\per\femto\barn} and
\unit{500}{\per\femto\barn}. Since the latter will give ten times as many events in
all channels, we reduce the listed statistical error accordingly to give a value
comparable to the small luminosity case. We also give the polarisation uncertainty (\ref{eqn:polerr}) for both estimates of the
polarisation error $\Delta P/P = \unit{0.25}{\%} $ and $
\unit{0.1}{\%}$. Finally we give the total errors after adding all combinations of
individual errors in quadrature.

On the signal side, we look at the different classes derived in section \ref{sec:finalxsects} with respect to their polarisation dependence. For comparison,
we use a common reference value of 500 events in the unpolarised case for an integrated luminosity of
\SI{50}{\per\femto\barn} and derive the corresponding number of
events for polarised input and potentially larger luminosity. Using the numbers in table \ref{tbl:bkgevts}, we
look at the ratio $N_{\text{S}} / \Delta N_{\text{B}}$ that has to be
maximised in order to get the strongest bound on the coupling. The results are shown in table \ref{tbl:sigoverbkgestimate}, where we give values only
for polarisation signs with the largest numbers for that ratio. 
In each combination we mark the positron polarisation that leads to the largest
value for the ratio. 

In the majority of cases, larger polarisation of the incoming leptons enhances the result. For high
statistics and taking the conservative value for the polarisation error, the systematic uncertainty
associated with increased polarisation
may be larger than the gain in the number of signal events, though. In those
cases, which appear only in scalar-- and left--coupling models, less polarised
beams may lead to better results. 
In regard to this analysis it is therefore clearly beneficial for the real \textsc{Ilc} to have a systematical uncertainty on the polarisation as small as possible, in particular for large $P^+$.  

\begin{table}
\centering
\begin{tabular}{r@{\quad}r@{\quad}c rr@{\qquad}c rr@{\qquad}c rrrr}
\toprule
IA type & $P^-/P^+$ & $N_{\text{S}}$ && $r_{50 P}$ & $r_{50 \tilde{P}}$ & $r_{500 P}$ &
$r_{500 \tilde{P}}$ \\
\midrule
Scalar &$\num[retainplus]{+0.8}/\num[retainplus]{+0.3}$   &  620  &&  1.41  &  1.85  &  \textbf{1.90}  &  3.90 \\
 &$\num[retainplus]{+0.8}/\num[retainplus]{+0.6}$   &  740  &&  \textbf{1.48}  &  \textbf{2.08}  &  1.86  &  \textbf{4.02} \\
\midrule
Vector &$\num[retainplus]{+0.8}/\num{-0.3}$   & 620  &&  1.82  &  2.04  &  3.21  &  5.34 \\
& $\num[retainplus]{+0.8}/\num{-0.6}$   &  740  &&  \textbf{2.41}  &  \textbf{2.54}  &  \textbf{5.36}
&  \textbf{7.40} \\
\midrule
Left & $\num{-0.8}/\num[retainplus]{+0.3}$   &  \num{1170}  &&  1.72 &  1.95  &  \textbf{2.95}  &  5.04 \\
& $\num{-0.8}/\num[retainplus]{+0.6}$   &  \num{1440}  &&  \textbf{1.78}  &  \textbf{2.16}  &  2.64  &  \textbf{5.07} \\
\midrule
Right &$\num[retainplus]{+0.8}/\num{-0.3}$   &  \num{1170}  &&  3.43  &  3.85  &  6.06  &  10.09 \\
& $\num[retainplus]{+0.8}/\num{-0.6}$   &  \num{1440}  &&  \textbf{4.69}  &  \textbf{4.95}  &  \textbf{10.43}  &  \textbf{14.40} \\
\bottomrule
\end{tabular}
\caption{Determination of the best ratio $r  \equiv 
  {N_{\text{S}}}/{\Delta N_{\text{B}}}$ with $\Delta N_{\text{B}}$ given by the different
  total errors determined in table \ref{tbl:bkgevts}. $N_{\text{S}}$ describes the
number of polarised signal events for the different classes described in section \ref{sec:finalxsects}
with a common reference value of 500 events for an integrated luminosity of \SI{50}{\per\femto\barn}. We only give the
results for the polarisation signs, which lead to the best ratios. We mark the
numbers which lead to the best signal to background ratio in bold.}
\label{tbl:sigoverbkgestimate}
\end{table}

\section[Increasing $\sqrt{s}$ to \SI{1}{\TeV}]{Increasing $\boldsymbol{\sqrt{s}}$ to \SI{1}{\TeV}}
\label{sec:1tev}

\begin{table}[b]
\centering
\begin{tabular}{r@{\qquad}r@{\quad}r@{\quad}r@{\quad}}
\toprule
$P^-/P^+$ & $\Pnu \Pnu \Pphoton$ & $\Pnu \Pnu \Pphoton \Pphoton$ & $\Ppositron \Pelectron$ \\
\midrule
\midrule
$0/0$ & 2677  & 268 & 304  \\
$\num[retainplus]{+0.8}/\num{-0.3}$ & 421 & 42  & 304  \\
$\num{-0.8}/\num[retainplus]{+0.3}$ & 6217 & 622  & 304  \\
\bottomrule
\end{tabular}
\caption{Simulated and modelled number of events in the different background
  sources after application of all selection criteria for $\sqrt{s} = \unit{1}{\TeV}$. The numbers are
  calculated for an integrated luminosity of \SI{1}{{\per\femto\barn}} in different
  polarisation settings.}
\label{tbl:neventsperscenario_1tev}
\end{table}

\begin{table}[t]
\centering
\begin{tabular}{r@{\quad}r@{\quad}c rr@{\qquad}c rr@{\qquad}c rrrr}
\toprule
$P^-/P^+$ & $N_{\text{B}}$ && $\Delta^\text{stat}_{50}$  & $\displaystyle\tilde{\Delta}^\text{stat}_{500}$ && $\Delta^\text{sys}_{P}$
& $\Delta^\text{sys}_{\tilde{P}}$  && $\Delta^\text{tot}_{50 P}$ & $\Delta^\text{tot}_{50 \tilde{P}}$ & $\displaystyle\tilde{\Delta}^\text{tot}_{500 P}$
& $\displaystyle \tilde{\Delta}^\text{tot}_{500 \tilde{P}}$ \\
\midrule
\midrule
$0/0$ & \num{162437} && 403 & 127 && 0 & 0 && 403 & 403 & 127 & 127 \\
\midrule
$\num[retainplus]{+0.8}$/$\num[retainplus]{+0.3}$ & \num{54649} && 234 & 74 && 380 & 152 && 446 & 279 & 387 & 169 \\
$\num[retainplus]{+0.8}$/$\num[retainplus]{+0.6}$ & \num{62791} && 251 & 79 && 469 & 188 && 531 & 314 & 476 & 203 \\
\midrule
$\num[retainplus]{+0.8}$/$\num{-0.3}$ & \num{38365} && 196 & 62 && 201 & 82 && 281 & 212 & 210 & 102 \\
$\num[retainplus]{+0.8}$/$\num{-0.6}$ & \num{30223} && 174 & 55 && 125 & 50 && 214 & 181 & 137 & 74 \\
\midrule
$\num{-0.8}$/$\num[retainplus]{+0.3}$ & \num{357173} && 598 & 189 && 428 & 171 && 735 & 622 & 468 & 255 \\
$\num{-0.8}$/$\num[retainplus]{+0.6}$ & \num{435979} && 660 & 209 && 612 & 245 && 900 & 704 & 647 & 322 \\
\midrule
$\num{-0.8}$/$\num{-0.3}$ & \num{199561} && 447 & 141 && 284 & 114 && 530 & 461 & 317 & 181 \\
$\num{-0.8}$/$\num{-0.6}$ & \num{120755} && 348 & 110 && 411 & 165 && 538 & 385 & 425 & 198 \\
\bottomrule
\end{tabular}
\caption{Total number of background events ($N_{\text{B}}$) and different
  error sources (see table \ref{tbl:bkgevts}) for $\sqrt{s} = \unit{1}{\TeV}$.}
\label{tbl:bkgevts_1tev}
\end{table}

\begin{table}[bt]
\centering
\begin{tabular}{r@{\qquad}rr@{\quad}rrrr}
\toprule
Model & $P^-/P^+$ & $N_{\text{S}}$ & $r_{50 P}$ & $r_{50 \tilde{P}}$ & $r_{500 P}$ &
$r_{500 \tilde{P}}$ \\
\midrule
\midrule
Scalar &$\num[retainplus]{+0.8}$/$\num[retainplus]{+0.3}$   &  620  &  1.39  &  2.22  &  1.60  &  3.7 \\
 &$\num[retainplus]{+0.8}$/$\num[retainplus]{+0.6}$   &  740  &  \textbf{1.39}  &  \textbf{2.36}  &  1.55  &  3.65 \\
\midrule
Vector &$\num[retainplus]{+0.8}$/$\num{-0.3}$   &  620  &  2.21  &  2.92  &  2.95  &  6.08 \\
& $\num[retainplus]{+0.8}$/$\num{-0.6}$   &  740  &  \textbf{3.46}  &  \textbf{4.09}  &  \textbf{5.40}
&  \textbf{10.00} \\
\midrule
Left & $\num{-0.8}$/$\num[retainplus]{+0.3}$   &  \num{1170}  &  1.59 &  1.88  &  2.50  &  \textbf{4.59} \\
& $\num{-0.8}$/$\num[retainplus]{+0.6}$   &  \num{1440}  &  \textbf{1.60}  &  \textbf{2.05}  &  2.23  &  4.47 \\
\midrule
Right &$\num[retainplus]{-0.8}$/$\num{-0.3}$   &  \num{1170}  &  4.16  &  5.52  &  5.57  &  11.47 \\
&$\num[retainplus]{-0.8}$/$\num{-0.6}$   &  \num{1440}  &  \textbf{6.73}  &  \textbf{7.96}  &  \textbf{10.51}  &  \textbf{19.46} \\
\bottomrule
\end{tabular}
\caption{Determination of the best ratio $r  \equiv 
  {\Delta N_{\text{B}}}/{N_{\text{S}}}$ (see
  table \ref{tbl:sigoverbkgestimate}) for $\sqrt{s} =
  \unit{1}{\TeV}$. }
\label{tbl:sigoverbkgestimate_1tev}
\end{table}

We also consider the possibility of a doubled center of mass
energy, which is currently discussed as a potential \textsc{Ilc} upgrade. This changes the previous analysis as follows:
\begin{itemize}
\item We generate events in a larger photon energy range $E_\gamma \in
  \left[\unit{8}{\GeV}, \unit{500}{\GeV}\right]$ and reduce it to the interval $\left[
      \unit{10}{\GeV}, \unit{450}{\GeV} \right]$ after performing the energy resolution
    smearing $\Delta E / E$. This again reduces background events from the
    $\PZzero$--resonance, which now is positioned at \unit{496}{\GeV}.
\item Dark matter signal processes can now be produced with masses up to $\SI{490}{\GeV}$.
\item We generate new monophoton neutrino events with the increased center of mass
  energy and model the diphoton neutrino events with the same conversion factor as in the low energy case. 
\item We use our previously modelled distribution for the Bhabha background
  and rescale it  by a factor of 1/4, taking into account that the full cross
  section for this \textsc{Qed}--driven process falls with $1/s$. 
\item We use, as a rough approximation, the same \textsc{Isr}-- and Beamstrahlung parameters in
\textsc{C}alc\textsc{hep}, efficiency factors and systematic error estimates. 
\end{itemize}

Tables \ref{tbl:neventsperscenario_1tev}-\ref{tbl:sigoverbkgestimate_1tev}
summarise again the number of background events per background scenario, the individual error
sources and the determination of the best polarisation setting for the
increased center of mass energy. In contrast to the Bhabha cross section that falls
almost completely according to $\sigma \propto 1/s$, the neutrino background gets a significant
contribution from t--channel $\PW$s which give  $s / m_W^2$ --terms in the
evaluation of the 
total cross section\footnote{These terms behave still regular for large $s$, since they in fact have a nontrivial logarithmic dependence on $m_W$ and $s$. This may lead to a locally increased cross section if $\sqrt{s}$ is doubled, but converges to $0$ for $\sqrt{s} \rightarrow \infty$.}.  The left--handed neutrino contribution therefore gets enhanced
whereas the Bhabha background becomes less dominant in some polarisation
channels. This leads to a larger relative polarisation error and therefore a
larger impact on the size of the background
fluctuation. Consequently, vector-- and right--coupling models receive stronger signal to background enhancement
for polarised input than in the $\sqrt{s} = \unit{500}{\GeV}$ case, whereas
the other models  suffer from the larger impact of systematics on the total
error and prefer a weaker polarisation. 
\section{Results}
\label{sec:ilcresults}
In order to understand the sensitivity of the \textsc{Ilc} to our various \textsc{Wimp} models, we perform a counting experiment.
We determine the total
number of background events along with the statistical and systematic fluctuation
$\Delta N_{\text{B}}$ and exclude coupling constants which would lead to a
larger number of signal events than \unit{1.64}{$\Delta N_{\text{B}}$}. This corresponds
 to the \unit{90}{\%} confidence interval of the background--only
assumption and is calculated with a \emph{Rolke} test \cite{TRolke, Root}. We only give results for an integrated luminosity of \SI{500}{{\per\femto\barn}}
and the systematic polarisation error set to $\Delta P/P = \unit{0.1}{\%}$. For
each benchmark model we choose the polarisation setting that leads to the best
signal to background ratio for the corresponding polarisation behaviour according to tables \ref{tbl:sigoverbkgestimate} and
\ref{tbl:sigoverbkgestimate_1tev}. Results for different parameters can be found by rescaling the coupling according to $G^\prime = G \sqrt{r^\prime / r}$ where $r$ denotes the signal to background uncertainty ratio given in tables \ref{tbl:sigoverbkgestimate} and \ref{tbl:sigoverbkgestimate_1tev}.

In figures  \ref{img:ilcbounds500gev} and \ref{img:ilcbounds} we show the derived bounds on the coupling
constants for \textsc{Ilc} center of mass energies of \unit{500}{\GeV} and \unit{1}{\TeV}. We mark the region that either violates the tree level approach with a
too large dimensionless coupling constant $g^2 > 4 \pi$ or by having a too small mediator
mass $M_\Omega \leq \sqrt{s}$ for the effective approach to be valid. Note
that the leading order in models with fermionic mediators has a different mass
dimension and therefore leads to a different definition  for the effective
coupling constant $G_\text{eff}$. Also, models with dimensionful fundamental couplings $g_\chi$ (SS and VS) do not have a simple perturbative $4 \pi$ bound. A proper unitarity and perturbativity analysis has to be performed in order to get correct bounds on \Geff in those cases, which is beyond the scope of this work.

If a model has no separate \emph{pseudoscalar}
or \emph{axialvector} result, it is identical to the corresponding \emph{scalar}/
\emph{vector} line due to identical cross section formulae. For masses far from
the threshold, the \textsc{Ilc}
is able to exclude coupling constants up to order
$\unit{1e-7}{\per\Square\GeV}$ or \\
$\unit{1e-4}{\per\GeV}$, depending on the mass dimension. This corresponds to a signal cross section of about \SI{0.3}{\femto\barn}. 

Significant exceptions arise for models
with vector dark matter: These tend to have very strong exclusion limits
  for small dark matter masses. This is caused by the $1/M_\chi^4$ dependence in the photon
  cross section, which leads to divergences for very small vector boson
  masses. It has been shown that only spontaneously broken
  gauge theories can lead to models with massive vector particles that are not
  divergent \cite{unitarity}. Therefore our initial fundamental model cannot be the full
  theory for all energies. Our effective approach restricts the energy to a
  maximum and in that case one can still find perturbatively valid results for
  mass ranges that do not violate unitary bounds. However, the perturbatively allowed mass range is difficult to evaluate, since such an analysis needs more information about the size
  of the individual couplings and the relation between the mass of the
  mediator and the dark matter mass itself. Here, a
  more detailled fundamental theory is needed to evaluate the breakdown of
  perturbation theory in this scenario. We therefore conclude that even though astrophysical analyses with simple vector dark matter models give reasonable results (see chapter \ref{chap:wmap}, \cite{DMChina2},$\ldots$), one has to be careful in interpreting these models for all mass ranges in collider studies.

At last we note that for models with fermionic operators the subleading
order again has a negligible effect, as can be seen from the nearly identical lines
for the \emph{low} and \emph{high} scenario.

\begin{figure}
\centering
\includegraphics[width=0.49\textwidth]{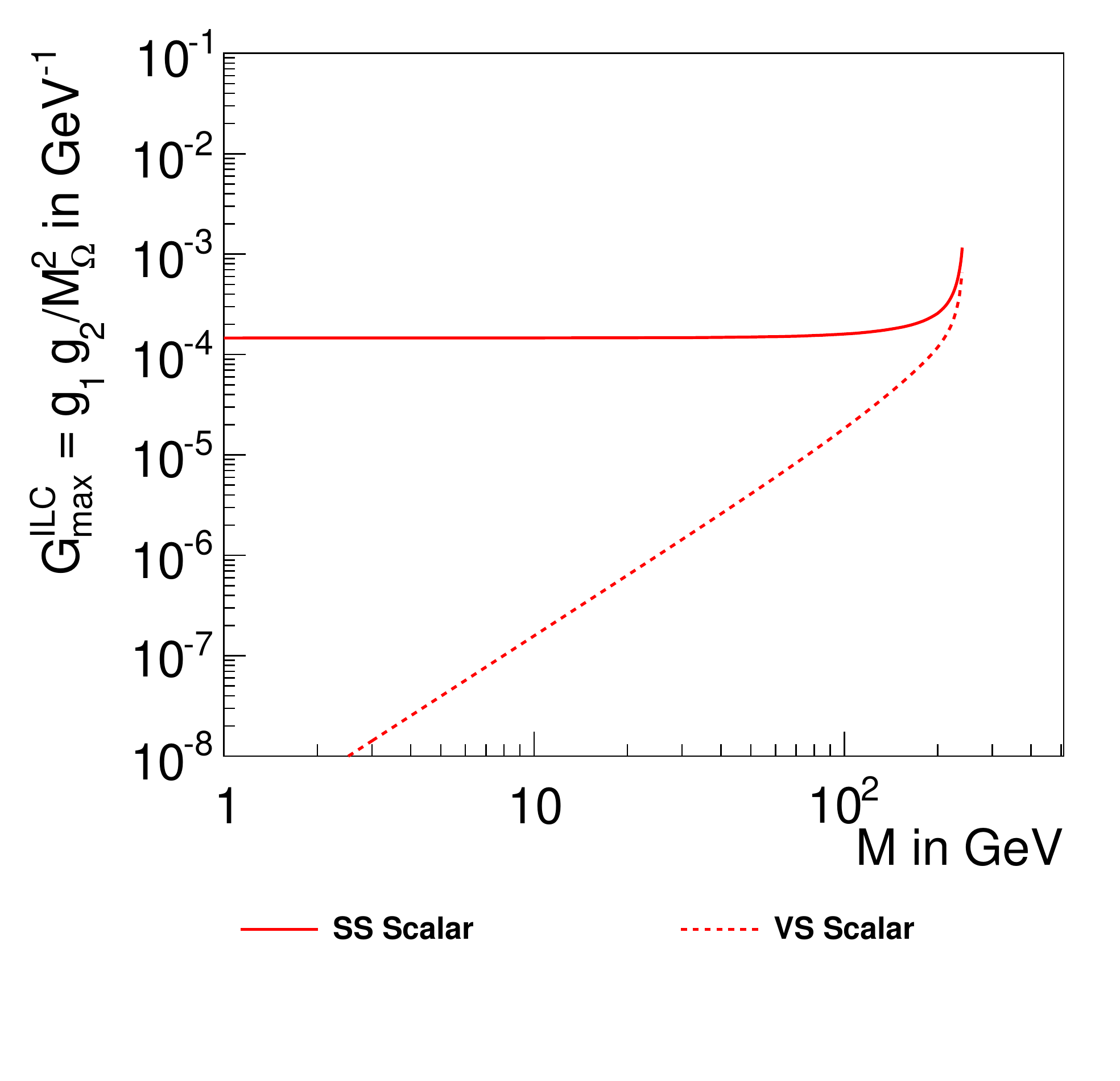} \hfill
\includegraphics[width=0.49\textwidth]{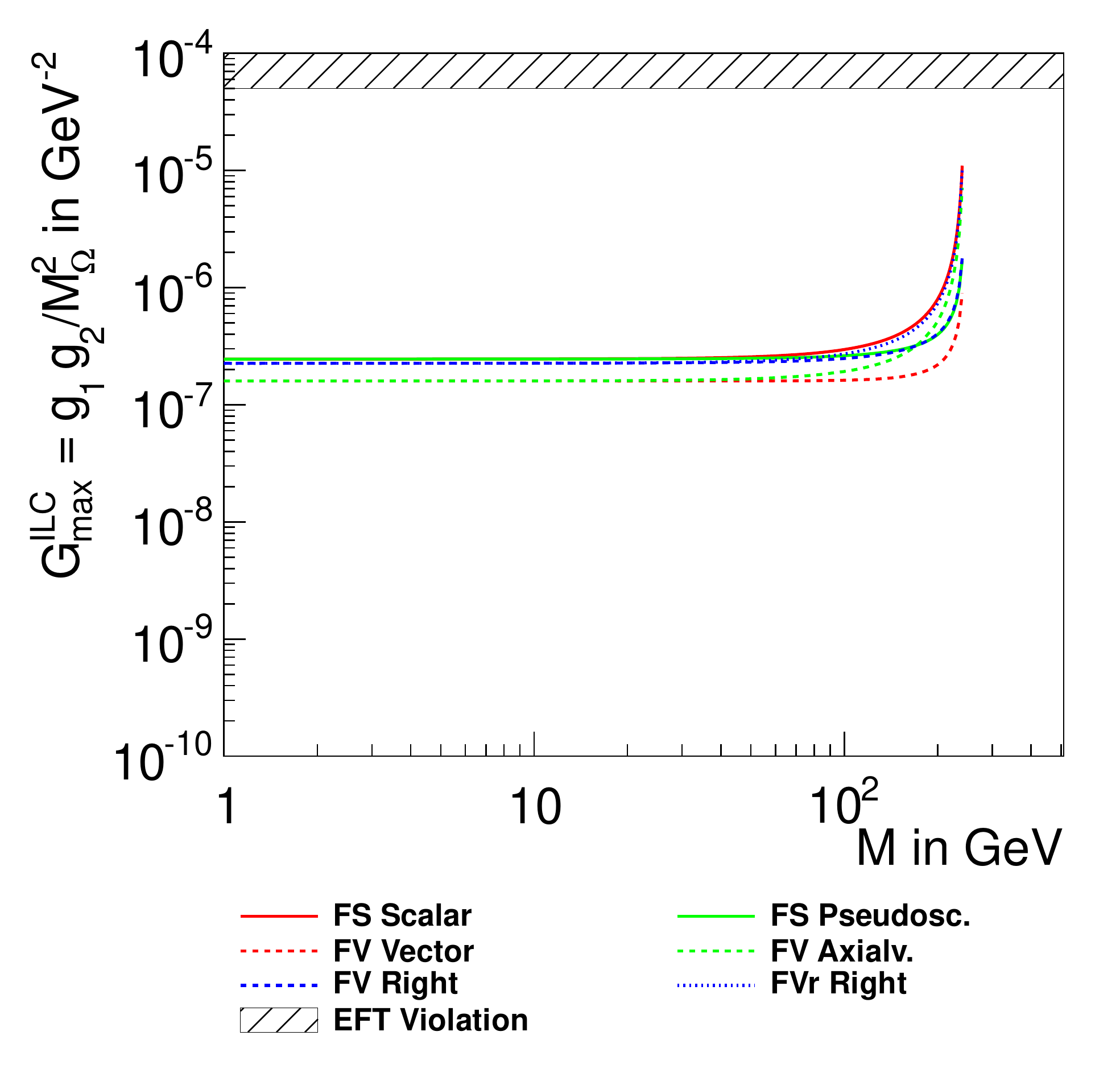} \\ \vspace{-0.25cm}
\includegraphics[width=0.49\textwidth]{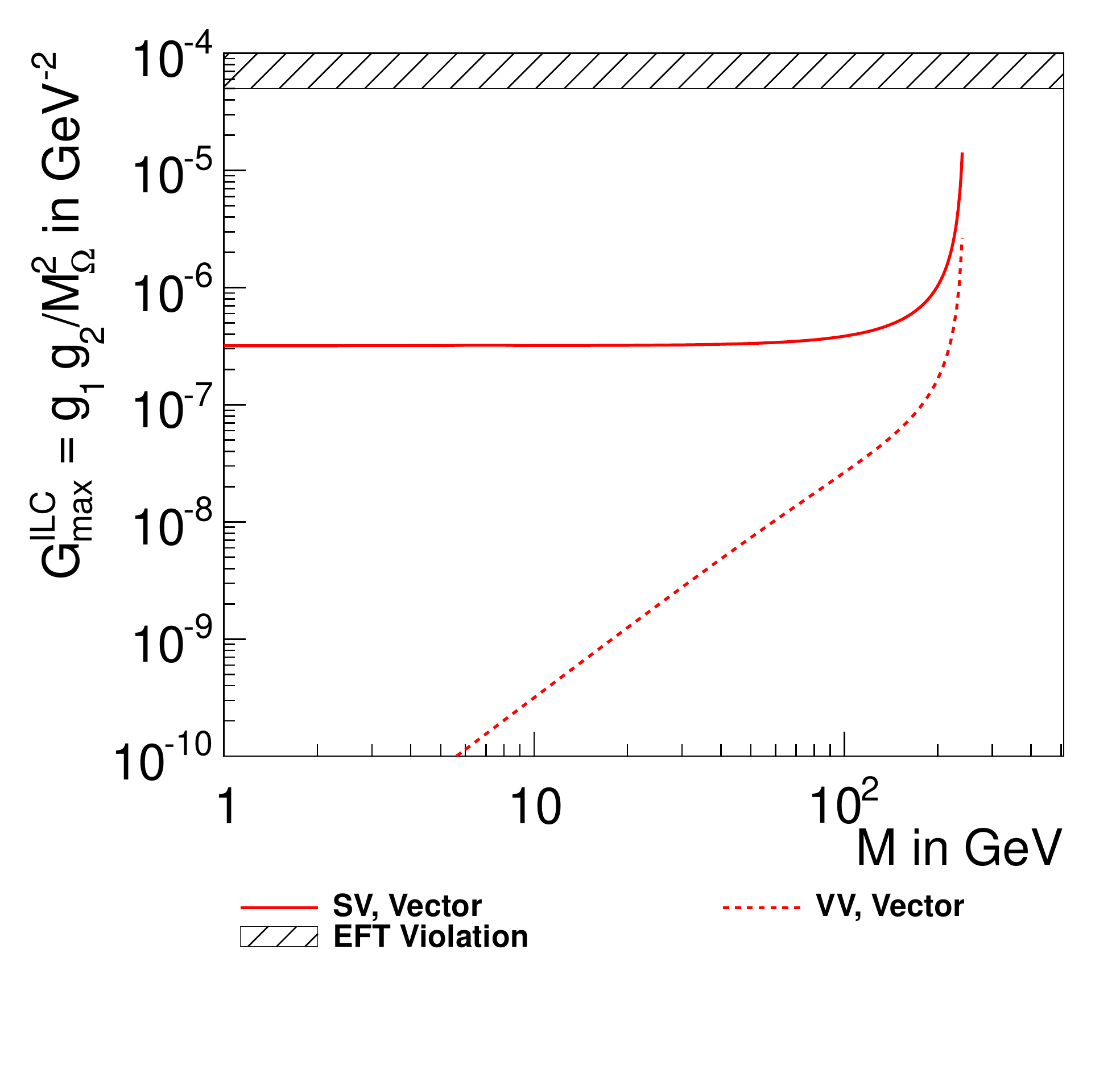} \hfill
\includegraphics[width=0.49\textwidth]{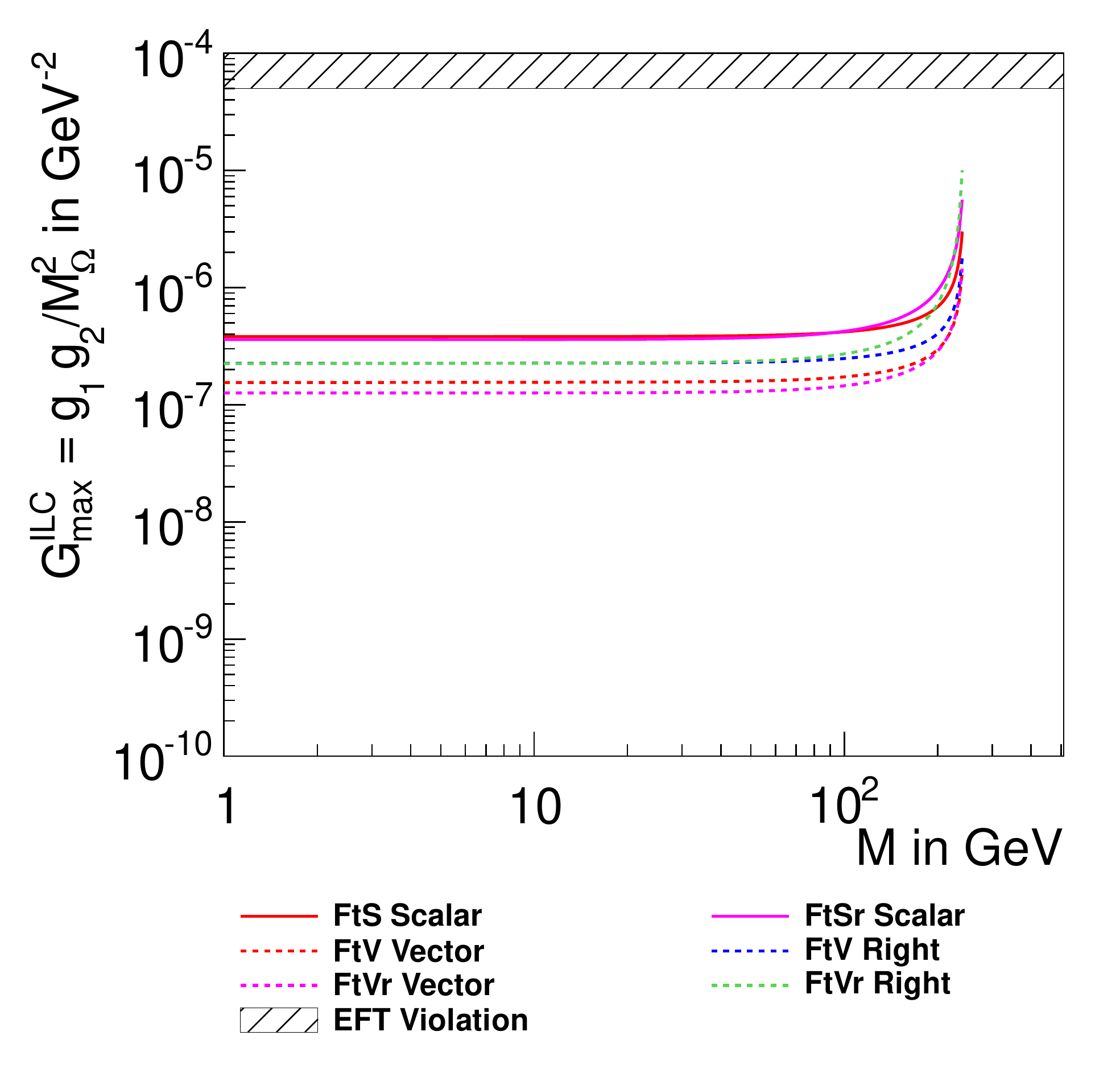} \\\vspace{-0.25cm}
\includegraphics[width=0.49\textwidth]{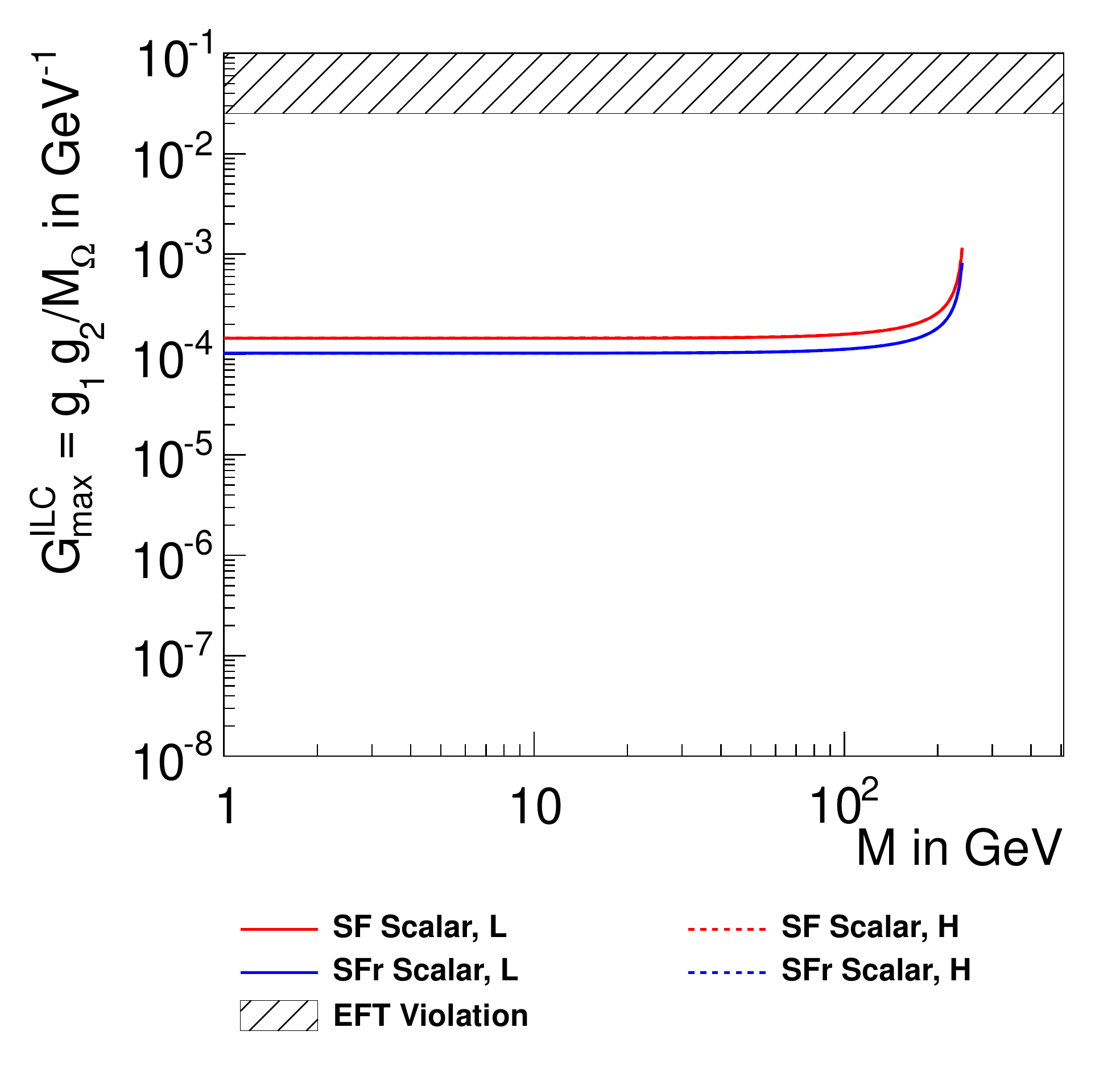} \hfill
\includegraphics[width=0.49\textwidth]{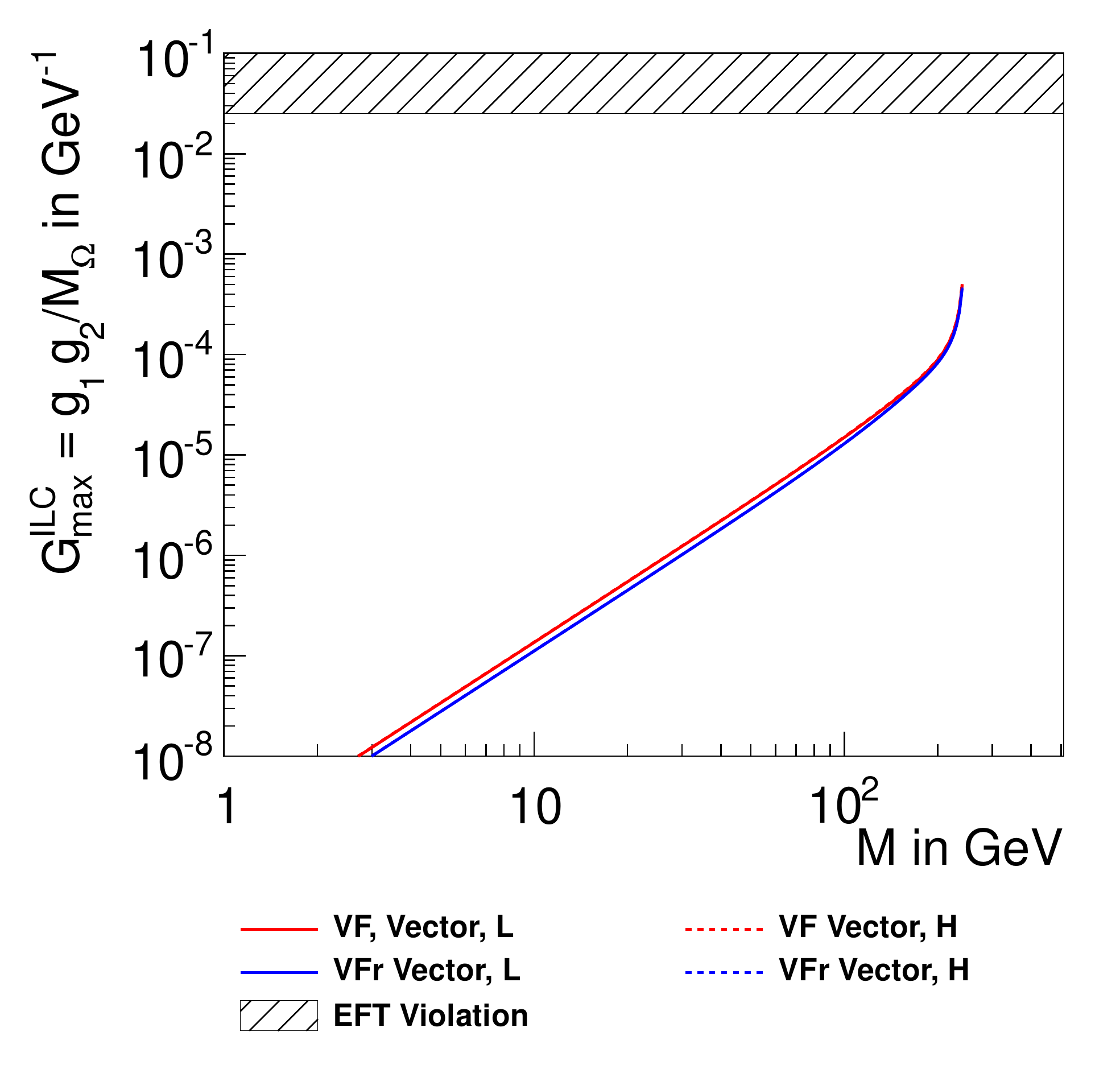} \\\vspace{-0.25cm}
\caption{\unit{90}{\%} exclusion limits on all effective couplings accessible at the \textsc{Ilc} with $\sqrt{s} =
  \unit{500}{\GeV}$.}
\label{img:ilcbounds500gev}
\end{figure}

\begin{figure}
\centering
\includegraphics[width=0.49\textwidth]{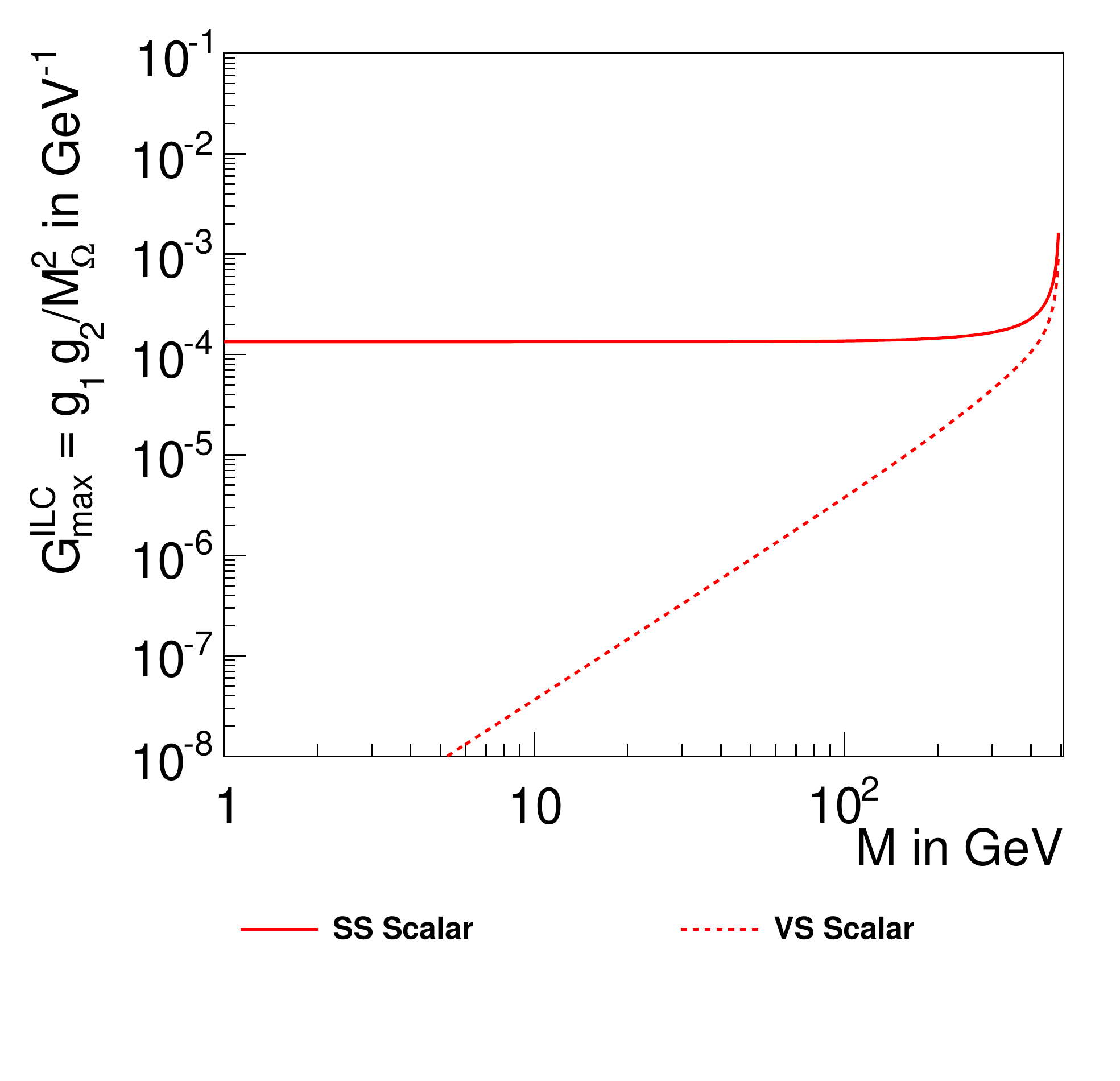} \hfill
\includegraphics[width=0.49\textwidth]{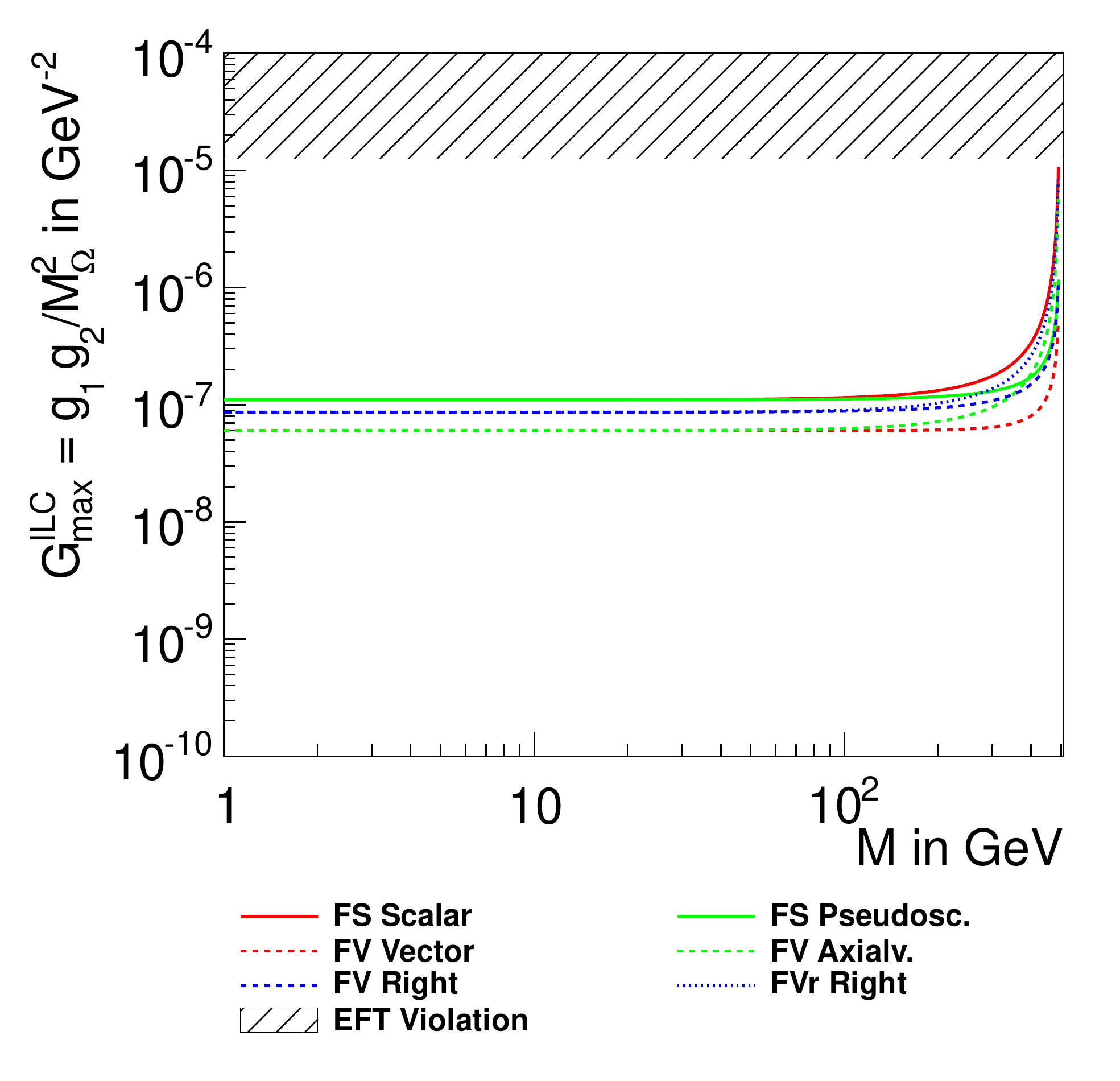} \\ \vspace{-0.25cm}
\includegraphics[width=0.49\textwidth]{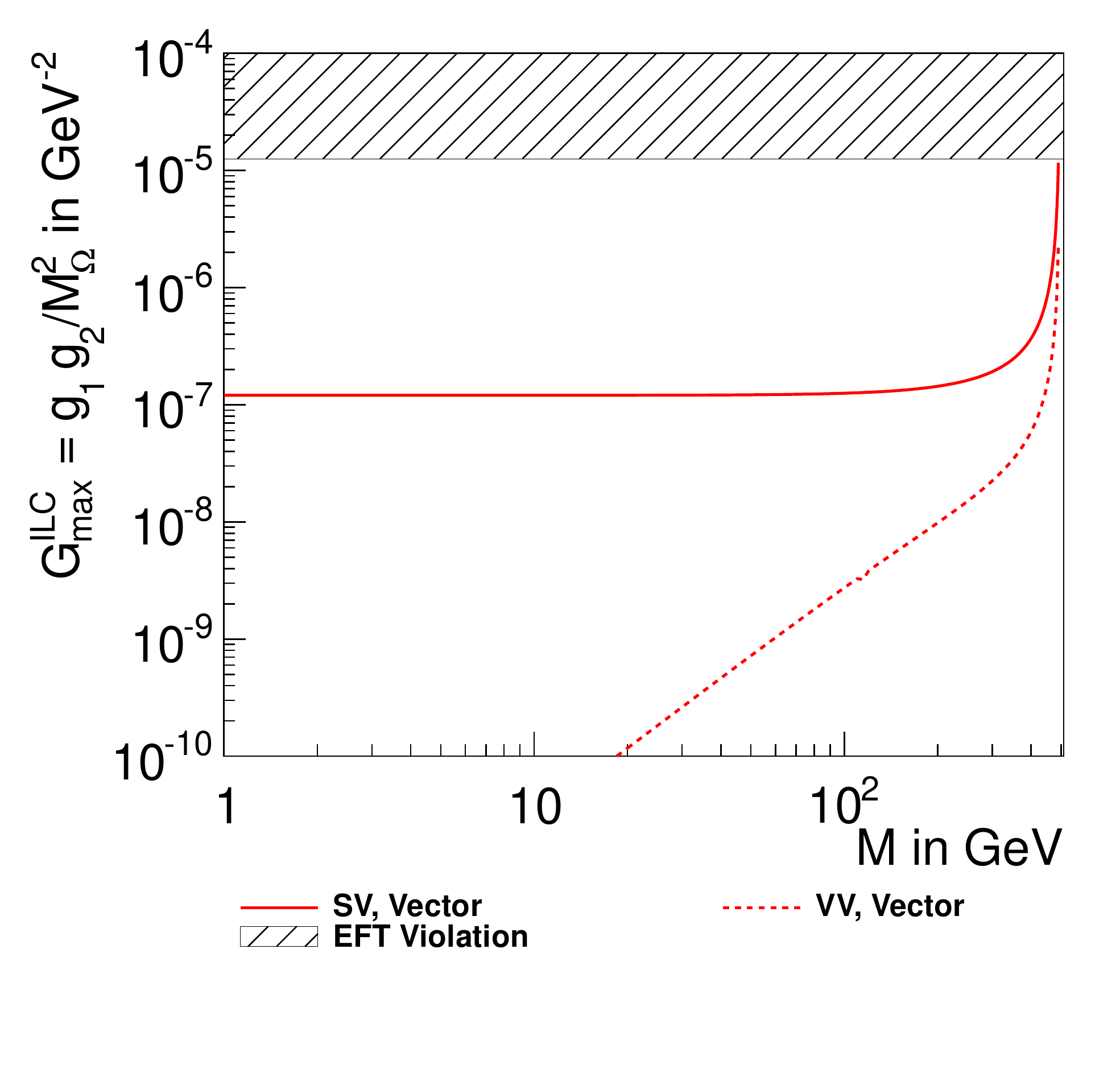} \hfill
\includegraphics[width=0.49\textwidth]{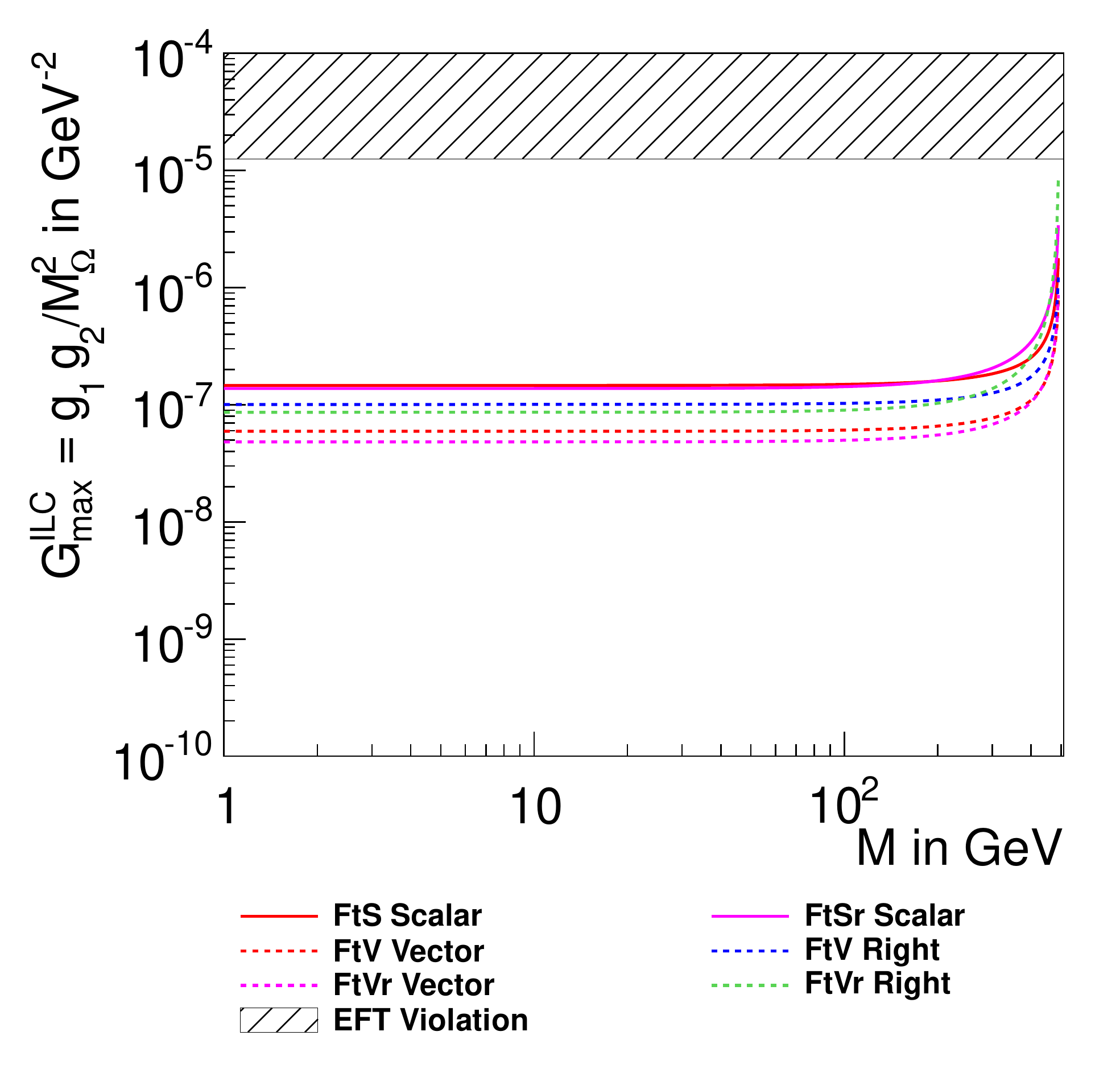} \\ \vspace{-0.25cm}
\includegraphics[width=0.49\textwidth]{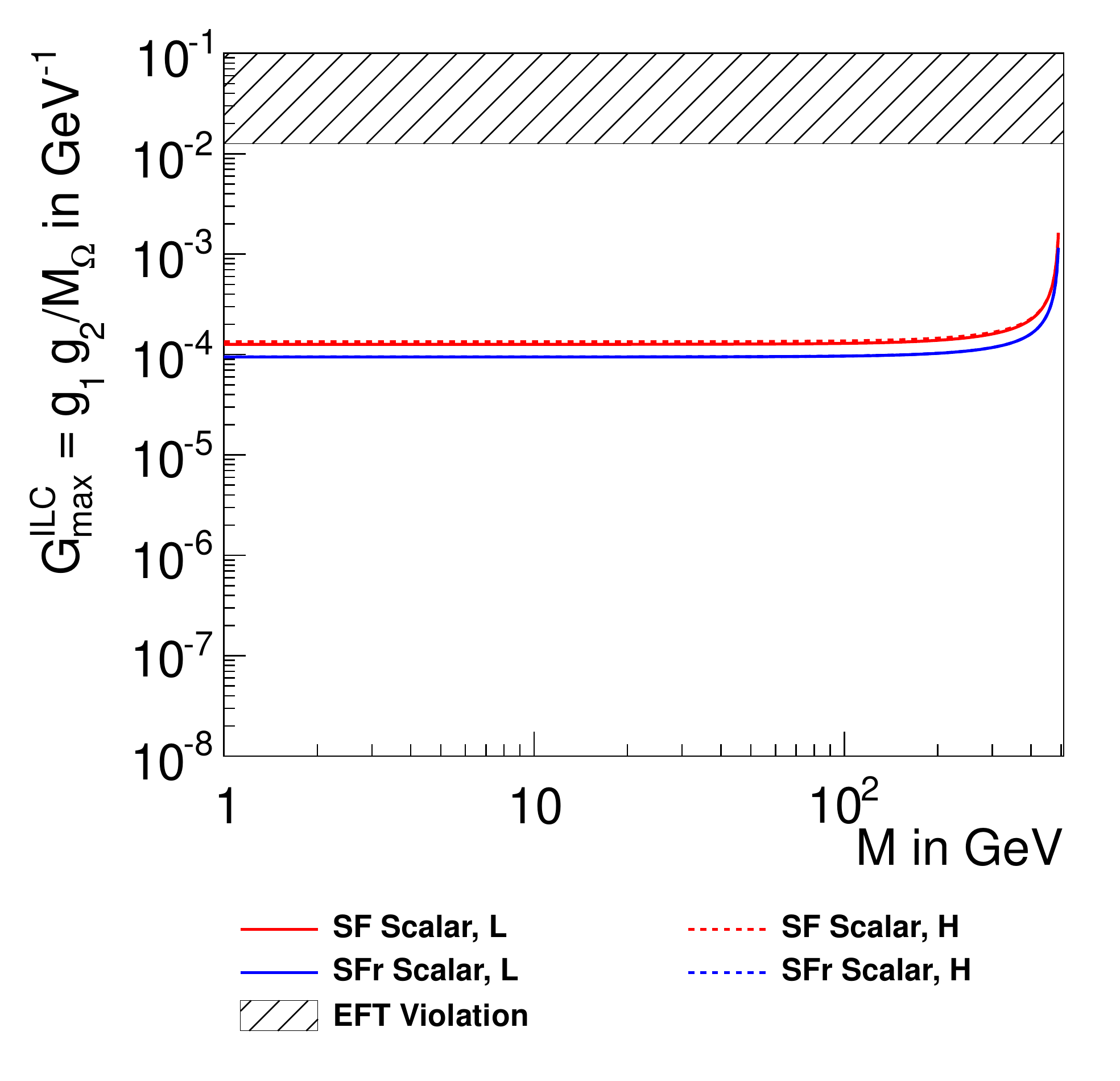} \hfill
\includegraphics[width=0.49\textwidth]{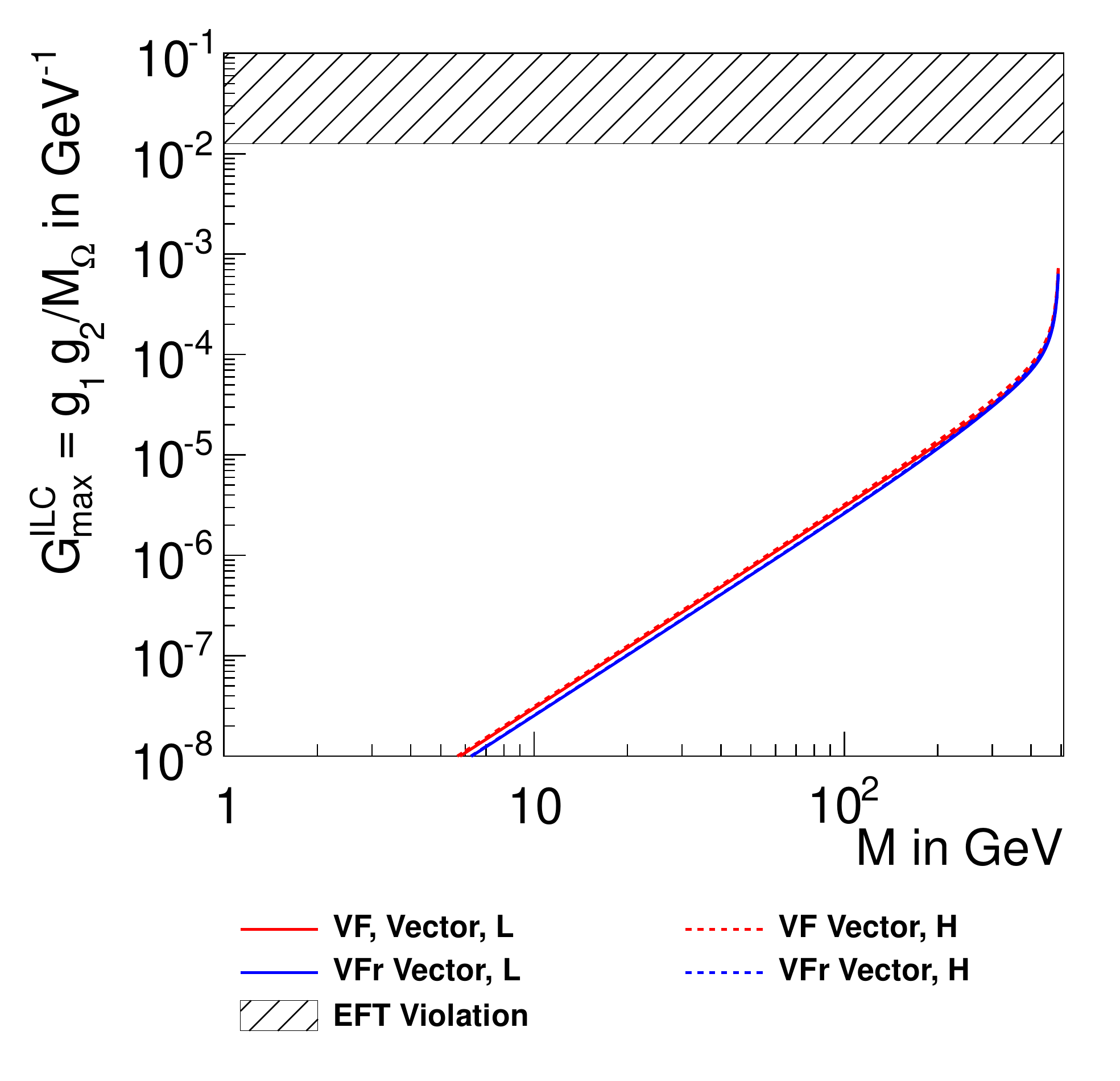}\\ \vspace{-0.25cm}
\caption{\unit{90}{\%} exclusion limits on all effective couplings accessible at the \textsc{Ilc} with $\sqrt{s} =
  \unit{1}{\TeV}$.}
\label{img:ilcbounds}
\end{figure}
\chapter{Limits on Direct Dark Matter Detection}
\label{chap:xenon}
The final step of the analysis of our \textsc{Wimp} interaction models is to compare the expected \textsc{Ilc} bounds on the effective coupling to the current direct detection limits on cross sections for dark matter--proton scattering $\sigma_P$. We use results from the \textsc{Xenon} experiment \cite{Xenon100}, which currently gives the strongest exclusion limits for dark matter masses above \SI{10}{\GeV}. In section \ref{sec:xenongeneral} we briefly summarise the main physics behind the \textsc{Xenon} project and how the current bounds on elastic proton scattering are established. We then describe in section \ref{sec:ilctoddxsect} how we can translate our \textsc{Ilc} and \textsc{Wmap} limits on \Geff into upper bounds on $\sigma_P$ in order to compare the excluded areas for all three experiments in section \ref{sec:xenonanalysis}.

\section{Direct Detection with \textsc{Xenon}}
\label{sec:xenongeneral}
\begin{figure}
\centering
\includegraphics[width=0.6\textwidth]{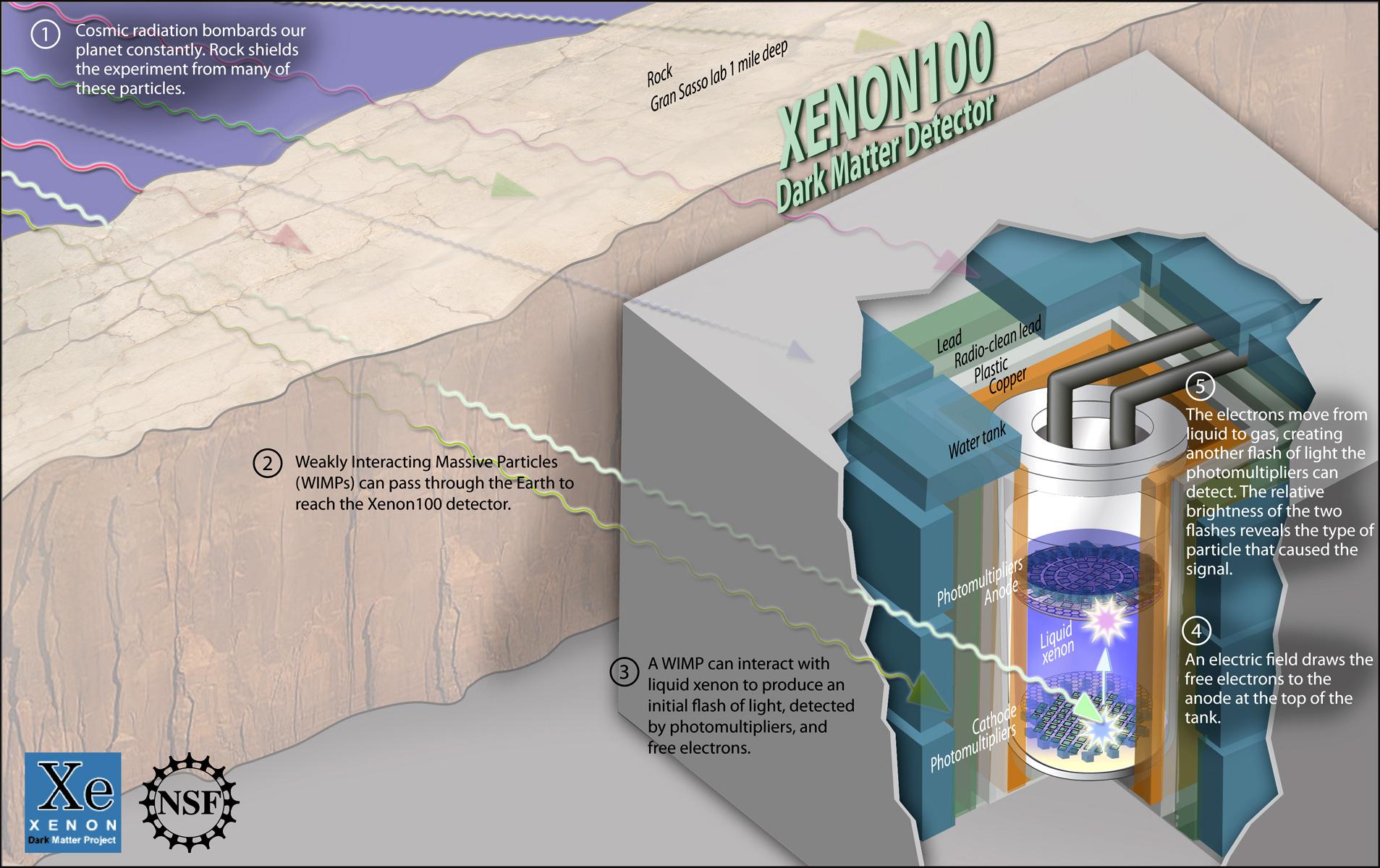}
\caption{Schematic picture of the \textsc{Xenon} experiment \cite{xenondesignbild}.}
\label{img:xenon}
\end{figure}
The \textsc{Xenon} Dark Matter Project aims to find \textsc{Wimp}s that are present in the vicinity of our Earth by looking for elastic scattering with an atom inside a tank of liquid xenon. The experiment is located at the Gran Sasso underground laboratory with a water equivalent depth at \SI{3100}{\meter} to reduce background from cosmic radiation. The project started in 2007 with its first phase, using \SI{15}{\kilo\gram} of liquid xenon and registering events for \SI{58.6}{days} \cite{Xenon10General}. Currently phase 2 is running with \SI{161}{\kilo\gram} and the most recent results given for \SI{225}{days} of data taking \cite{Xenon100General}. The collaboration is already designing the final phase 3 experiment, increasing the total xenon mass to around \SI{1000}{\kilo\gram} \cite{Xenon1TGeneral}.

Whenever a dark matter particle happens to interact with one of the atoms inside the tank, excitation light will be emitted and measured by photo tubes, thus obtaining information about the recoil energy $E_R$ of the scattered nucleus. Let $R$ be the event rate per unit mass of the detector. The differential rate with respect to $E_R$ can be expressed as
\begin{align}
\left. \abl{R}{E_R} \right|_{\text{observed}} \hspace{-0.8cm}= R_0 S(E_R) F^2(E_R) I,
\end{align}
where $R_0$ is the total event rate. $S(E_R)$ denotes the spectral function, which takes into account energy shifts like Doppler effects due to the earth's movement, detection efficiencies and resolution effects. $F(E_R)$ denotes form factor corrections due to the intrinsic structure of the nucleus, which is extracted from the interaction cross section
\begin{align}
\sigma \equiv \frac{\sigma_0}{4 \mu^2 v^2} \int \de q^2 F^2(q) \label{eqn:formfactor}
\end{align}
with $\sigma_0$ denoting the cross section at zero momentum transfer. Here,
 \begin{align}
\mu \equiv M_\chi M_N / (M_\chi + M_N) \label{eqn:themu}
\end{align}
denotes the reduced mass of the \textsc{Wimp}--nucleus system. $I$ describes an energy independent interaction function taking into account general modifying factors like coherent enhancements. Analytical estimates for those functions are given in \cite{XenonCalc}.

Due to the non-observation of a significant number of signal events, \textsc{Xenon} is able to set upper bounds on $\sigma_0$. In general these bounds are given for two classes of interactions: Firstly for the case of spin--independent scattering, the dark matter particle can interact with all nucleons inside the xenon--atom, which enhances the individual nucleon cross section by a factor of $A^2 \approx \num{3e3}$. Since nucleon spins tend to anti--align inside the nucleus when occupying the lowermost energy state, this enhancement does not occur in the case of spin--dependent interactions.

\section{Proton Scattering via a Tree--Level Interaction with Quarks}
\label{sec:ilctoddxsect}
\textsc{Xenon} gives exclusion bounds on the
dark matter--proton cross sections, but starting from a fundamental interaction
theory, one usually only knows the interaction Lagrangian of dark matter with quarks. To
compare with experiment, one has to translate this information into matrix
elements for dark matter proton interaction by using nuclear form factors. We will now show a method to do this
for a large class of interactions which has been discussed in \cite{DMClass} for a small subset of the models we consider here.

\subsection*{Rewriting the Matrix Element}
We start with a general four--particle--interaction of the type $G_q \bar{q} \Gamma q
\bar{\chi} \Gamma \chi$ between the dark matter particles $\chi$
and the quark fields $q$. For now, the operator $\Gamma$ is completely general and may even by different in the quark and the dark matter bilinear. We also allow a coupling constant $G_q$ that may
be different for each quark type, for example as it would be in a Yukawa--like scenario. The matrix element is then given by the transition amplitude from incoming to
outgoing states. In our case this is  an incoming dark matter particle $\chi$
with 4--momentum $p$, a proton $P$ at rest with 4--momentum $k$ and the respective
outgoing particles with 4--momenta $p^\prime$ and $k^\prime$:
\begin{align}
i \mathcal{M} \left(p+k\rightarrow p^\prime+k^\prime\right) \left( 2 \pi \right)^4 \delta^{4}\left(p+k-p^\prime-k^\prime \right)  =
  \langle \chi(p^\prime) P(k^\prime) \mid i T \mid \chi(p) P(k) \rangle.
\end{align}
To leading order the $T$--matrix elements are obtained from the interaction
  Lagrangian. Since our operator is already factorised into a quark and a dark matter bilinear, we  can separate the quark transition from the quark--nucleon matrix element:
\begin{align}
\langle \chi(p^\prime) P(k^\prime) \mid T \mid \chi(p) P(k) \rangle 
&=  \int \de^4 x  \sum_q G_q \ \langle \chi(p^\prime) \mid \bar{\chi}(x)
\Gamma \chi(x)\mid \chi(p) \rangle \langle P(k^\prime) \mid \bar{q}(x) \Gamma
  q(x) \mid  P(k) \rangle. \nonumber
\end{align}
We use the translation operator, generated by the momentum operator $\hat{P}$,
to shift the fields to the interaction point $x=0$. 
\begin{align}
\langle \chi(p^\prime) \mid \bar{\chi}(x) \Gamma \chi(x)\mid \chi(p) \rangle &=
\langle \chi(p^\prime) \mid e^{i \hat{P} x} \bar{\chi}(0) e^{-i \hat{P}
  x} \Gamma e^{i \hat{P} x} \chi(0) e^{-i \hat{P} x}\mid \chi(p) \rangle \\
&= e^{i (p^\prime-p) x} \langle \chi(p^\prime) \mid  \bar{\chi}(0) \Gamma  \chi(0)
\mid \chi(p) \rangle. \\
\intertext{We do the same for the proton matrix element. Due to the small
  velocity of the free dark matter particles, the momentum transfer during the
  scattering process will be small compared to the masses of the dark matter particle and the nucleus. It is therefore reasonable to consider
  all incoming and outgoing particles to be at rest:}
\langle P(k^\prime) \mid \bar{q}(x) \Gamma q(x)\mid P(k) \rangle &= e^{i
  (k^\prime-k) x} \langle P(0) \mid  \bar{q}(0) \Gamma  q(0)
\mid P(0) \rangle.
\end{align}
Putting the pieces together and performing the space integral $\int \de^4 x \
e^{i p x} = (2 \pi)^4 \delta^4(p)$, we receive the following approximate
formula for the matrix element:
\begin{align}
\mathcal{M} \approx \sum_q G_q \ \langle \chi \mid  \bar{\chi} \Gamma \chi
\mid \chi \rangle \langle P \mid  \bar{q} \Gamma q \mid P \rangle.
\end{align}
\subsection*{Non--Relativistic Dark Matter}
We first want to evaluate the matrix elements for non--relativistic dark matter in case of different operators $\Gamma$. In case of a scalar \textsc{Wimp}, external fields trivially give $1$ and momenta can be approximated as $p^\mu = M_\chi \delta^\mu_0$. This leads to the following matrix elements:
\begin{align}
\left< \chi | \ \chi^\dagger \chi \ | \chi \right> &= 1, \\
i \left< \chi |\  \chi^\dagger \partial_\mu\chi \ | \chi \right> &= M_\chi \delta_\mu^0, \label{eqn:derivinmatrix}\\
i \left< \chi |\  \chi \partial_\mu\chidag \ | \chi \right> &= - M_\chi \delta_\mu^0.
\end{align}
For fermion dark matter, external fields give the free spinor functions $u(p)$ (see appendix \ref{sec:appdirac}). In the non--relativistic case we can use their explicit representation to calculate the bilinears as follows:
\begin{align}
\Big< \chi^{s^\prime} \Big| \ \bar{\chi} \chi \ \Big| \chi^s \Big> &= \bar{u}^{s^\prime}(0) u^s(0) = 2 M_\chi \delta^{s s^\prime}, \\
\Big< \chi^{s^\prime} \Big| \ \bar{\chi} \gamma^5 \chi \ \Big| \chi^s \Big> &=0, \label{eqn:axialcurrent} \\
\Big< \chi^{s^\prime} \Big| \ \bar{\chi} \gamma^\mu \chi \ \Big| \chi^s \Big> &= 2 M_\chi \delta^{s s^\prime} \delta^\mu_0, \label{eqn:vectorcurrent}\\
\Big< \chi^{s^\prime} \Big| \ \bar{\chi} \gamma^\mu \gamma^5 \chi \ \Big| \chi^s \Big> &= 2 M_\chi \delta^\mu_i \xi^{\dagger s^\prime} \sigma^i \xi^s, \label{eqn:axialvectorcurrent} \\
\Big< \chi^{s^\prime} \Big| \ \bar{\chi} \sigma^{\mu \nu} \chi \ \Big| \chi^s \Big> &= 2 M_\chi \delta^\mu_i \delta^\nu_j \epsilon^{ijk} \xi^{\dagger s^\prime} \sigma_k \xi^s.
\end{align}

The $\sigma^{\mu \nu}$ term may appear in t--channel interactions which can be reformulated into a sum of s--channel operators by using Fierz' identities. However, this sum of operators will always include either an additional scalar or vector interaction which both are spin--independent. As we argued in the last paragraph, spin independent interactions are always enhanced by multiple orders of magnitude, such that spin--dependent contributions can always be neglected as soon as they appear in a sum with spin--independent interactions. This is why we do not discuss tensor interactions any further from here on.

In the final step we want to evaluate matrix elements for vector dark matter, for which we need the corresponding polarisation vectors $\epsilon_\mu^s(k)$ for the external fields. These are defined by the following properties:
\begin{align}
\epsilon^s_\mu(k) \epsilon^{s^\prime\mu}(k) = -\delta^{s s^\prime}, \qquad k^\mu \epsilon^s_\mu(k)  = 0. \label{eqn:app:ddxsects:1}
\end{align}
For a massive particle at rest, we have $k^\mu = (M, 0, 0, 0)$ such that we can easily give three vectors that fulfil (\ref{eqn:app:ddxsects:1}):
\begin{align}
\epsilon_\mu^1 = \begin{pmatrix}
0 \\ -1 \\ 0 \\ 0
\end{pmatrix}, \qquad
\epsilon_\mu^2 = \begin{pmatrix}
0 \\ 0 \\ -1 \\ 0
\end{pmatrix}, \qquad
\epsilon_\mu^3 = \begin{pmatrix}
0 \\ 0 \\ 0 \\ -1
\end{pmatrix}.
\end{align}
In short form we may write $\epsilon_\mu^s = g_{\mu s}$, which allows us to write matrix elements for vector particles as
\begin{align}
\Big< \chi^i \Big| \ \chi^\dagger_\mu \chi_\nu \ \Big| \chi^j \Big> &= \epsilon^i_{\mu} \epsilon^j_{\nu} = g_{\mu i} g_{\nu j}
\end{align}
plus cases with additional derivatives $\partial_\rho = M_\chi \delta_\rho^0$.

\subsection*{Quark--Nucleon Form Factors}
We now continue with the evaluation of the low energy quark--proton matrix elements \linebreak
$\left< P \mid  \bar{q} \Gamma q \mid P \right> $. These can approximately be evaluated through lattice calculations, effective nuclear theories and nucleon scattering experiments. We will need results for $\Gamma \in \left[ 1, \gamma^5, \gamma^\mu, \gamma^5 \gamma^\mu \right]$ to give matrix elements for all benchmark models defined in table \ref{tbl:constraints}. We use the standard normalisation for external fields in non--relativistic field--theories $\displaystyle \langle P | P \rangle \equiv 2 M_P \langle \tilde{P} | \tilde{P} \rangle$ and give the matrix elements in terms of the new states $| \tilde{P} \rangle$.  If not mentioned explicitly, spin indices combine to $\delta^{s s^\prime}$. Numerical values for the different form factors we use in this section can be found in appendix \ref{sec:constants} and are taken from \cite{fnumbers, deltanumbers}.

The scalar quark current counts the number of valence-- and sea--quark pairs inside the proton. Up--, down-- and strange--quarks contribute to that number directly, whereas other quark types contribute only virtually through a gluon loop with a numerically different form:
\begin{align}
\langle \tilde{P} \mid\bar{q} q\mid \tilde{P} \rangle &\approx 
\left\{
     \begin{array}{lr}
\frac{M_p}{m_q} f_q^p \qquad &\text{for u--, d-- and s--quark} \\
\frac{M_p}{m_q} \frac{2}{27} \left(1 - f_u^p - f_d^p - f_s^p \right) \qquad & \text{for c--, b-- and t--quark}
     \end{array}
   \right.  \label{eqn:scalarcurrent} \\
\intertext{Axial quark currents always vanish in the non--relativistic limit [see (\ref{eqn:axialcurrent})]:}
\langle \tilde{P} \mid\bar{q} \gamma^5 q\mid \tilde{P} \rangle &\approx 0 .\\
\intertext{The vector current only couples to the up and down quarks inside the proton. This is analogous to the photon that mainly couples to the charge constituents of the nucleon, i.e. the valence quarks.}
\langle \tilde{P} \mid\bar{q} \gamma^\mu q\mid \tilde{P} \rangle &\approx \left\{
     \begin{array}{lr}
2 \delta^\mu_0 \qquad& \text{for u--quark,} \\
  \delta^\mu_0 \qquad& \text{for d--quark.} \\
0 \qquad& \text{else}
     \end{array}
\right. \label{eqn:vectorcurrentmatrixelement}\\
\intertext{The axialvector current couples spin dependently [see (\ref{eqn:axialvectorcurrent})]. It therefore counts the spin--contribution for all (valence and sea) quarks, which is only significantly large for up--, down-- and strange--quarks. This current is the only one that leads to spin--dependent interactions.}
\langle \tilde{P}^{s^\prime} \mid\bar{q} \gamma^\mu \gamma^5 q\mid \tilde{P^s} \rangle &\approx \Delta_q^N \delta^\mu_i \xi^{\dagger s^\prime} \sigma^i \xi^s.
\end{align}

\subsection*{Elastic Cross Section}
Knowing the matrix element, the total cross section for non--relativistic particles at zero momentum transfer is then given as follows:
\begin{align}
\sigma_0 = \frac{1}{4 \pi (M_P + M_\chi)^2} \frac{1}{2 (2 s_\chi + 1)} \sum_{\text{spins}} \left| \mathcal{M}_{if} \right|^2.
\end{align} 
The $s_\chi$ term averages the number of incoming spin configurations, which is $2$ for the proton and $2 s_\chi + 1$ for the dark matter particle of model--dependent spin. We give the final solutions for all benchmark models in appendix \ref{sec:nonrelmatrixelements}.
%
%
%
%
%

\section{Proton Scattering via Loop Interactions with Leptons}
\begin{figure}
\centering
\includegraphics[width=0.3\textwidth]{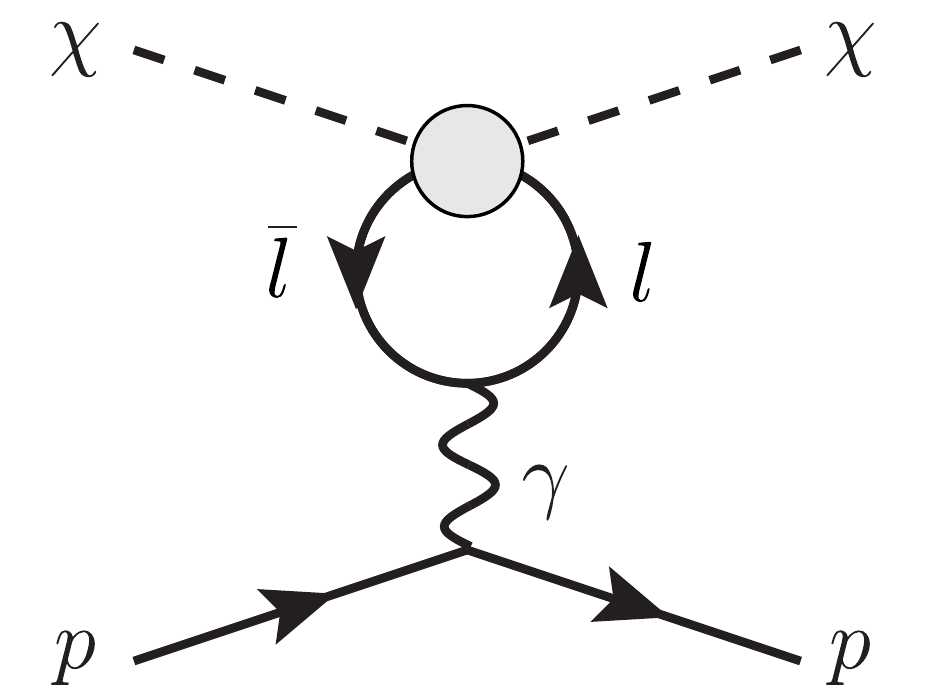}
\caption{Feynman--Diagram for lepton--only dark matter--proton interaction with the SV Vector model as an example.}
\label{img:loop}
\end{figure}

If we forbid quark--couplings, as discussed in section \ref{sec:benchmarks}, tree level scattering between dark matter and the proton cannot occur. In some cases, an interaction with the proton is still possible by higher order loop diagrams including a virtual lepton pair that couples electromagnetically according to figure \ref{img:loop} \cite{loopstatements}. However, this interaction will only give significant cross sections for a small subset of models: Axial vector currents will always give zero cross section at loop order because they lead to traces with an odd number of $\gamma^5$ matrices. Those either vanish directly or give terms proportional to $\epsilon^{\mu \nu \rho \sigma}$ that cannot contract with four independent Lorentz--vectors in an unpolarised $2 \rightarrow 2$ process and therefore will definitely give zero. Models with scalar currents need at least two photons to give traces with an even number of $\gamma^\mu$--matrices, so they can only happen at two loop order and are therefore negligible. 

Therefore, dark matter that couples to leptons only can only give significant proton--dark matter cross sections if they include vector currents. To be exact, an s--channel vector coupling $\bar{\psi} \gamma^\mu \psi$ is sufficent, such that all t--channel interactions have to be taken into account as long as their Fierz reformulation includes a vector part. 

In all these cases, the matrix elements from the tree level calculation get a loop--correction factor and change to\footnote{Note that there is a sign difference in \cite{LEPShinesLight}. We are confident in our result since it behaves regularly for $q \rightarrow 0$}. 
\begin{align}
\mathcal{M} &\approx \frac{\alpha_\text{em}}{9 \pi q^2} \left( q^2 g^{\mu \nu} - q^\mu q^\nu \right)  \hspace{-0.3cm} \sum_{l\ =\ e\text{, }\mu\text{, }\tau}  \hspace{-0.3cm} G^l \ \langle \chi \mid  \bar{\chi} \Gamma_\mu \chi
\mid \chi \rangle \langle P \mid  \bar{P} \gamma_\nu P \mid P \rangle \times F(q^2, m_l), \label{eqn:loopM}\\
F(q^2, m_l) &\equiv  \frac{1}{q^2} \left[12 m_l^2 + 5 q^2 -\left(12 m_l^2 + 6 q^2\right)  \beta_q \arcoth \beta_q  - 3 q^2 \ln \frac{m_l^2}{\Lambda^2}\right],
\end{align}
where we defined $\beta_q \equiv \sqrt{1-4m_l^2/q^2}$. $\Lambda$ gives the renormalisation scale and in a simple cutoff regularisation scheme it is defined as the scale at which new physics appears. Within the effective approach, this scale is given by the mass of the heavy mediator, which is why we assume $\Lambda$ to be \SI{1}{\TeV}. The actual value has only negligible impact on the result, as long as it is set around the \TeV{} scale. $m_l$ denotes the mass of the virtual lepton and $q$ the exchanged 4--momentum. In this scattering process, $q$ is always space--like ($q^2 < 0$) such that $\beta_q$ is real and the loop factor can be evaluated for all values of $m_l$ and $q^2$. We follow the approximation in \cite{LEPShinesLight} and conservatively assume a maximal scattering angle, leading to an angle--independent approximation $q^2 \approx - 4 \mu^2 v_\chi^2$. The reduced \textsc{Wimp} nucleus mass $\mu$ is given by (\ref{eqn:themu}) and we choose $v_\chi$ to be the local escape velocity of about \SI{500}{\km \per \second} for a \textsc{Wimp} in a typical dark matter halo.  

Even though we have worked with $q^2 \approx 0$ up to now and neglected it at tree level, it is necessary to include it within this calculation: The values for $\sigma_0$ given by the \textsc{Xenon}--collaboration are experimentally evaluated from the form factor corrected cross section $\sigma$ in (\ref{eqn:formfactor}) under the assumption that the interaction itself is momentum--independent and couples to all nucleons coherently. This was a reasonable approximation due to $|q^2| \ll M_\chi, M_N$, but it does not work for the new loop factor $F(q^2, m_l)$, since $m_l^2$ can be of the same order as $|q^2|$. We have to manually cancel this new $q$---dependence in the \textsc{Xenon}-data we use in order to compare like with like by rescaling their results as follows:
\begin{align}
\sigma_p^{\text{\textsc{Xenon}, Loop}} = \sigma_p^{\text{\textsc{Xenon}}} \left(\frac{\sum_l f(\tilde{q}^2, m_l)}{\sum_l f(q^2, m_l)} \cdot \frac{A}{Z} \right)^2 , \label{eqn:xenoncorrection}
\end{align}
with a different scattering momentum $\tilde{q} \equiv q(\mu = M_\chi M_P / (M_\chi + M_P))$. The $A/Z$ factor takes into account that the measured total cross section only needs to be divided among the protons, not all nucleons.
Our theoretical cross sections are now calculated by using protons as the external particles and using the non--relativistic vector current (\ref{eqn:vectorcurrent}) to evaluate the proton bilinear $\bar{\psi} \gamma^\mu \psi$. One also needs the non--relativistic time--component of the momentum exchange $q^0 \approx M_\chi - M_\chi = 0$. Apart from the new loop factor, the calculation is analogous to the tree level case, such that the cross sections look similar. We can conveniently relate the loop-- to the tree--results as follows: As argued before, only vector currents lead to non--vanishing loop--level results. After examining the results of the tree level calculation (see appendix \ref{sec:nonrelmatrixelements}), one finds that due to (\ref{eqn:vectorcurrentmatrixelement}), this current always lead to factors of either $B_P \equiv 2 G_\text{eff}^u + G_\text{eff}^d$ or $\tilde{B}_P \equiv B^p M_\chi + 2 G_\text{eff}^u m_u + G_\text{eff}^d m_d$. This term has to be replaced now by the loop contribution according to (\ref{eqn:loopM}), whereas terms from other currents have to vanish. We can summarise this prescription as follows:
\begin{align}
\sigma_0^\text{Loop} &= \frac{\alpha^2_\text{em}}{81 \pi^2}  \cdot
 \left\{
     \begin{array}{lr}
\displaystyle \left[\sum_l G_\text{eff}^l F(q^2, m_l) \right]^2  \left. \sigma_0^\text{Tree}\right|_{B_P = 1, F_P=D_P=0} \qquad& \text{if $\sigma_0^\text{Tree}$ contains $B_p$}, \\
\displaystyle \left[\sum_l (m_l+M_\chi) G_\text{eff}^l F(q^2, m_l) \right]^2 \left. \sigma_0^\text{Tree}\right|_{\tilde{B}_P = 1, F_P=D_P=0}  \qquad& \text{if $\sigma_0^\text{Tree}$ contains $\tilde{B}_p$}, \\
\quad & \quad \\
0 \qquad& \text{else.}
     \end{array}
\right.
\end{align}
\section{Combined Analysis}
\label{sec:xenonanalysis}
The full list of combined maximum
exclusion limits for dark matter--proton interaction at \textsc{Wmap}, \textsc{Xenon}
and the \textsc{Ilc} can be found in figures \ref{img:totalbounds1}-\ref{img:totalbounds14} listed in the appendix. We will give a subset of examples here to focus on particular properties.

\begin{figure}
\centering
 \includegraphics[width=0.49\columnwidth]{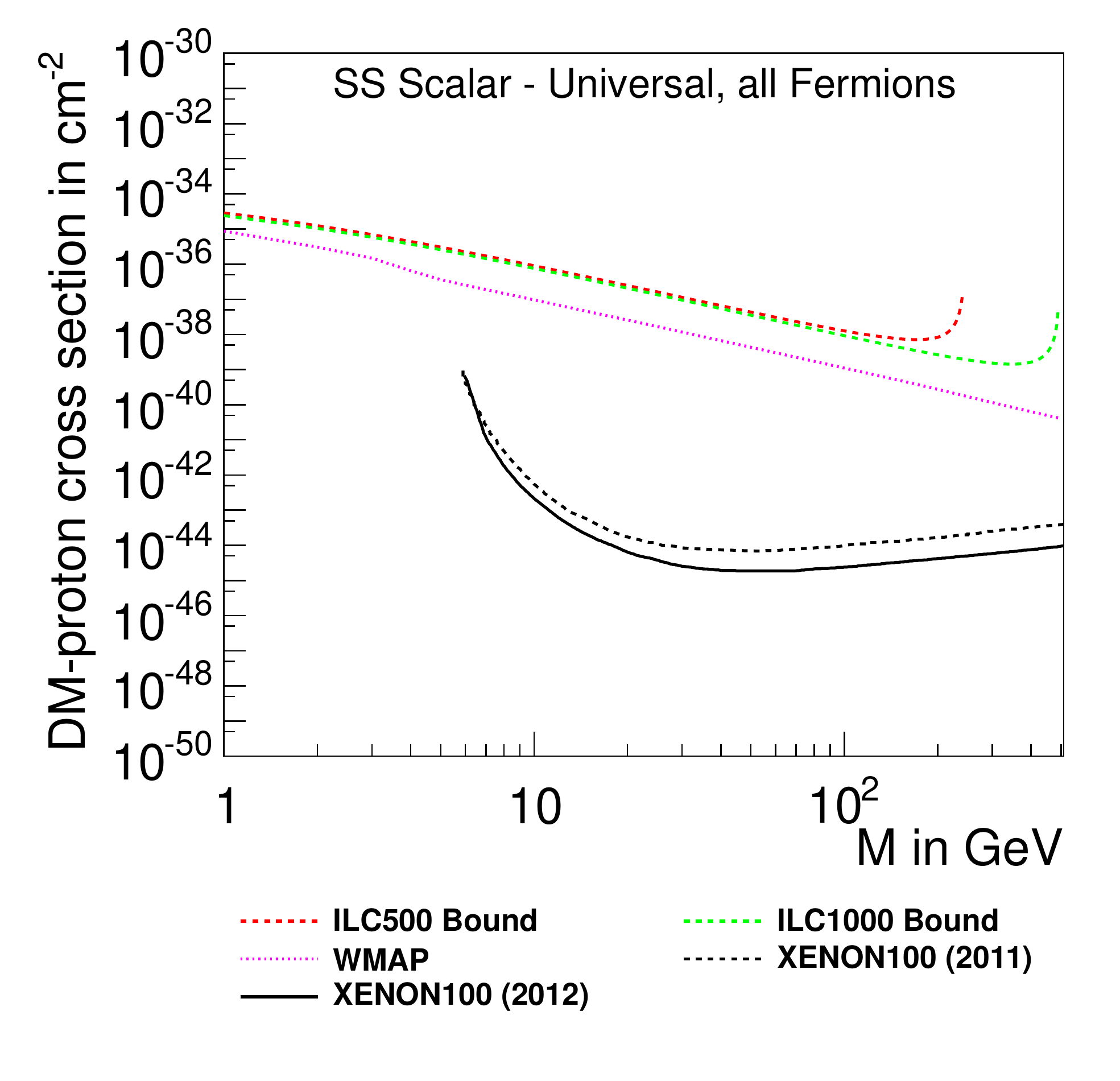} \hfill
 \includegraphics[width=0.49\columnwidth]{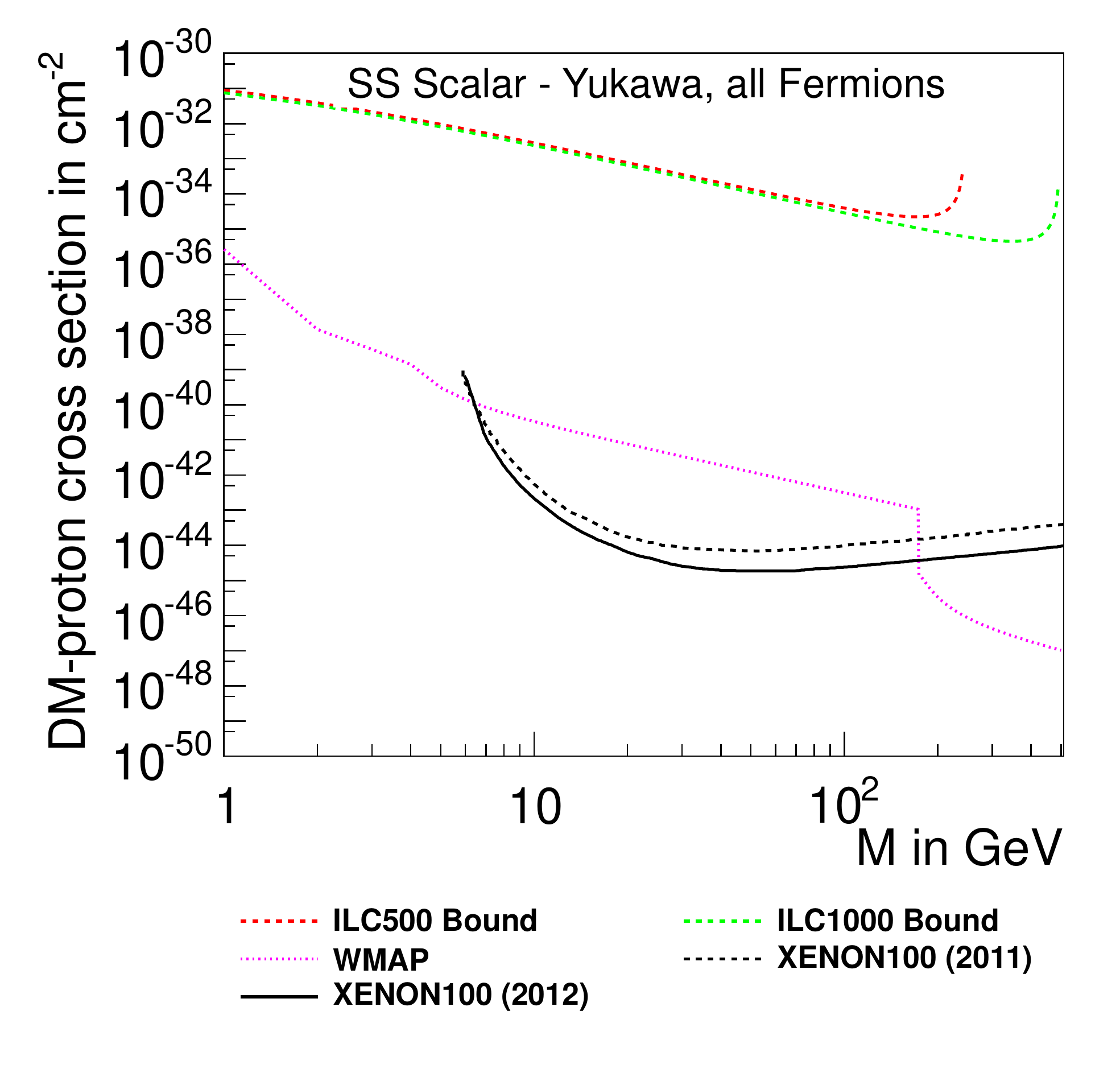} \\
 \includegraphics[width=0.49\columnwidth]{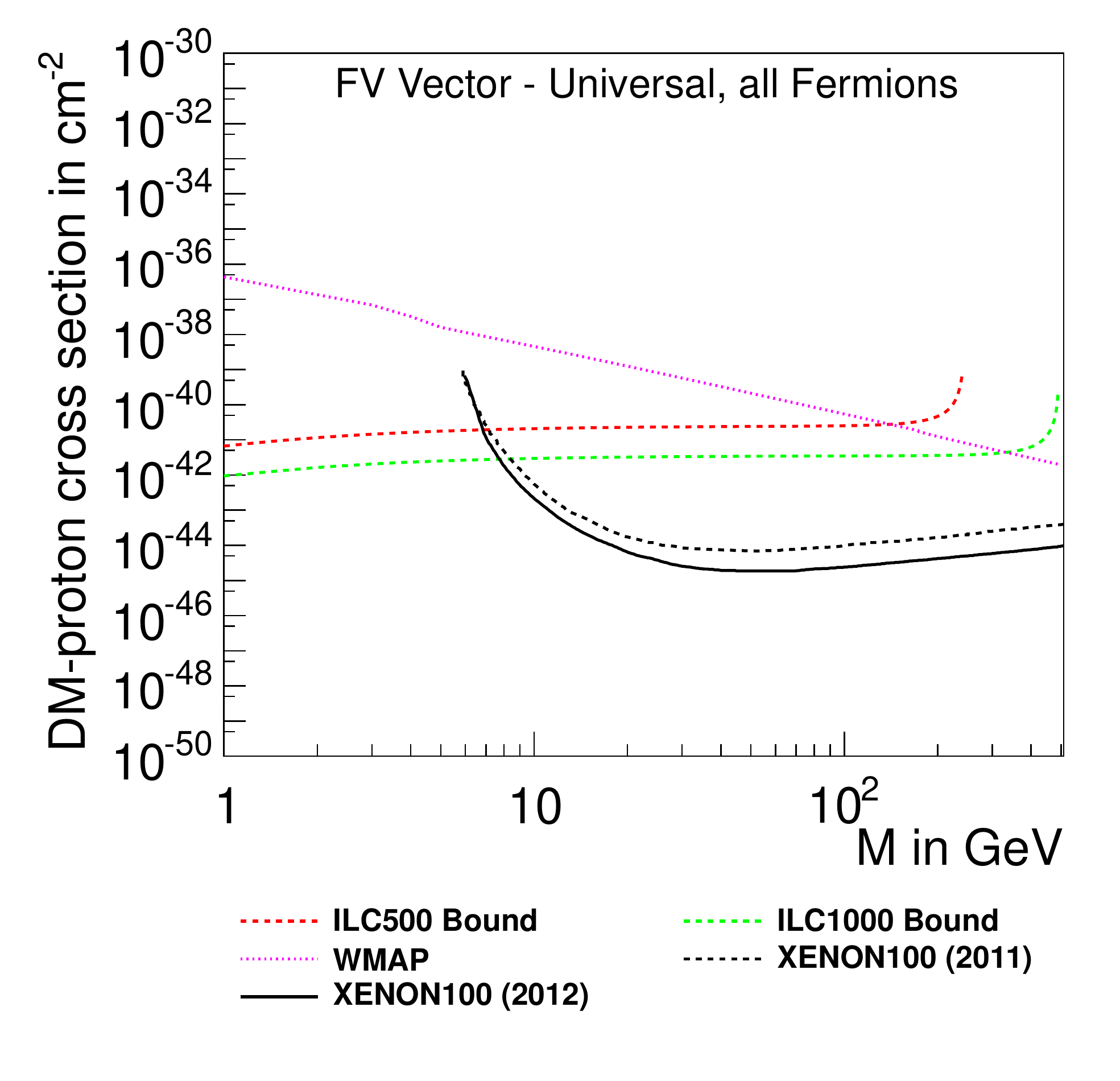} \hfill
 \includegraphics[width=0.49\columnwidth]{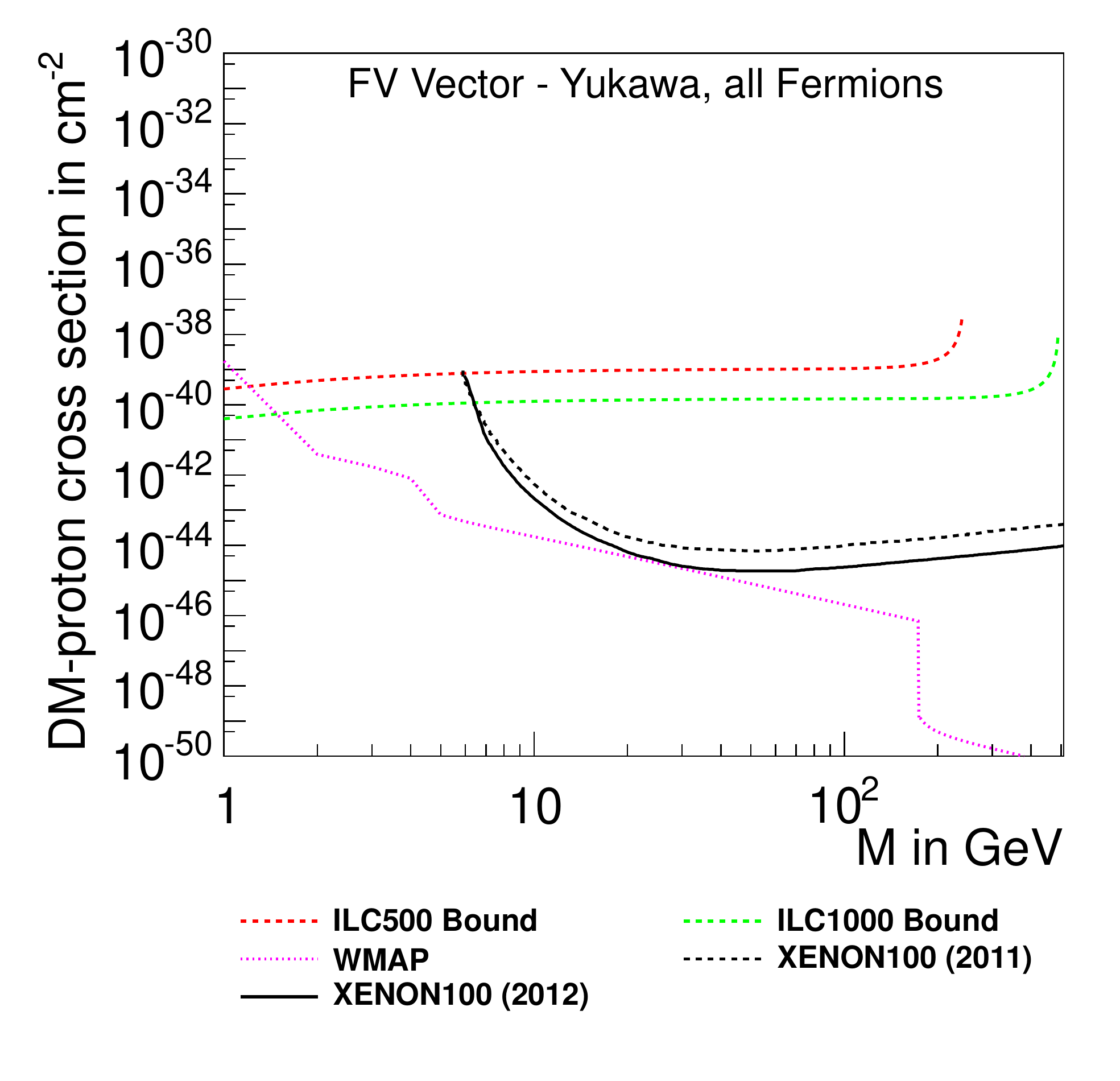} \\
 \includegraphics[width=0.49\columnwidth]{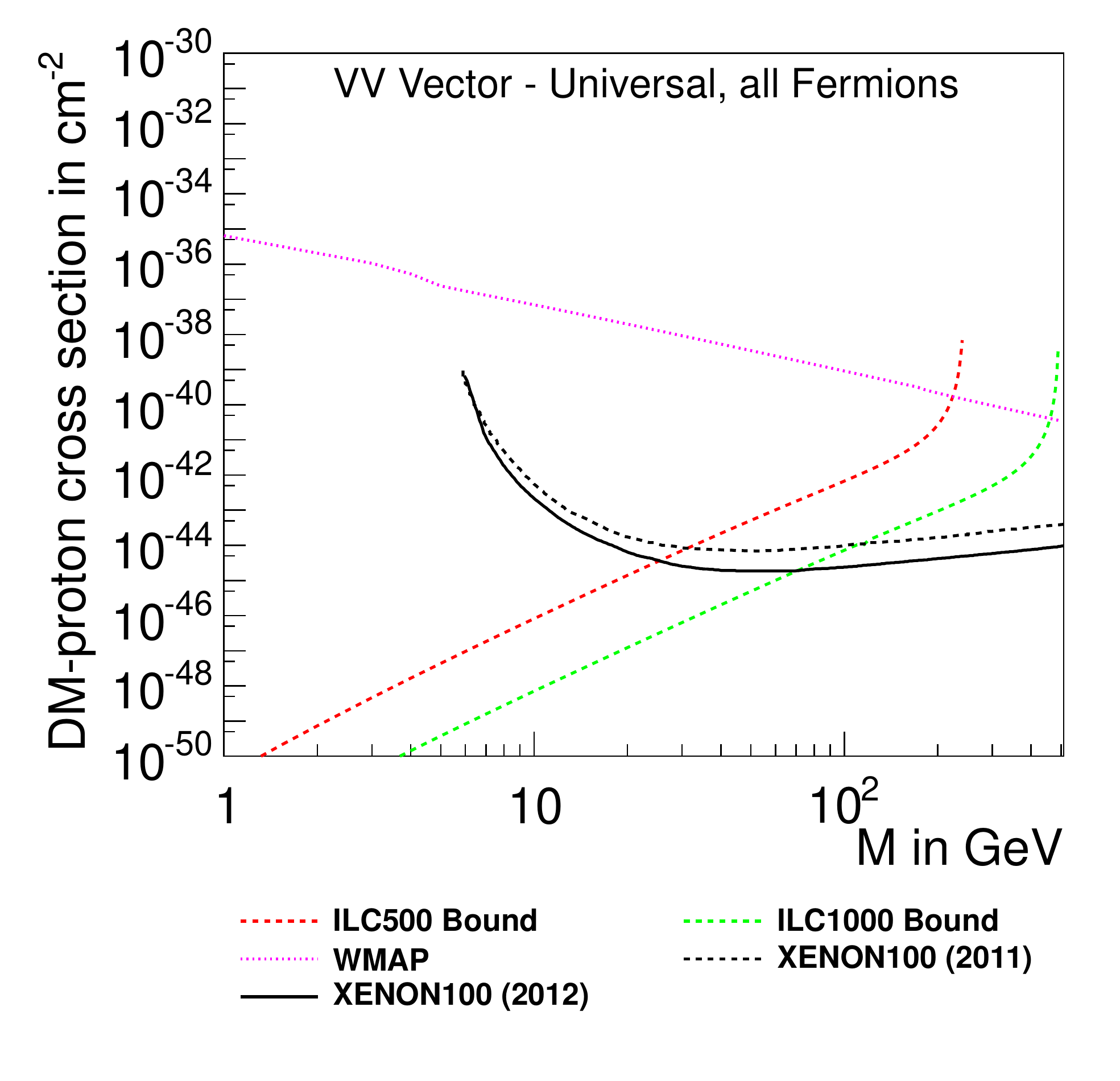} \hfill
 \includegraphics[width=0.49\columnwidth]{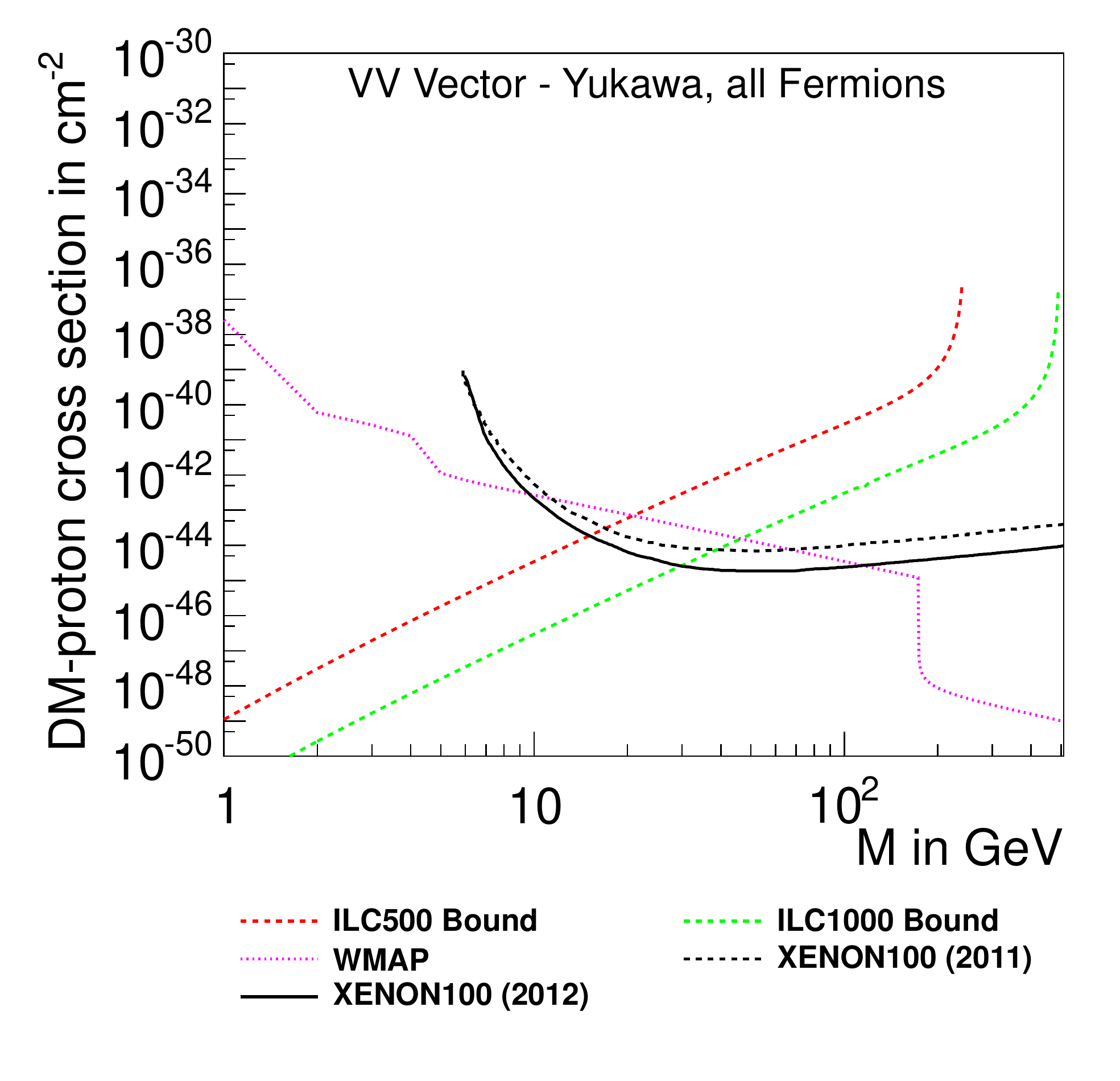}
\caption{Combined \unit{90}{\%} exclusion limits on the spin independent dark matter proton cross
   section from \textsc{Ilc}, \textsc{Wmap} and \textsc{Xenon}. For a subset of models we compare universal or Yukawa--like scenarios.}
\label{fig:comparison1}
\end{figure}
In general, \textsc{Xenon} gives an upper bound on the cross section under the assumption that if the coupling was larger, one would see a signficiant excess. On the other hand, \textsc{Wmap} limits $\sigma_0$ from below by arguing that if the coupling was smaller, dark matter would be to abundant and the universe would not be flat. So from that point of view, only areas below the \textsc{Xenon}-- and above the \textsc{Wmap} exclusion lines are allowed. As can be seen in figure \ref{fig:comparison1}, these two experiments already forbid dark matter masses in the range from about \SI{10}{\GeV} to \SI{500}{\GeV} for a large class of models with universal coupling to the Standard Model. So if we assume one of these models and a \textsc{Wimp} mass  within that interval, we need to include 
additional mechanisms that change the relic density estimate and allow for smaller couplings. As a short example this could be achieved if dark matter may annihilate into further invisible relativistic particles that do not interact with the Standard Model sector. This could bring the relic abundance down to the neccessary value even for small couplings but could neither be probed at the \textsc{Ilc} nor at \textsc{Xenon}.

Taking the \textsc{Ilc} into account, figure \ref{fig:comparison1} shows the model--dependent behaviour of the excluded cross section for different dark matter masses. Most of the models we analysed have a relatively flat mass dependence with somewhat different behaviours on the high mass threshold. The exclusion limits for the \textsc{Ilc} are of the order of the \textsc{Xenon} bounds near the threshold of $M_\chi = \SI{10}{\GeV}$ and improve them for smaller masses. For larger masses, the \textsc{Ilc} cannot compete with the strong limits from direct detection. A slightly different shape can be seen for models with dimensionful coupling constants $g_\chi$, which receive an additional $1/M^2_\chi$ dependence in the direct detection cross section (see appendix \ref{sec:nonrelmatrixelements}). This leads to weaker limits in the small mass region which become stronger for larger masses. However, the general exclusion strength is still too weak to improve the direct detection limits for large masses in all cases. Finally, for models with vector dark matter we have seen that the \textsc{Ilc} cross section includes a unique $1/M_\chi^4$ prefactor. This leads to a better exclusion for the direct detection cross section for smaller masses even if the coupling constant has dimension of mass. However, we argued before that these models have to be taken with caution because of their divergent behaviour in the production cross section for small \textsc{Wimp} masses.

\begin{figure}
\centering
 \includegraphics[width=0.49\columnwidth]{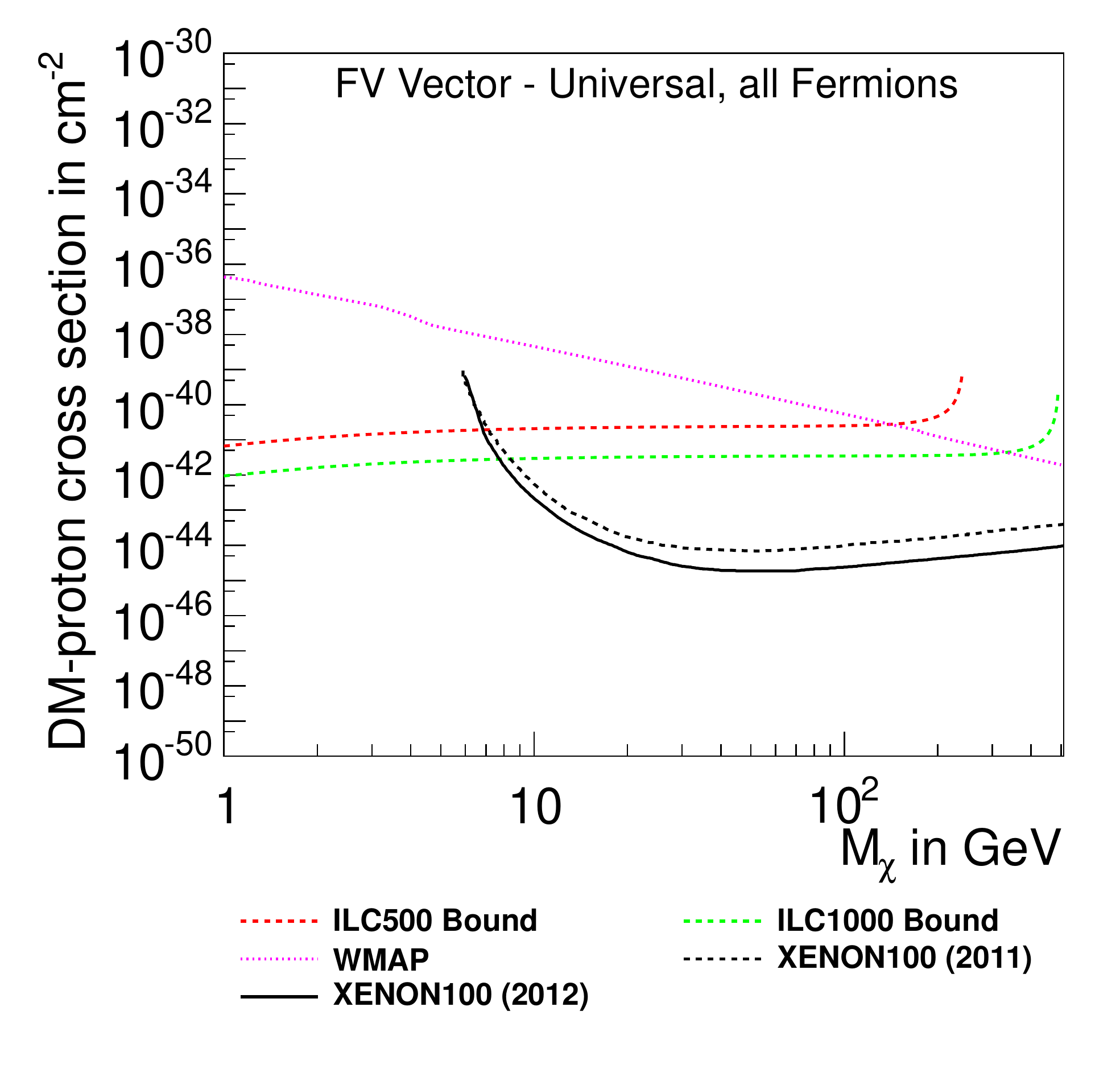} \hfill
 \includegraphics[width=0.49\columnwidth]{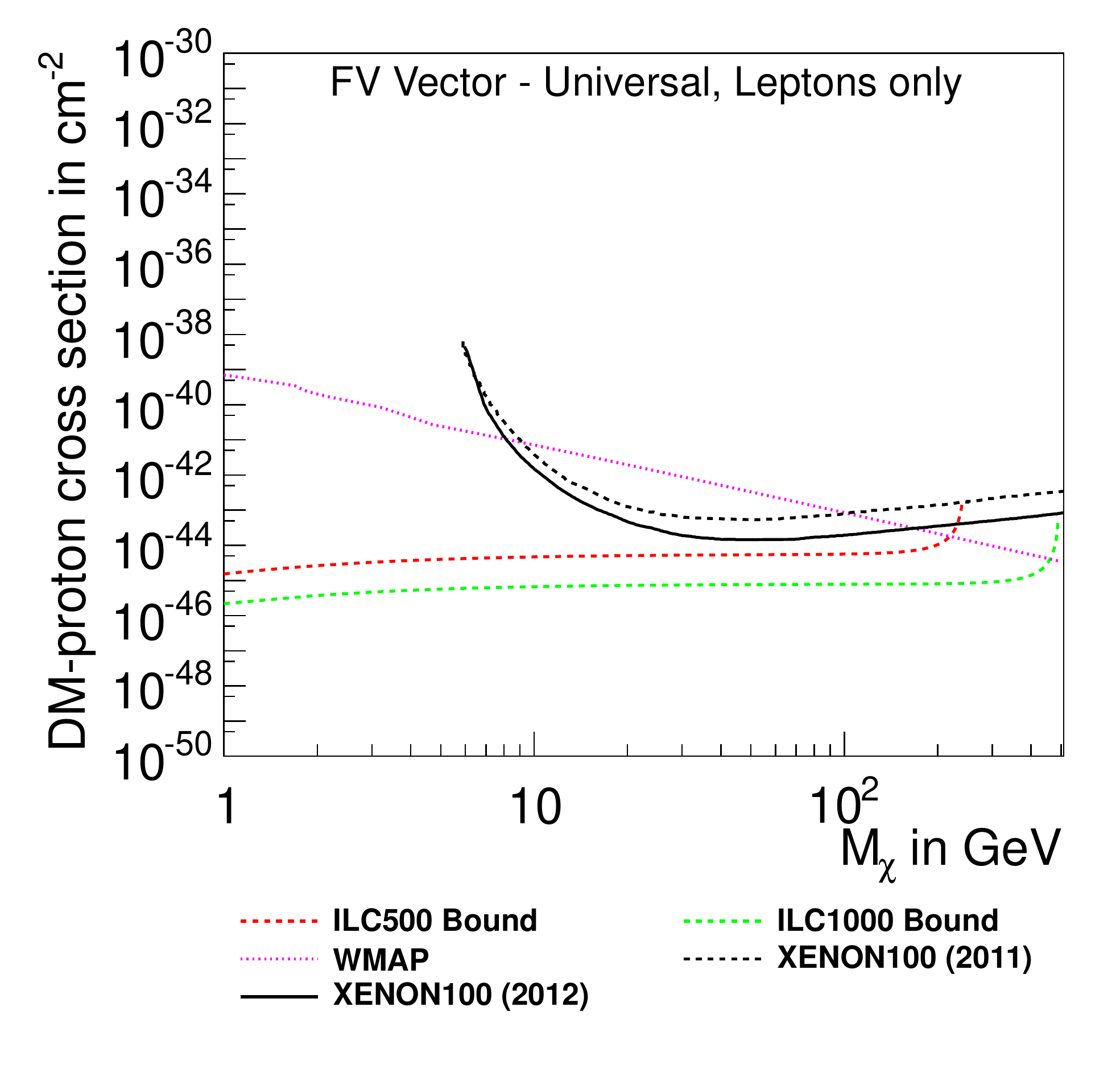}
\caption{Combined \unit{90}{\%} exclusion limits on the spin independent dark matter proton cross
   section from \textsc{Ilc}, \textsc{Wmap} and \textsc{Xenon}. For the FV Vector model, we compare the case with coupling to all Standard Model fermions to a scenario with lepton--only interactions.}
\label{fig:comparison2}
\end{figure}
The FV Vector model can be used to compare with previous collider studies, since this particular model has also been analysed for \textsc{Lep} \cite{LEPShinesLight}, Tevatron \cite{TevatronDarkMatter} and the \textsc{Lhc} \cite{CMSDM}. Direct comparison shows that the expected \textsc{Ilc} exclusion bounds can improve the current collider limits on $\sigma_0$ by up to three orders of magnitude.

If we go from a universal scenario to one including Yukawa--like couplings, \textsc{Ilc} results get significantly weakened by about two orders of magnitude. Since the electron mass is smaller than quark masses, pair production at an electron--positron collider is less likely than elastic scattering with hadronic objects at direct detection experiments. Hence, this leads to a reduced exclusion power. On the other hand, the relic density limits become weaker which leads to a larger allowed dark matter mass range for most models. This effect is largest as soon as annihilation into top--quarks, which have the strongest coupling in this scenario, is kinematically allowed (see discussion in chapter \ref{chap:wmap}).

\begin{figure}
\centering
 \includegraphics[width=0.49\columnwidth]{nopamela_FVVectorplotter_protonxsect_universal} \hfill
 \includegraphics[width=0.49\columnwidth]{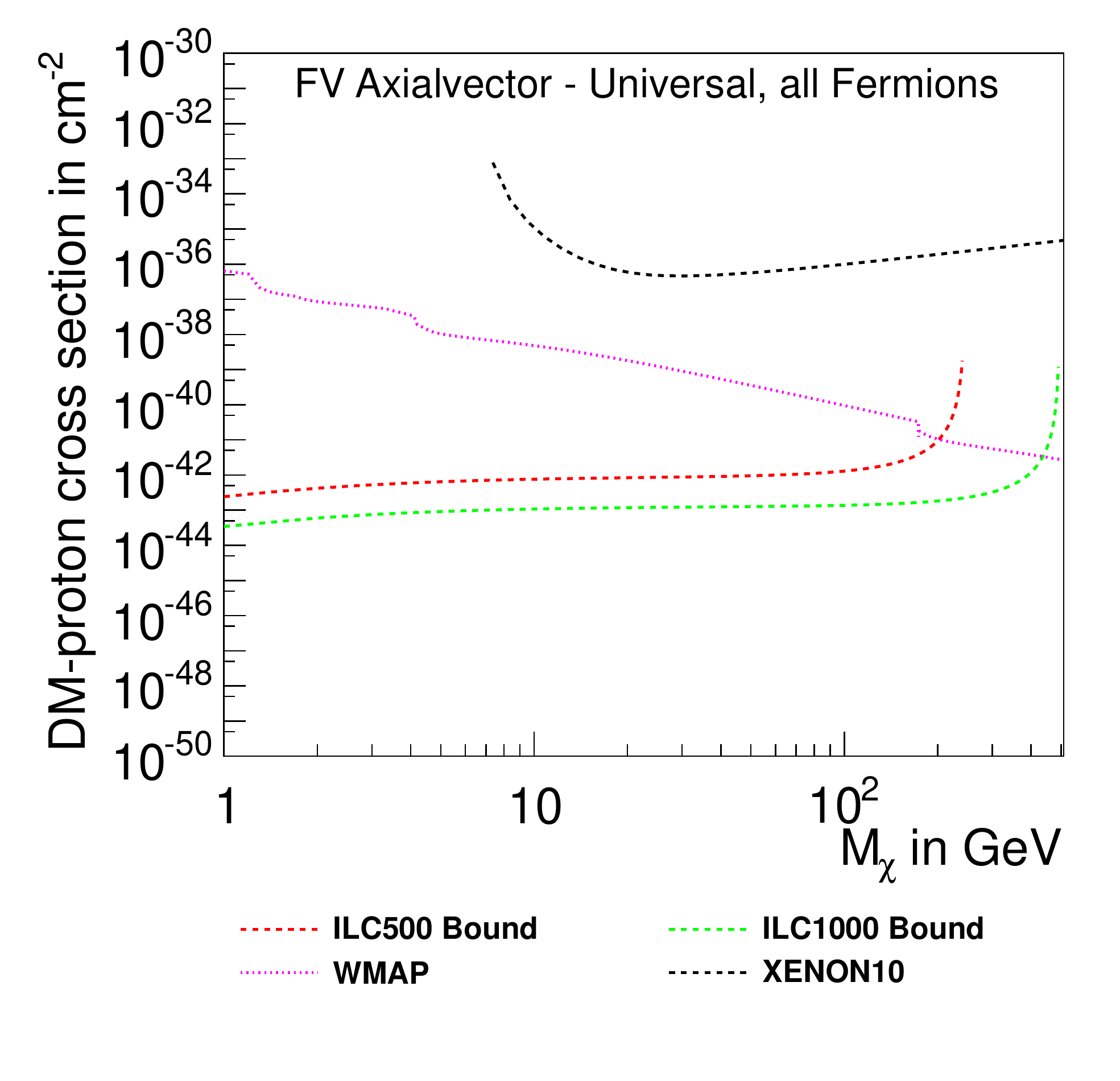} 
\caption{Combined \unit{90}{\%} exclusion limits on the dark matter proton cross
   section from \textsc{Ilc}, \textsc{Wmap} and \textsc{Xenon}. We compare spin--dependent and spin--independent limits for different FV models.}
\label{fig:comparison3}
\vspace{-0.4cm}
\end{figure}

We show results for the spin--dependent interaction in figure \ref{fig:comparison3}, which is only given by the FV Axialvector model. There were no official exclusion limits for spin--dependent interactions given by the \textsc{Xenon}100 experiment when this thesis was completed (October 2012). Hence we only use the result of its predecessor \textsc{Xenon}10, with the remark that the results will probably be improved in the phase 2 results by up to three orders of magnitude. However, it still can be expected that the \textsc{Ilc} will give the leading exclusion bounds on the proton cross section over the whole accessible dark matter mass range up to \SI{500}{\GeV}.

\begin{figure}
\centering
 \includegraphics[width=0.49\columnwidth]{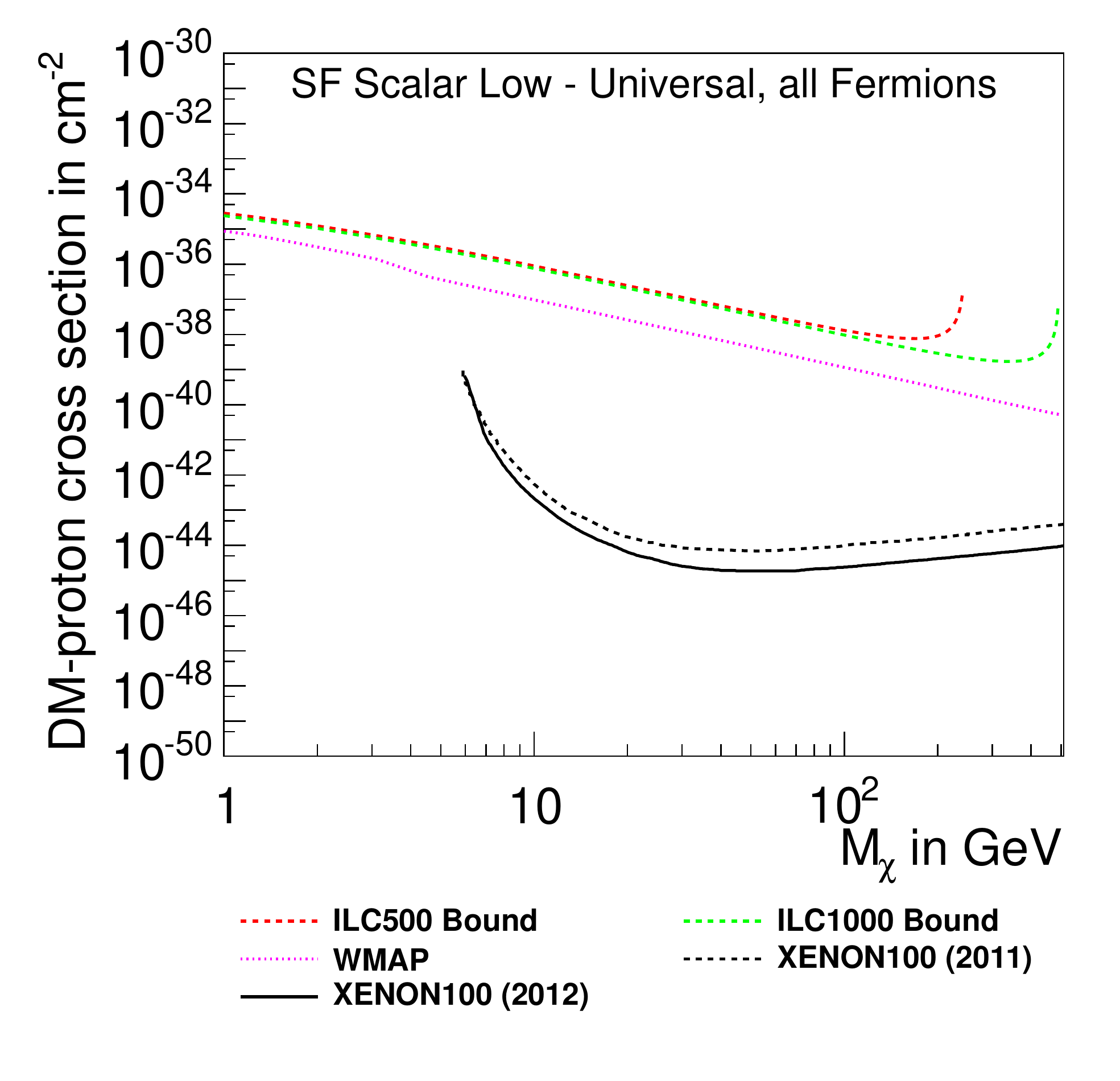} \hfill
 \includegraphics[width=0.49\columnwidth]{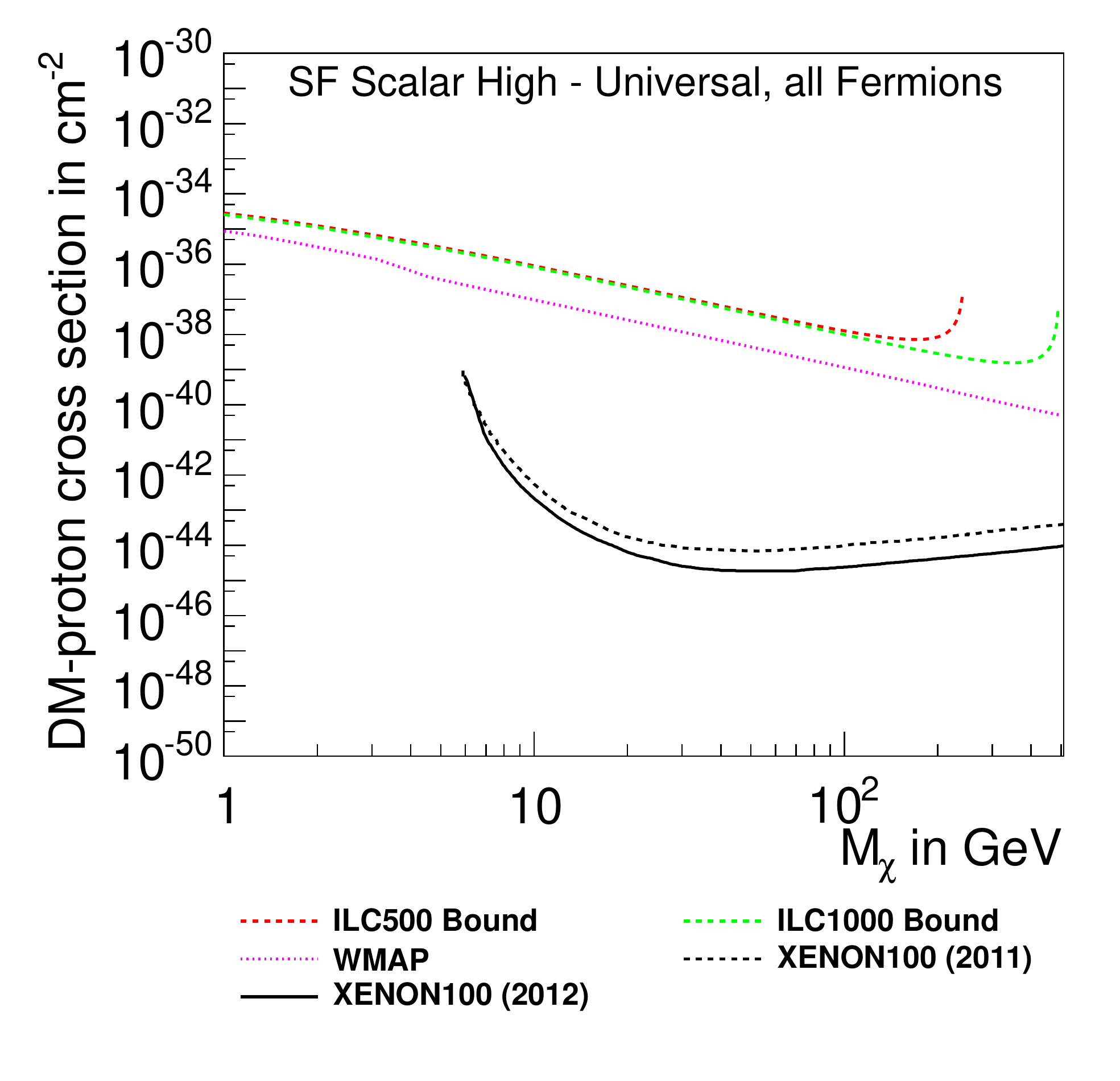} \\ 
 \includegraphics[width=0.49\columnwidth]{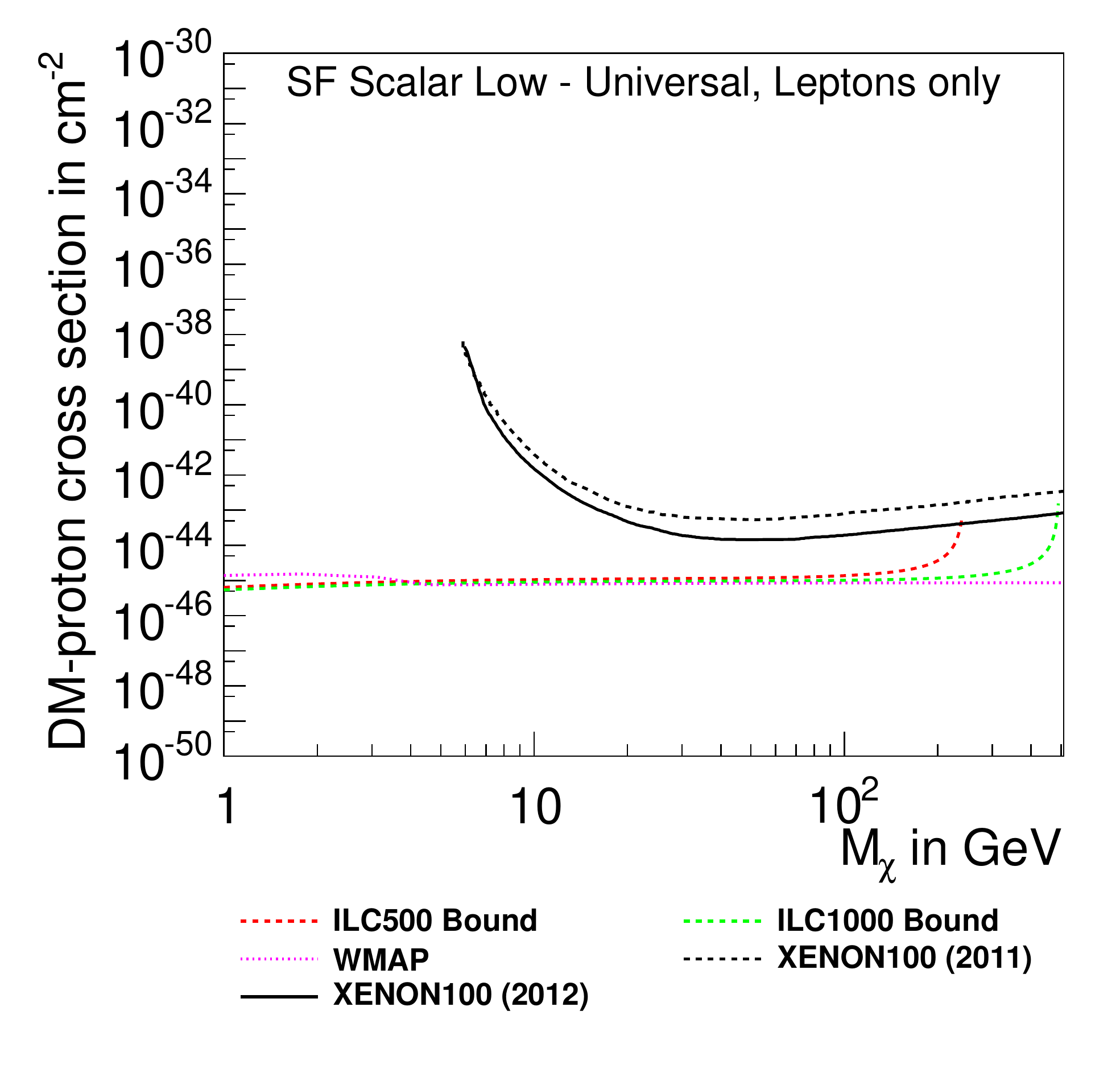} \hfill
 \includegraphics[width=0.49\columnwidth]{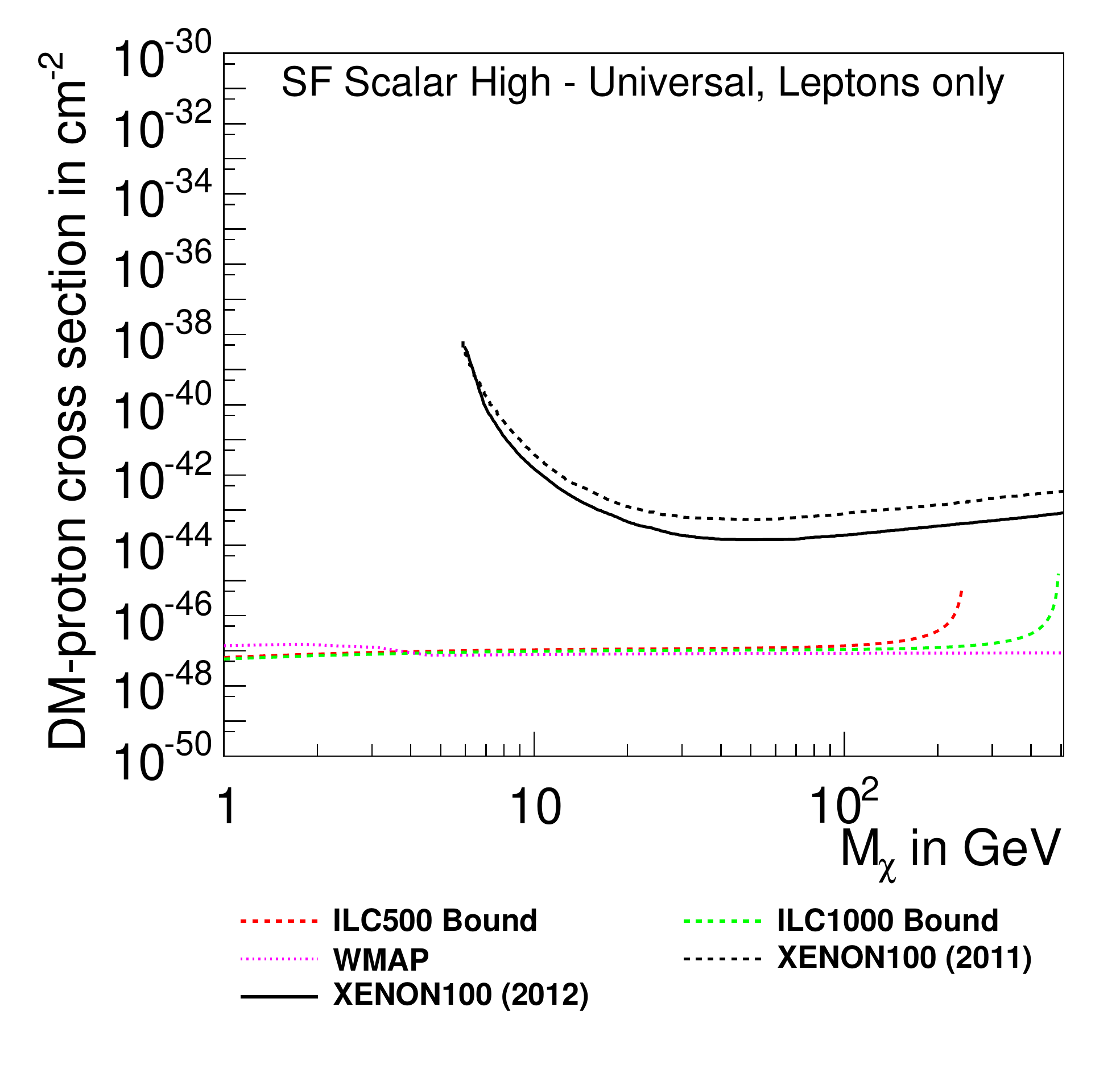} 
\caption{Combined \unit{90}{\%} exclusion limits on the spin independent dark matter proton cross
   section from \textsc{Ilc}, \textsc{Wmap} and \textsc{Xenon}. For the SF Scalar model, we compare the case with coupling to all Standard Model fermions to a scenario with lepton--only interactions for different masses of the mediator.}
\label{fig:comparison22}
\end{figure}
Finally In figures \ref{fig:comparison2} and \ref{fig:comparison22}, we compare results for tree level quark scattering to limits for loop induced scattering to leptons only. The \textsc{Xenon}--results are not only weakened by the correction factor in (\ref{eqn:xenoncorrection}) but also the \textsc{Ilc} limits get a strong enhancement by multiple orders of magnitude. This is due to the much smaller direct detection cross section caused by the loop factor and the appearance of $\alpha_\text{em}^2 / 81 \pi^2$. Models with fermion mediators get an even stronger enhancement, caused by the fact that the $1/M_\Omega$ terms in the operator are scalar--like whereas the $1/M_\Omega^2$ terms show a vector--like structure. Tree--level interactions can occur through a scalar quark current (\ref{eqn:scalarcurrent}) inside the nucleon, but loop induced scattering needs a vector current to allow for photon--interaction. From that it follows that only the sub-leading term in the effective operator induces loop interactions, leading to an additional $1/M_\Omega^2$ suppression of the scattering cross section. In that case, the actual value of the mediator mass $M_\Omega$ is needed in order to know the exact additional suppression going from the tree level to the loop interaction. This can be seen by the significantly different exclusion lines for \emph{Low} or \emph{high} suppression scales in the SF models, defined in table \ref{tbl:constraints}.
For other models with only two--loop or no interaction at all, the \textsc{Ilc} will always give the strongest exclusion limits, since direct detection is not sensitive as soon as the expected interaction rate is negligibly small.

In general we can conclude that for most dark matter models the \textsc{Ilc} is expected to give the strongest accessible exclusion limits for cold dark matter lighter than \SI{10}{\GeV}. For larger masses, direct detection still prevails, unless quark coupling is forbidden or coherence is lost due to spin--dependent interactions. In those cases, the \textsc{Ilc} will give the strongest bounds on the dark matter proton cross section over the whole accessible mass range.
\chapter{Conclusions}
\label{chap:conclusions}
In this thesis we have looked at a complete list of models with one dark matter particle of specific spin that couples via different interaction mechanisms by one heavy mediator to a pair of standard model fermions. We discussed how in general the operators can be derived by using a well--defined renormalisable fundamental theory and formally integrating out the heavy particle from the path integral. This gave us a list of models which could easily be analysed in various experimental scenarios, since they only consisted of a simple new 4--particle--vertex with a new coupling parameter \Geff{} and the \textsc{Wimp} mass $M_\chi$.

We started by constraining these models to give the right relic density for dark matter, measured by the \textsc{Wmap} experiment, according to standard big bang cosmology. By allowing further unknown contributions to $\Omega_\text{DM}^0$, this analysis gave a minimum value on the effective coupling strength $G_\text{eff}$. In general the bound is stronger the heavier the dark matter particle is assumed to be. Models show different behaviour as soon as  the mass $M_\chi$ passes thresholds to allow for annihilation into heavy quarks or leptons. For the exclusion limits for heavy dark matter it is important to which Standard Model fermions the dark matter is allowed to couple in general, and in particular if different particles couple with the same strength or proportionally to their respective mass. Different assumptions and parameter values can change the exclusion limits by multiple orders of magnitude.

As a next step, we analysed the pair creation of \textsc{Wimp}s at the International Linear Collider with an additional hard photon in the final state by using polarised incoming leptons. We numerically discussed the impact of beam, detector and background effects. It could be shown that under the assumption that no signal event is measured, signal cross sections down to around \SI{0.3}{\femto \barn} can be excluded to \SI{90}{\%} confidence level for a large \textsc{Wimp} mass range. Different models generally show a similar behaviour with exclusion limits not differing by more than a factor of 2. However, peculiar divergent behaviour for small \textsc{Wimp} masses could be shown for models including vector dark matter, showing that these models cannot be simply analysed in an effective theory without enhancing the underlying fundamental model to a spontaneously broken gauge symmetry. This has not been seen in previous studies, since effective vector dark matter has only been analysed with respect to astrophysical interpretations, where these divergences do not occur.

Regarding the experimental requirements of the collider, we could show that using the polarisation of the incoming particles may enhance the signal to background ratio. This is done by filtering chiral Standard Model neutrino events from the background, which at the same time introduces a large systematic error due to fluctuations in the experimental polarisation value. We showed that as soon as enough data is taken to make the statistical uncertainty small, systematics may nullify any gain of significance for larger polarisations if the error $\Delta P/P$ is too large. Considering a potentially larger positron polarisation should therefore be strongly linked to the discussion of a smaller experimental uncertainty on that value. A doubled center of mass energy always leads to a better exclusion potential by not only enlarging the accessible dark matter mass range from about \SI{240}{\GeV} to \SI{490}{\GeV} but also by reducing the dominant Bhabha--background and therefore leading to a better signal to background ratio. 

Finally we translated our derived \textsc{Wmap} and \textsc{Ilc} bounds on \Geff{} into limits on elastic dark matter proton scattering at zero momentum transfer $\sigma_P^0$ with the intention of comparing them to results by the \textsc{Xenon} collaboration. This shows that the International Linear Collider may give the strongest bounds for dark matter masses below \SI{5}{\GeV}, also compared to previous collider studies at \textsc{Lep}, Tevatron and the \textsc{Lhc}, and of competitive order to \textsc{Xenon} in the range up to \SI{10}{\GeV}. It is not able to compete with direct detection experiments for larger masses, as long as the direct \textsc{Wimp} proton interaction is not forbidden by any additional mechanism. If one forbids the interaction of dark matter with quarks, \textsc{Wimps} can only scatter with nucleons by photon coupling to a virtual lepton loop. In that case, the translated \textsc{Ilc} bounds get strongly enhanced due to additional loop suppression factors, such that it will improve the direct detection limits. This is also the case for interactions that couple spin--dependently, since in that case interaction does not take place coherently with all protons and neutrons in the nucleus.

All in all it could be shown that the International Linear Collider may contribute important information to the dark matter puzzle by giving the strongest exclusion statements for \textsc{Wimp}s in the low--\GeV{} mass region, or even beyond for scenarios in which direct detection is not sensitive.

\section*{Outlook}
In this thesis, we formulated fundamental theories and translated them into effective models in order to reduce the number of free parameters. It would be interesting to know in general how low the mediator mass can go such that the effective approach still gives accurate results compared to using the full fundamental theory. In particular it may be asked how differences for low mass mediators manifest themselves in the different exclusion limits we derived. This has been  partially done in a \textsc{Lep} analysis in \cite{LEPShinesLight} but only on the collider level with a small subset of models.

In addition it is yet unclear how the peculiar divergent \textsc{Ilc} results for vector dark matter may change if one assumes a fully unitary theory. The current results show that these models receive the strongest exclusion limits for production processes, but it is unclear whether this statement survives as soon as the $M_\chi \rightarrow 0$ divergence is regularised.

\sisetup{range-phrase=-}
For this thesis we only looked at the \emph{exclusion} potential of the \textsc{Ilc}. This naturally raises the question how sensitive the collider would be to \emph{discover} a potential dark matter candidate and how precise mass and coupling strength of the \textsc{Wimp} could be measured. This question has been partially analysed for a more simplified interaction in \cite{BartelsThesis, BartelsList, BartelsList2, BartelsList3, BartelsList4}. The authors quote that unpolarised \textsc{Wimp} pair production cross section down to \SI{25}{\femto\barn} could be discovered with $5\sigma$ sensitivity, assuming a total integrated luminosity of \SI{500}{\per\femto\barn} and a center of mass energy of $\sqrt{s} = \SI{500}{\GeV}.$ Signal cross sections above that value could be measured with an accuracy of $2 - 5$ \unit{}{\femto\barn}. \textsc{Wimp} masses of the order \SI{100}{\GeV} can be measured with a resolution of $0.5 - 3$\unit{}{\%}. Smaller masses in the few--\GeV{} range have been analysed in the related context of light neutralinos in Supersymmetry \cite{Conley:2010jk}, from which one can expect resolutions of around $\unit{2}{\GeV}$. We assume that the accuracy will be of the same order for out set of dark matter models.

As it has been discussed, analyses of monophoton events do generally suffer from the unknown shape of Bhabha--background. Compared to the neutrino background this is particularly bad since the Bhabha contribution cannot be reduced by means of beam polarisation. If one knew the detailed spectrum after consideration of all detector effects, one would be able to promote the \textsc{Ilc} exclusion statements into a shape dependent analysis with a much stronger exclusion power. Regarding the expected discovery sensitivity of the \textsc{Ilc}, one could even use signal data to not only find but also distinguish between different effective models. This could be done by looking at the different threshold behaviour for large photon energies as well as the signal's response on changing the incoming lepton polarisation. However, this is not possible without knowing the shape of the dominant background sources. It remains uncertain whether this can be estimated accurately without having a physically built detector at hand to compare the simulation with.

\pagebreak
\appendix
\chapter{Definitions and Parameters}
\section{General Conventions}
We follow the conventions in \cite{Peskin}. Throughout this thesis we use natural units with $\hbar = c = k_\text{B} = 1$. The metric tensor is defined as $g_{\mu \nu} = \text{diag}(1, -1, -1, -1)$. Fourier transformations and delta-distributions are defined as follows:
\begin{align}
f(x) &= \int \frac{\de^4 k}{(2 \pi)^4}  \ \tilde{f}(k) e^{-i k \cdot x}, \\
\tilde{f}(k) &= \int \de^4 x \ f(x) e^{i k \cdot x}, \\
\int \de^4 x \ e^{i k \cdot x} &= (2 \pi)^4  \delta^4 (k), \\
\int \de^4 x \  \delta^4 (x) &= 1.
\end{align}
\section{Dirac--Algebra and Spinor--Identities}
\label{sec:appdirac}
We now want to give a short list of definitions and properties that are needed for various calculations within this thesis.
The standard basis for the Pauli matrices reads as follows:
\begin{align}
\sigma^1 = \begin{pmatrix} 0 & 1 \\ 1 & 0 \end{pmatrix}, \qquad
\sigma^2 = \begin{pmatrix} 0 & -i \\ i & 0 \end{pmatrix}, \qquad
\sigma^3 = \begin{pmatrix} 1 & 0 \\ 0 & -1 \end{pmatrix}. \qquad
\end{align}
We combine them to the four vectors $\sigma^\mu \equiv (\mathbb{1}, \sigma^i)$, $\bar{\sigma}^\mu \equiv (\mathbb{1}, -\sigma^i)$. The following identities then hold:
\begin{align}
\sigma^\mu + \bar{\sigma^\mu} &= 2 \delta^\mu_0 \mathbb{1}, \\
\sigma^\mu - \bar{\sigma^\mu} &= 2 \delta^\mu_i \sigma^i,\\
\sigma^i \sigma^j &= \delta^{ij} \mathbb{1} + i \epsilon^{ijk} \sigma_k, \\
\text{tr}(\sigma^i) &= 0, \\
\text{tr}(\sigma^i \sigma^j) &= 2 \delta^{ij}.
\end{align}
We use the chiral representation for the Dirac matrixes $\gamma^\mu$ 
\begin{align}
\gamma^\mu = \begin{pmatrix} 0 & \sigma^\mu \\ 
                            \bar{\sigma}^\mu & 0 
\end{pmatrix},
\qquad \gamma^5 = \begin{pmatrix} -\mathbb{1} & 0 \\ 
                            0 & \mathbb{1}
\end{pmatrix}
\end{align}
and define $\slashed{a} \equiv a^\mu \gamma_\mu$ as well as $\sigma^{\mu \nu} \equiv i/2 (\gamma^\mu \gamma^\nu - \gamma^\nu \gamma^\mu)$. They fulfil the following relations:
\begin{align}
\gamma^\mu \gamma^\nu + \gamma^\nu \gamma^\mu &= 2 g^{\mu \nu}, \\
\gamma^\mu \gamma^\nu \gamma^\rho &= g^{\mu \nu} \gamma^\rho + g^{\nu \rho} \gamma^\mu - g^{\mu \rho} \gamma^\nu - i \epsilon^{\mu \nu \rho \sigma} \gamma_\sigma \gamma_5.
\end{align}
Any general spinor $\psi(x)$ that solves the Dirac equation $(i \slashed{\partial} - m) \psi(x) = 0$ can be written as a linear combination of positive and negative frequency solutions $u^s$ and $v^s$ that fulfil $(\slashed{p} - m) u^s(p) = (\slashed{p} + m) v^s(p) = 0$. In this representation, they can be formulated in terms of normalised two--component spinors $\xi^s$ as follows:
\begin{align}
u^s(p) = \begin{pmatrix}
\sqrt{p \cdot \sigma} \ \xi^s \\
\sqrt{p \cdot \bar{\sigma}} \ \xi^s
\end{pmatrix}, \qquad 
v^s(p) = \begin{pmatrix}
\sqrt{p \cdot \sigma} \ \xi^s \\
- \sqrt{p \cdot \bar{\sigma}} \ \xi^s
\end{pmatrix}.
\end{align}
Moreover we define $\bar{u} \equiv u^\dagger \gamma^0$. We can evaluate the spin indices in $\xi^s$ and $u^s(p)$ with the following relations:
\begin{align}
\xi^{\dagger s} \xi^{s^\prime} &= \delta^{s s^\prime}, \\
\sum_{s, s^\prime} \left| \xi^{\dagger s} \mathcal{G}(1, \gamma^\mu, \gamma^5, \gamma^\mu \gamma^5, \Sigma^{\mu \nu}) \xi^{s^\prime} \right|^2 &= \text{tr}(\mathcal{G} \mathcal{G}^*) \\
\sum_{s} u^s(p) \bar{u}^s(p) &= \slashed{p} + m, \\
\sum_{s} v^s(p) \bar{v}^s(p) &= \slashed{p} - m.
\end{align}
Products of general spinor bilinears can be rewritten by using Fierz' transformations \cite{DreinerSPinors}. The most important identities we need are
\begin{align}
\psib \chi \chib \psi &= -\frac{1}{4} \left[ \psib \psi \chib \chi + \psib \gaf \psi \chib \gaf \chi  + \psib \gamu \psi \chib \gamuu \chi - \psib \gamu \gaf \psi \chib \gamuu \gaf \chi \right. \nonumber \\
& \left. \hspace{0.5\textwidth} + \frac{1}{2} \psib \smunu \psi \chib \smunuu \chi \right], \\
\psib \gaf \chi \chib \gaf \psi &= -\frac{1}{4} \left[ \psib \psi \chib \chi + \psib \gaf \psi \chib \gaf \chi  - \psib \gamu \psi \chib \gamuu \chi + \psib \gamu \gaf \psi \chib \gamuu \gaf \chi \right. \nonumber \\
& \left. \hspace{0.5\textwidth} + \frac{1}{2} \psib \smunu \psi \chib \smunuu \chi \right], \\
\psib \gamu \chi \chib \gamuu \psi &= -\left[ \psib \psi \chib \chi - \psib \gaf \psi \chib \gaf \chi  -\frac{1}{2} \psib \gamu \psi \chib \gamuu \chi -\frac{1}{2} \psib \gamu \gaf \psi \chib \gamuu \gaf \chi \right], \\
\psib \gamu \gaf \chi \chib \gamuu \gaf \psi &= -\left[- \psib \psi \chib \chi + \psib \gaf \psi \chib \gaf \chi  -\frac{1}{2} \psib \gamu \psi \chib \gamuu \chi -\frac{1}{2} \psib \gamu \gaf \psi \chib \gamuu \gaf \chi \right].
\end{align}
We define real spinors through the general Majorana condition $\psi = C \bar{\psi}^T$ with the charge conjugation matrix $C$ which fulfils the following identities \cite{Denner:1992vza}:
\begin{align}
C^\dagger &= C^{-1}, \\ C^T &= -C, \\ C\Gamma_i^TC^{-1} &= \eta_i \Gamma_i, \label{eqn:theetai}\\
u^s(p) &= C (\bar{v}^s(p))^T, \\  v^s(p) &= C (\bar{u}^s(p))^T, \\
C (\slashed{p}-m)^{-1} C^{-1} &= (-\slashed{p} - m)^{-1}.
\end{align}
There is no summation over $i$ in (\ref{eqn:theetai}) and $\eta_i$ is defined as 
\begin{align}
\eta_i = \left\{
     \begin{array}{rl}
1 \qquad& \text{for } \Gamma_i = 1, \gamma^5 \text{ or }\gamma^{\mu} \gamma^5 \\
  -1 \qquad& \text{for } \Gamma_i = \gamma^\mu \text{ or } \sigma^{\mu \nu}
     \end{array}
\right.
\end{align}
\pagebreak
\section{Physical Constants}
\label{sec:constants}
\begin{table}[H]
\begin{center}
\begin{tabular*}{\textwidth}{@{\extracolsep{\fill}}ll S[tabformat=1.3e2, tabnumalign=left] @{\hspace{-0.9cm}} s l}
\toprule
\multicolumn{5}{c}{General Constants} \\
\midrule
fine structure constant & $\alpha(0)$ & 137.036&& \\
                                        & $\alpha({1}{\TeV})$ & 125.229& {\cite{alpha}}&\\
reduced Planck's constant &$\hbar$& 6.582e-22&\MeV \second& \\
speed of light in vacuum & $c$ &2.998e8&\metre \per \second& \\
Boltzmann konstant & $k_\text{B}$ & 8.617e-5 & \eV \per \kelvin& \\
\midrule
\multicolumn{5}{c}{Particle Masses} \\
\midrule
proton mass & $m_P$ & 938.3&\hspace{-1.7cm}\GeV& \\
electron mass & $m_e$ & 511.0&\hspace{-1.7cm}\keV& \\
muon mass &$m_\mu$&105.7&\hspace{-1.7cm}\MeV& \\
tau mass &$m_\tau$&1.8&\hspace{-1.7cm}\GeV& \\
up quark mass &$m_u$&2.3&\hspace{-1.7cm}\MeV& \\
down quark mass &$m_d$&4.8&\hspace{-1.7cm}\MeV& \\
strange quark mass &$m_s$&95&\hspace{-1.7cm}\MeV& \\
charm quark mass &$m_c$&1.3&\hspace{-1.7cm}\GeV& \\
bottom quark mass &$m_b$&4.2&\hspace{-1.7cm}\GeV& \\
top quark mass &$m_t$&173.5&\hspace{-1.7cm}\GeV& \\
$W^\pm$ boson mass & $m_W$ & 80.4&\hspace{-1.7cm}\GeV& \\
$Z^0$ boson mass & $m_Z$ & 91.2&\hspace{-1.7cm}\GeV& \\
\midrule
\multicolumn{5}{c}{Nuclear Form Factors} \\
\midrule
scalar contribution (u--quarks) & $f^p_u$ & 0.020&\multirow{3}{*}{$\Biggr\}$ \cite{fnumbers}}&  \\
scalar contribution (d--quarks) & $f^p_d$ & 0.026&& \\
scalar contribution (s--quarks) & $f^p_s$ & 0.118&& \\
spin contribution (u--quarks) & $\Delta_u^p$ & -0.427&\multirow{3}{*}{$\Biggr\}$ \cite{deltanumbers}}& \\ 
spin contribution (d--quarks) & $\Delta_d^p$ & 0.842&& \\ 
spin contribution (s--quarks) & $\Delta_s^p$ & -0.085&& \\ 
\midrule
\multicolumn{5}{c}{Astrophysical Constants} \\
\midrule
gravitational constant  & $G_\text{N}$ &6.674e-11&\meter\cubed\per\kilo\gram\per\second\Square& \\
Planck mass & $m_\text{Pl}$&1.221e19&\GeV& \\
Hubble constant &$H_0$&71.0&\kilo\meter\per\second& \\
little $h$ & $h$ & 0.710 & & \\
critical density &$\rho_c$&5.312e11&\GeV\per\centi\meter& \\
baryonic matter density & $\Omega_\text{B}$&0.05&& \\
dark matter density &$\Omega_\text{DM}$&0.22&& \\
dark energy density &$\Omega_\Lambda$&0.73&& \\
\bottomrule
\end{tabular*}
\caption{List of all numerical values for the various physical parameters that are used throughout this thesis. If not mentioned otherwise, they are taken from \cite{PDG}. Since we only give qualitative results, we do not perform a thorough error analysis, which is why we do not show the corresponding experimental uncertainties.}
\end{center}
\end{table}

\chapter{Dark Matter Interaction Cross Sections}
\section{Annihilation}
\label{app:annixsects}
We give the leading expansion terms in the thermally averaged annihilation cross section times velocity $\langle \sigma v \rangle = a + b v^2$, which we need for the relic density analysis in chapter \ref{chap:wmap}.  We abbreviate $\lambda \equiv m_f / M_\chi \leq 1$, with $m_f$ denoting the mass of the final state fermion which, due to kinematic reasons, cannot be larger than the \textsc{Wimp} mass. 
\allowdisplaybreaks
\begin{align}
a_{SS}&=\frac{\sqrt{1-\lambda ^2} \left(g_a^2-g_s^2 \left(\lambda ^2-1\right)\right)}{4 \pi  M_\Omega^4}\\
b_{SS}&=\frac{ \left(g_s^2 \left(-5 \lambda ^4+7 \lambda ^2-2\right)+g_a^2 \left(3 \lambda ^2-2\right)\right)}{32 \pi  \sqrt{1-\lambda ^2} M_\Omega^4} \\[3.5ex]  
a_{SF}&=\frac{\left(1-\lambda ^2\right)^{3/2}}{M_\Omega^2   \sqrt{1-\lambda ^2} } \left[-\frac{\lambda   M_\chi \left(g_s^4-g_a^4\right)}{2 \pi  M_\Omega}+\frac{\lambda ^2 M_\chi^2 \left(g_s^2+g_a^2\right)^2}{4 \pi  M_\Omega^2}+\frac{ \left(g_s^2-g_a^2\right)^2}{4 \pi} \right]\\
b_{SF}&= \frac{ \left(5 \lambda ^4-7 \lambda ^2+2\right) }{ M_\Omega^2} \left[\frac{\lambda M_\chi \left(g_s^4-g_a^4\right)}{16 \pi M_\Omega}-\frac{\left(5 \lambda ^4-7 \lambda ^2+2\right) \left(g_s^2-g_a^2\right)^2}{32 \pi}\right]+ \nonumber \\
&\quad \frac{\sqrt{1-\lambda ^2} M_\chi^2 \left(15 \lambda ^4 \left(g_s^2+g_a^2\right)^2+4\left(1- \lambda ^2\right) \left(g_s^4+6 g_s^2 g_a^2+g_a^4\right)\right)}{96 \pi  M_\Omega^4} \\[3.5ex]  
a_{SFr}&=\frac{ \left(1-\lambda ^2\right)^{3/2}}{M_\Omega^2} \left[-\frac{2 \lambda  M_\chi \left(g_s^4-g_a^4\right)}{\pi  M_\Omega}+\frac{\lambda ^2 M_\chi^2 \left(g_s^2+g_a^2\right)^2}{\pi  M_\Omega^2}+\frac{\left(g_s^2-g_a^2\right)^2}{\pi}\right]\\
b_{SFr}&= \frac{\left(5 \lambda ^4-7 \lambda ^2+2\right)}{M_\Omega^2} \left[\left(\frac{\lambda   M_\chi \left(g_s^4-g_a^4\right)}{4 \pi  \sqrt{1-\lambda ^2} M_\Omega}-\frac{\lambda ^2 M_\chi^2 \left(g_s^2+g_a^2\right)^2}{8 \pi  \sqrt{1-\lambda ^2} M_\Omega^2}-\frac{\left(g_s^2-g_a^2\right)^2}{8 \pi  \sqrt{1-\lambda ^2}}\right) \right]\\[3.5ex]  
a_{SV} &= 0 \\
b_{SVv}&=\frac{\sqrt{1-\lambda ^2} M_\chi^2 v^2 \left(4 \left(g_l^2+g_r^2\right)-\lambda ^2 \left(g_l^2-6 g_l g_r+g_r^2\right)\right)}{48 \pi  M_\Omega^4} \\[3.5ex]  
 a_{FS} &= \frac{g_a^2 \sqrt{1-\lambda ^2} M_\chi^2 \left(g_a^2-g_s^2 \left(\lambda ^2-1\right)\right)}{2 \pi  M_\Omega^4} \\
 b_{FS} &= \frac{M_\chi^2 \left(2 g_s^4 \left(\lambda ^2-1\right)^2+g_s^2 g_a^2 \left(-3 \lambda ^4+\lambda ^2+2\right)+g_a^4 \lambda ^2\right)}{16 \pi  \sqrt{1-\lambda ^2} M_\Omega^4} \\[3.5ex]  
 a_{FV} &= \frac{\sqrt{1-\lambda ^2} M_\chi^2 \left(4 \lambda ^2 g_l^2 g_r^2+\left(g_l^2+g_r^2\right) (g_l+g_r)^2\right)}{8 \pi  M_\Omega^4}\\
 b_{FV} &= \frac{M_\chi^2}{192 \pi  \sqrt{1-\lambda ^2} M_\Omega^4}  \Big[2 \lambda ^4 \left(g_l^4-6 g_l^3 g_r+32 g_l^2 g_r^2-6 g_l g_r^3+g_r^4\right)  \nonumber\\
& \quad -\lambda ^2 \left(g_l^4-30 g_l^3 g_r+50 g_l^2 g_r^2-30 g_l g_r^3+g_r^4\right)+ 2 \left(g_l^2+g_r^2\right) \left(g_l^2-6 g_l g_r+g_r^2\right)\Big]  \\[3.5ex]  
 a_{FVr} &=\frac{\lambda ^2 \sqrt{1-\lambda ^2} M_\chi^2 (g_l-g_r)^4}{8 \pi  M_\Omega^4} \\
 b_{FVr} &= \frac{M_\chi^2 (g_l-g_r)^2 \left(\left(19 \lambda ^4-32 \lambda ^2+16\right) (g_l^2 + g_r^2)+6 \lambda ^2 \left(8-9 \lambda ^2\right) g_l g_r \right)}{192 \pi  \sqrt{1-\lambda ^2} M_\Omega^4}  \\[3.5ex]  
 a_{FtS} &=\frac{\sqrt{1-\lambda ^2} M_\chi^2 \left(g_s^2 (\lambda -1)-g_a^2 (\lambda +1)\right)^2}{8 \pi  M_\Omega^4} \\
 b_{FtS} &= \frac{M_\chi^2}{192 \pi  \sqrt{1-\lambda ^2} M_\Omega^4}  \Big(2 g_s^2 g_a^2 \left(-13 \lambda ^4+11 \lambda ^2+2\right) + \nonumber \\ & \quad  g_s^4 (\lambda -1)^2 (\lambda  (17 \lambda +16)+2)+g_a^4 (\lambda +1)^2 (\lambda  (17 \lambda -16)+2)\Big) \\[3.5ex]  
 a_{FtSr} &= \frac{\sqrt{1-\lambda ^2} M_\chi^2 \left(g_a^2 (\lambda -1)-g_s^2 (\lambda +1)\right)^2}{8 \pi  M_\Omega^4} \\
 b_{FtSr} &= \frac{M_\chi^2}{192 \pi  \sqrt{1-\lambda ^2} M_\Omega^4} \Big(-2 g_s^2 g_a^2 \left(\lambda ^4+13 \lambda ^2-14\right) + \nonumber \\
&\quad g_s^4 (\lambda +1)^2 (\lambda  (29 \lambda -40)+14) + g_a^4 (\lambda -1)^2 (\lambda  (29 \lambda +40)+14)\Big) \\[3.5ex]  
 a_{FtV} &= \frac{\sqrt{1-\lambda ^2} M_\chi^2 \left(g_l^4+2 \lambda  g_l^3 g_r+2 \left(\lambda ^2+2\right) g_l^2 g_r^2+2 \lambda  g_l g_r^3+g_r^4\right)}{8 \pi  M_\Omega^4}\\
 b_{FtV} &= \frac{M_\chi^2}{192 \pi  \sqrt{1-\lambda ^2} M_\Omega^4} \Big(\left(2 \lambda ^4-\lambda ^2+2\right) (g_l^4 +g_r^4)+6 \lambda  \left(3 \lambda ^2-2\right) (g_l^3 g_r + g_r^3 g_l)+ \nonumber \\ 
& \quad 6 \left(9 \lambda ^4-10 \lambda ^2+4\right) g_l^2 g_r^2\Big) \\[3.5ex]  
 a_{FtVr} &= \frac{\sqrt{1-\lambda ^2} \left(\lambda  M_\chi \left(g_l^2+g_r^2\right)-4 g_l M_\chi g_r\right)^2}{8 \pi  M_\Omega^4} \\
 b_{FtVr} &= \frac{M_\chi^2}{192 \pi  \sqrt{1-\lambda ^2} M_\Omega^4} \Big(\left(19 \lambda ^4-32 \lambda ^2+16\right) (g_l^4+g_r^4) + \nonumber \\ & \quad 24 \lambda  \left(2-3 \lambda ^2\right) (g_l^3 g_r + g_l g_r^3)+6 \left(25 \lambda ^4-32 \lambda ^2+16\right) g_l^2 g_r^2\Big) \\[3.5ex]  
 a_{VS} &= \frac{\sqrt{1-\lambda ^2} \left(g_a^2-g_s^2 \left(\lambda ^2-1\right)\right)}{12 \pi  M_\Omega^4} \\
 b_{VS} &= \frac{\left(g_s^2 \left(-7 \lambda ^4+5 \lambda ^2+2\right)+g_a^2 \left(\lambda ^2+2\right)\right)}{288 \pi  \sqrt{1-\lambda ^2} M_\Omega^4} \\[3.5ex]  
 a_{VF} &= \frac{\left(1-\lambda ^2\right)^{3/2}}{M_\Omega^2} \left[\frac{\lambda  g_l M_\chi g_r \left(g_l^2+g_r^2\right)}{12 \pi  M_\Omega}+\frac{5  g_l^2 g_r^2}{36 \pi}+ \right. \nonumber \\ & \quad \left.\frac{ M_\chi^2 \left(8 \left(g_l^4+g_r^4\right)+\lambda ^2 \left(g_l^4+18 g_l^2 g_r^2+g_r^4\right)\right)}{144 \pi  M_\Omega^2}\right] \\
 b_{VF} &= -\frac{\lambda  \left(19 \lambda ^4-41 \lambda ^2+22\right) g_l M_\chi g_r \left(g_l^2+g_r^2\right)}{288 \pi  \sqrt{1-\lambda ^2} M_\Omega^3}+ \frac{\sqrt{1-\lambda ^2} \left(25 \lambda ^2+6\right) g_l^2 g_r^2}{288 \pi  M_\Omega^2}+ \nonumber \\
 & \quad \frac{\sqrt{1-\lambda ^2} M_\chi^2 \left(-144 \lambda ^2 \left(g_l^4+g_r^4\right)+248 \left(g_l^4+g_r^4\right)+\lambda ^4 \left(31 g_l^4+270 g_l^2 g_r^2+31 g_r^4\right)\right)}{3456 \pi  M_\Omega^4} \\[3.5ex]  
 a_{VFr} &= \frac{  \left(1-\lambda ^2\right)^{3/2}}{M_\Omega^2} \left[\frac{\lambda g_l M_\chi g_r \left(g_l^2+g_r^2\right)}{9 \pi  M_\Omega}+\frac{g_l^2 g_r^2}{3 \pi}- \right. \nonumber \\ & \quad \left. \frac{ M_\chi^2 \left(\lambda ^2 \left(g_l^4-14 g_l^2 g_r^2+g_r^4\right)-8 \left(g_l^4+g_r^4\right)\right)}{36 \pi  M_\Omega^2}\right] \\
 b_{VFr} &= -\frac{\lambda  \left(\lambda ^4+\lambda ^2-2\right) g_l M_\chi g_r \left(g_l^2+g_r^2\right)}{72 \pi  \sqrt{1-\lambda ^2} M_\Omega^3}+\frac{\sqrt{1-\lambda ^2} \left(7 \lambda ^2+2\right) g_l^2 g_r^2}{72 \pi  M_\Omega^2} \nonumber \\
 & \quad \frac{\sqrt{1-\lambda ^2} M_\chi^2 \left(\left(-23 \lambda ^4+66 \lambda ^2+32\right) (g_l^4 + g_r^4)+6 \lambda ^2 \left(27 \lambda ^2-2\right) g_l^2 g_r^2\right)}{864 \pi  M_\Omega^4} \\[3.5ex]  
 a_{VV} &= 0 \\
 b_{VV} &= 
 \frac{\sqrt{1-\lambda ^2} M_\chi^2 \left(4 \left(g_s^2+g_a^2\right)-\lambda ^2 \left(g_s^2-6 g_s g_a+g_a^2\right)\right)}{16 \pi  M_\Omega^4} 
\end{align}
\section{Radiative Pair Production}
\label{app:2t3appendix}
General results for the differential photon cross section in the proccess $\Ppositron \Pelectron \rightarrow \chi \chi \gamma$ we need in chapters \ref{chap:chichigamma} and \ref{chap:ilcanalysis} are given in table  \ref{tbl:2t3crosssections}. We list all the additional terms here, which appear in the analytic cross section formulae but not in the Weizsäcker--Williams solution:
\begin{align}
A_\text{SF} &= \frac{ \left(1 - V_{x \theta} \right)}{4 M^2_\Omega}
\frac{\hat{s}}{1-x}  \left[(g_s+g_a)^4 C_R + (g_s-g_a)^4 C_L \right] \\
A_\text{SFr} &= \frac{\alpha}{8 \pi}
\frac{\hat{s}}{M_\Omega^2} \frac{x}{1-x}  \left[(g_s+g_a)^4 C_R + (g_s-g_a)^4 C_L \right] \\
A_\text{FtS} &= \frac{(1-V_{x \theta})}{4} \left[ C_S (\hat{s} - 4 M_\chi^2) + \frac{1}{1 - x} C_S (2 M_\chi^2 + \hat{s}) \right] \\
A_\text{VF} & = 20 G^2_{lr} C_S (1-V_{x \theta}) \frac{x }{1-x} (\hat{s}^2 + 4 M_\chi^2 \hat{s}-8  M_\chi^4  )+  \frac{(g_L^4 C_L + g_R^4 C_R)}{M_\Omega^2 } \Big[  \nonumber \\
&\quad -\frac{1}{32}\frac{x^4 \sin^2(2 \theta)}{(x-1)^2
    ((x-1)^2+1)}\hat{s} ( 3 \hat{s}^2 + 26 M_\chi^2 \hat{s} - 32 M_\chi^4)  \nonumber \\
&\quad+ 6 \frac{x}{((x-1)^2+1)} \hat{s} ( \hat{s}^2  + 7 M_\chi^2 \hat{s} - 24 M_\chi^4 )  \nonumber \\
&\quad- \frac{1}{4} (1-V_{x \theta}) (21 \hat{s}^3 + 282 M_\chi^2 \hat{s}^2 - 1144 M_\chi^4 \hat{s}  + 160 M_\chi^6)\nonumber \\
&\quad+\frac{3}{2}\frac{(1-V_{x \theta})}{(1-x)}  \hat{s} (\hat{s}^2-28 M_\chi^2 \hat{s}+ 16 M_\chi^4)\nonumber \\
&\quad+\frac{1}{4} \frac{(1-V_{x \theta})}{(1-x)^2} \hat{s} (7 \hat{s}^2-126 M_\chi^2 \hat{s}+ 32 M_\chi^4)\nonumber \\
&\quad\left. +\frac{(1-V_{x \theta})}{(1-x)^3} \hat{s} (\hat{s}^2 +2 M_\chi^2  \hat{s}+6 M_\chi^4) \right] \\ 
A_\text{VFr} & = \frac{(g_L^4 C_L + g_R^4 C_R)}{M_\Omega^2 } \Big[ - \frac{1}{32} \frac{x^4 \sin^2(2 \theta)}{(x-1)^2
    ((x-1)^2+1)} \hat{s}(  \hat{s}^2 + 32 M_\chi^2 \hat{s} -24 M_\chi^4    )  \nonumber \\
&\quad+ 2 \frac{x}{((x-1)^2+1)} \hat{s} (  \hat{s}^2 + 12 M_\chi^2 \hat{s} +  56 M_\chi^4  )  \nonumber \\
&\quad- \frac{1}{4}(1-V_{x \theta}) (7\hat{s}^3 + 144 M_\chi^2 \hat{s}^2 -168 M_\chi^4 \hat{s}+1280 M_\chi^6)\nonumber \\
&\quad+\frac{1}{2}\frac{(1-V_{x \theta})}{(1-x)}  \hat{s} (\hat{s}^2 -48 M_\chi^2 \hat{s}+ 56 M_\chi^4)\nonumber \\
&\quad+\frac{1}{4}\frac{(1-V_{x \theta})}{(1-x)^2} \hat{s} (9 \hat{s}^2 -272 M_\chi^2 \hat{s}+104 M_\chi^4)\nonumber \\
&\quad\left. +2 \frac{(1-V_{x \theta})}{(1-x)^3} \hat{s} (\hat{s}^2+2 M_\chi^2 \hat{s}+ 6 M_\chi^4) \right]
\end{align}

\section{Elastic Scattering}
\label{sec:nonrelmatrixelements}
We give the elastic proton scattering cross section we use in the direct detection analysis in chapter \ref{chap:xenon} for all benchmark models defined in table \ref{tbl:constraints}. We use the common abbreviations
\begin{align}
\mu &\equiv \frac{M_P M_\chi}{M_P + M_\chi}, \\
F_P &\equiv \sum_{q=u,d,s} \frac{M_P}{m_q} G_\text{eff}^q f_q^P + \frac{2}{27} \sum_{q=c,b,t} \frac{M_P}{m_q} G_\text{eff}^q  \left(1 - \sum_{q=u,d,s} f_q^p \right),\\
B_P &\equiv 2 G_\text{eff}^u + G_\text{eff}^d,\\
\tilde{B}^p &\equiv B^p M_\chi + 2G_\text{eff}^u m_u + G_\text{eff}^d m_d, \\
D_P &\equiv \sum_q G_\text{eff}^q \Delta_q^p. \\
\intertext{The cross sections can then be given as follows:}
\sigma_{\text{SS Scalar}} &= \frac{\mu^2}{4 \pi M_\chi^2}  F_P^2 \\
\sigma_{\text{SS Pseudosc.}} &= 0 \\
\sigma_{\text{SF Scalar/Pseudosc.}} &= \frac{\mu^2}{4 \pi} \left(\pm F_P + \frac{\tilde{B}_P }{M_\Omega}\right)^2\\
\sigma_{\text{SV Vector}} &= \frac{\mu^2}{\pi} B_P^2 \\
\sigma_{\text{SV Axialv.}} &= 0 \\
\sigma_{\text{SV Chiral}} &= \frac{\mu^2}{4 \pi} B_P^2 \\[3.5ex]
\sigma_{\text{FS Scalar}} &= \frac{\mu^2}{\pi} F_P^2 \\
\sigma_{\text{FS Pseudosc.}} &= 0 \\
\sigma_{\text{FV Vector}} &= \frac{\mu^2}{\pi} B_P^2 \\
\sigma_{\text{FV Axialv.}} &= \frac{3 \mu^2}{\pi} D_P^2 \\
\sigma_{\text{FV Chiral}} &= \frac{\mu^2}{16 \pi}  B_P^2 \\[3.5ex]
\sigma_{\text{FtS Scalar}} &= \frac{\mu^2}{16 \pi} \left(F_P + B_P \right)^2 \\
\sigma_{\text{FtS Pseudosc.}} &= \frac{\mu^2}{16 \pi} \left(F_P - B_P \right)^2 \\
\sigma_{\text{FtV Vector}} &= \frac{\mu^2}{\pi} \left(F_P - \half B_P \right)^2 \\
\sigma_{\text{FtV Axialv.}} &= \frac{\mu^2}{\pi} \left(F_P + \half B_P \right)^2 \\
\sigma_{\text{FtV Chiral}} &= \frac{\mu^2}{16 \pi} B_P^2 \\[3.5ex]
\sigma_{\text{VS Scalar}} &= \frac{\mu^2}{4 \pi M_\chi^2} F_P^2 \\
\sigma_{\text{VS Pseudosc..}} &= 0 \\
\sigma_{\text{VF Vector/Axialv.}} &= \frac{\mu^2}{4 \pi} ( F_P \mp \frac{\tilde{B}_P}{M_\Omega})^2 \\
\sigma_{\text{VV Vector}} &= \frac{\mu^2}{\pi} B_P^2 \\
\sigma_{\text{VV Axialv.}} &= 0 \\
\sigma_{\text{VV Chiral}} &= \frac{\mu^2}{4 \pi} B_P^2
\end{align}

\paragraph{Real Fields}
In case of real fields, the matrix element receives an additional contribution with the now undistinguishable $\chi$ and $\chi^\dagger$ fields interchanged. As explained in section \ref{sec:benchmarks}, this will double any scalar $F_P$ and axial--vector $D_P$ contribution, whereas vector--like $B_P$ terms vanish.
\chapter{Full List of Combined Exclusion Plots}
\begin{figure}[H]
\centering
 \includegraphics[width=0.475\columnwidth]{SSScalarplotter_protonxsect_universal} \hfill
 \includegraphics[width=0.475\columnwidth]{SSScalarplotter_protonxsect_yukawa} \\
 \includegraphics[width=0.475\columnwidth]{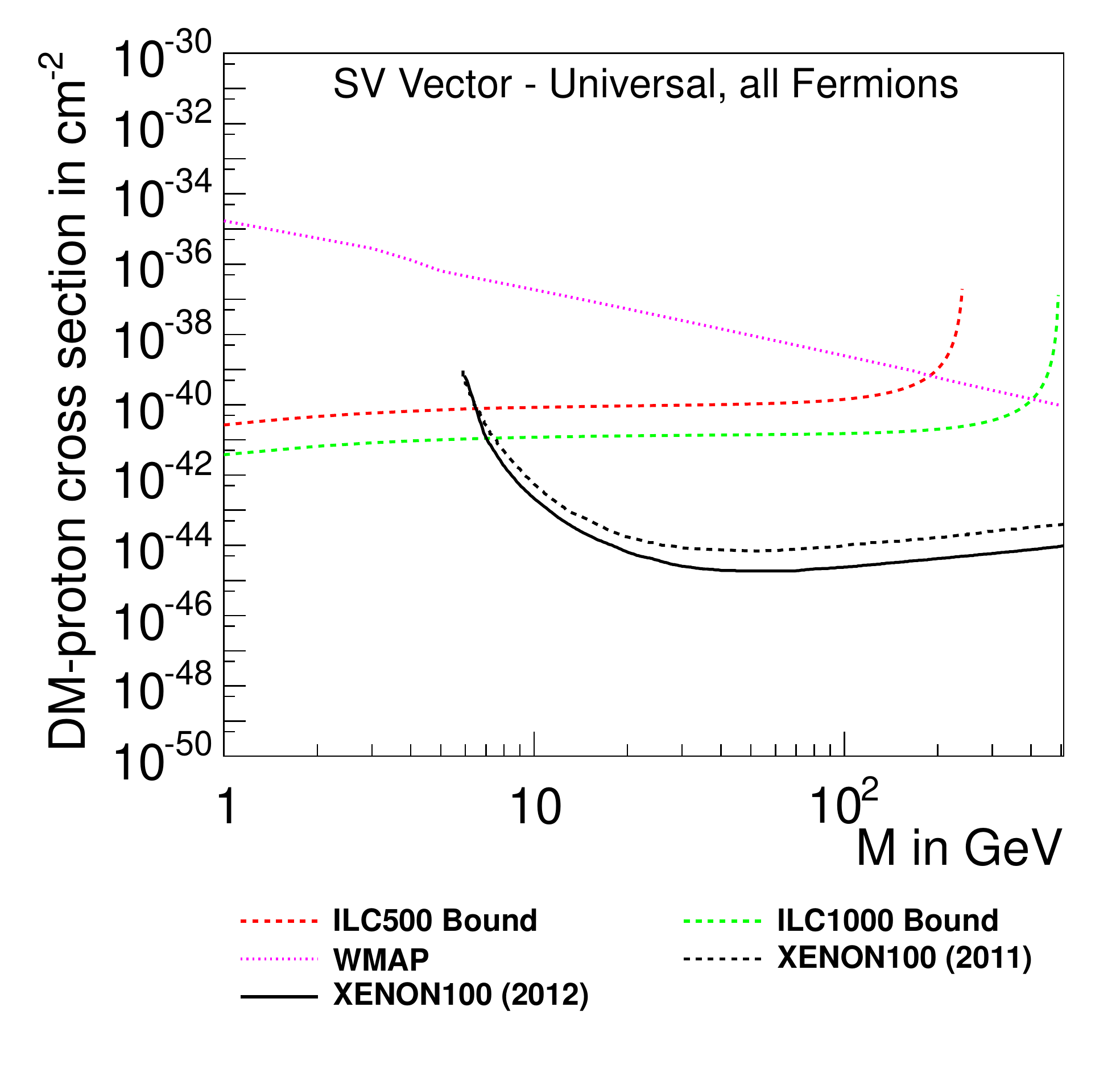} \hfill
 \includegraphics[width=0.475\columnwidth]{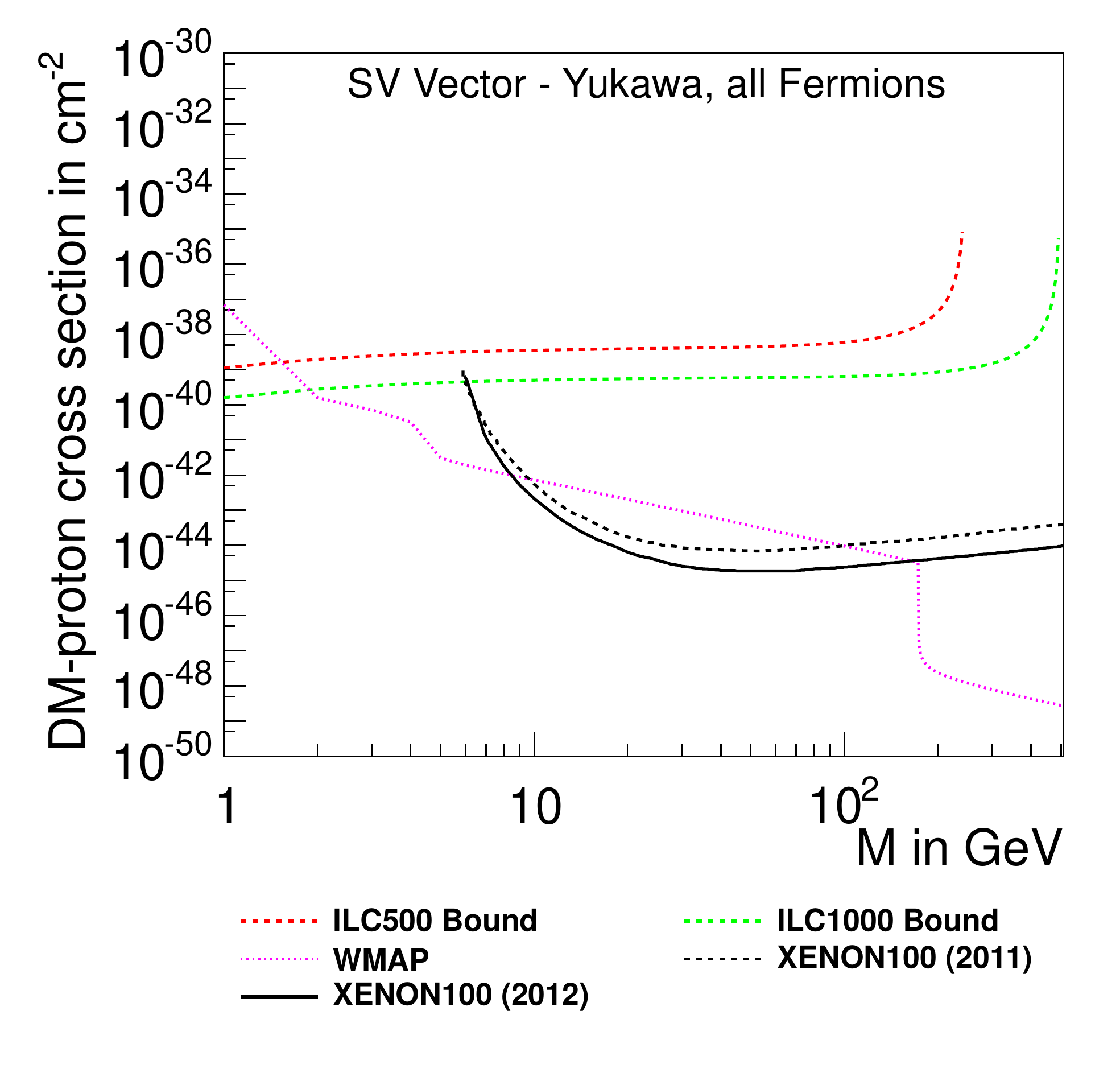}
 \caption{Combined \unit{90}{\%} exclusion limits on the \textbf{spin independent} dark matter proton cross
   section from \textsc{Ilc}, \textsc{Wmap} and \textsc{Xenon} for some \textbf{scalar dark matter} models with \textbf{s--channel scalar or vector coupling} to \textbf{all Standard Model fermions}.}
 \label{img:totalbounds1}
 \end{figure}

\begin{figure}[H]
\centering
 \includegraphics[width=0.475\columnwidth]{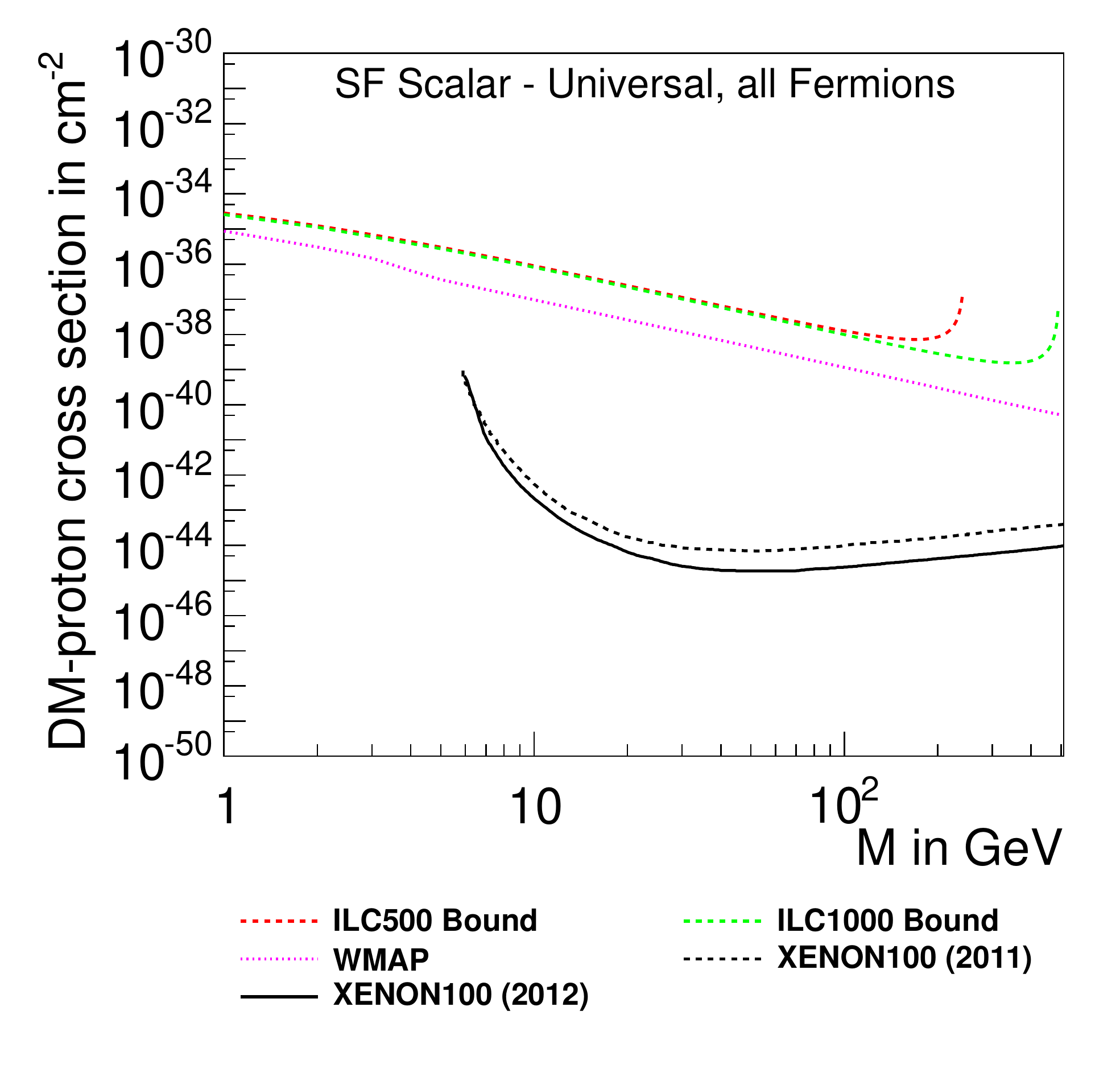} \hfill
 \includegraphics[width=0.475\columnwidth]{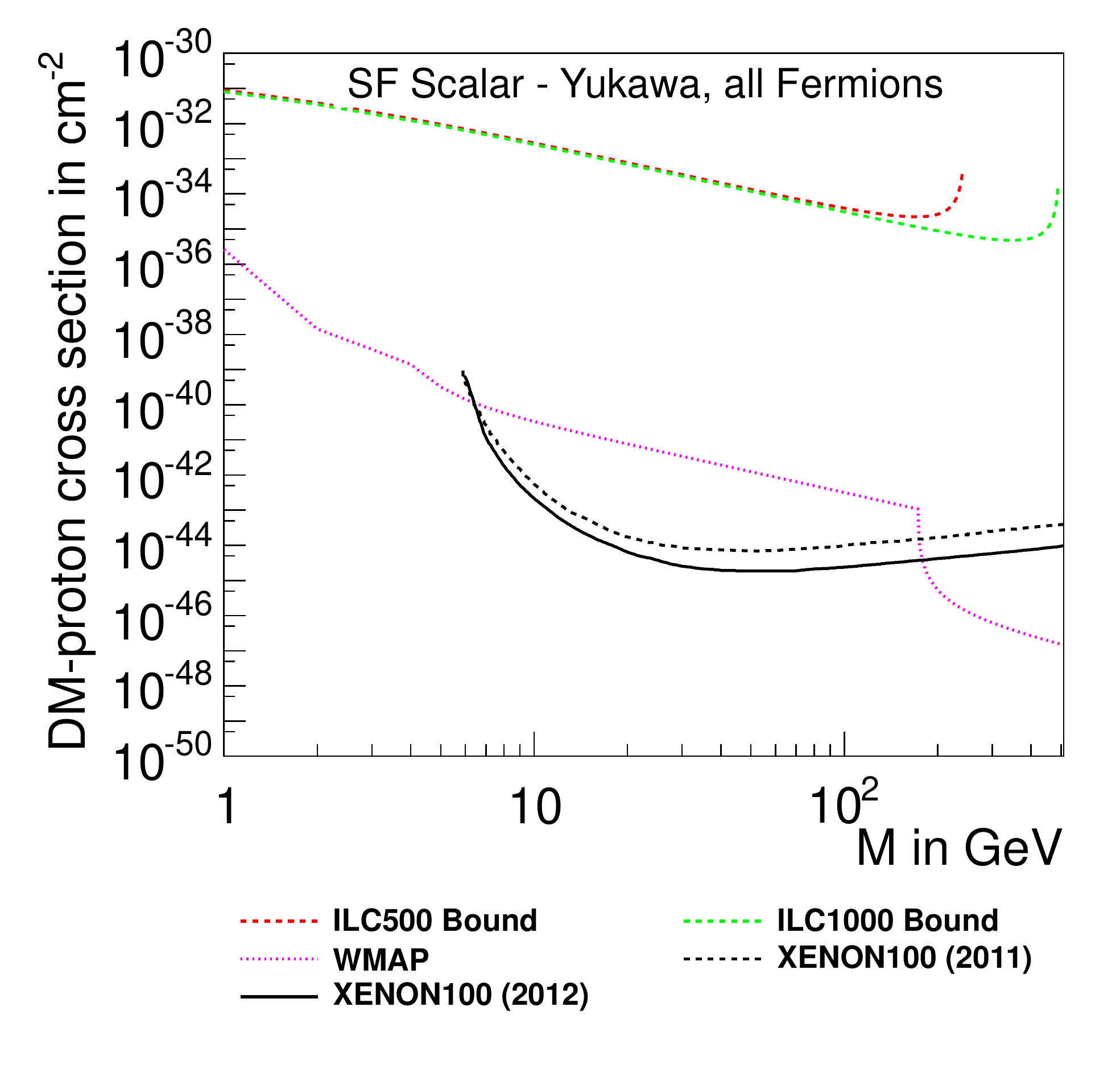} \\
 \includegraphics[width=0.475\columnwidth]{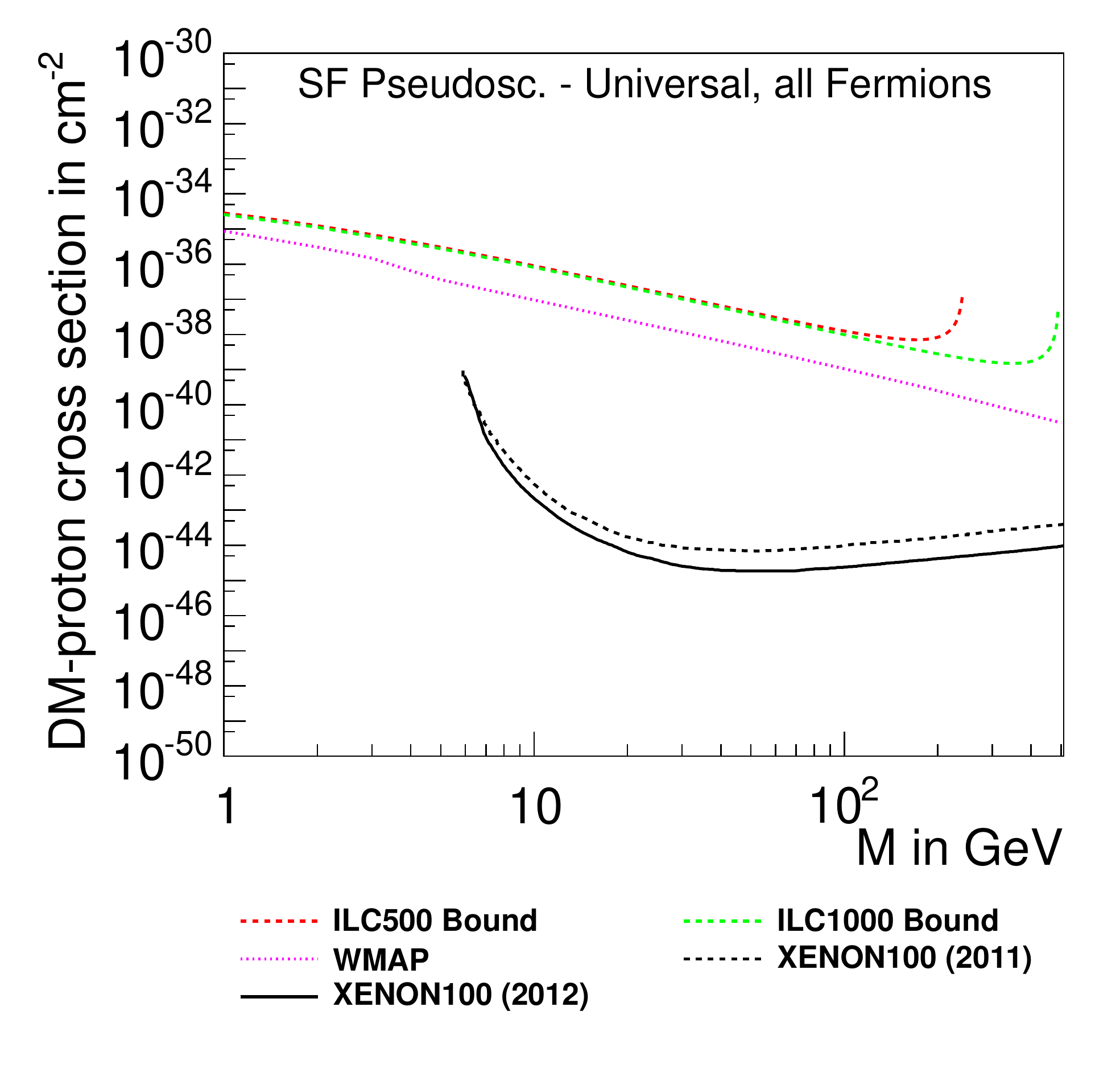} \hfill
 \includegraphics[width=0.475\columnwidth]{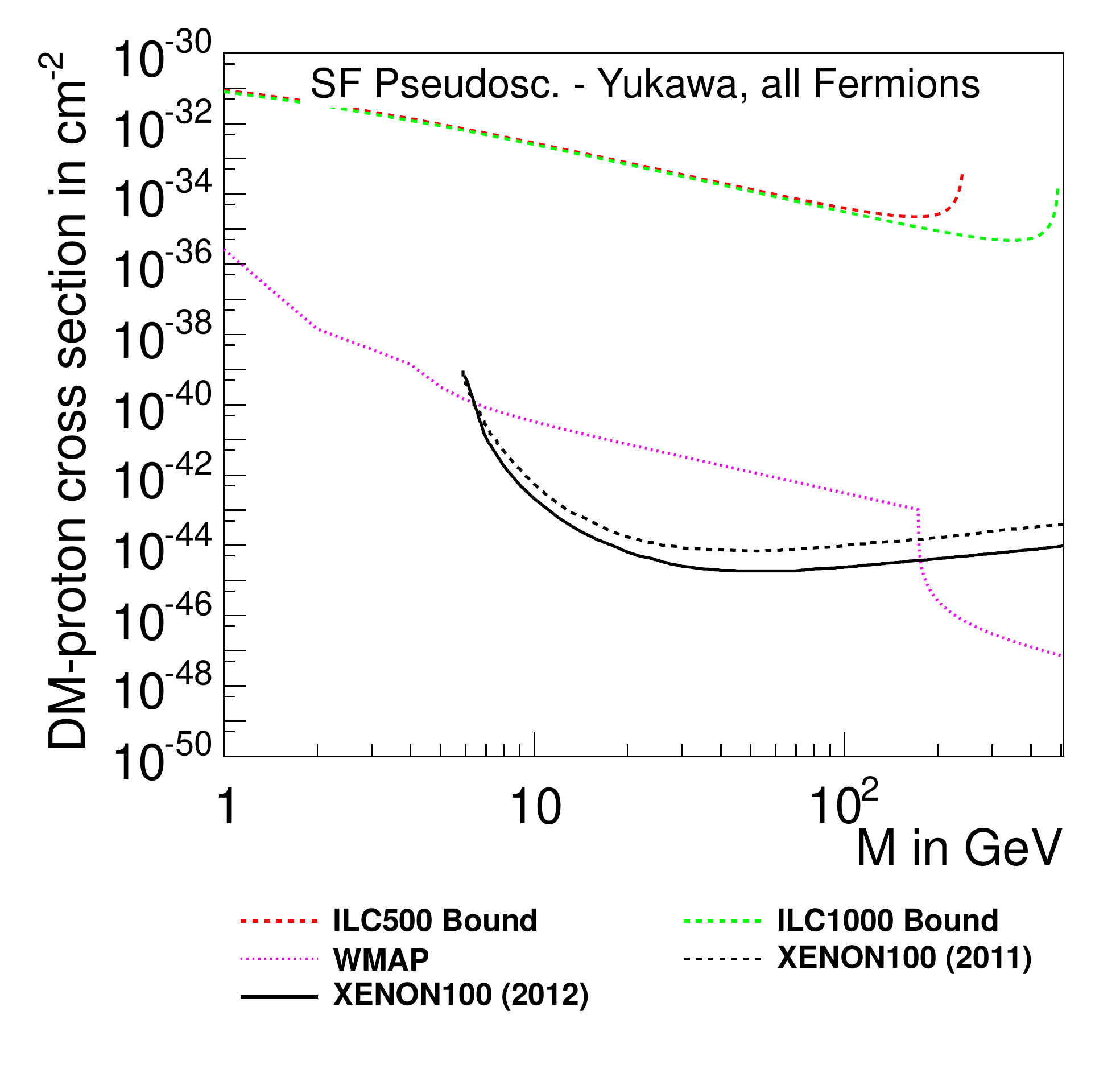} \\
 \includegraphics[width=0.475\columnwidth]{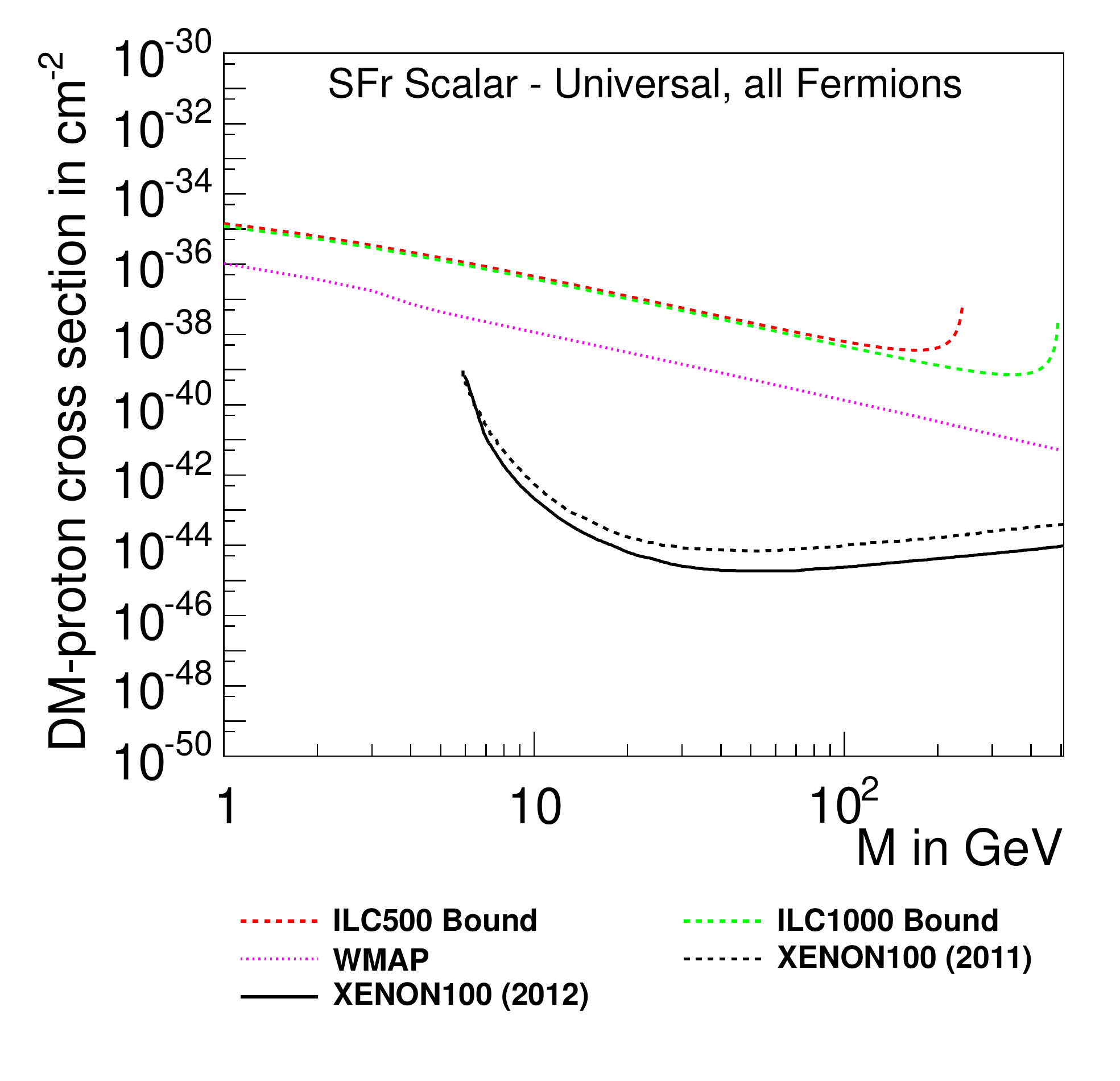} \hfill
 \includegraphics[width=0.475\columnwidth]{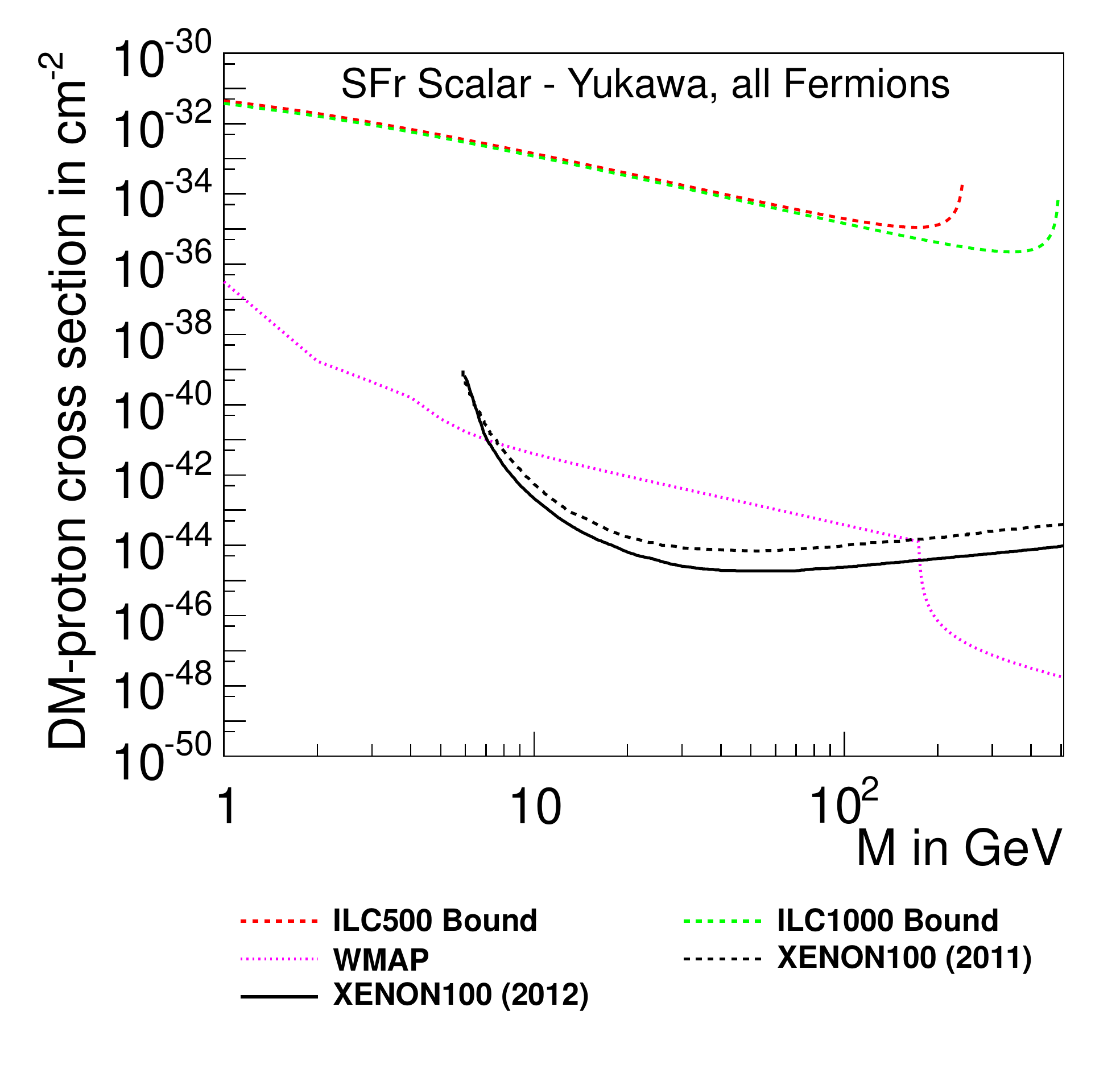}
 \caption{Combined \unit{90}{\%} exclusion limits on the \textbf{spin independent} dark matter proton cross
   section from \textsc{Ilc}, \textsc{Wmap} and \textsc{Xenon} for some \textbf{scalar dark matter} models with \textbf{t--channel fermion coupling} to \textbf{all Standard Model fermions}.}
 \label{img:totalbounds2}
 \end{figure}

\begin{figure}[H]
\centering
 \includegraphics[width=0.475\columnwidth]{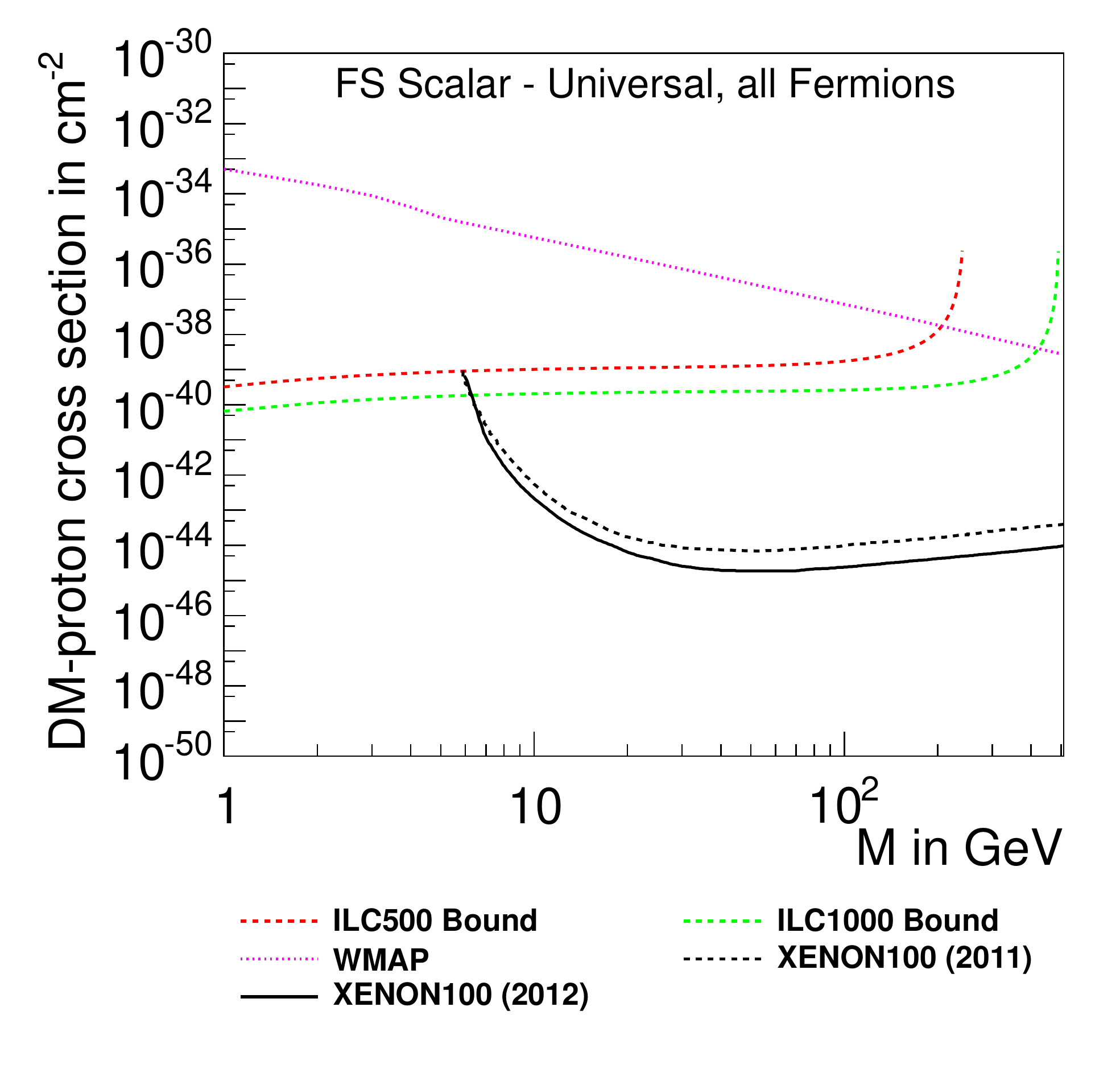} \hfill
 \includegraphics[width=0.475\columnwidth]{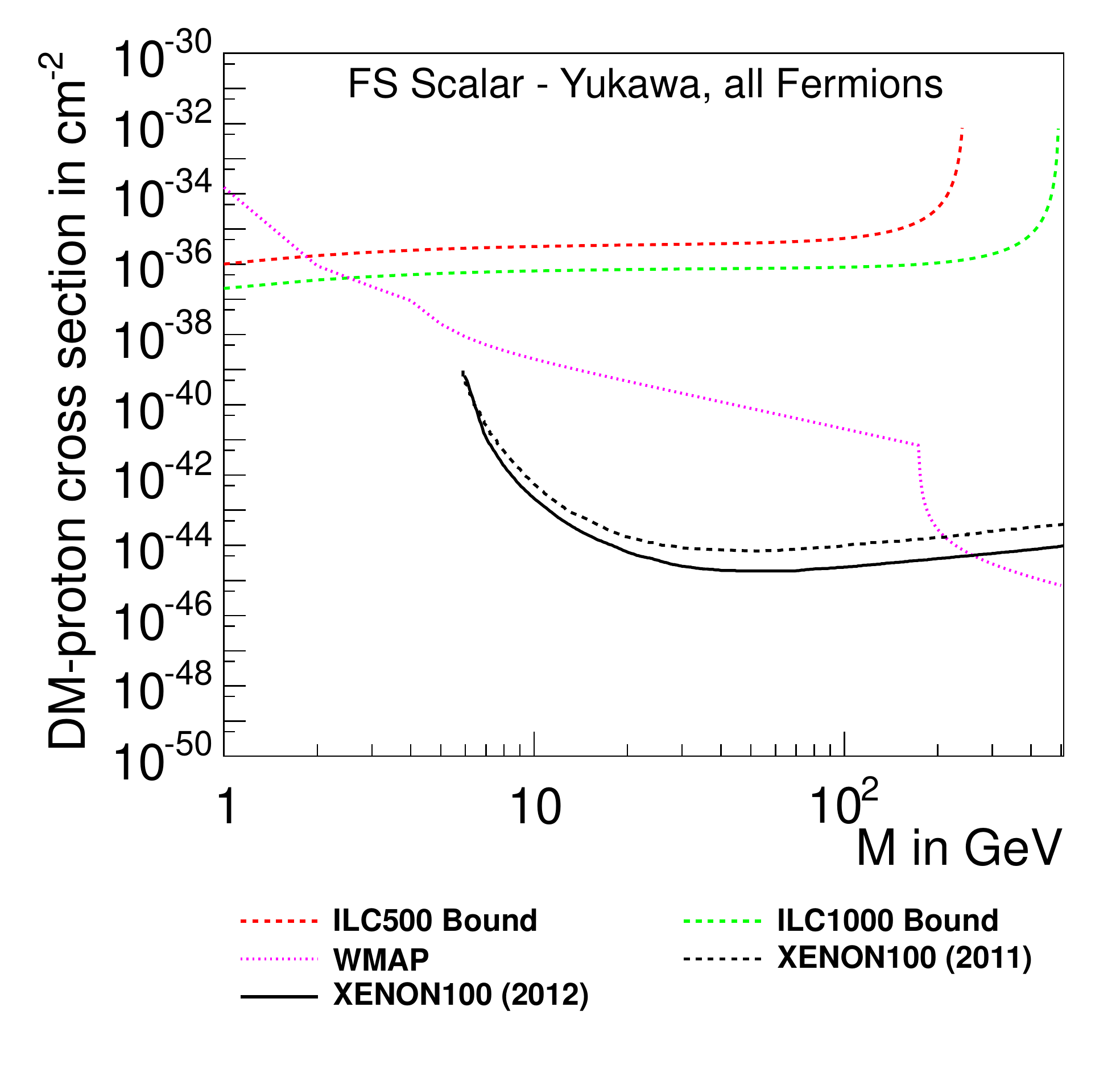} \\
 \includegraphics[width=0.475\columnwidth]{FVVectorplotter_protonxsect_universal} \hfill
 \includegraphics[width=0.475\columnwidth]{FVVectorplotter_protonxsect_yukawa} \\
 \includegraphics[width=0.475\columnwidth]{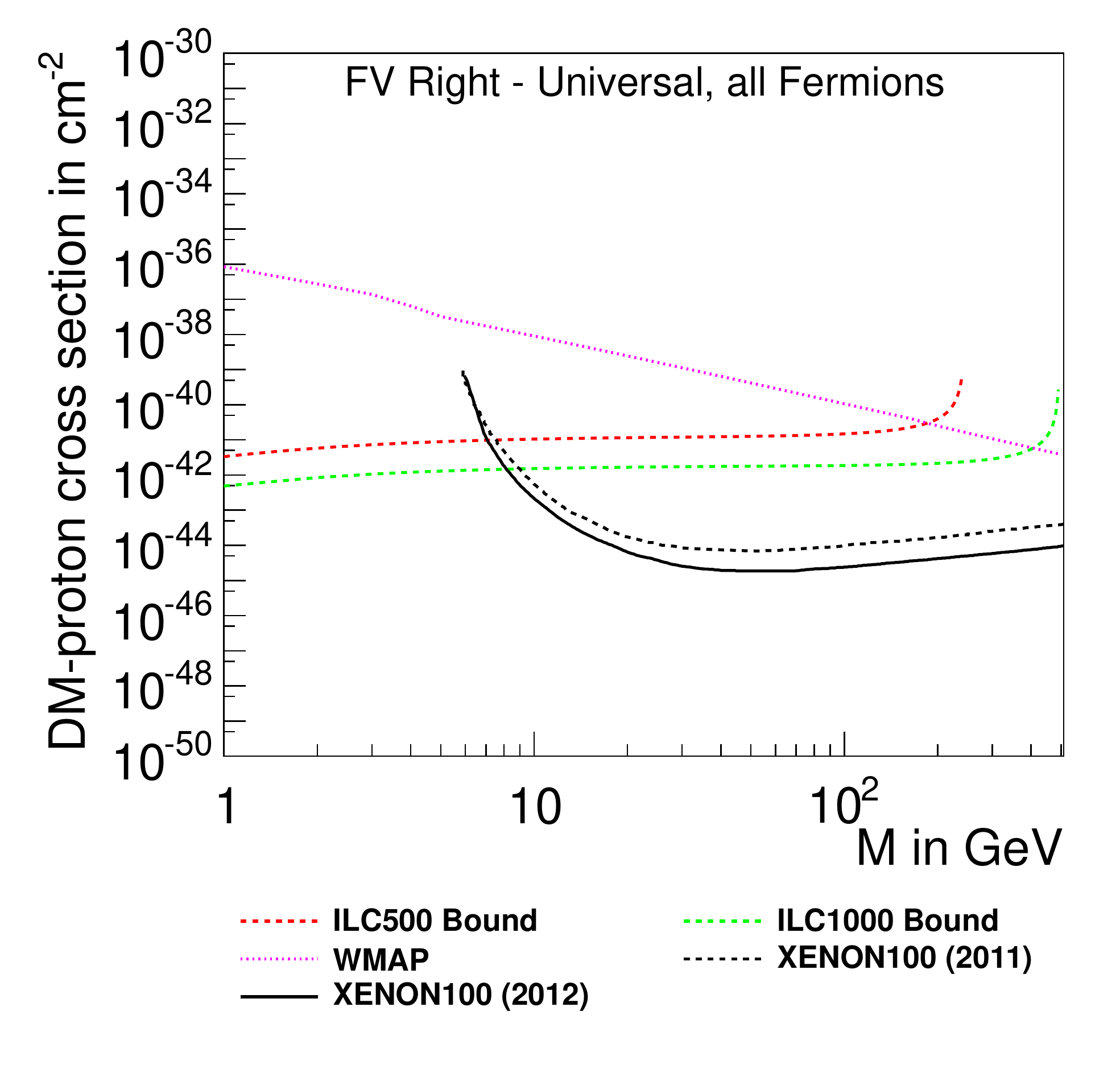} \hfill
 \includegraphics[width=0.475\columnwidth]{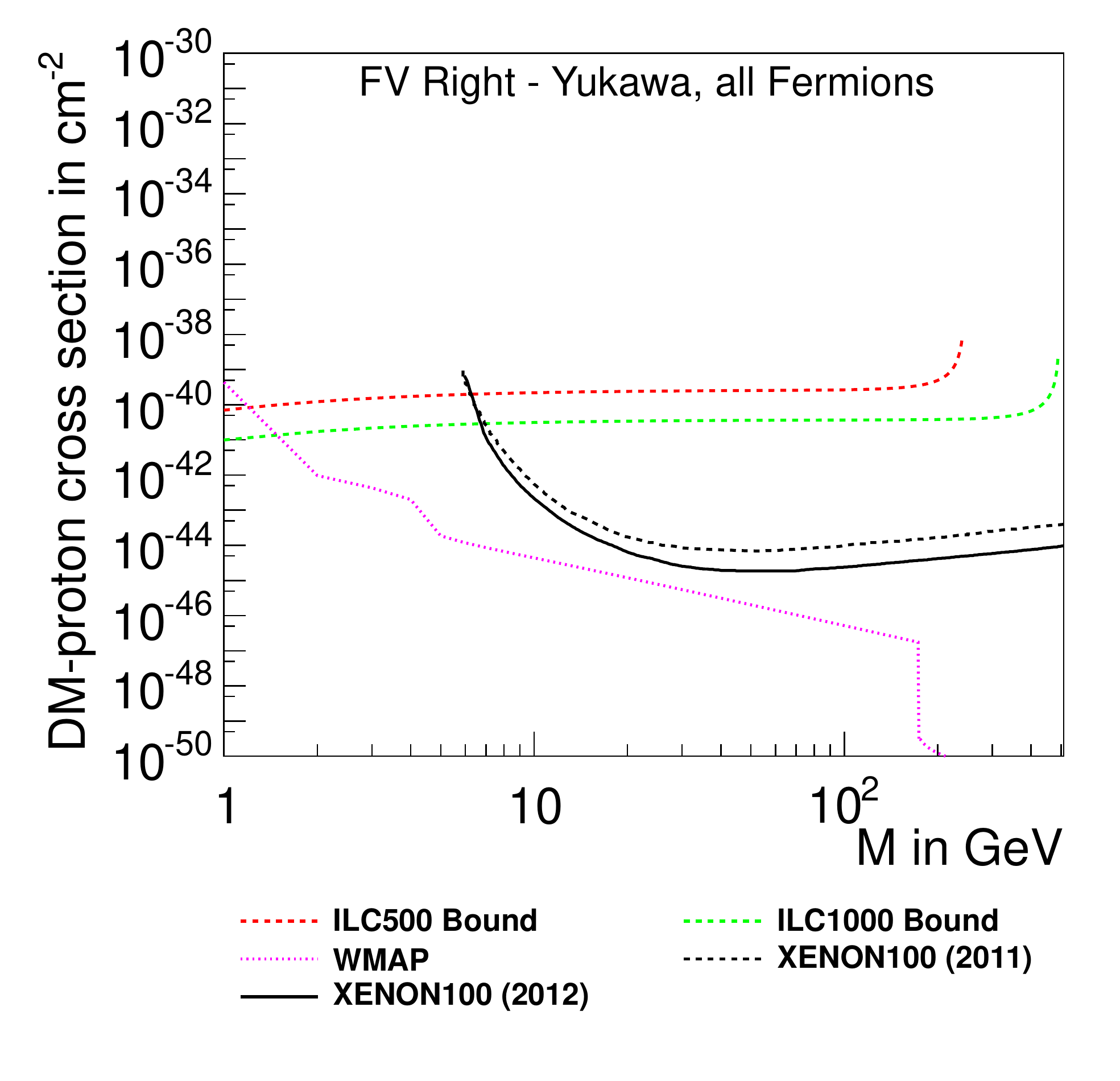}
 \caption{Combined \unit{90}{\%} exclusion limits on the \textbf{spin independent} dark matter proton cross
   section from \textsc{Ilc}, \textsc{Wmap} and \textsc{Xenon} for some \textbf{fermion dark matter} models with \textbf{s--channel scalar or vector coupling} to \textbf{all Standard Model fermions}.}
 \label{img:totalbounds3}
 \end{figure}
\begin{figure}[H]
\centering
 \includegraphics[width=0.475\columnwidth]{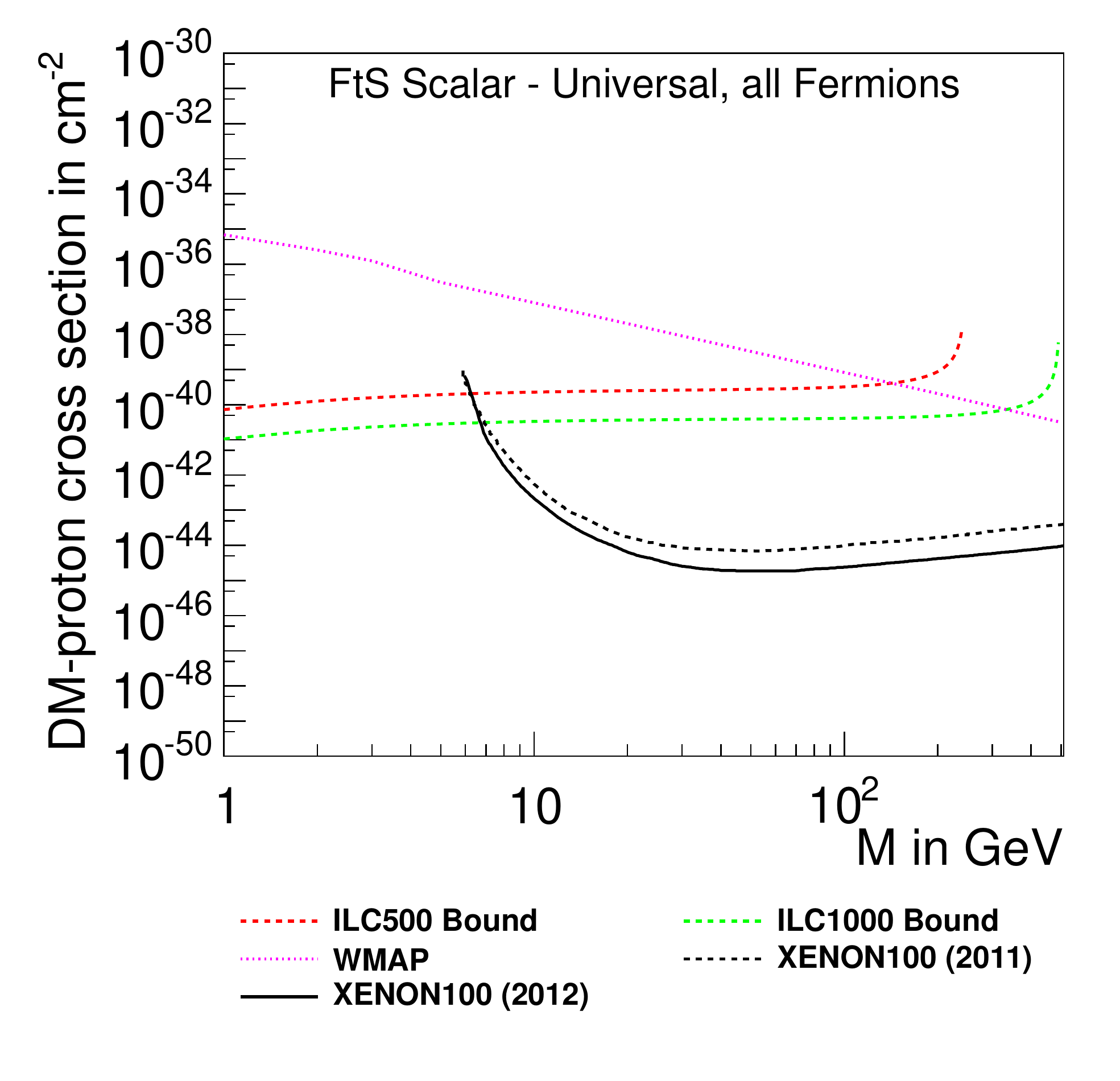} \hfill
 \includegraphics[width=0.475\columnwidth]{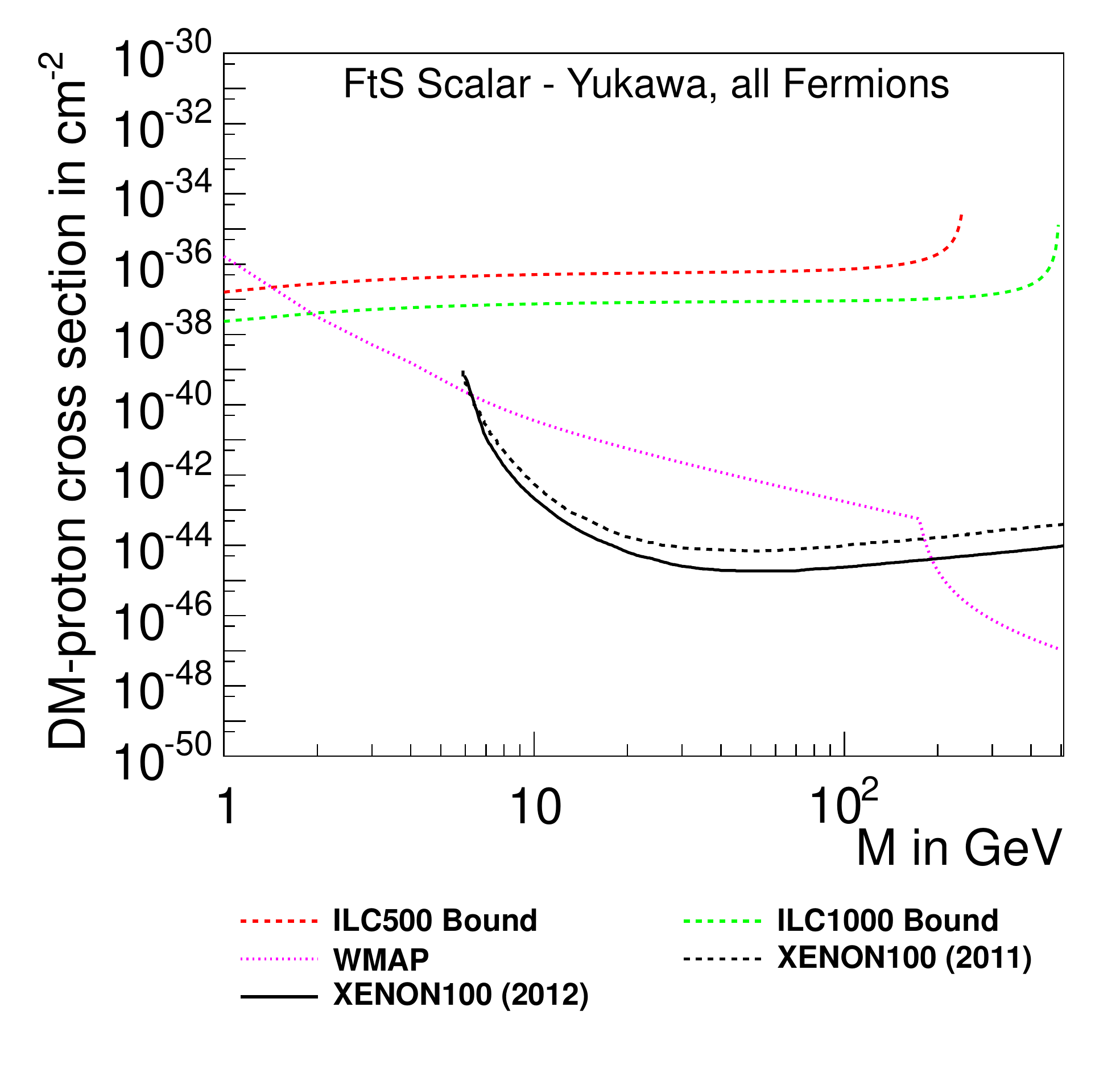} \\
 \includegraphics[width=0.475\columnwidth]{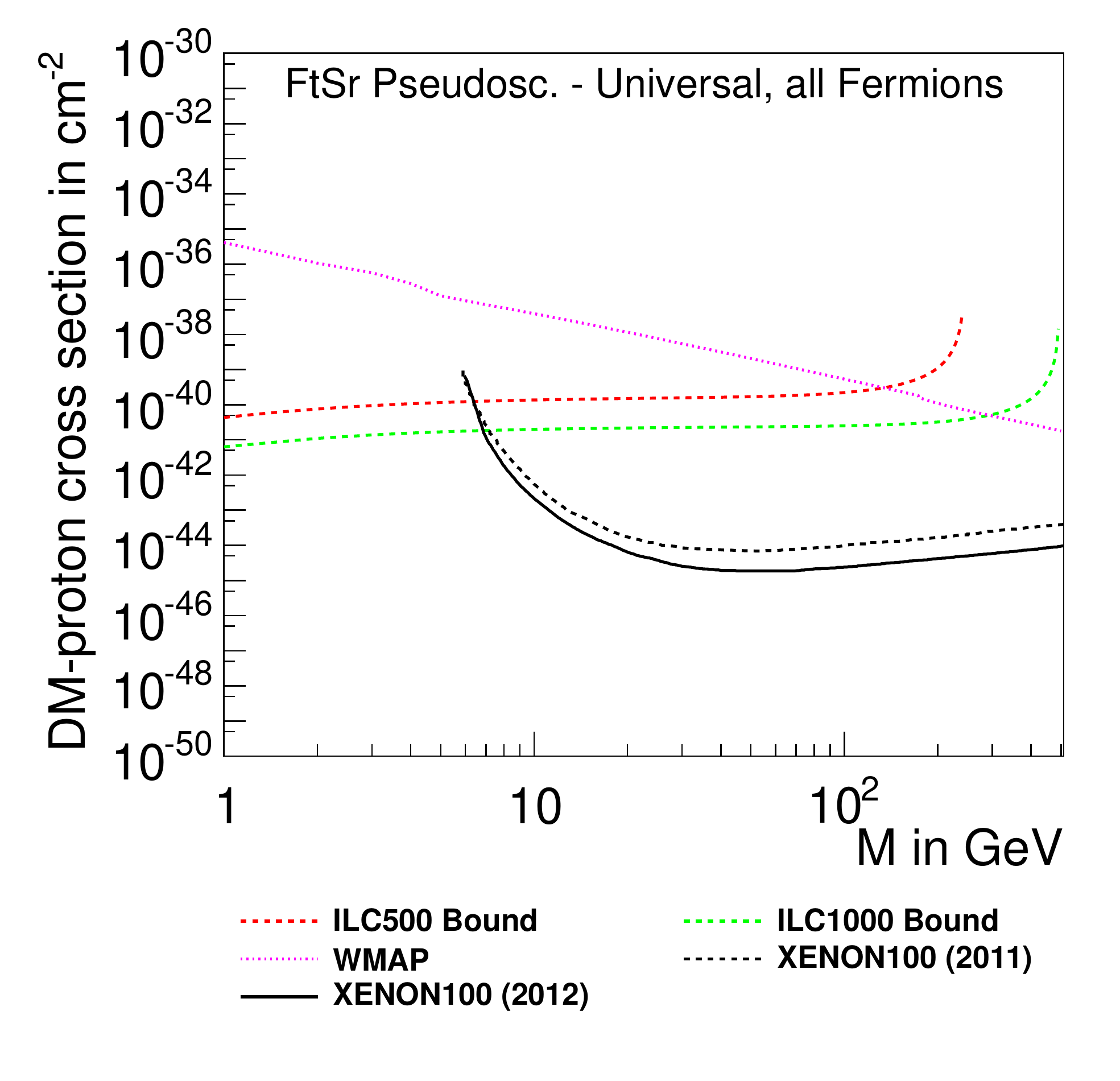} \hfill
 \includegraphics[width=0.475\columnwidth]{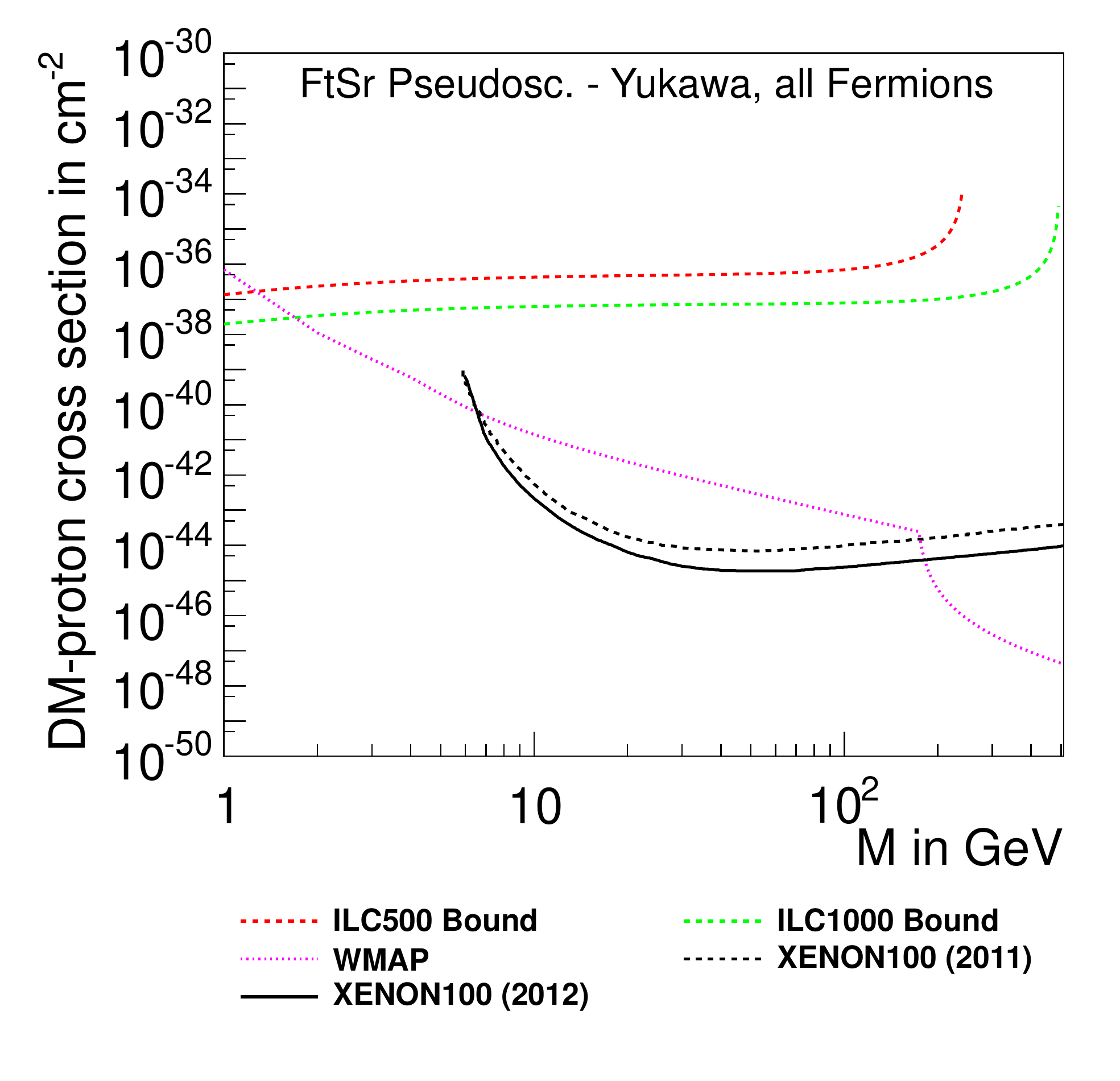} \\
 \includegraphics[width=0.475\columnwidth]{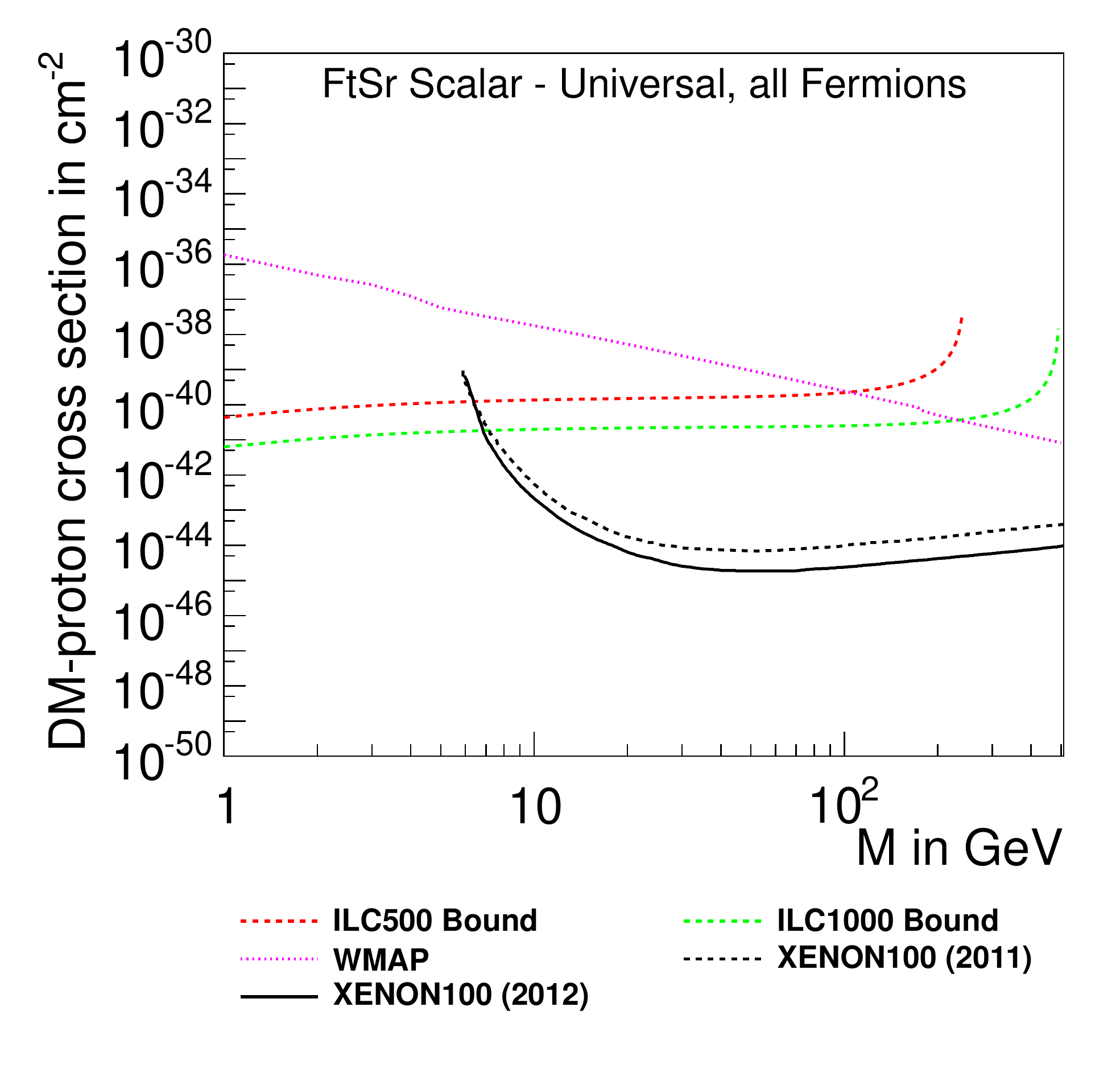} \hfill
 \includegraphics[width=0.475\columnwidth]{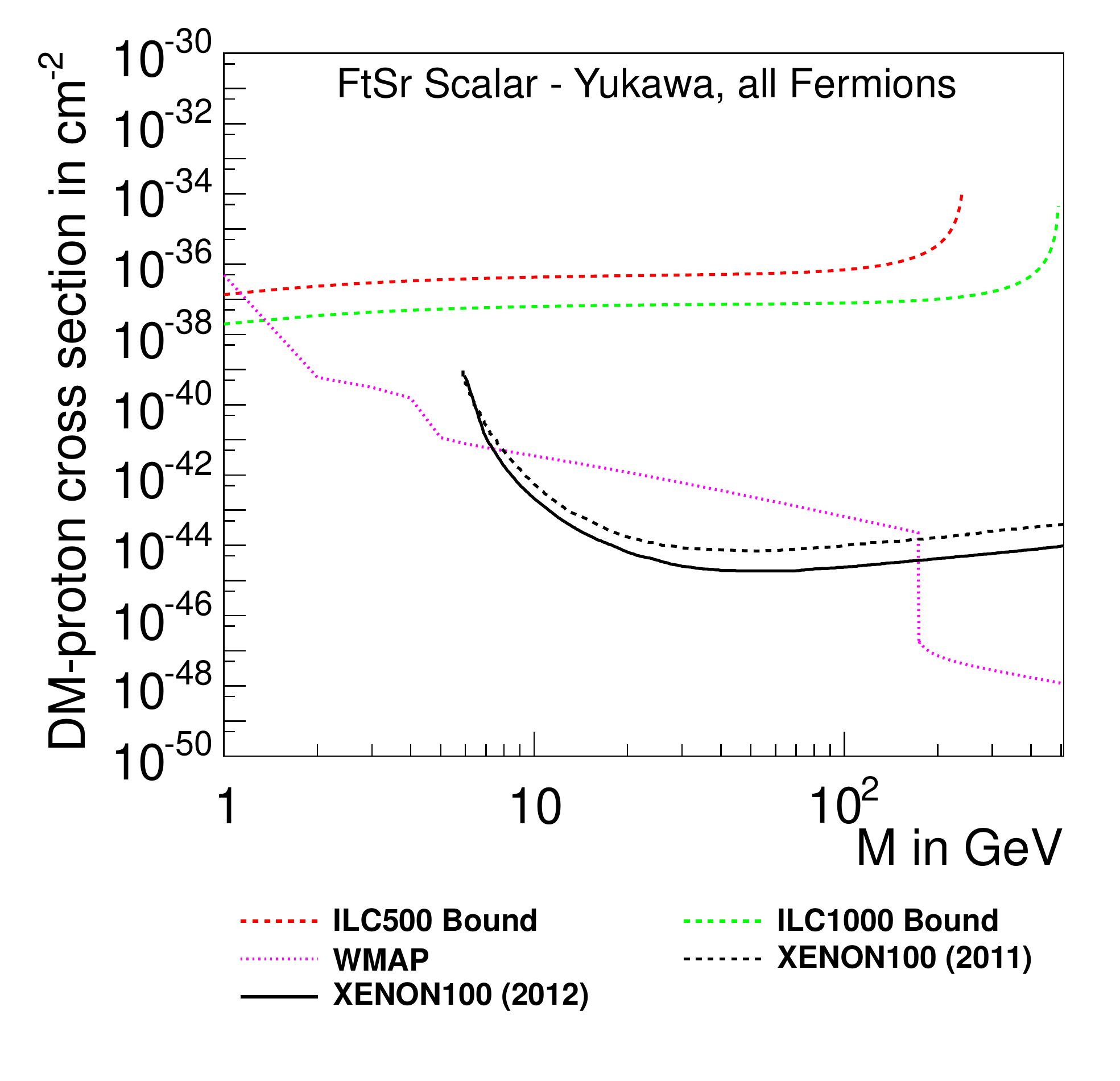}
 \caption{Combined \unit{90}{\%} exclusion limits on the \textbf{spin independent} dark matter proton cross
   section from \textsc{Ilc}, \textsc{Wmap} and \textsc{Xenon} for some \textbf{fermion dark matter} models with \textbf{t--channel scalar coupling} to \textbf{all Standard Model fermions}.}
 \label{img:totalbounds4}
 \end{figure}

\begin{figure}[H]
\centering
 \includegraphics[width=0.475\columnwidth]{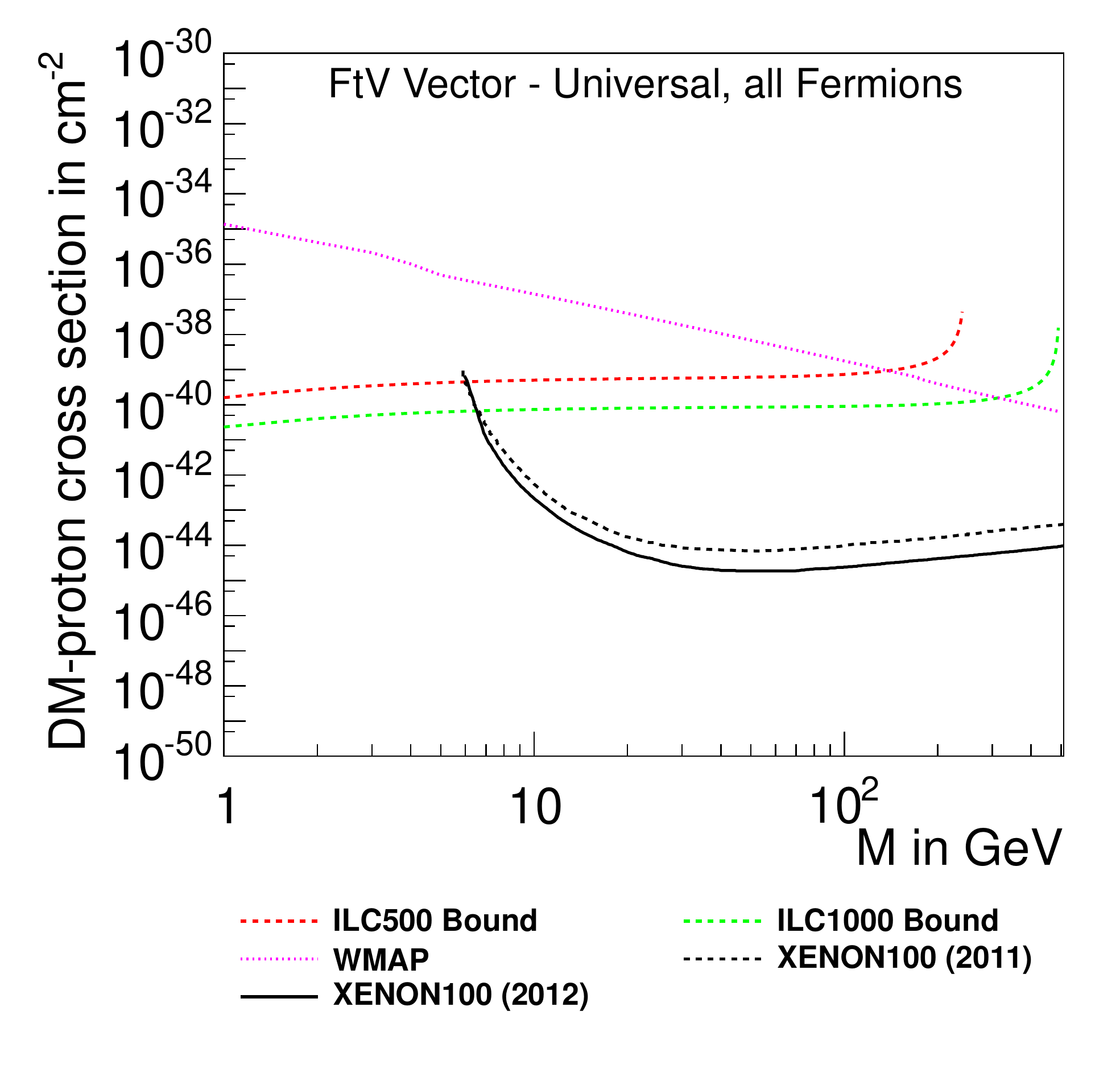} \hfill
 \includegraphics[width=0.475\columnwidth]{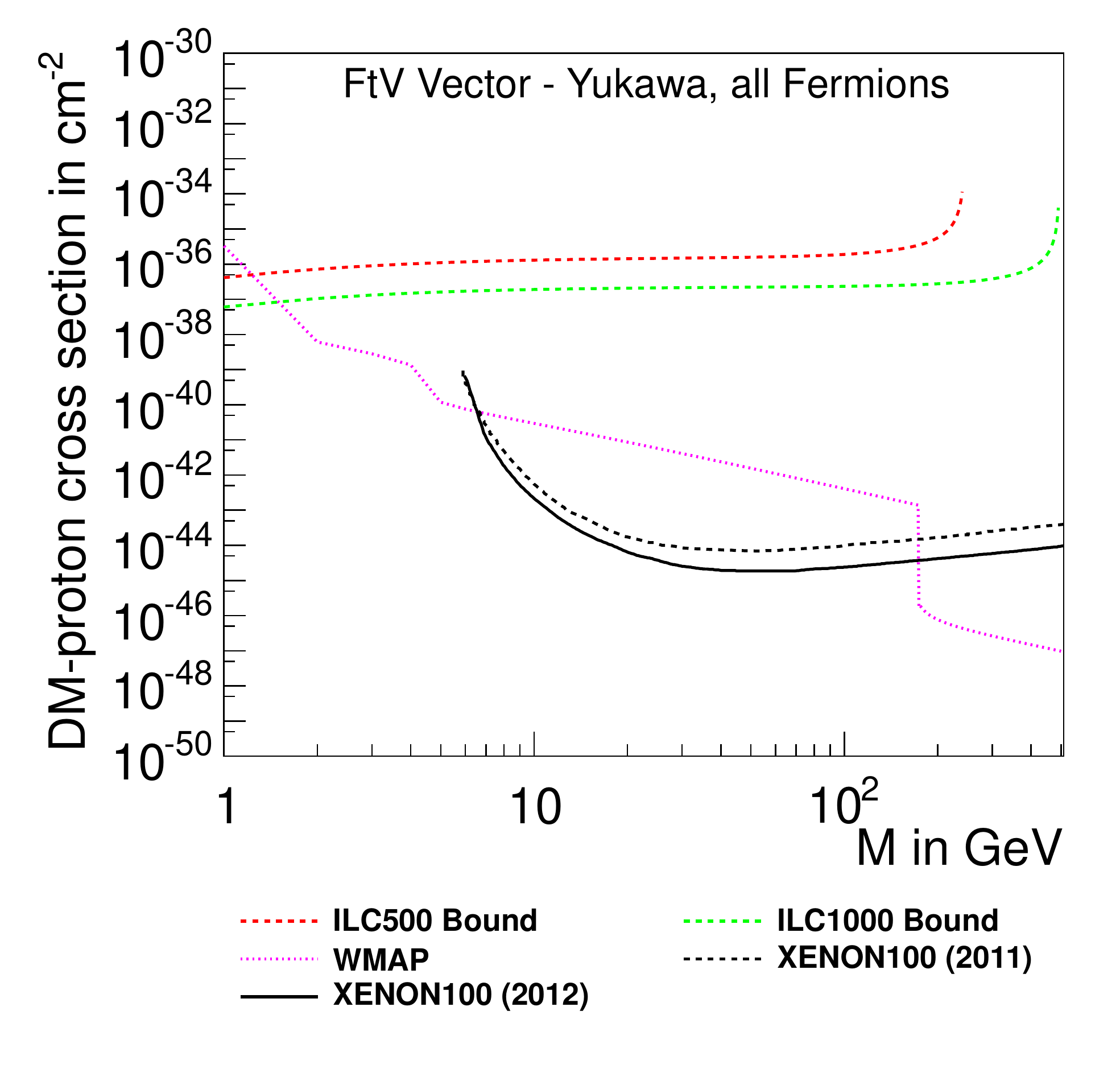} \\
 \includegraphics[width=0.475\columnwidth]{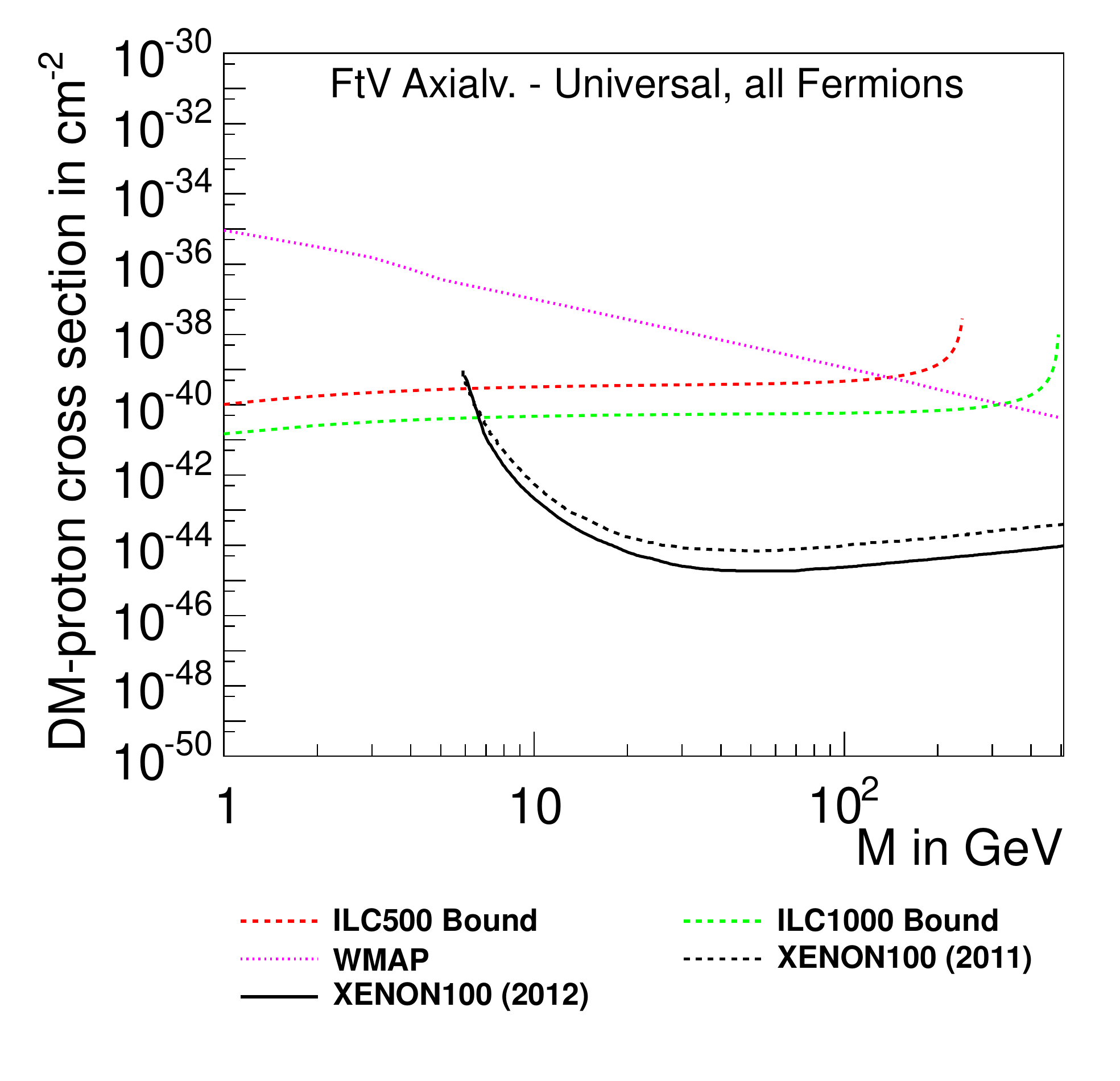} \hfill
 \includegraphics[width=0.475\columnwidth]{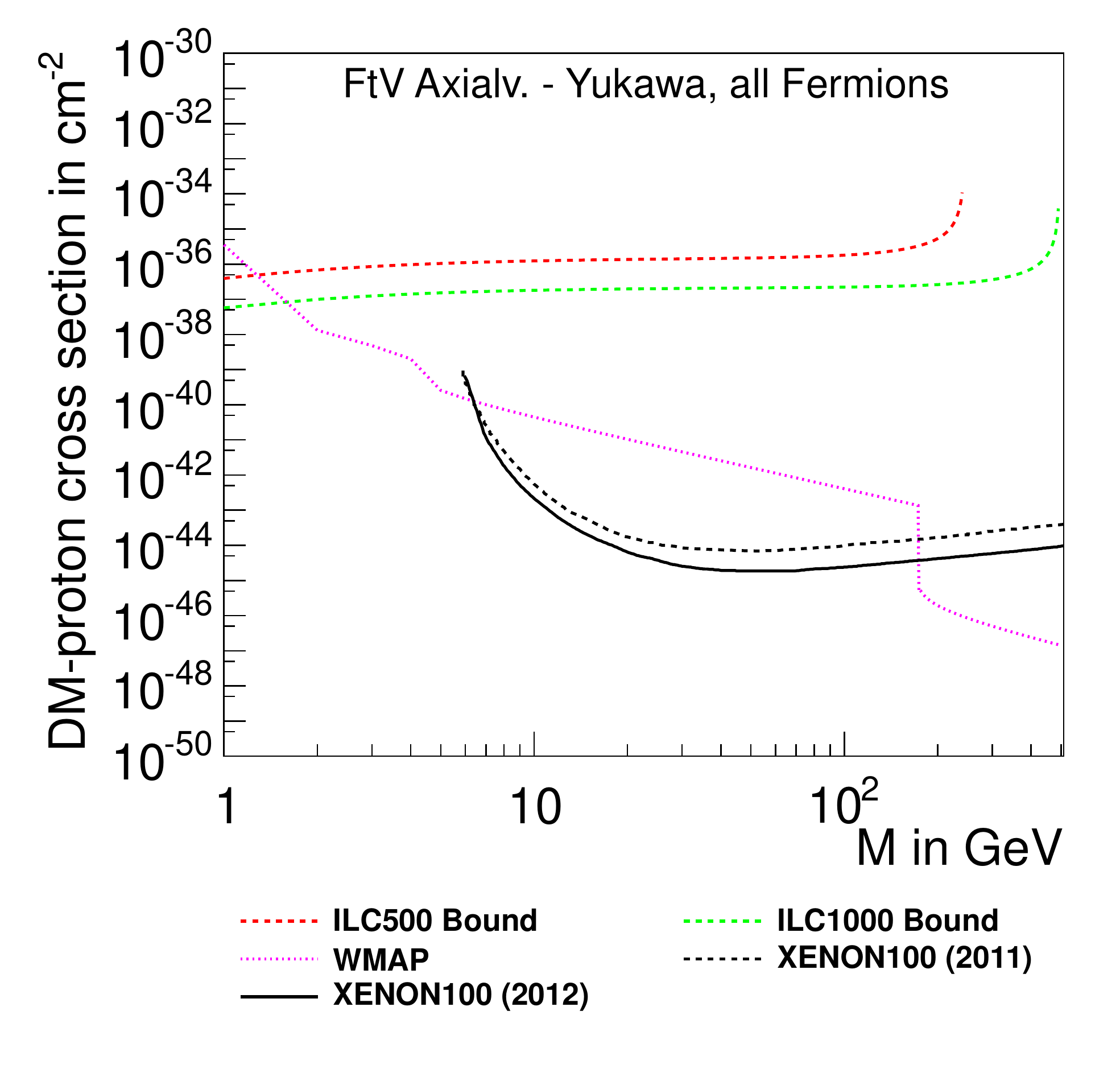} \\
 \includegraphics[width=0.475\columnwidth]{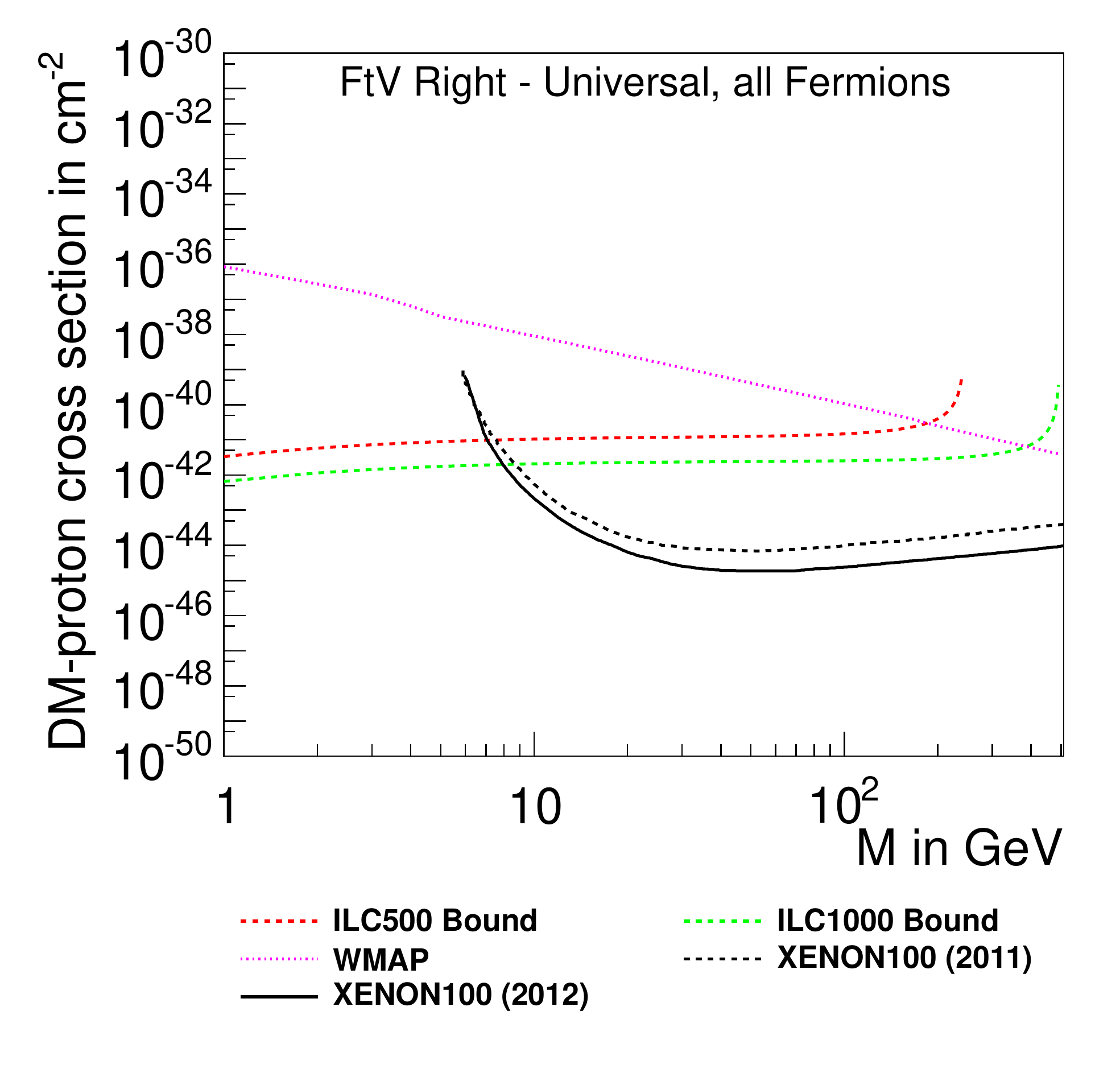} \hfill
 \includegraphics[width=0.475\columnwidth]{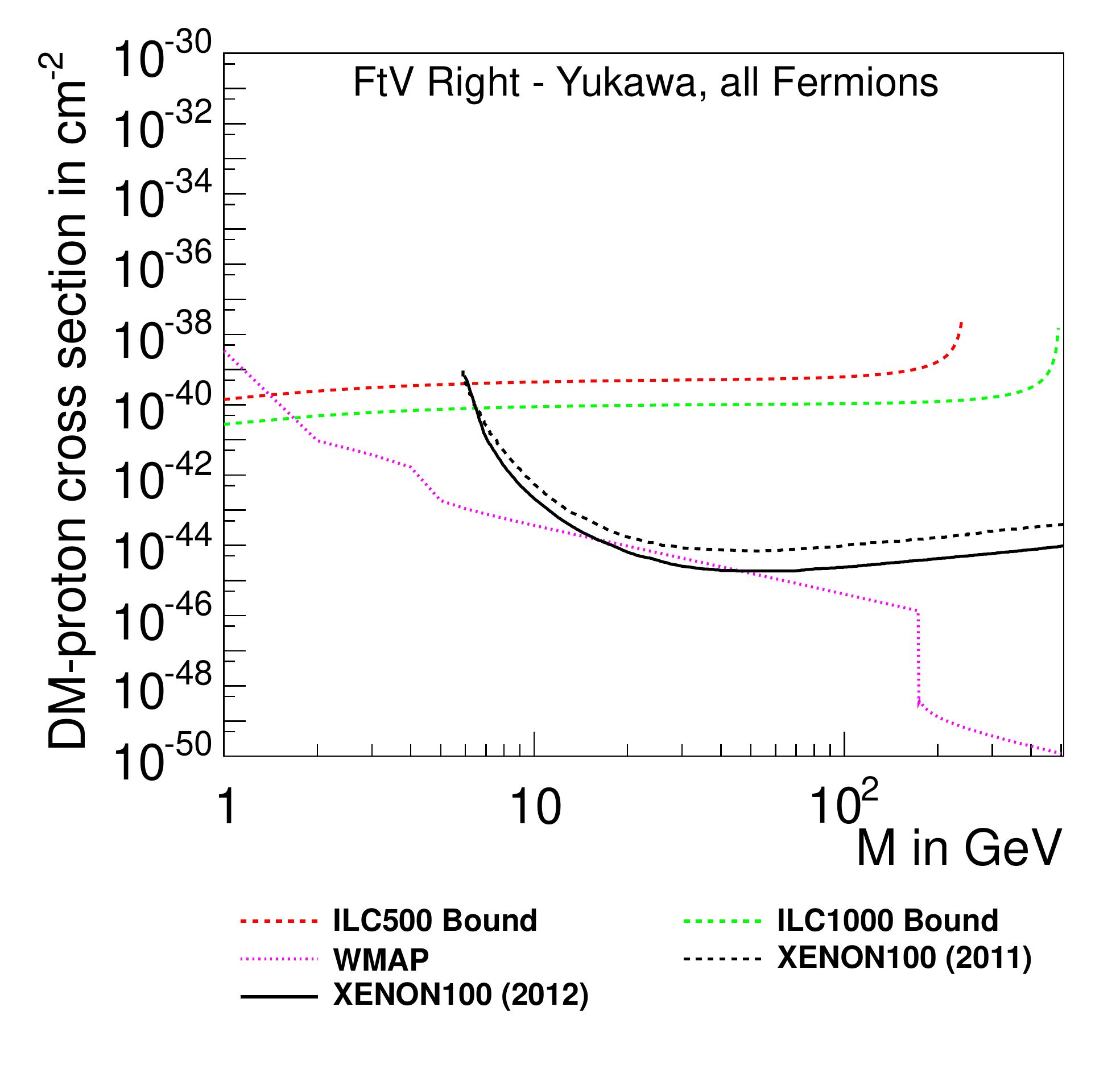}
 \caption{Combined \unit{90}{\%} exclusion limits on the \textbf{spin independent} dark matter proton cross
   section from \textsc{Ilc}, \textsc{Wmap} and \textsc{Xenon} for some \textbf{fermion dark matter} models with \textbf{t--channel vector coupling} to \textbf{all Standard Model fermions}.}
 \label{img:totalbounds5}
 \end{figure}
\begin{figure}[H]
\centering
 \includegraphics[width=0.475\columnwidth]{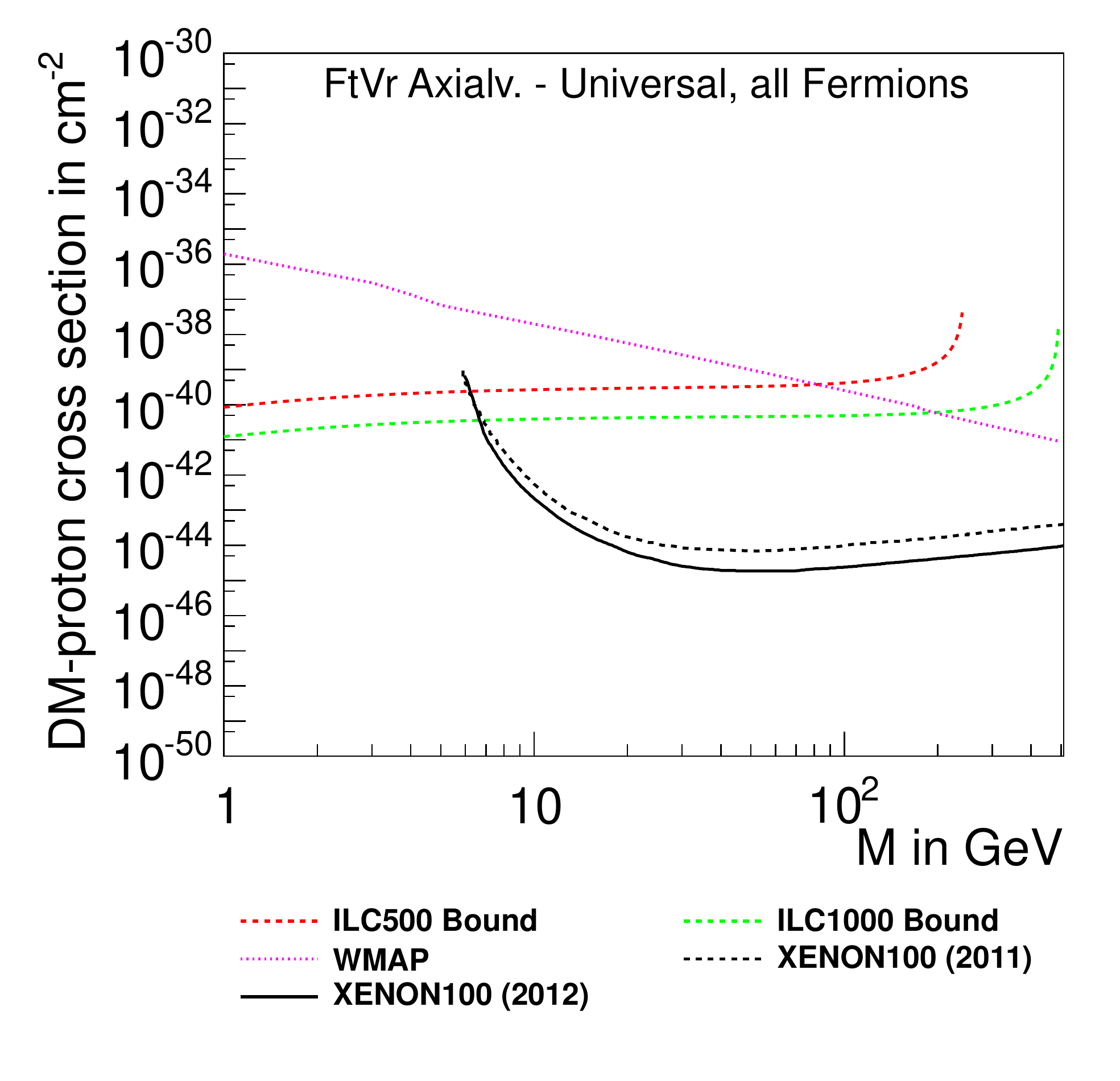} \hfill
 \includegraphics[width=0.475\columnwidth]{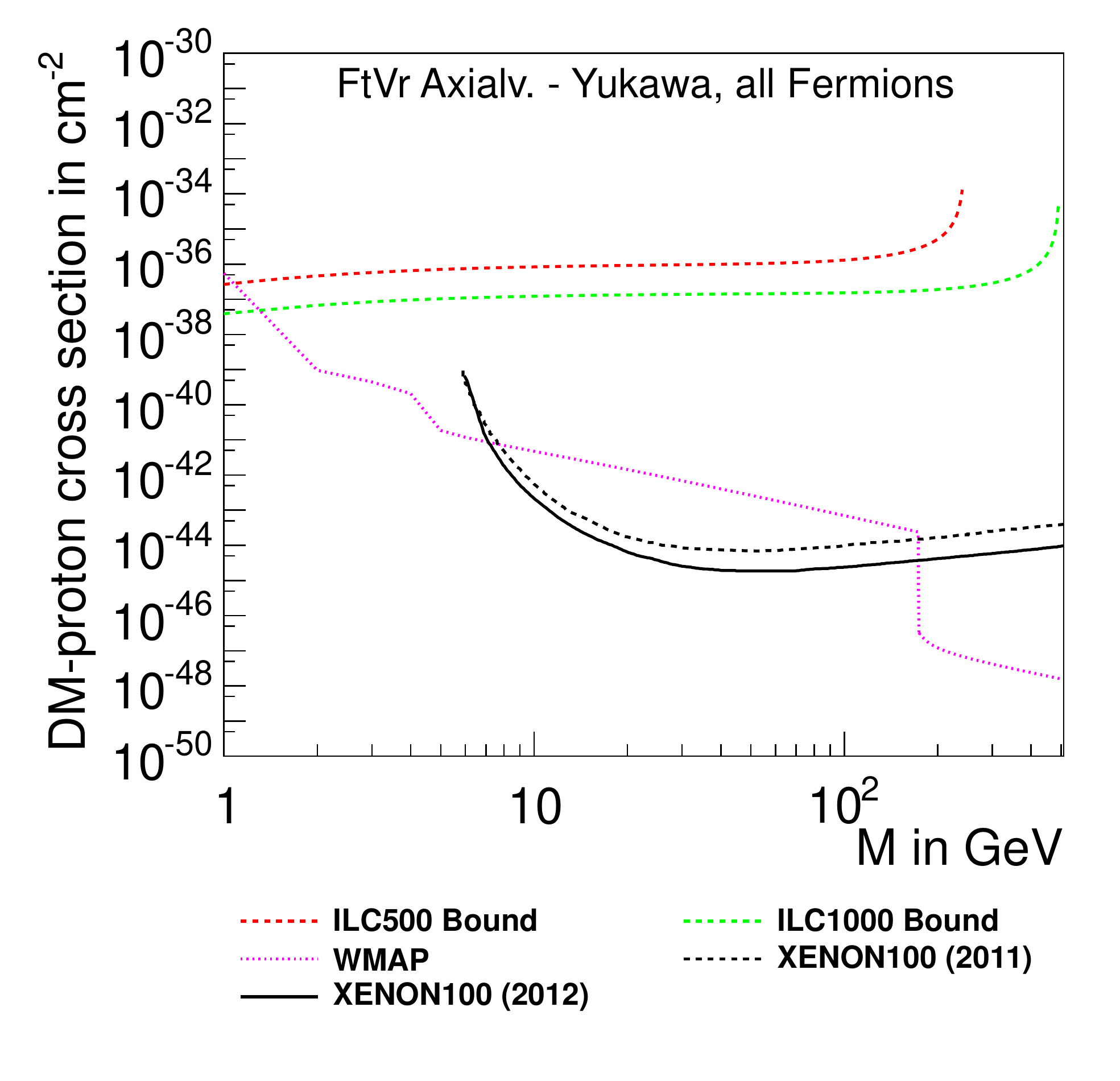} \\
 \includegraphics[width=0.475\columnwidth]{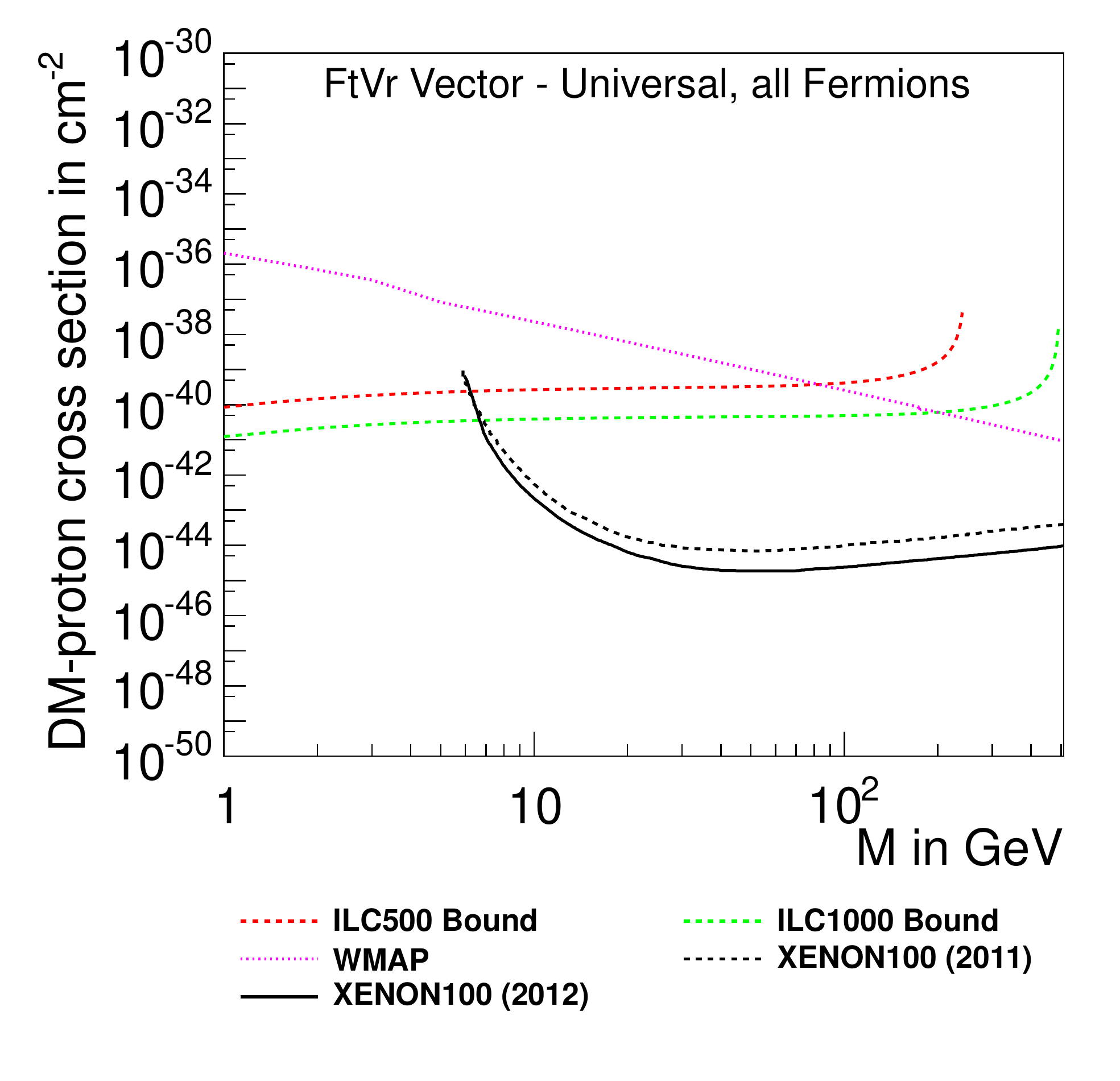} \hfill
 \includegraphics[width=0.475\columnwidth]{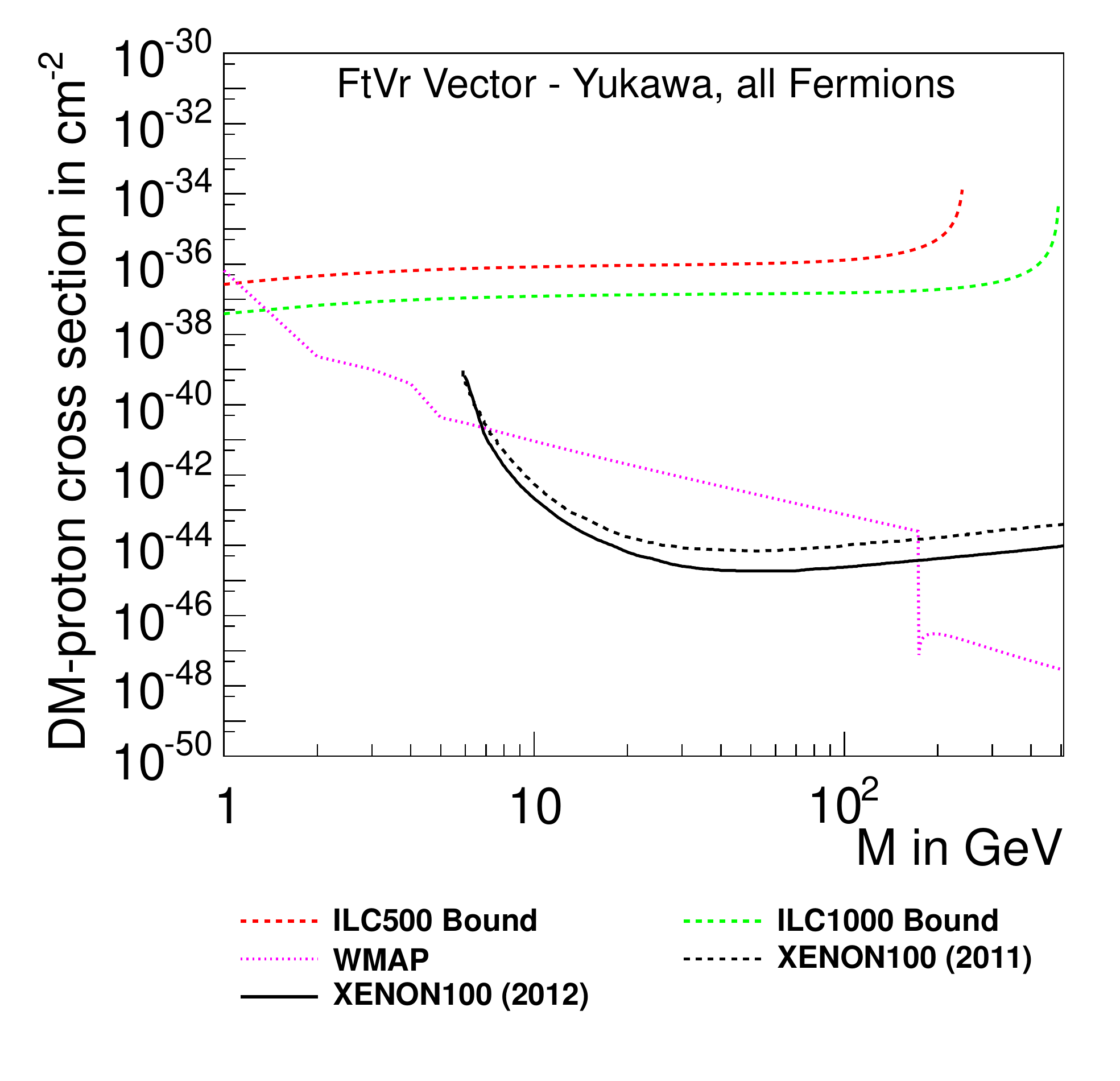}
 \caption{Combined \unit{90}{\%} exclusion limits on the \textbf{spin independent} dark matter proton cross
   section from \textsc{Ilc}, \textsc{Wmap} and \textsc{Xenon} for some \textbf{fermion dark matter} models with \textbf{t--channel vector coupling} to \textbf{all Standard Model fermions}.}
 \label{img:totalbounds6}
 \end{figure}

\begin{figure}[H]
\centering
 \includegraphics[width=0.475\columnwidth]{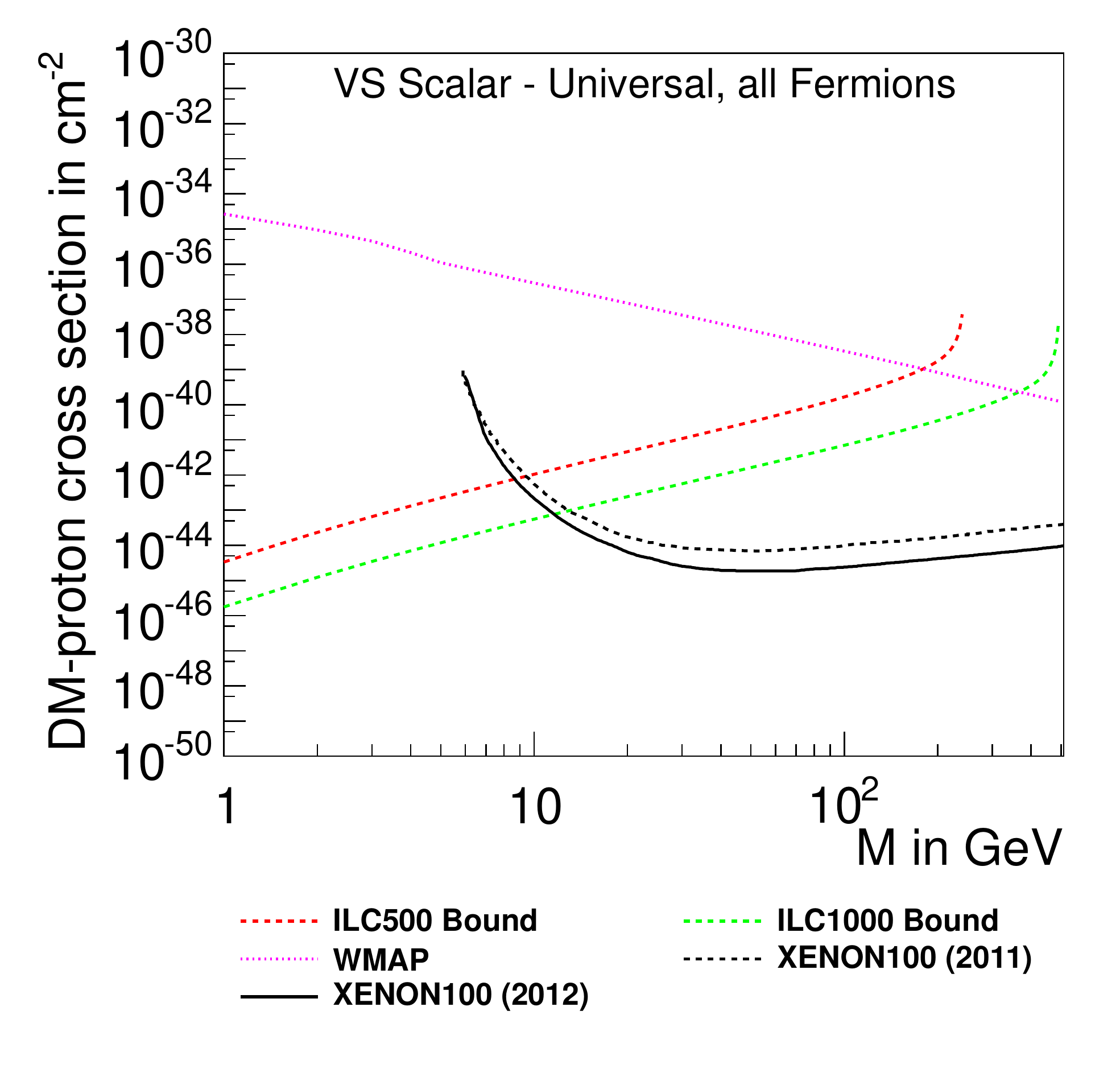} \hfill
 \includegraphics[width=0.475\columnwidth]{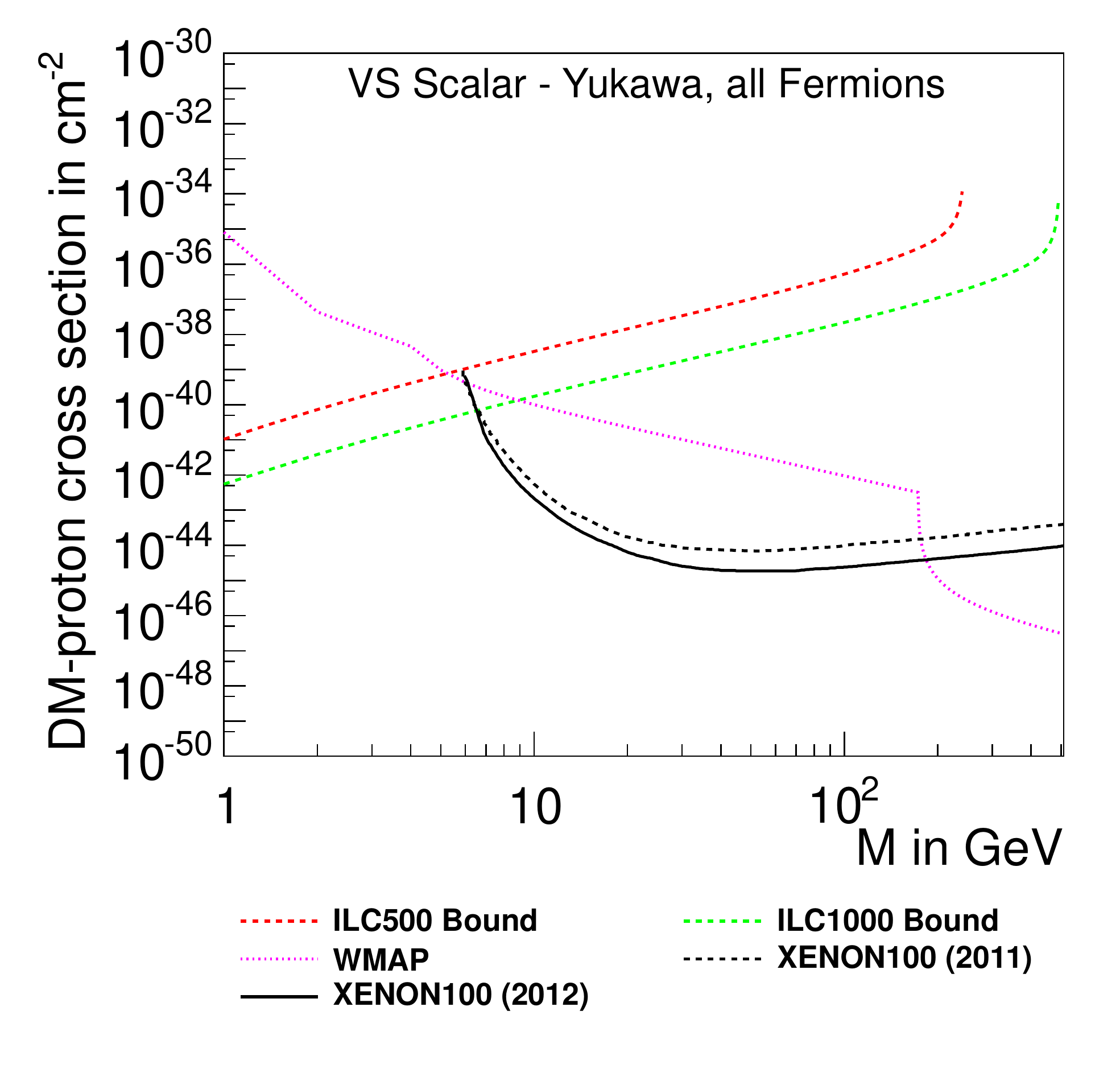} \\
 \includegraphics[width=0.475\columnwidth]{VVVectorplotter_protonxsect_universal} \hfill
 \includegraphics[width=0.475\columnwidth]{VVVectorplotter_protonxsect_yukawa}
 \caption{Combined \unit{90}{\%} exclusion limits on the \textbf{spin independent} dark matter proton cross
   section from \textsc{Ilc}, \textsc{Wmap} and \textsc{Xenon} for some \textbf{vector dark matter} models with \textbf{s--channel scalar or vector coupling} to \textbf{all Standard Model fermions}.}
 \label{img:totalbounds7}
 \end{figure}
\begin{figure}[H]
\centering
 \includegraphics[width=0.475\columnwidth]{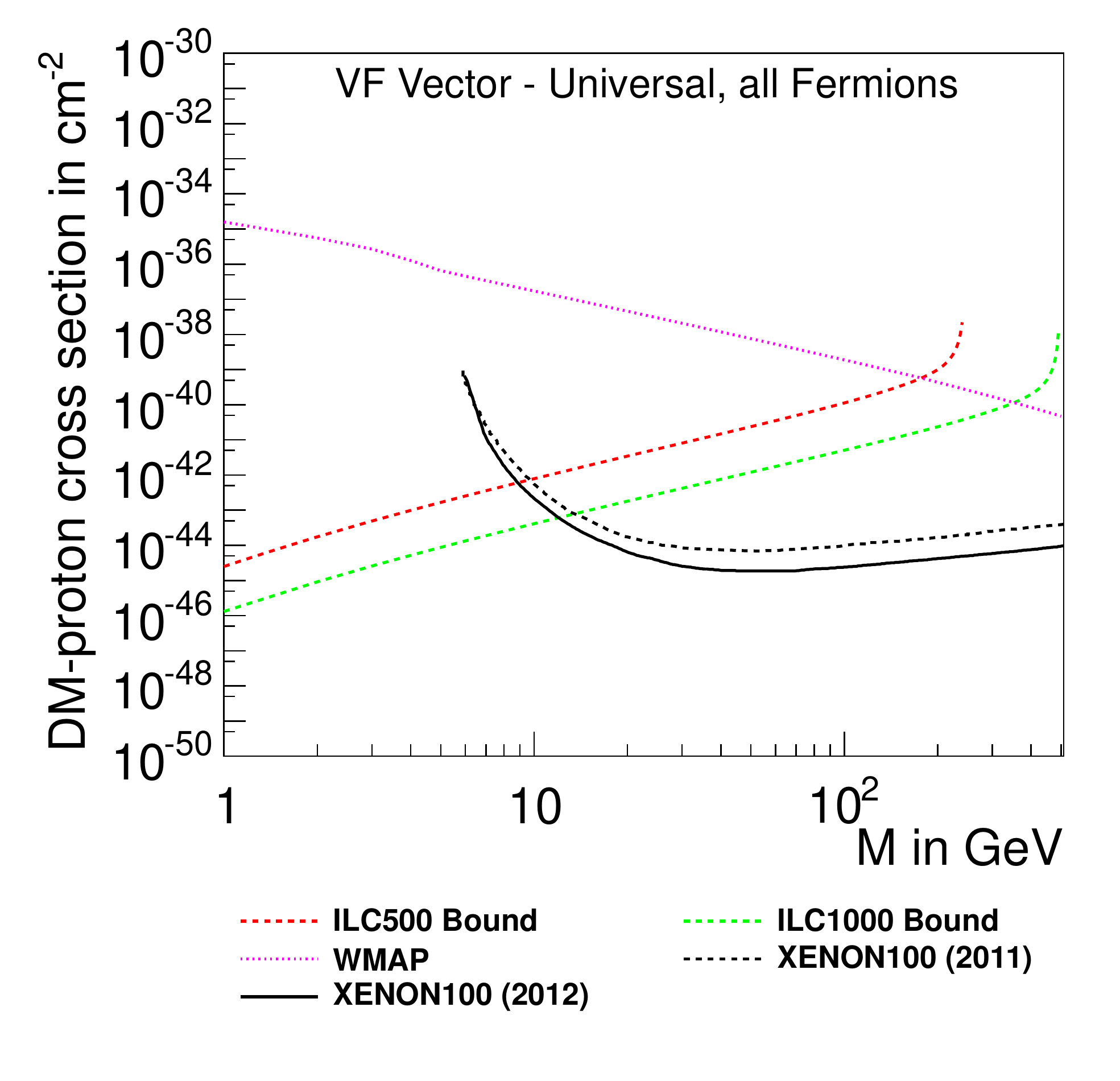} \hfill
 \includegraphics[width=0.475\columnwidth]{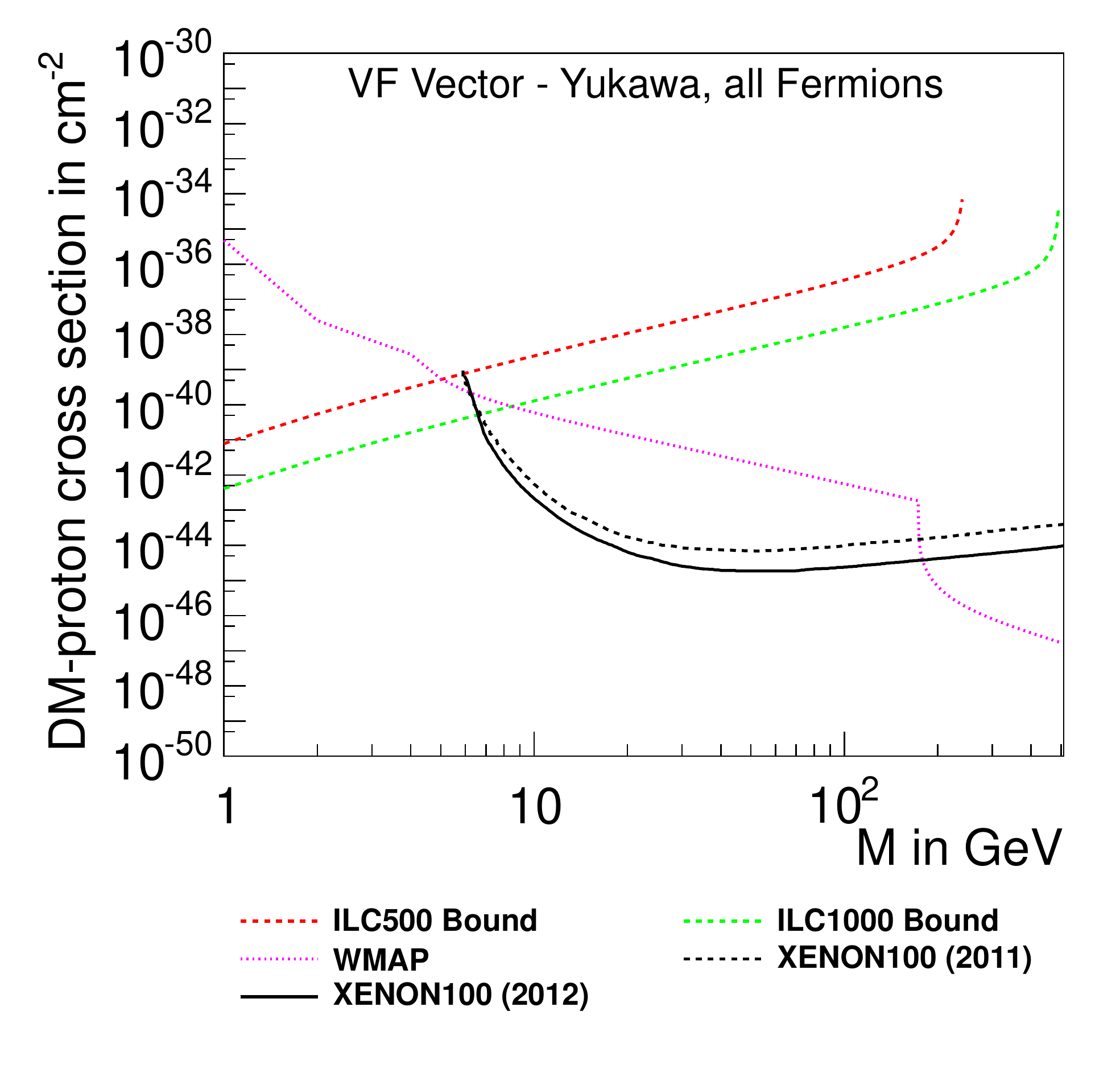} \\
 \includegraphics[width=0.475\columnwidth]{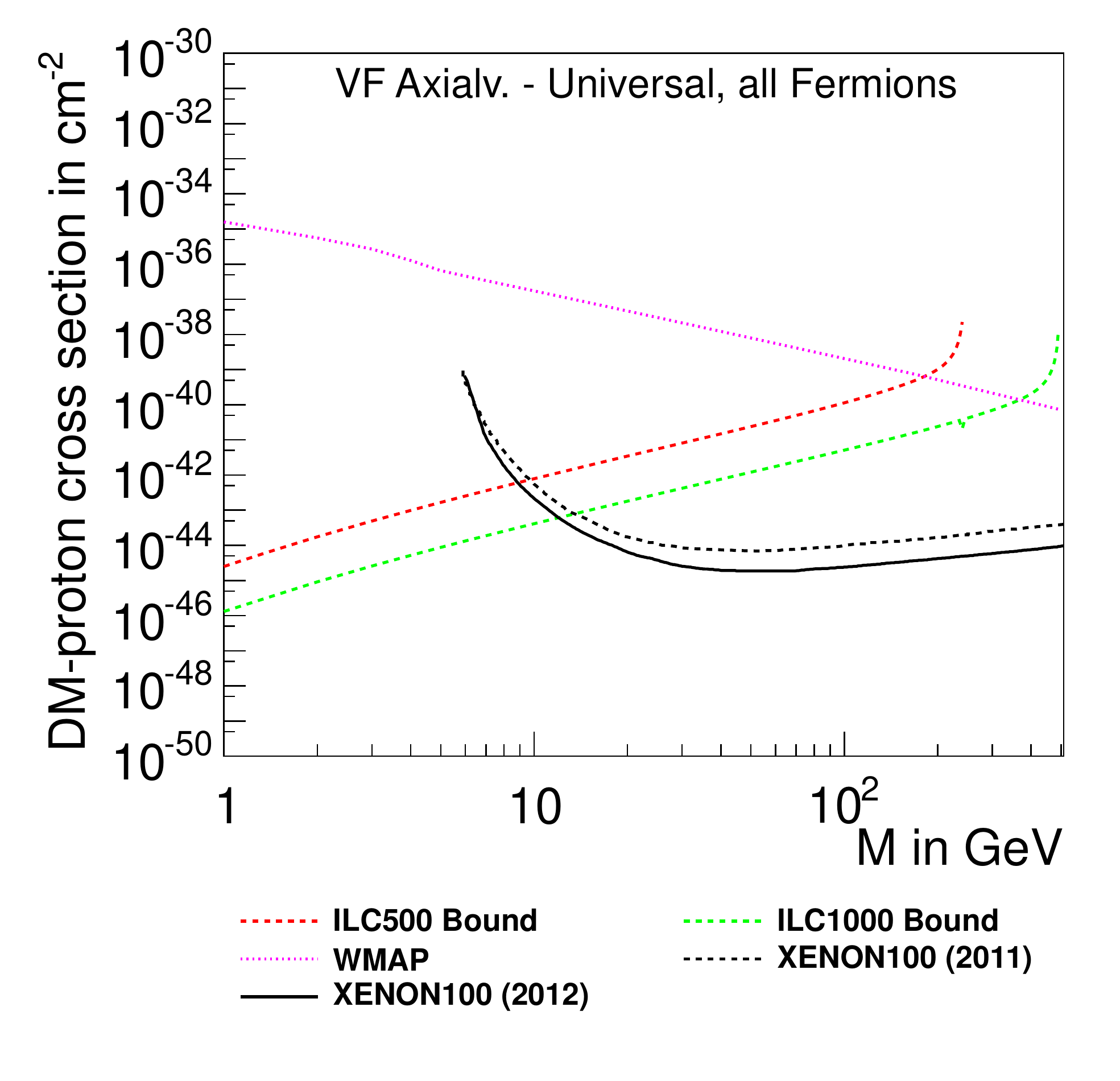} \hfill
 \includegraphics[width=0.475\columnwidth]{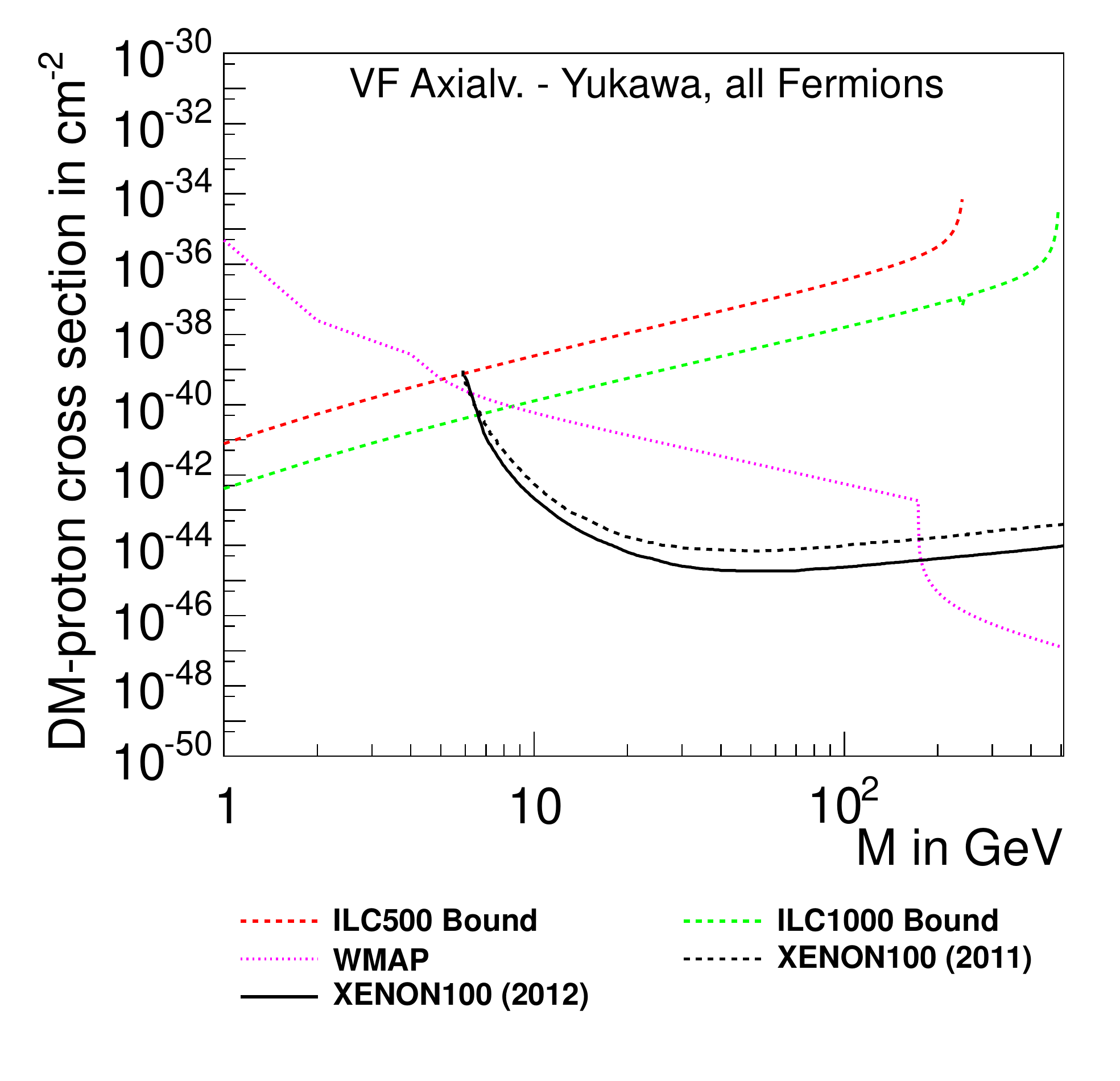} \\
 \includegraphics[width=0.475\columnwidth]{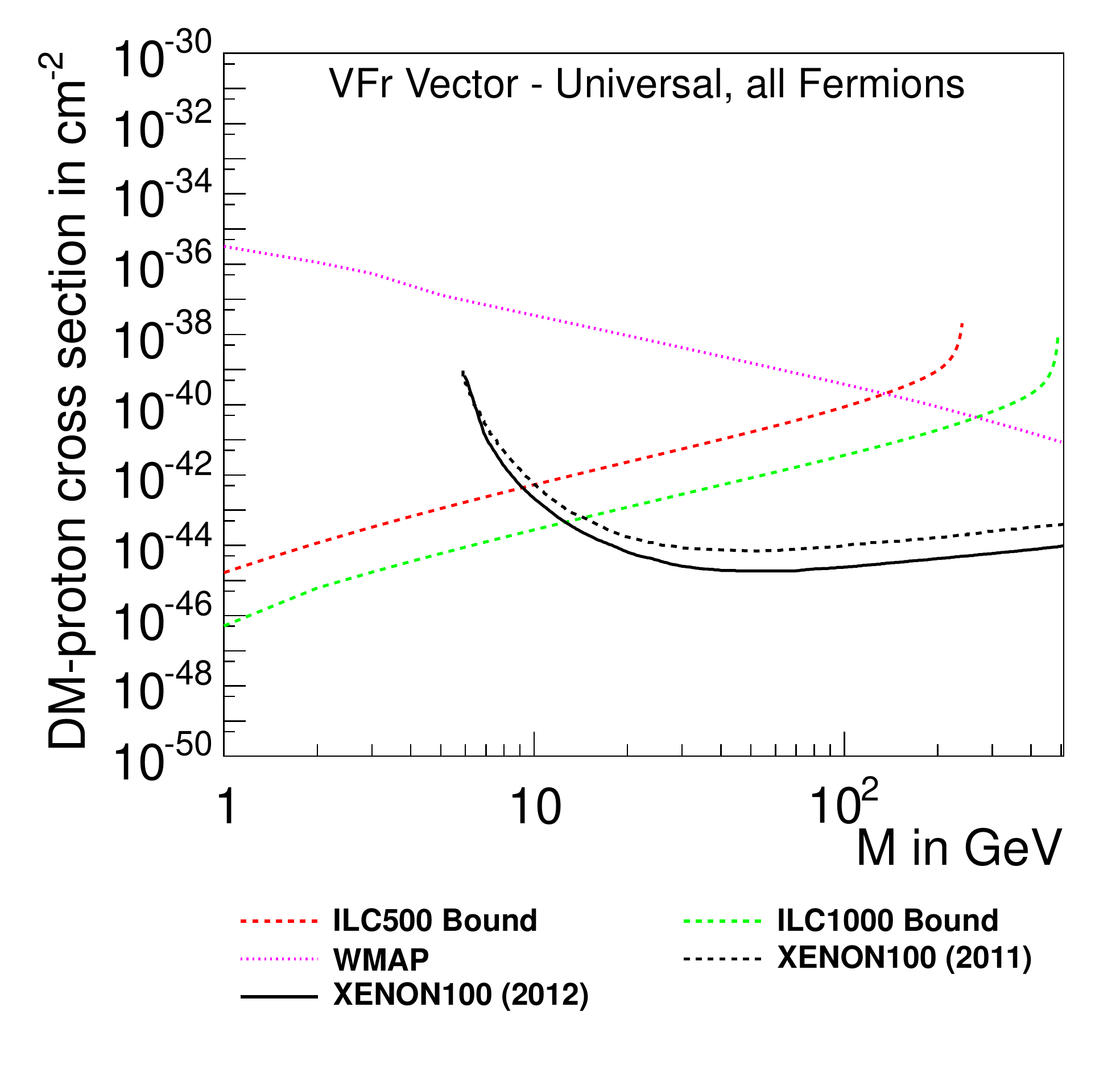} \hfill
 \includegraphics[width=0.475\columnwidth]{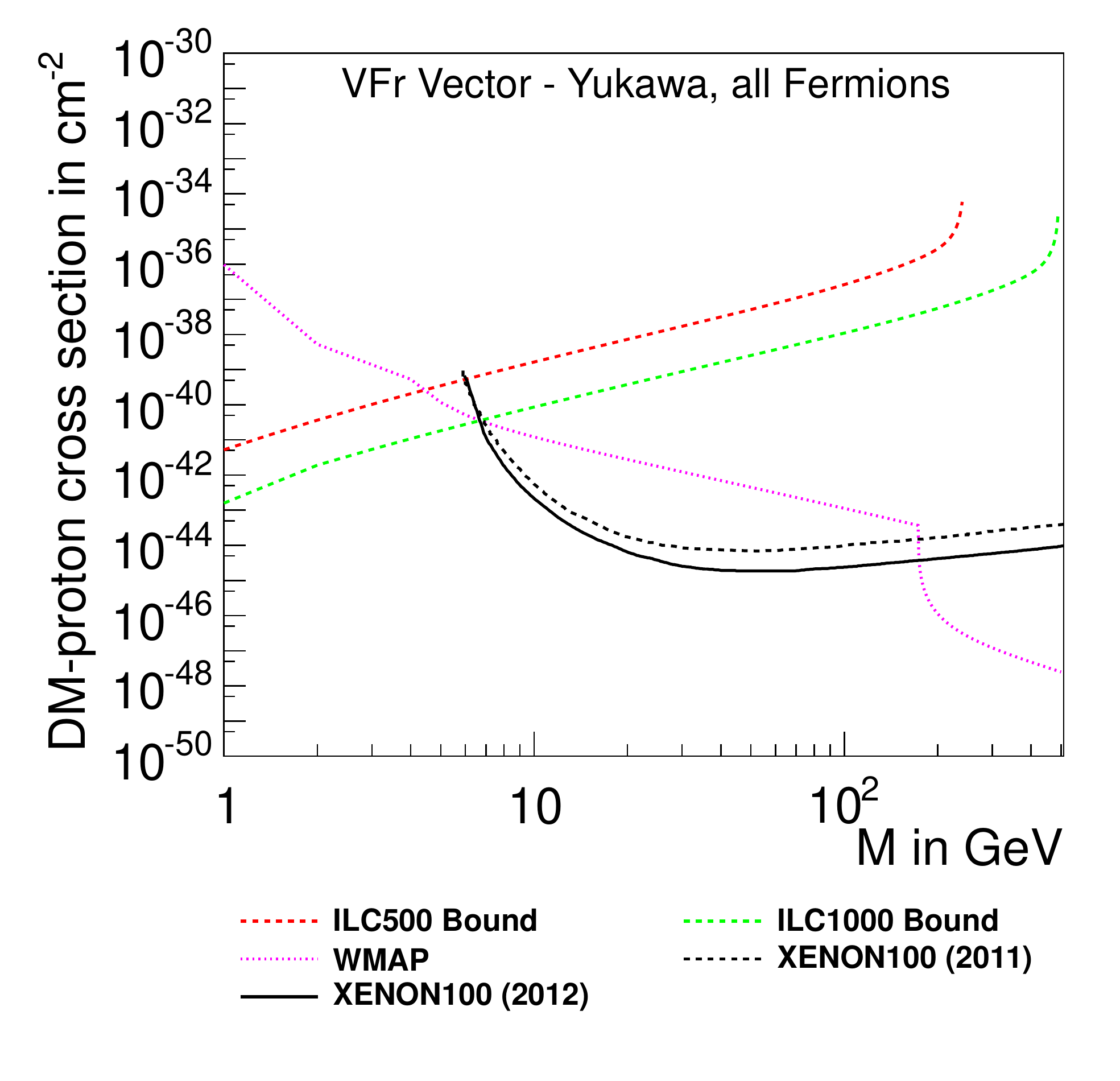}
 \caption{Combined \unit{90}{\%} exclusion limits on the \textbf{spin independent} dark matter proton cross
   section from \textsc{Ilc}, \textsc{Wmap} and \textsc{Xenon} for some \textbf{vector dark matter} models  with \textbf{t--channel fermion coupling} to \textbf{all Standard Model fermions}.}
 \label{img:totalbounds8}
 \end{figure}

\begin{figure}[H]
\centering
 \includegraphics[width=0.475\columnwidth]{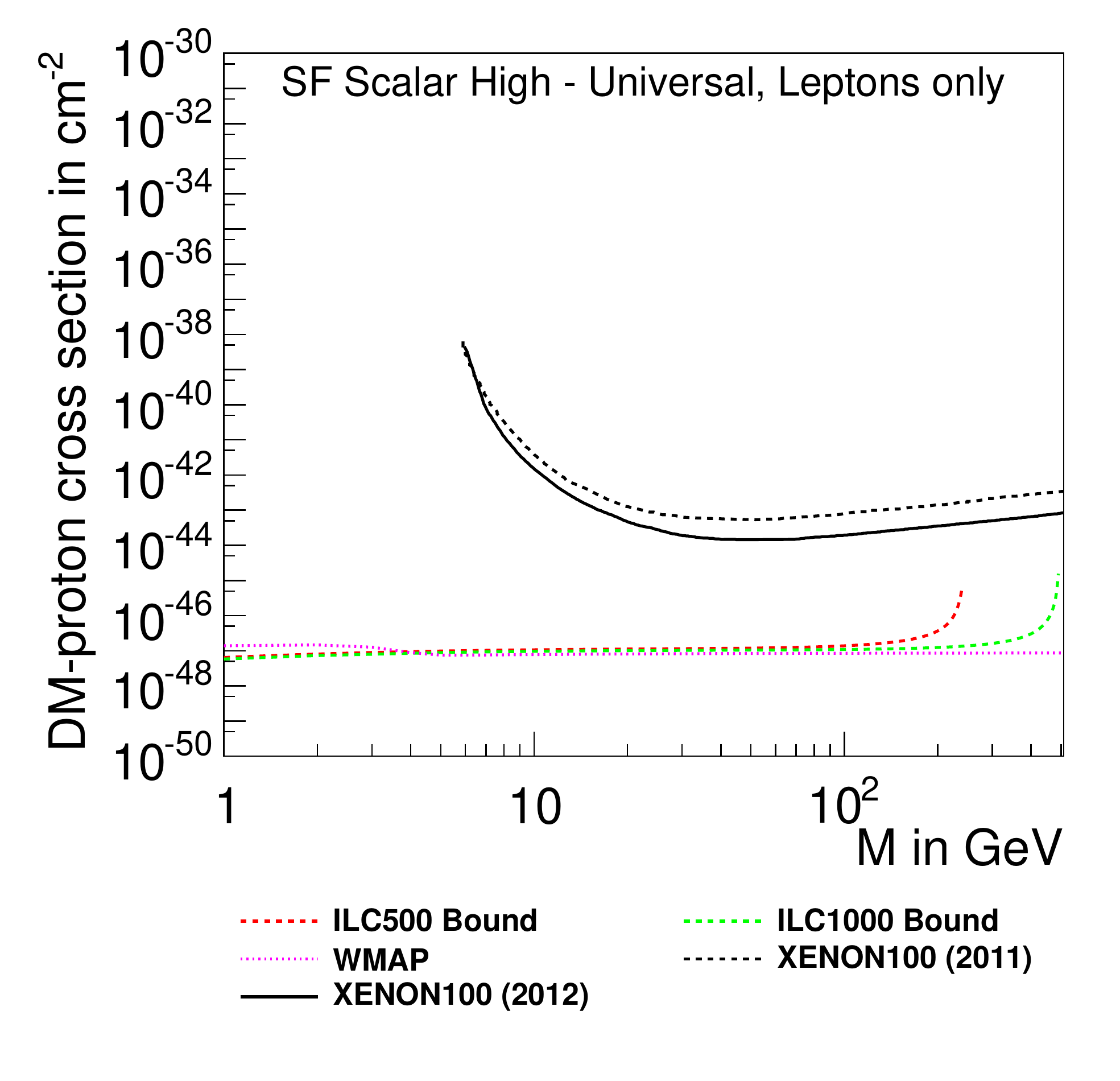} \hfill
 \includegraphics[width=0.475\columnwidth]{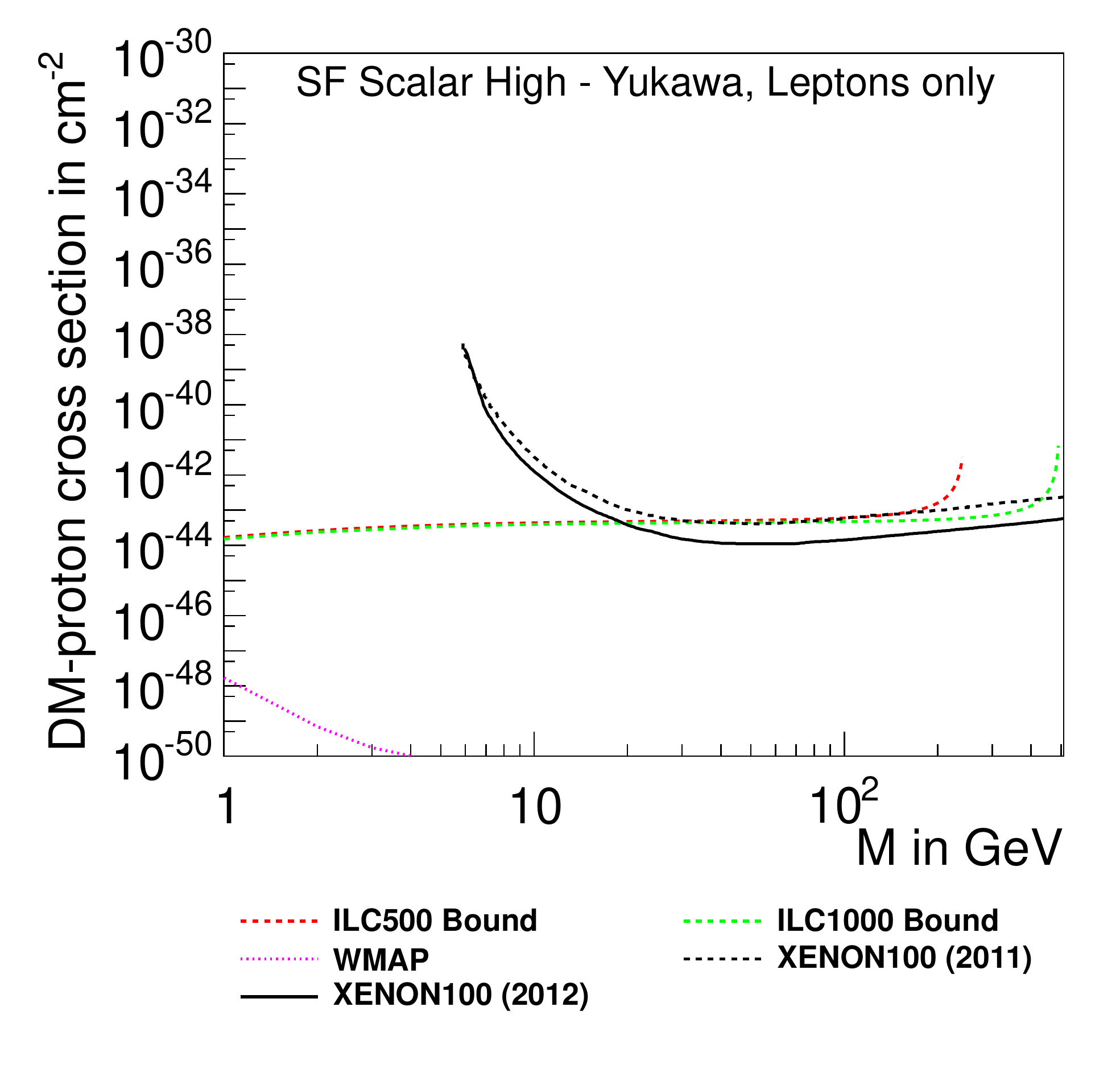} \\
 \includegraphics[width=0.475\columnwidth]{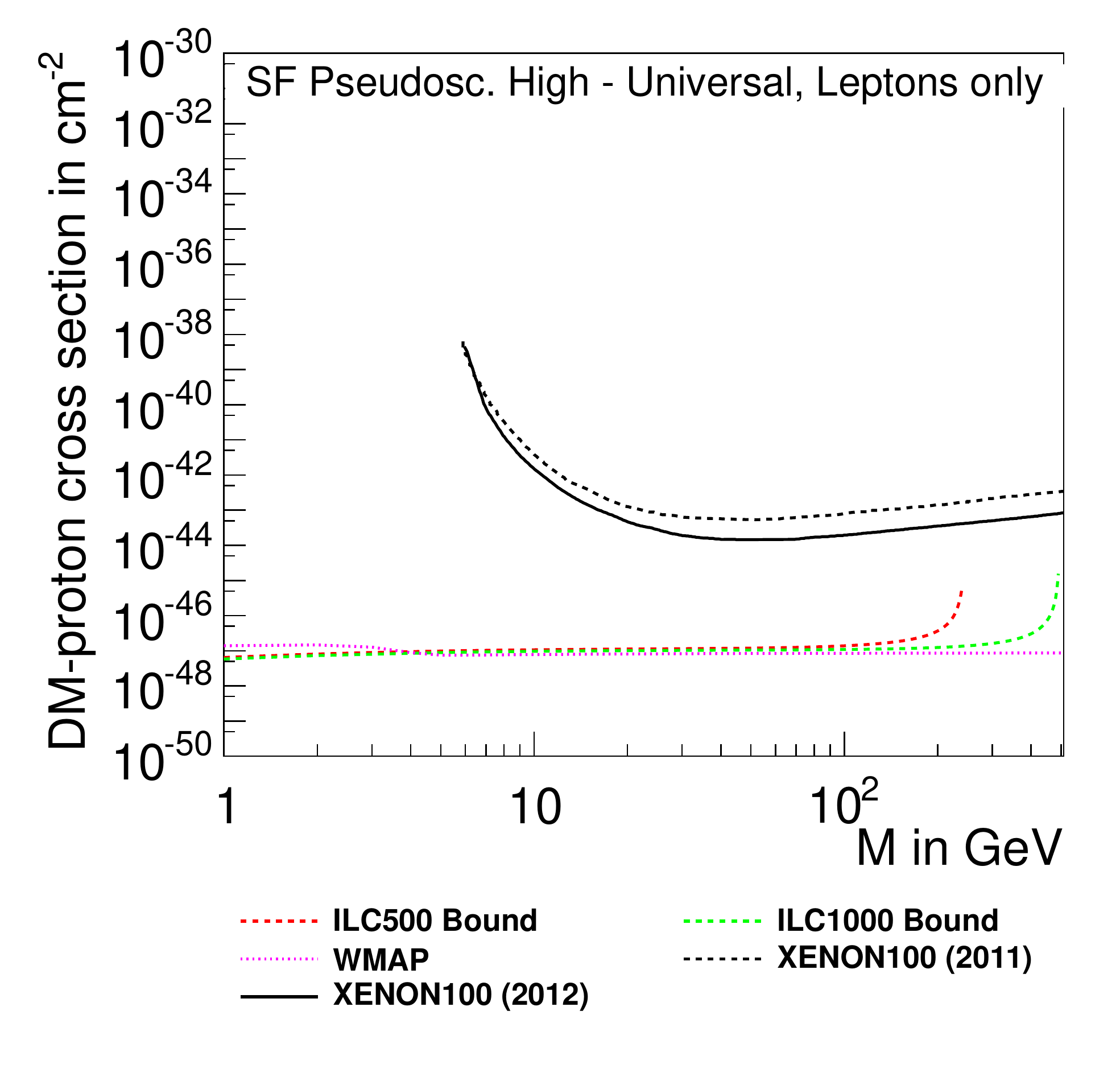} \hfill
 \includegraphics[width=0.475\columnwidth]{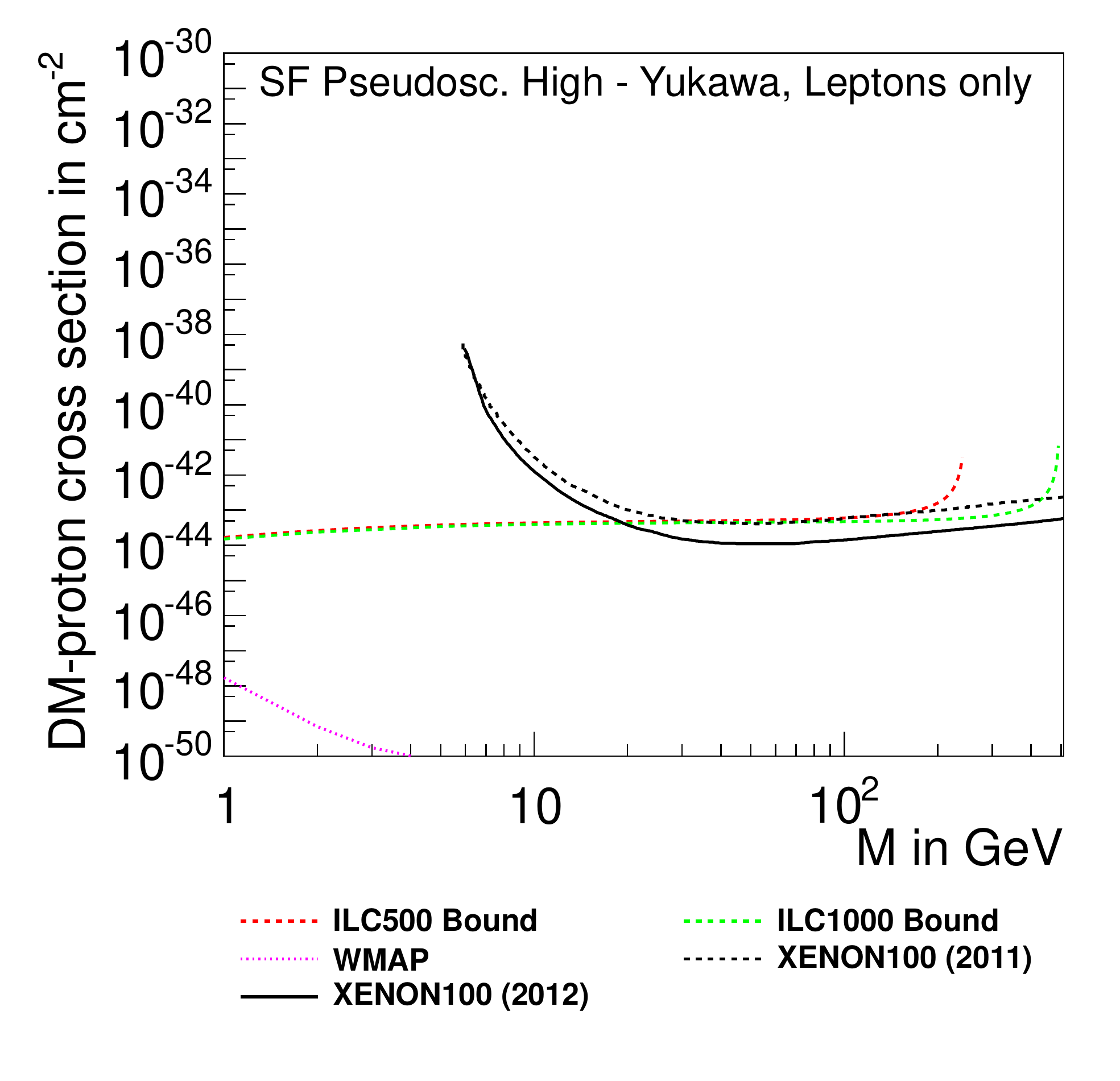} \\
 \includegraphics[width=0.475\columnwidth]{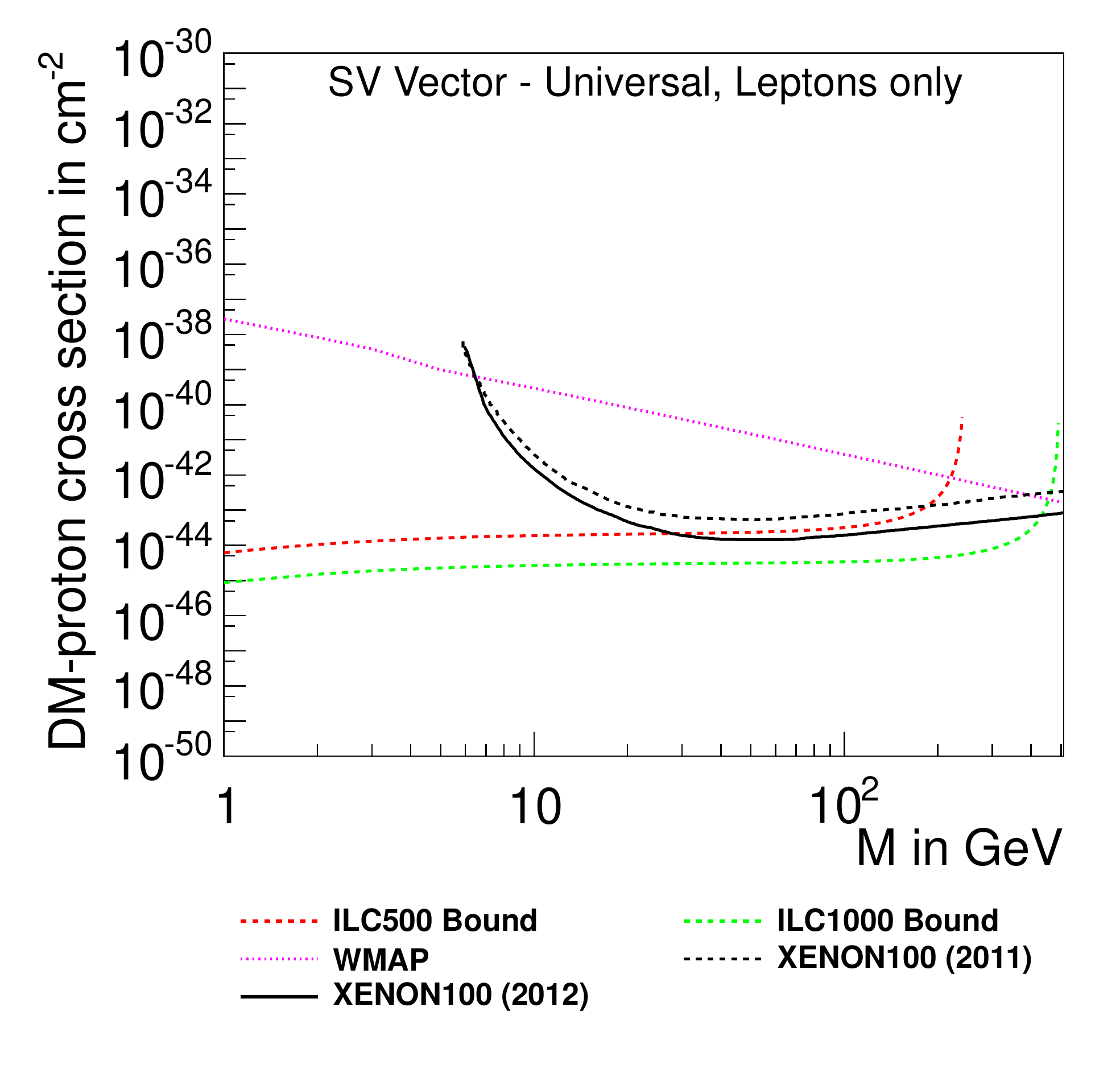} \hfill
 \includegraphics[width=0.475\columnwidth]{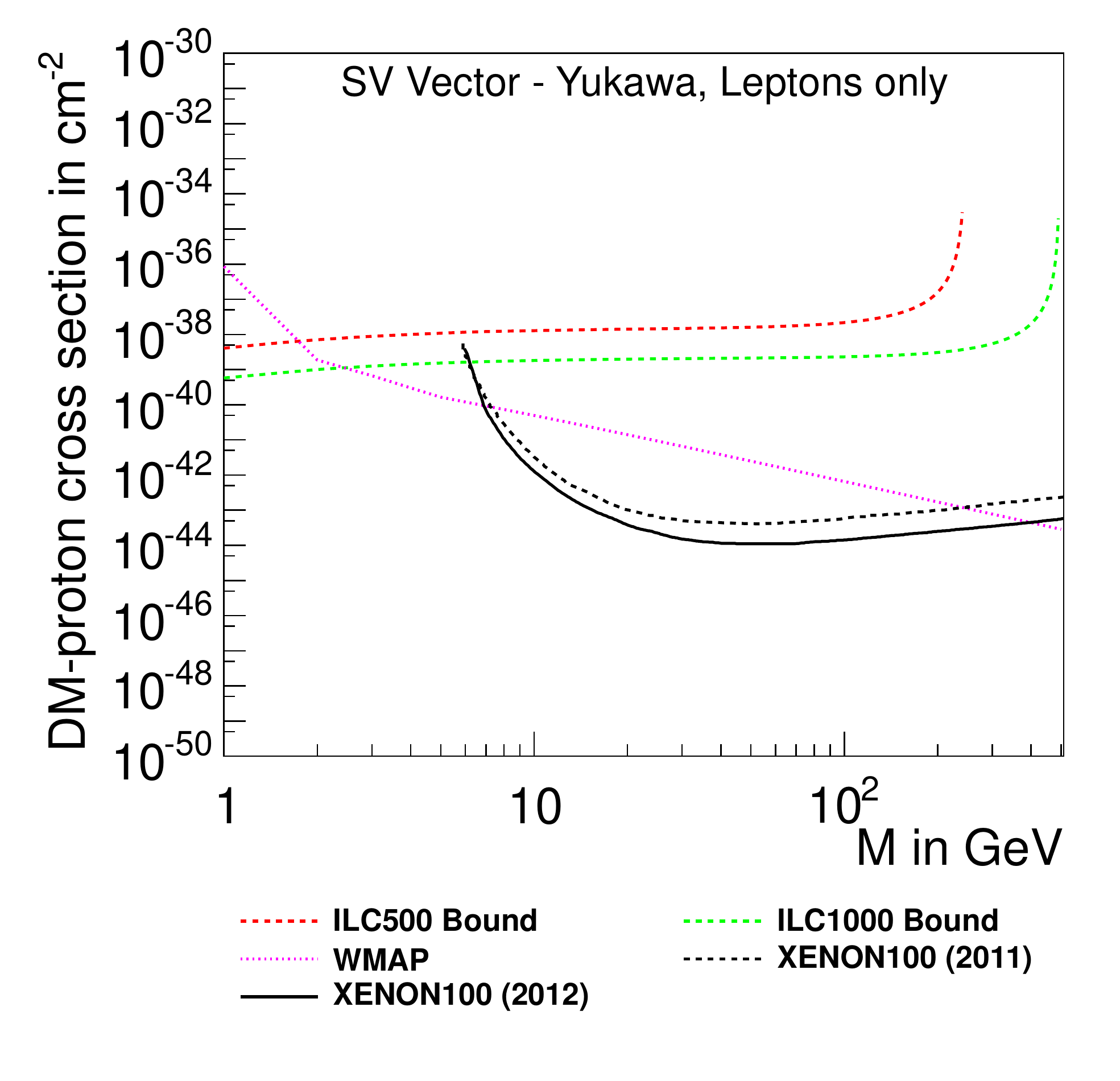}
 \caption{Combined \unit{90}{\%} exclusion limits on the \textbf{spin independent} dark matter proton cross
   section from \textsc{Ilc}, \textsc{Wmap} and \textsc{Xenon} for some \textbf{scalar dark matter} models with \textbf{t--channel fermion or s--channel vector coupling} to \textbf{leptons only}.}
 \label{img:totalbounds9}
 \end{figure}
\begin{figure}[H]
\centering
 \includegraphics[width=0.475\columnwidth]{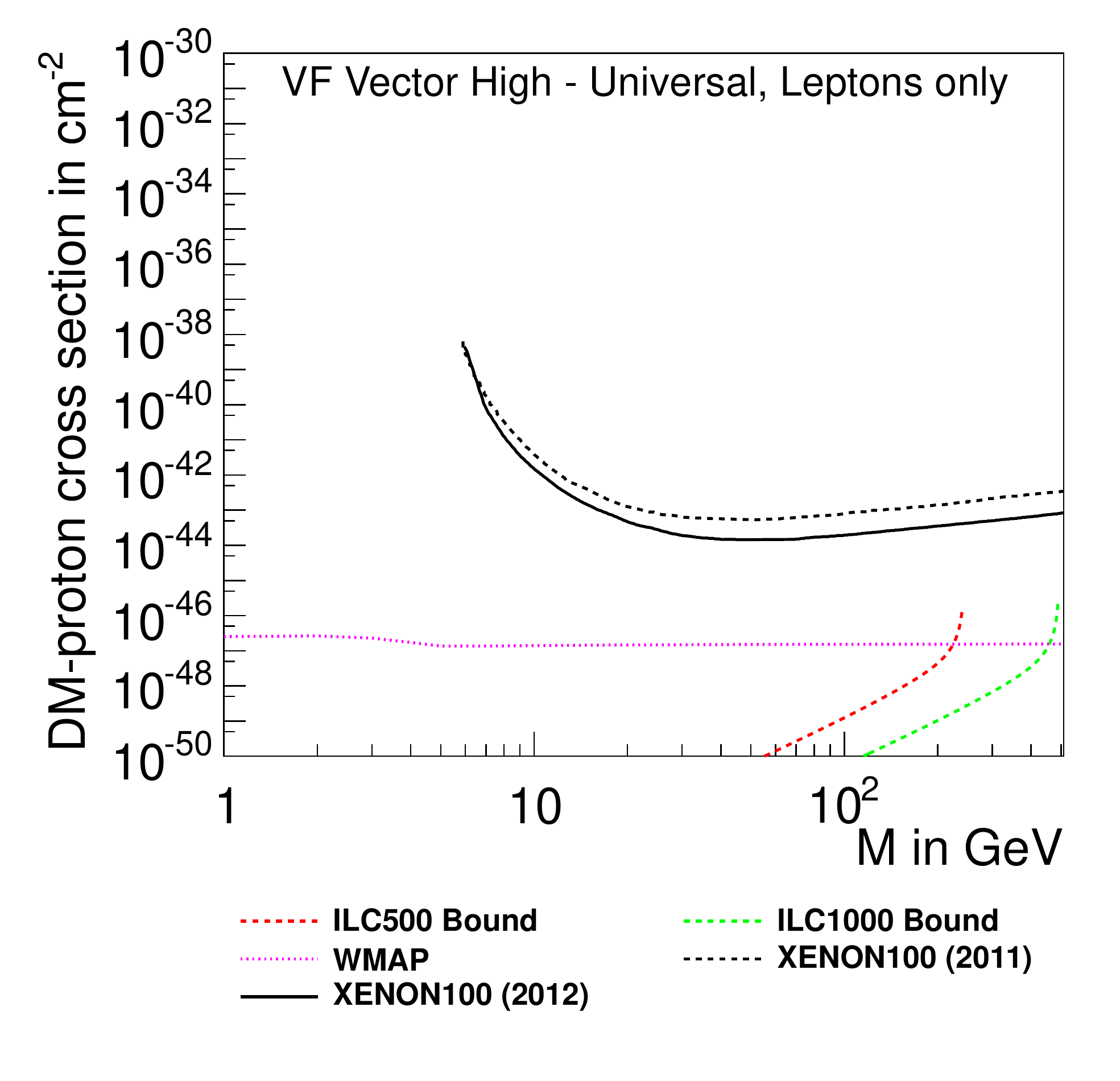} \hfill
 \includegraphics[width=0.475\columnwidth]{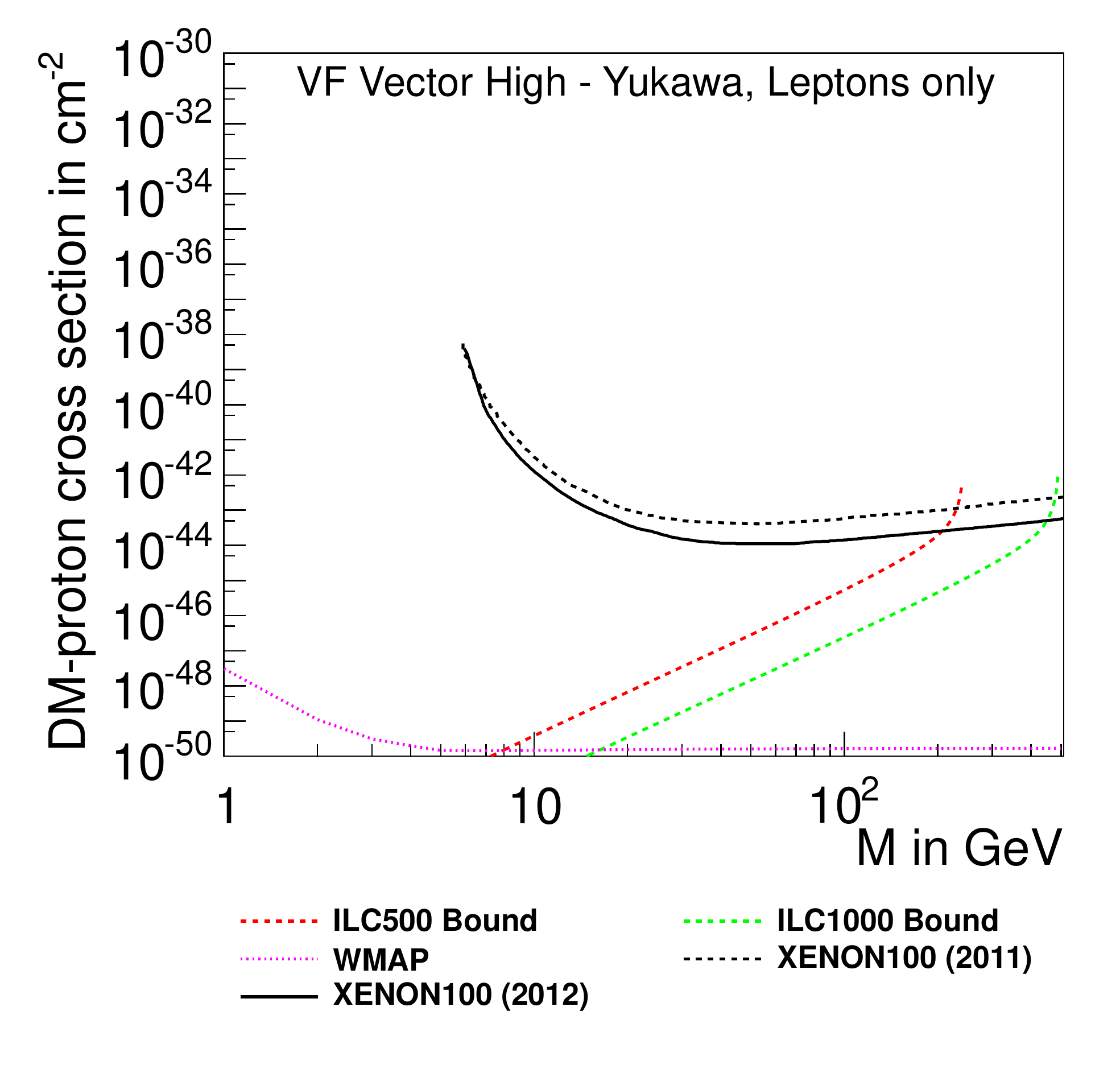} \\
 \includegraphics[width=0.475\columnwidth]{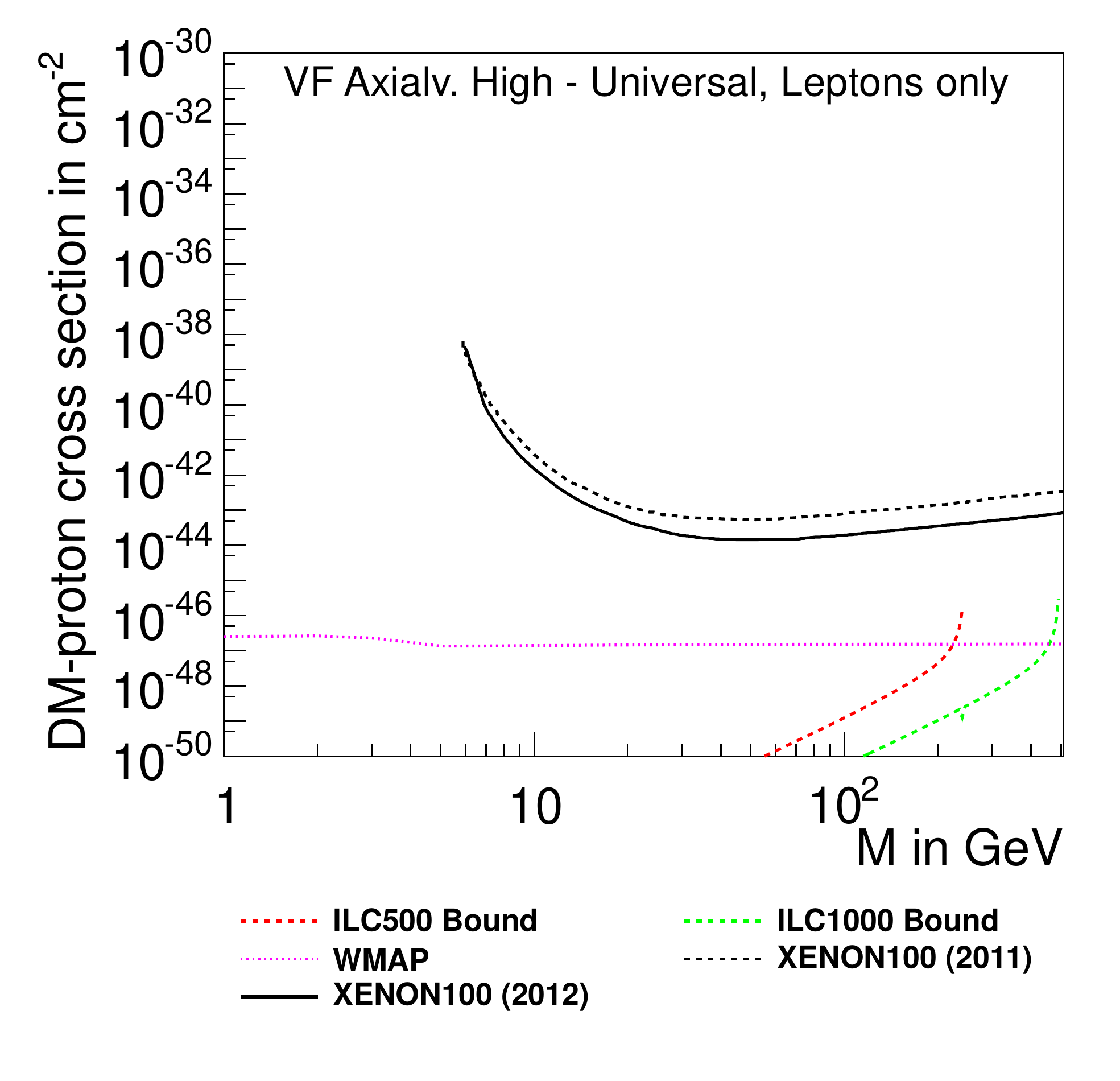} \hfill
 \includegraphics[width=0.475\columnwidth]{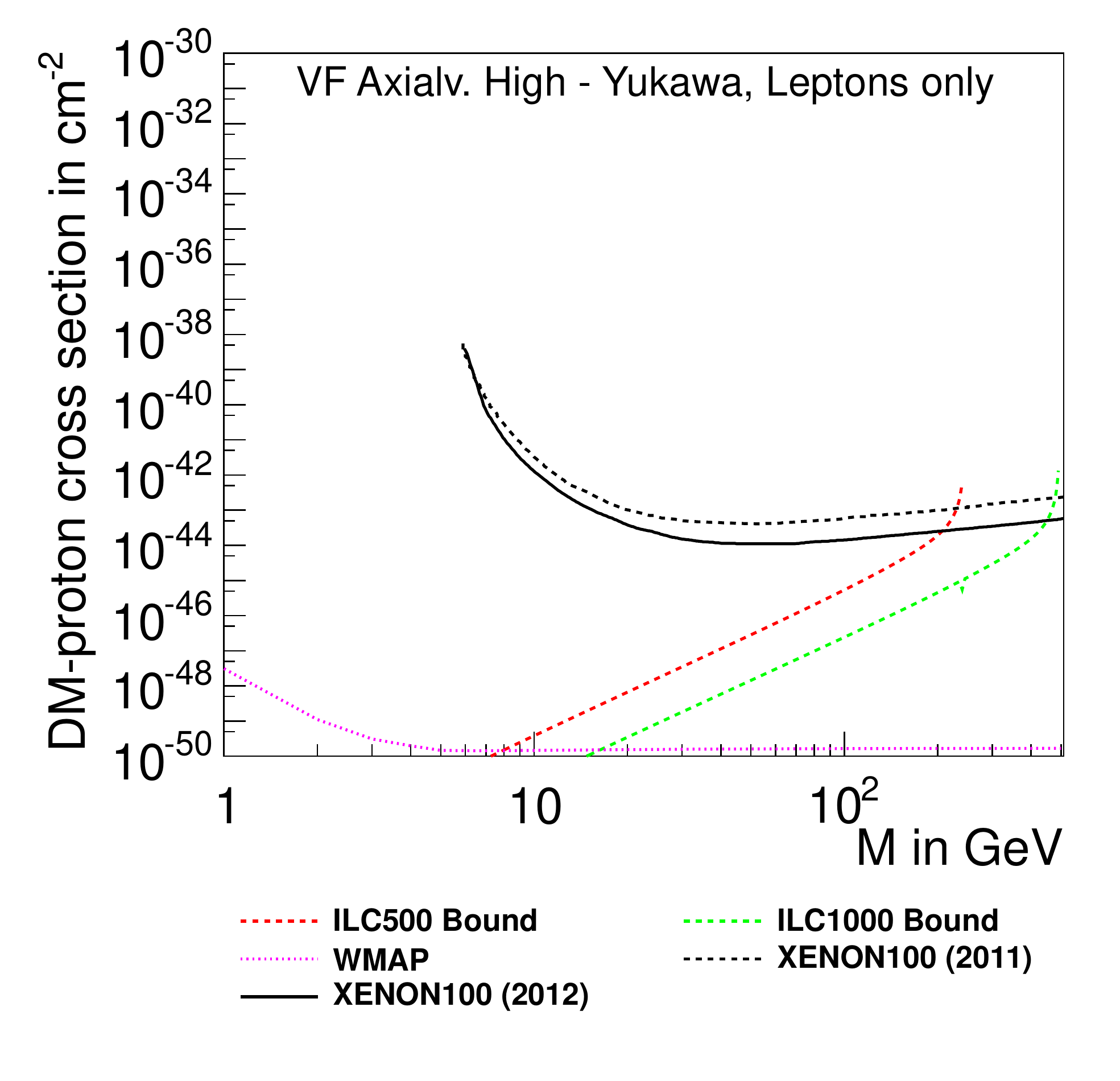} \\
 \includegraphics[width=0.475\columnwidth]{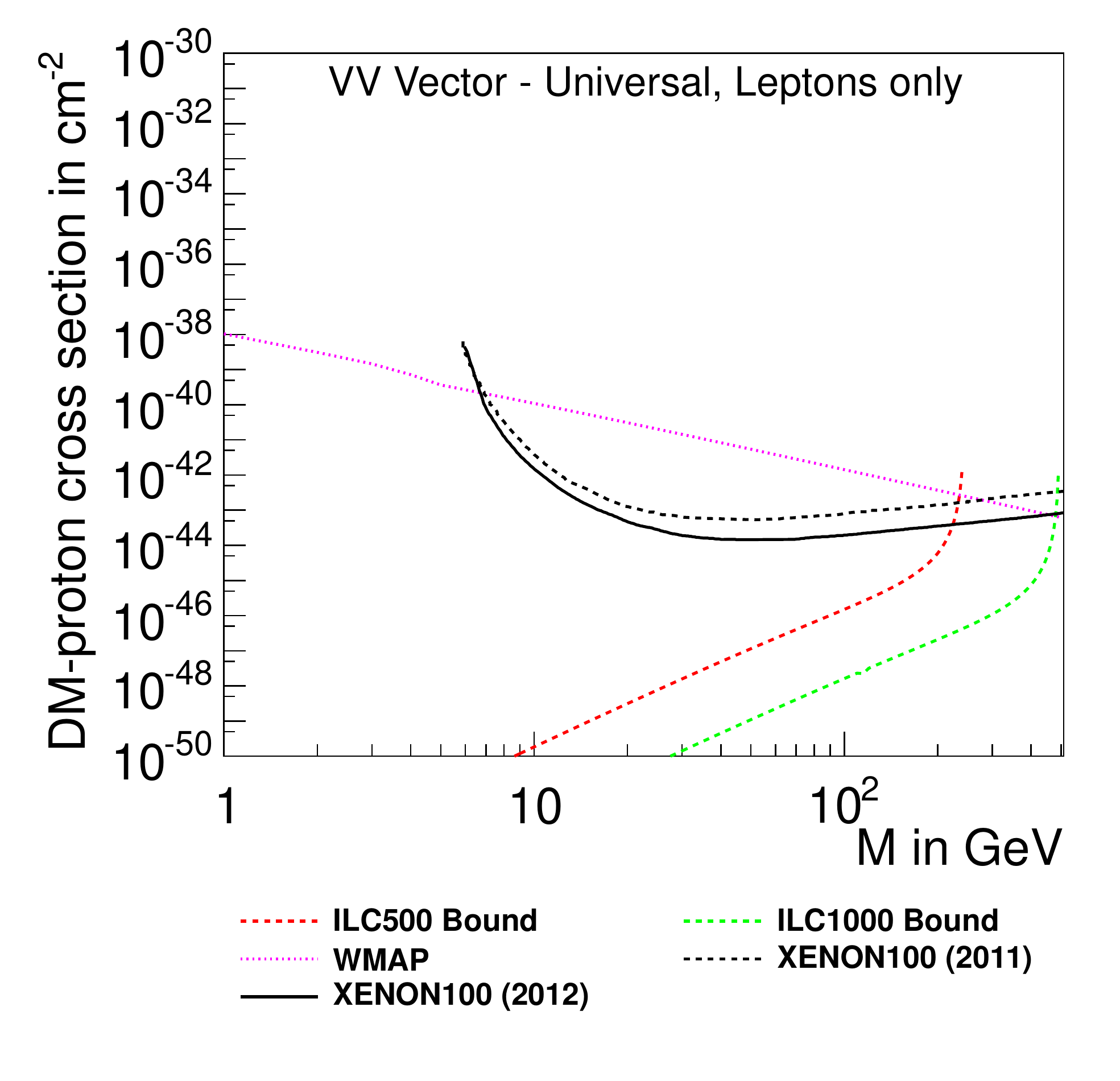} \hfill
 \includegraphics[width=0.475\columnwidth]{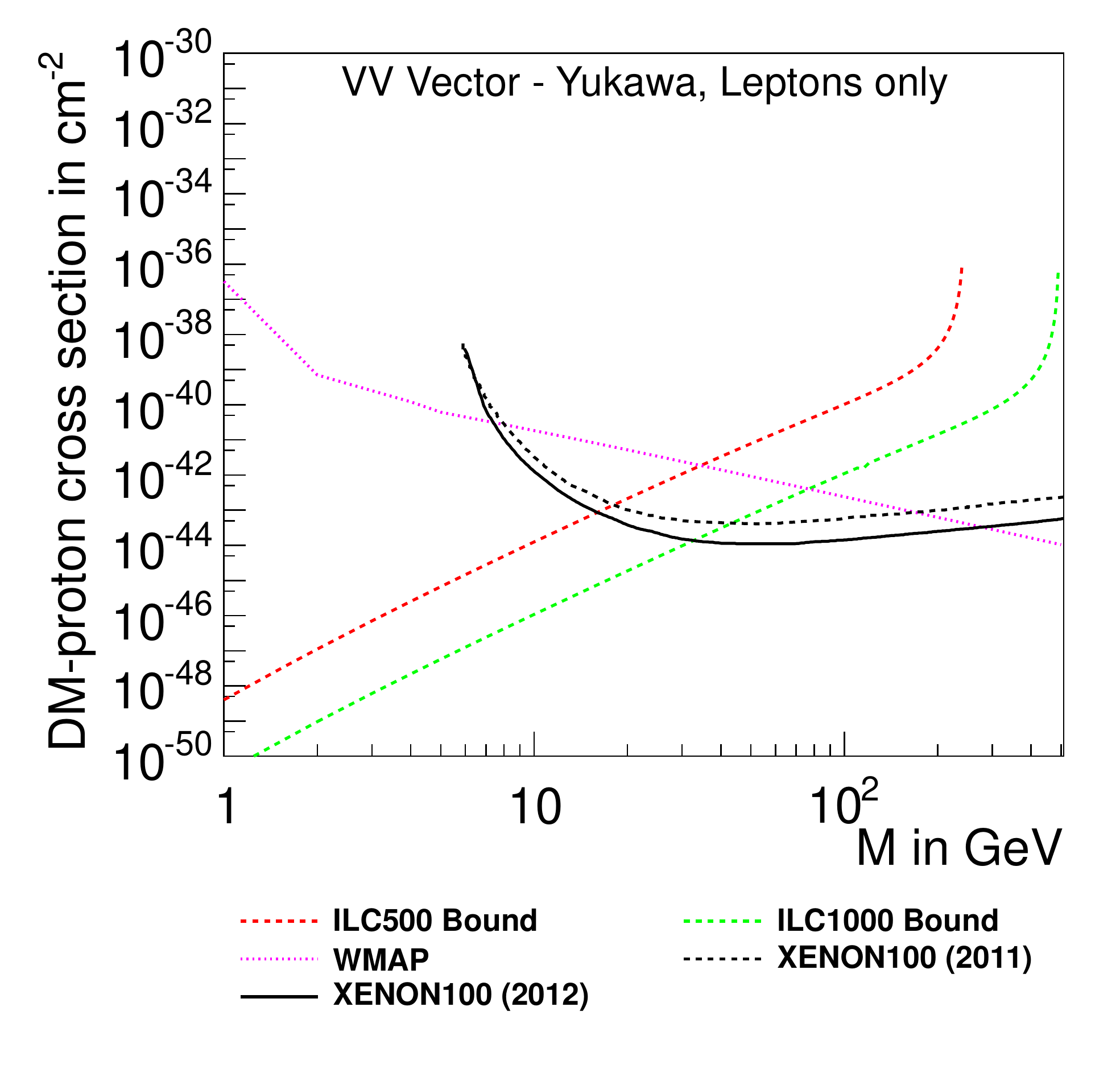}
 \caption{Combined \unit{90}{\%} exclusion limits on the \textbf{spin independent} dark matter proton cross
   section from \textsc{Ilc}, \textsc{Wmap} and \textsc{Xenon} for some \textbf{vector dark matter} models with \textbf{t--channel fermion or s--channel vector coupling} to \textbf{leptons only}.}
 \label{img:totalbounds10}
 \end{figure}

\begin{figure}[H]
\centering
 \includegraphics[width=0.475\columnwidth]{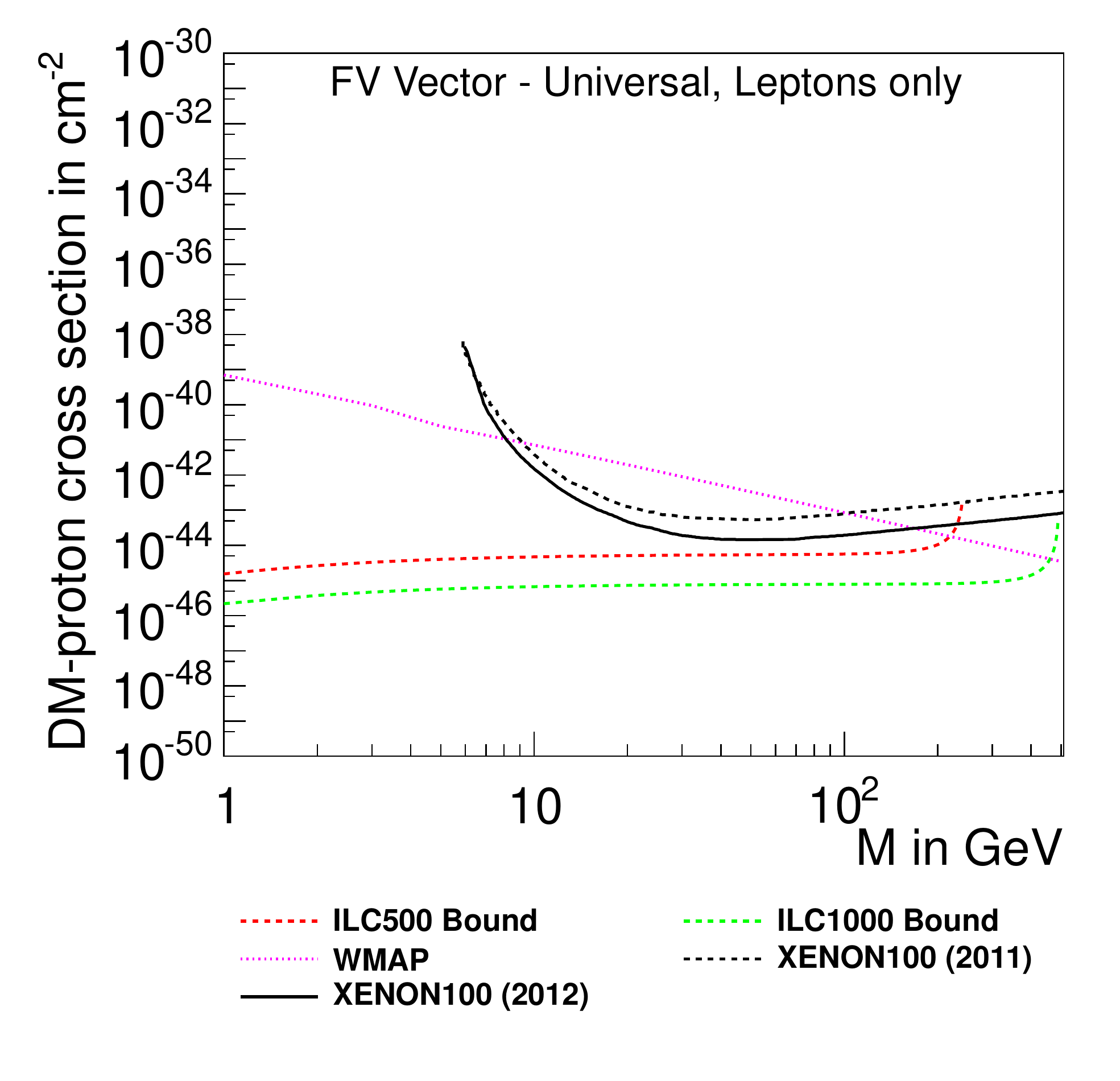} \hfill
 \includegraphics[width=0.475\columnwidth]{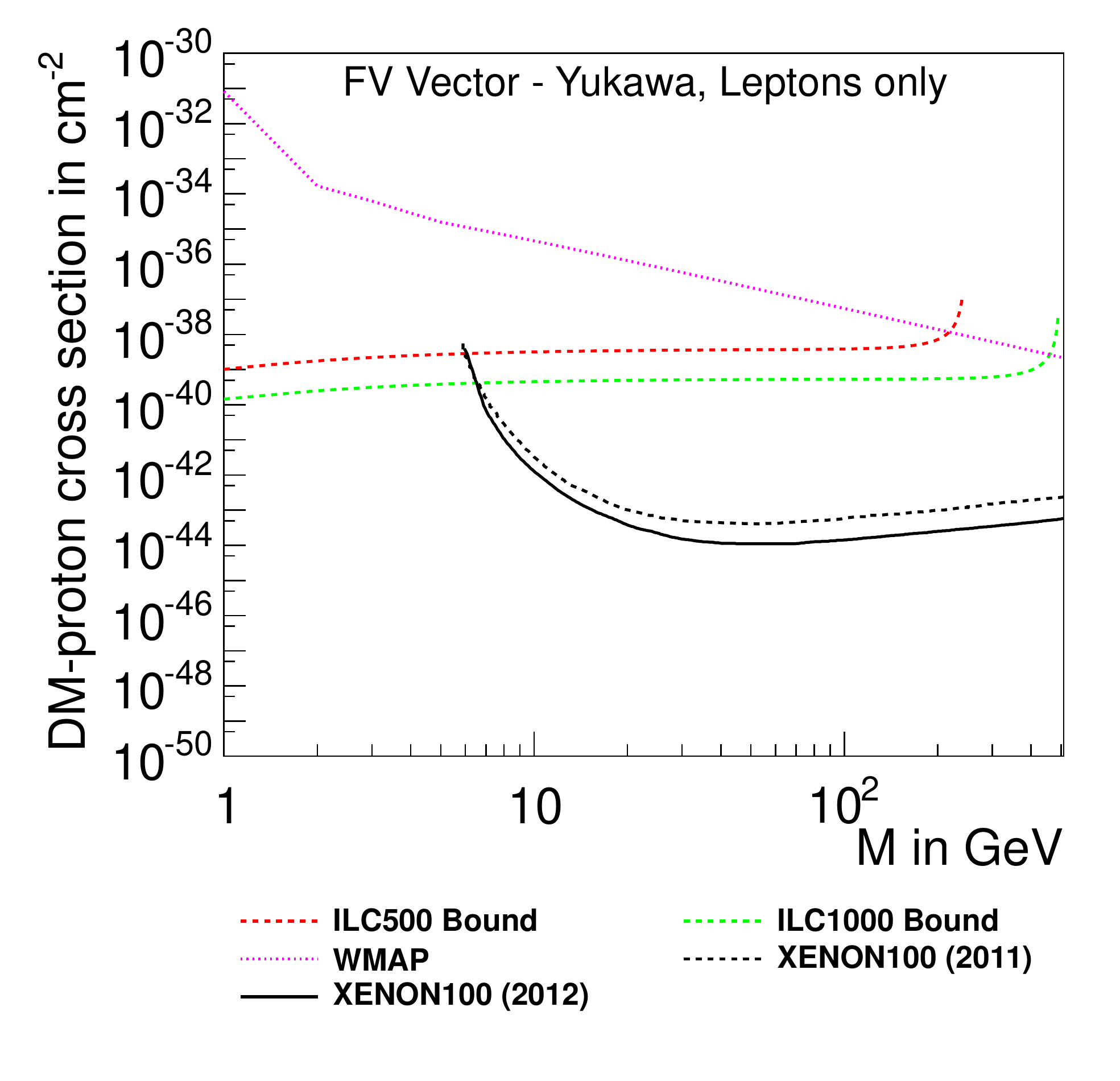} \\
 \includegraphics[width=0.475\columnwidth]{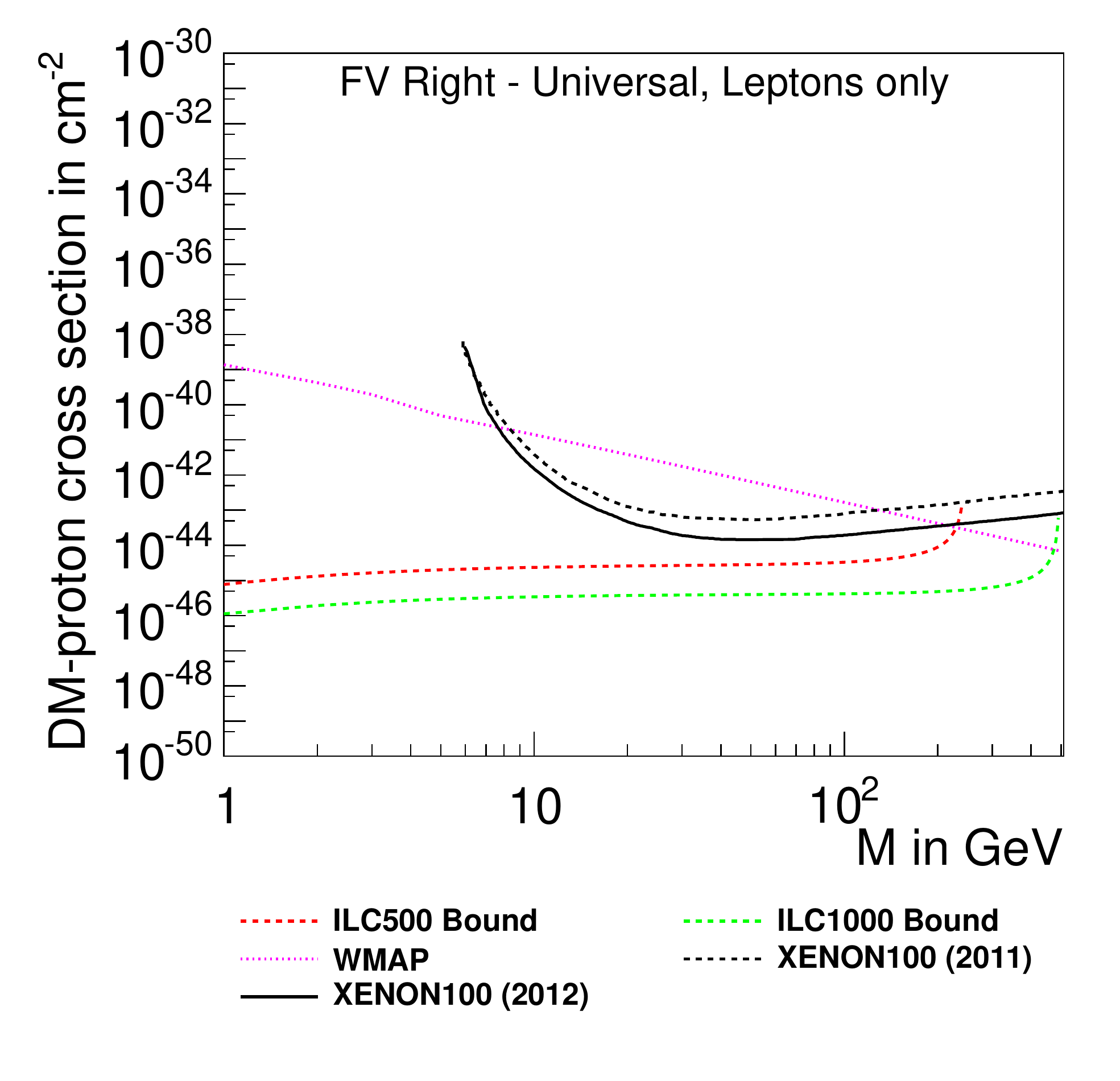} \hfill
 \includegraphics[width=0.475\columnwidth]{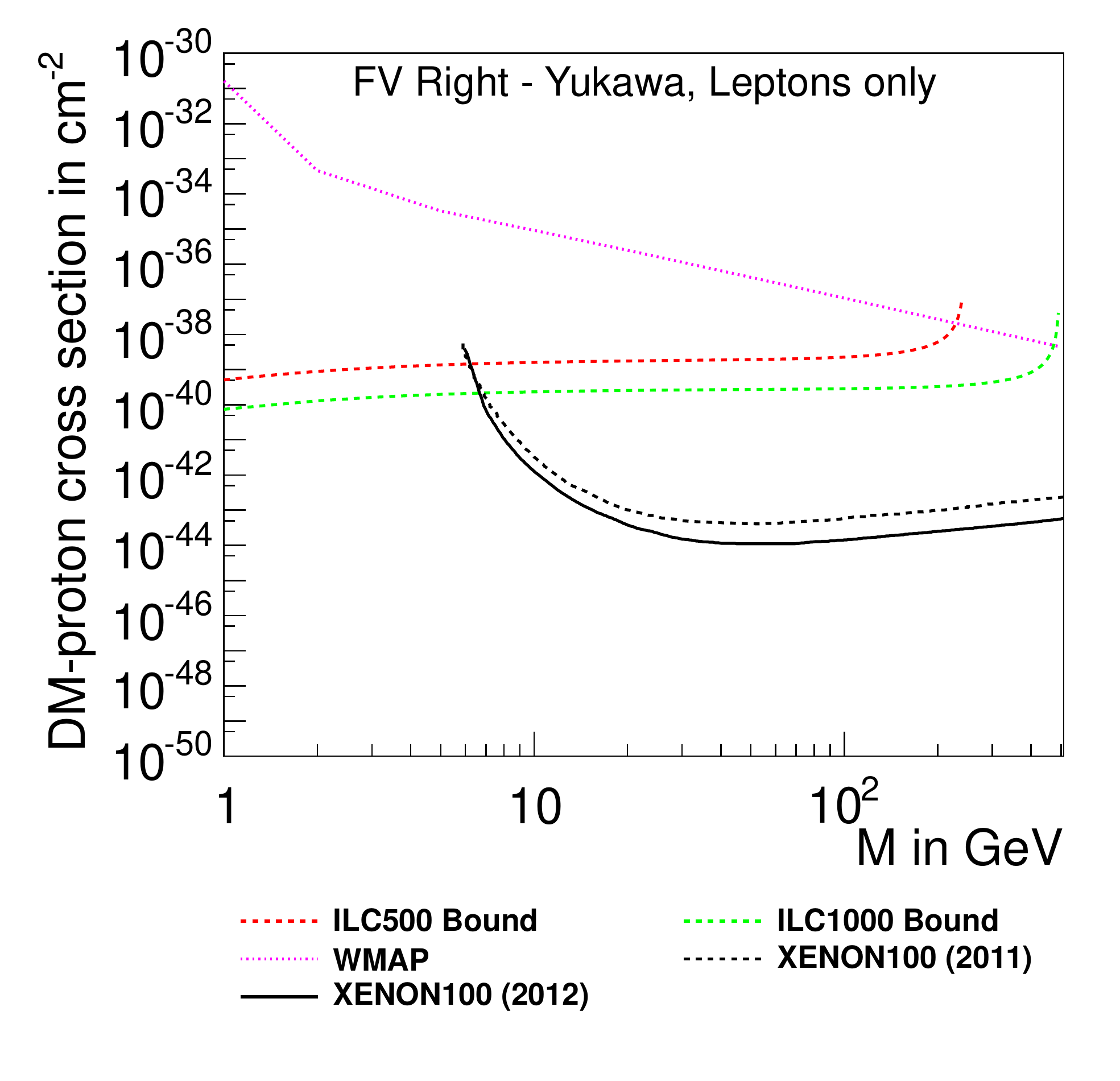}
 \caption{Combined \unit{90}{\%} exclusion limits on the \textbf{spin independent} dark matter proton cross
   section from \textsc{Ilc}, \textsc{Wmap} and \textsc{Xenon} for some \textbf{fermion dark matter} models with  \textbf{s--channel vector coupling} to \textbf{leptons only}.}
 \label{img:totalbounds11}
 \end{figure}
\begin{figure}[H]
\centering
 \includegraphics[width=0.475\columnwidth]{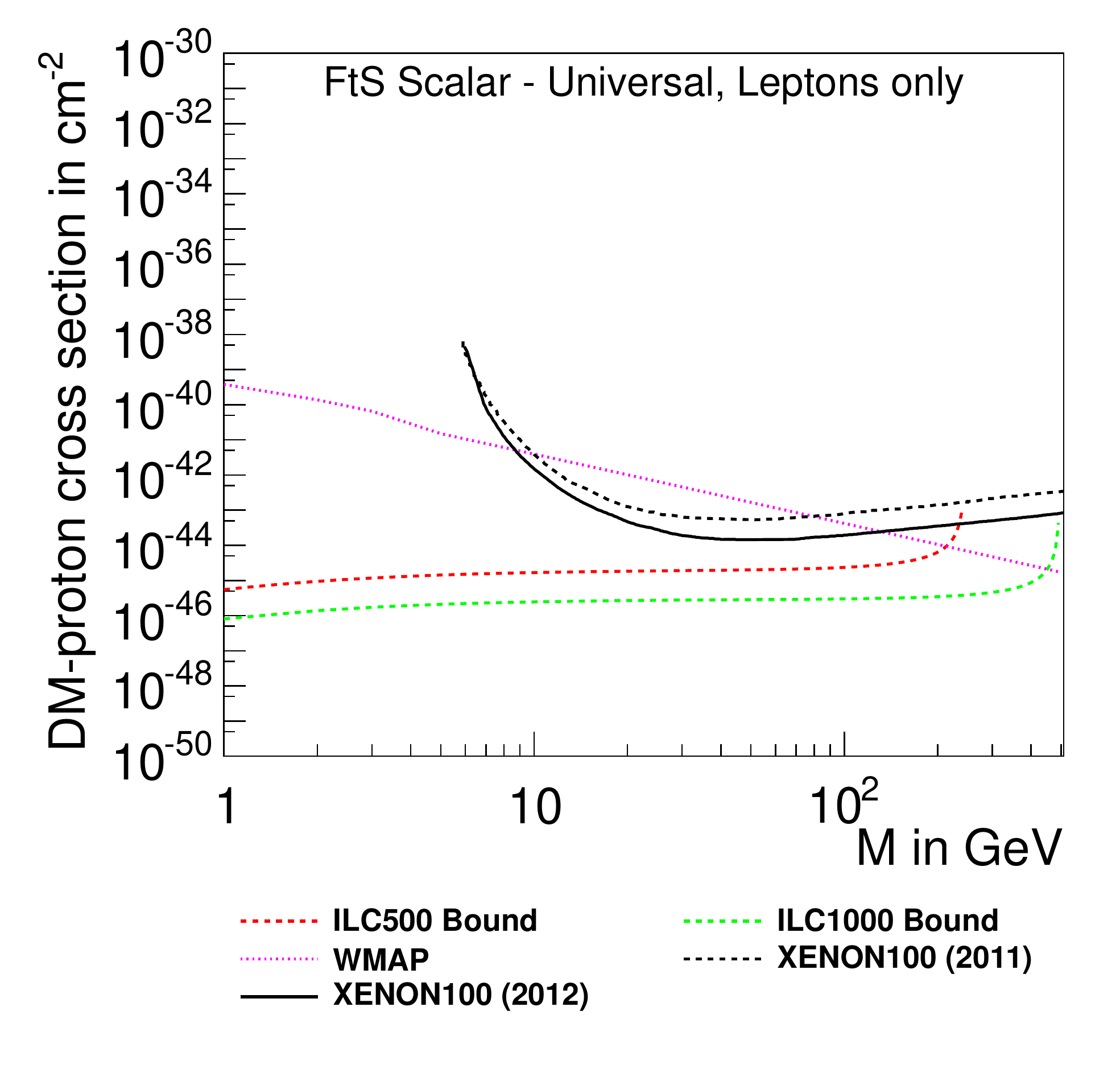} \hfill
 \includegraphics[width=0.475\columnwidth]{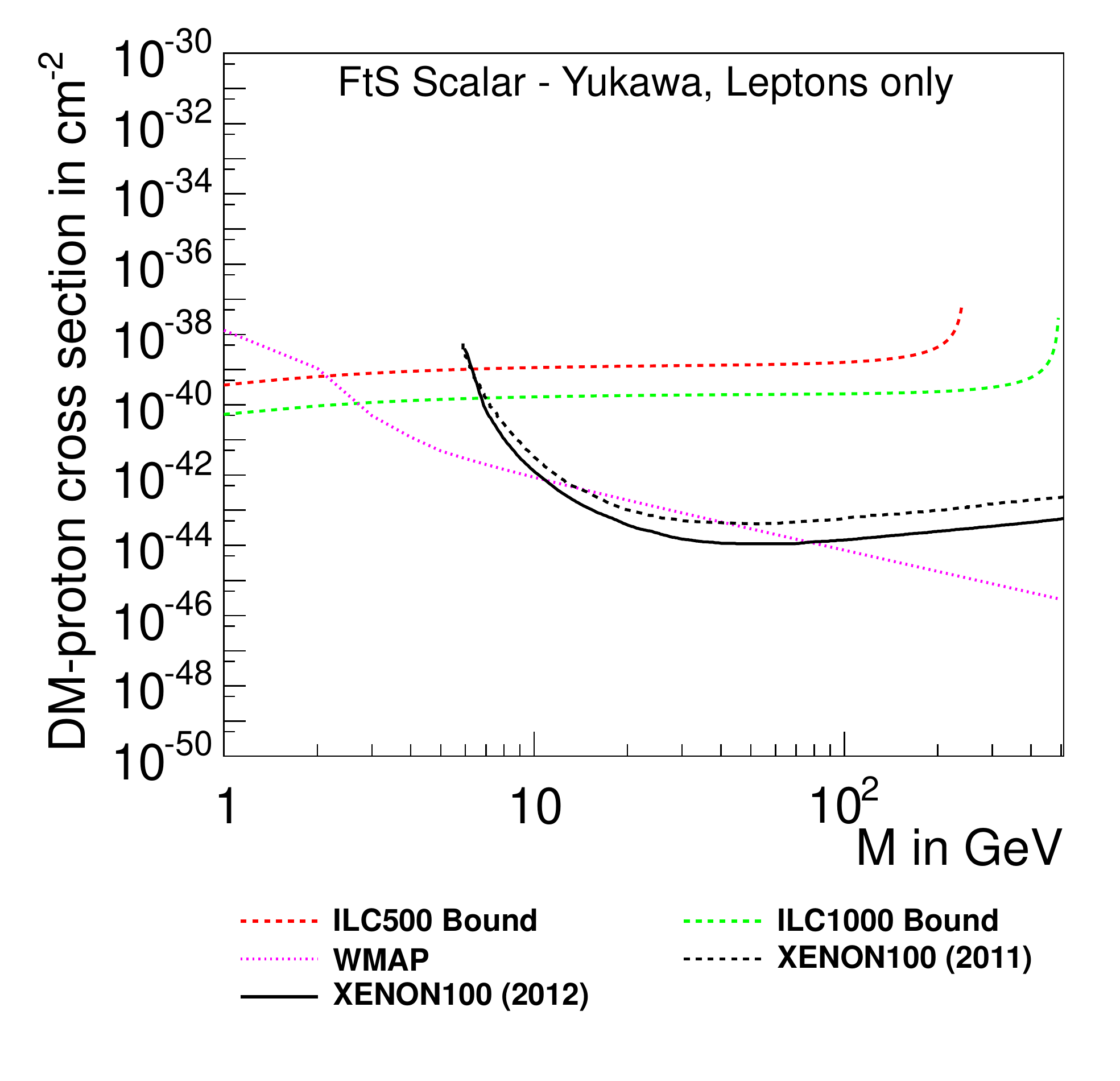} \\
 \includegraphics[width=0.475\columnwidth]{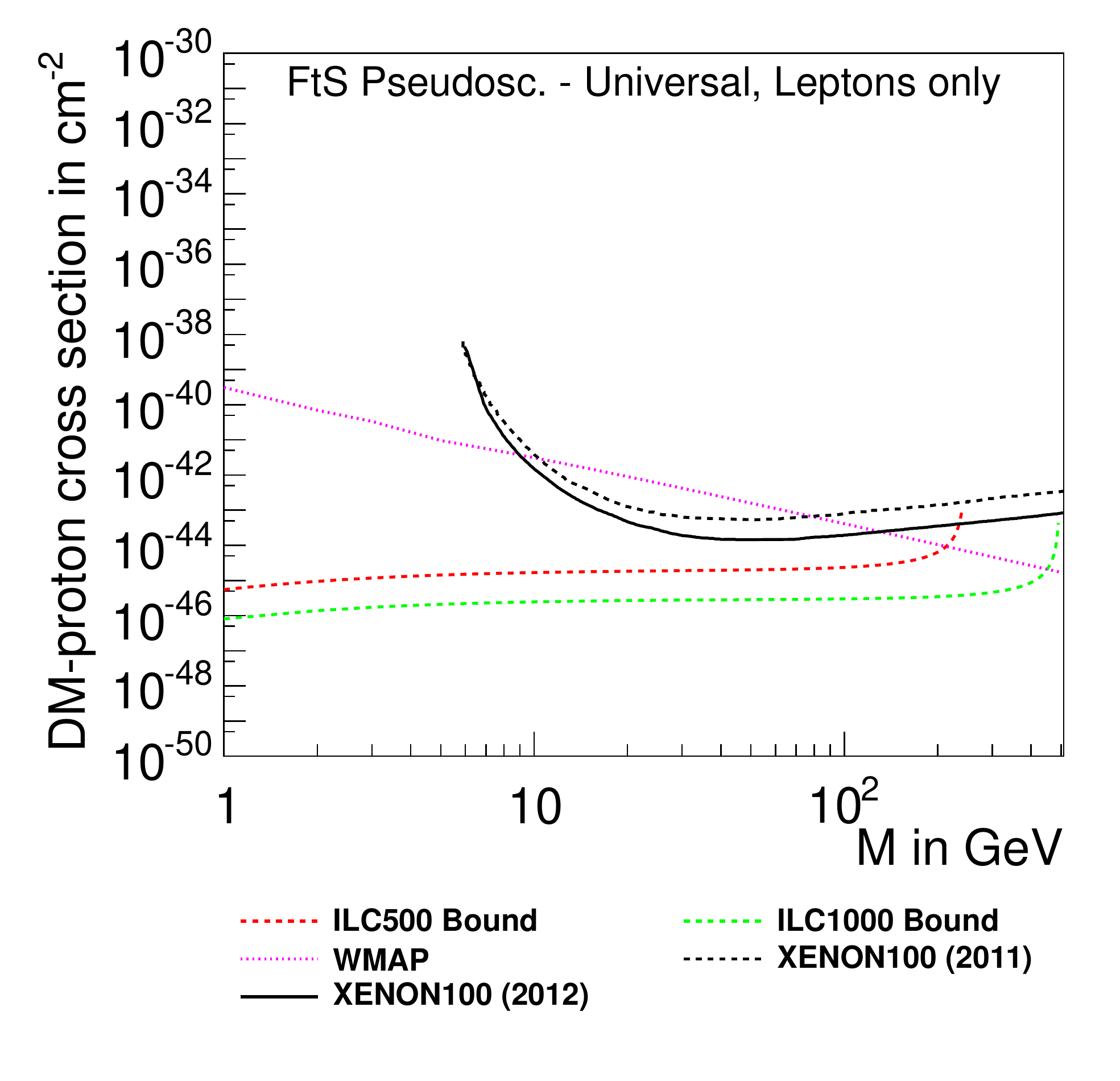} \hfill
 \includegraphics[width=0.475\columnwidth]{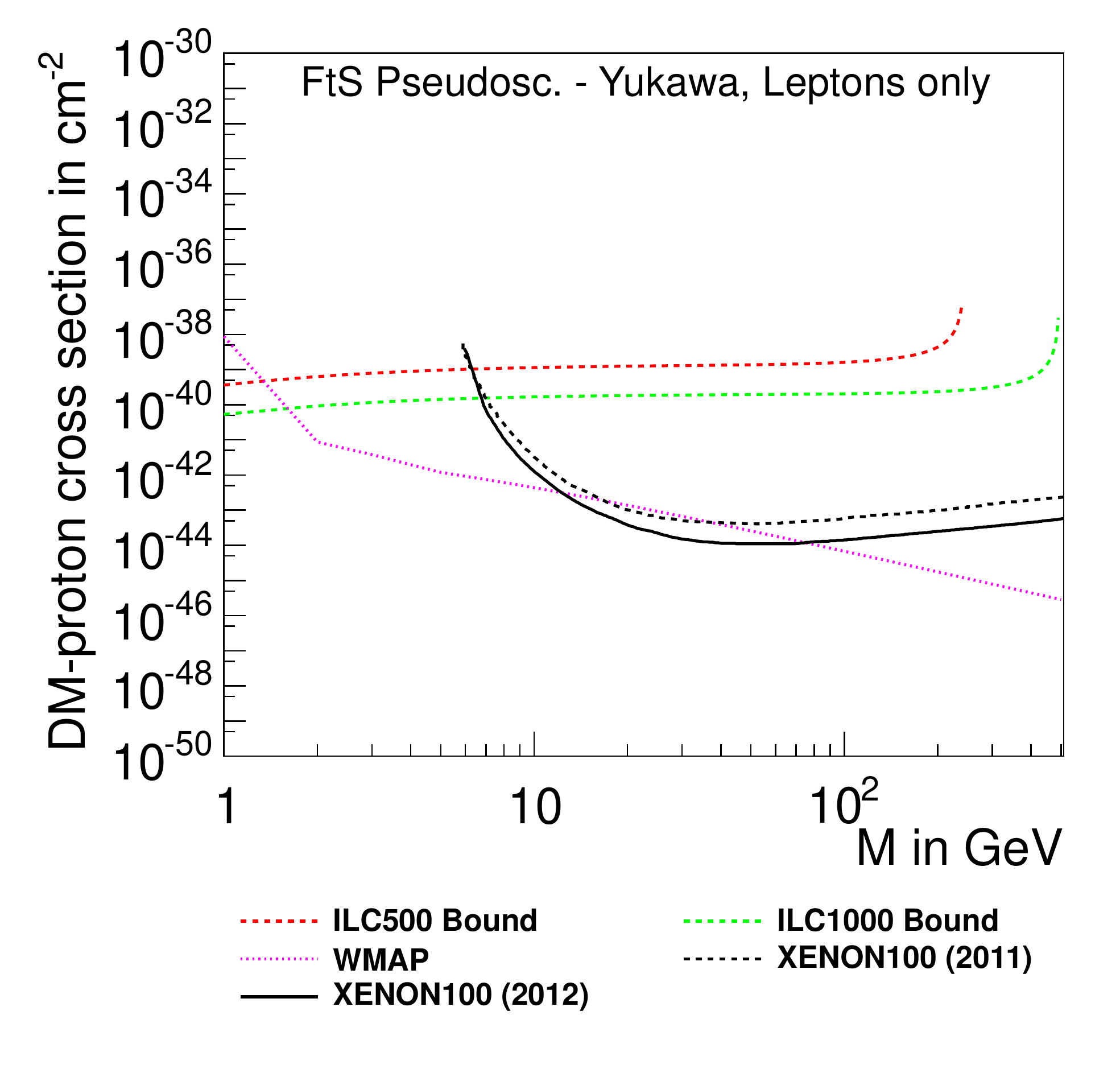}
 \caption{Combined \unit{90}{\%} exclusion limits on the \textbf{spin independent} dark matter proton cross
   section from \textsc{Ilc}, \textsc{Wmap} and \textsc{Xenon} for some \textbf{fermion dark matter} models  with \textbf{t--channel scalar coupling} to \textbf{leptons only}.}
 \label{img:totalbounds12}
 \end{figure}

\begin{figure}[H]
\centering
 \includegraphics[width=0.475\columnwidth]{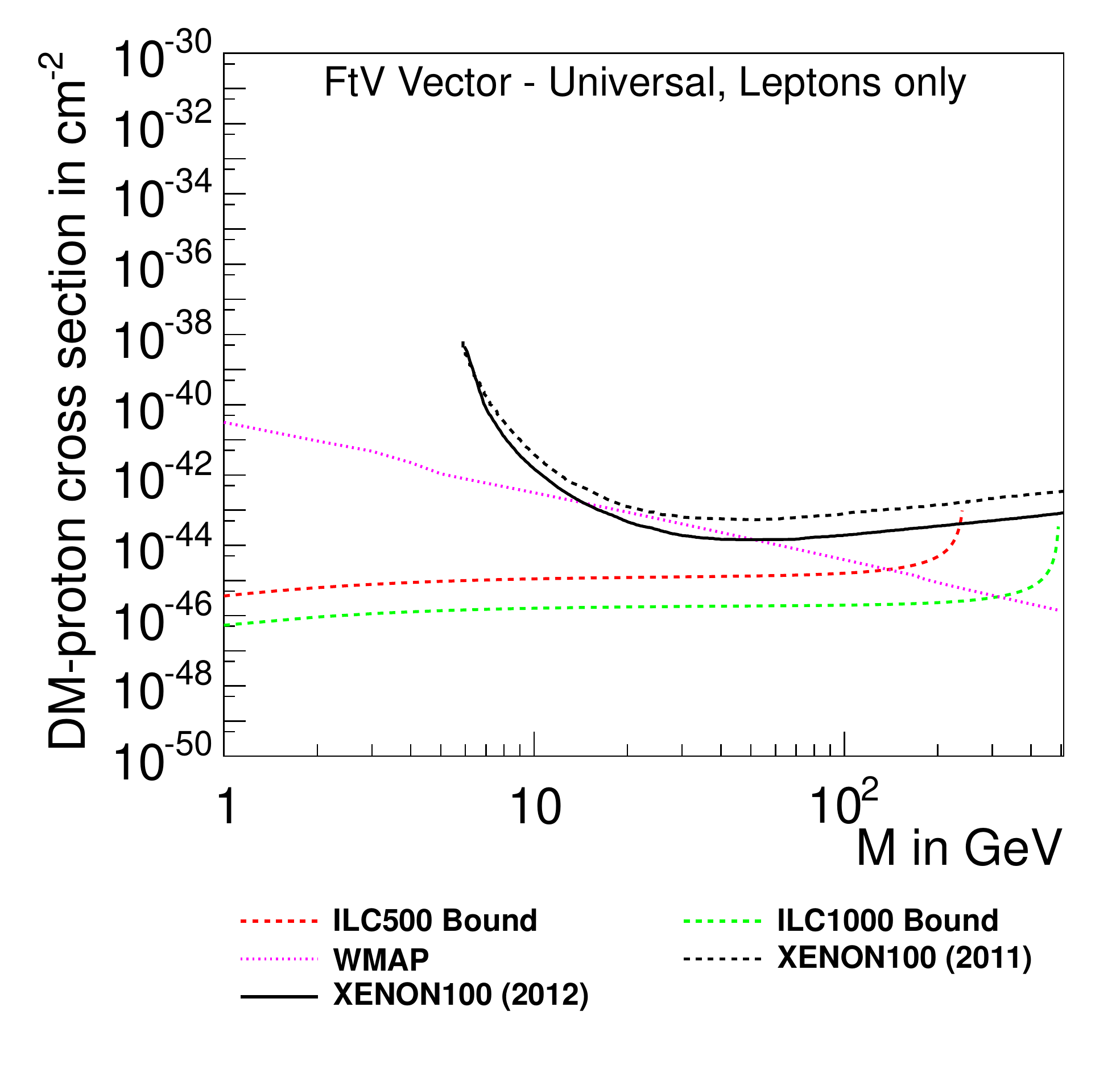} \hfill
 \includegraphics[width=0.475\columnwidth]{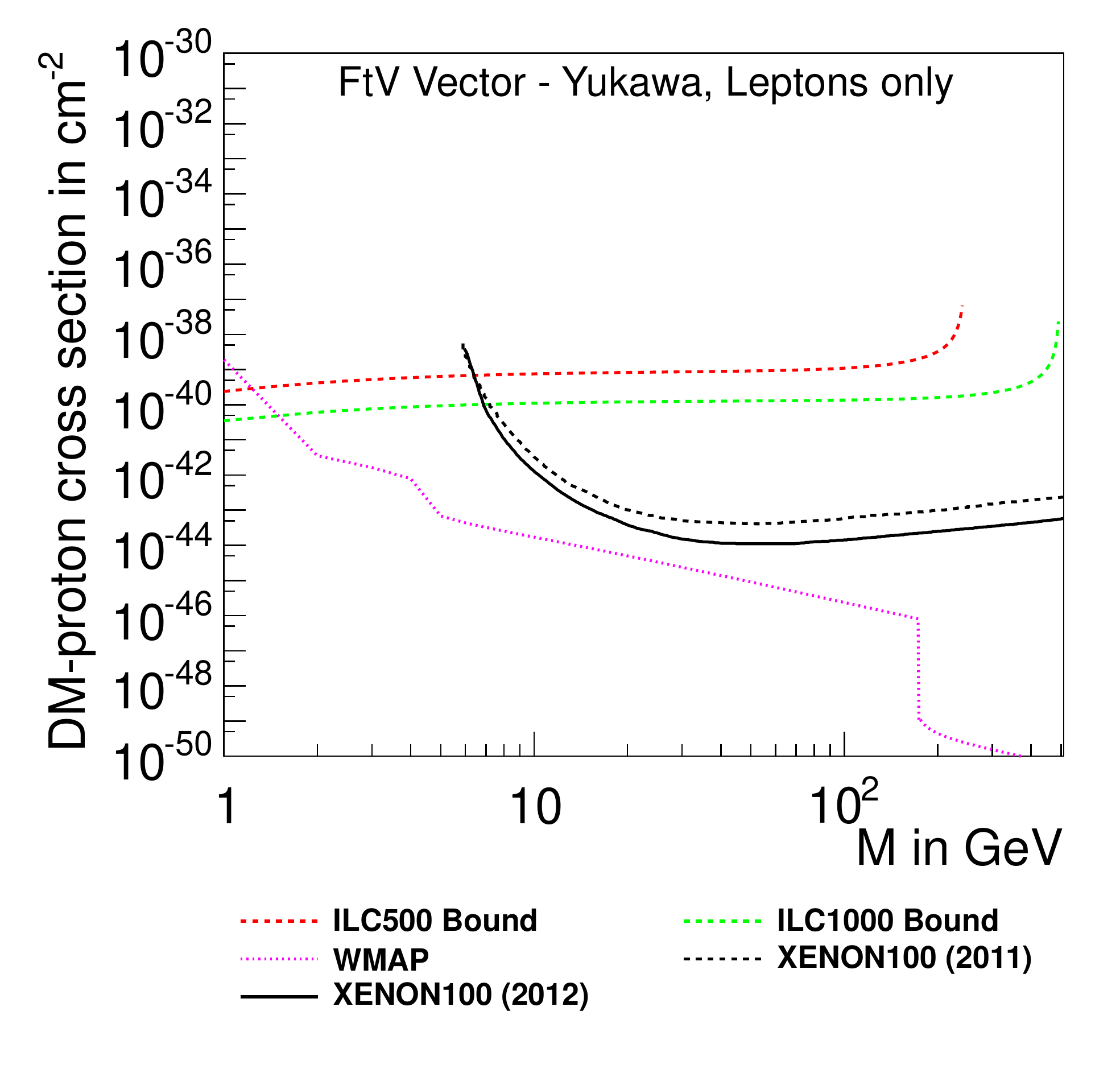} \\
 \includegraphics[width=0.475\columnwidth]{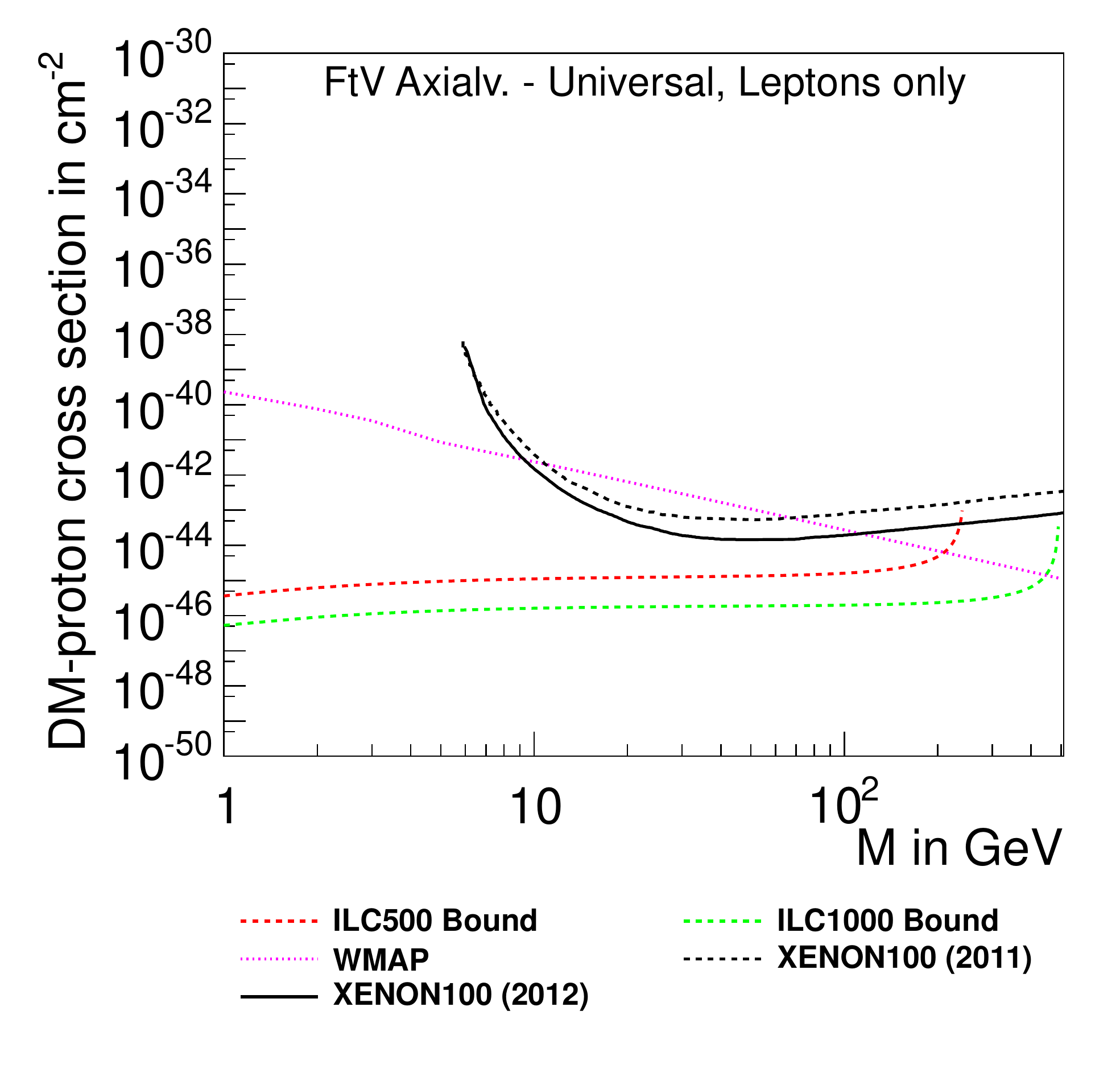} \hfill
 \includegraphics[width=0.475\columnwidth]{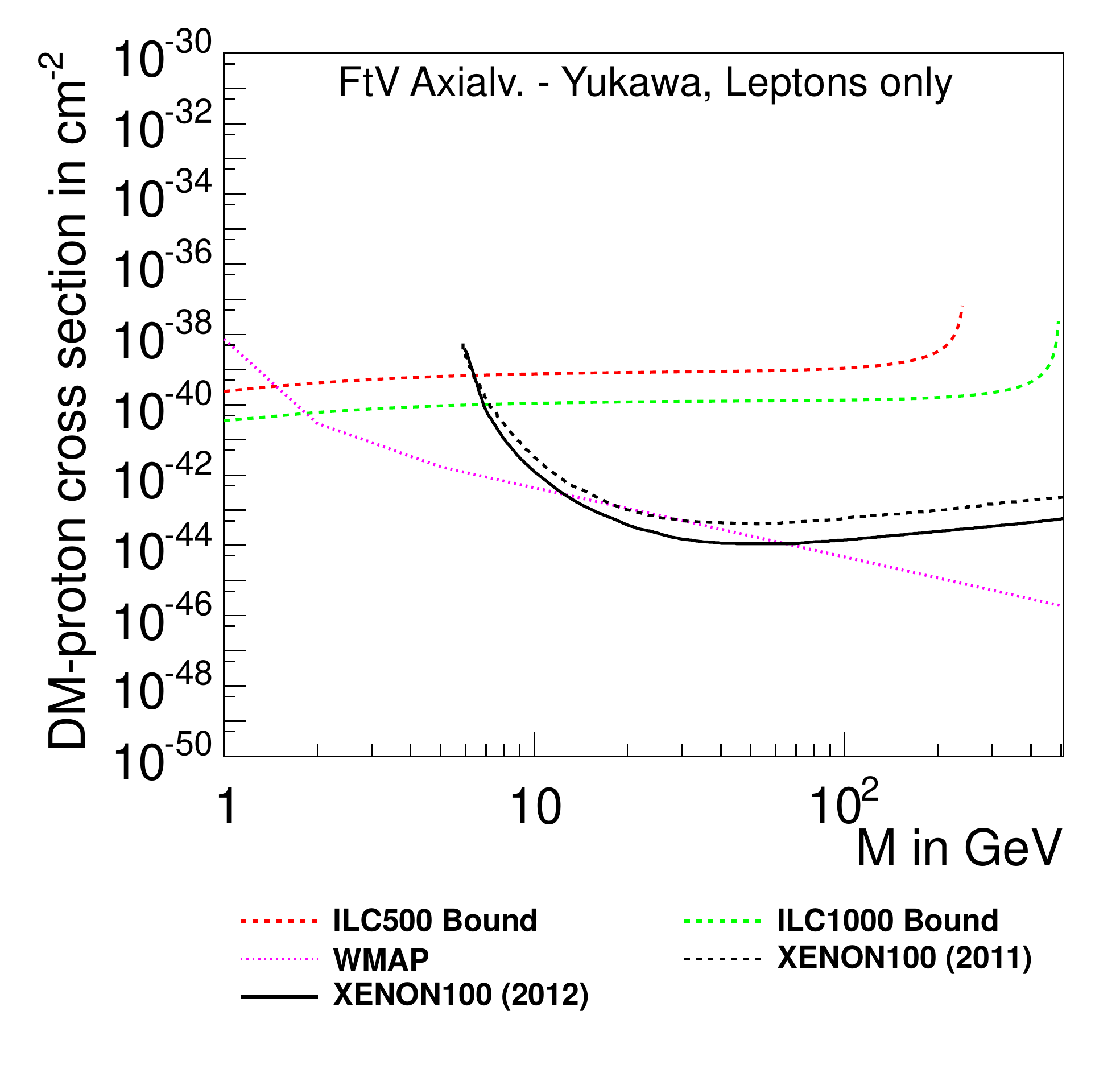} \\
 \includegraphics[width=0.475\columnwidth]{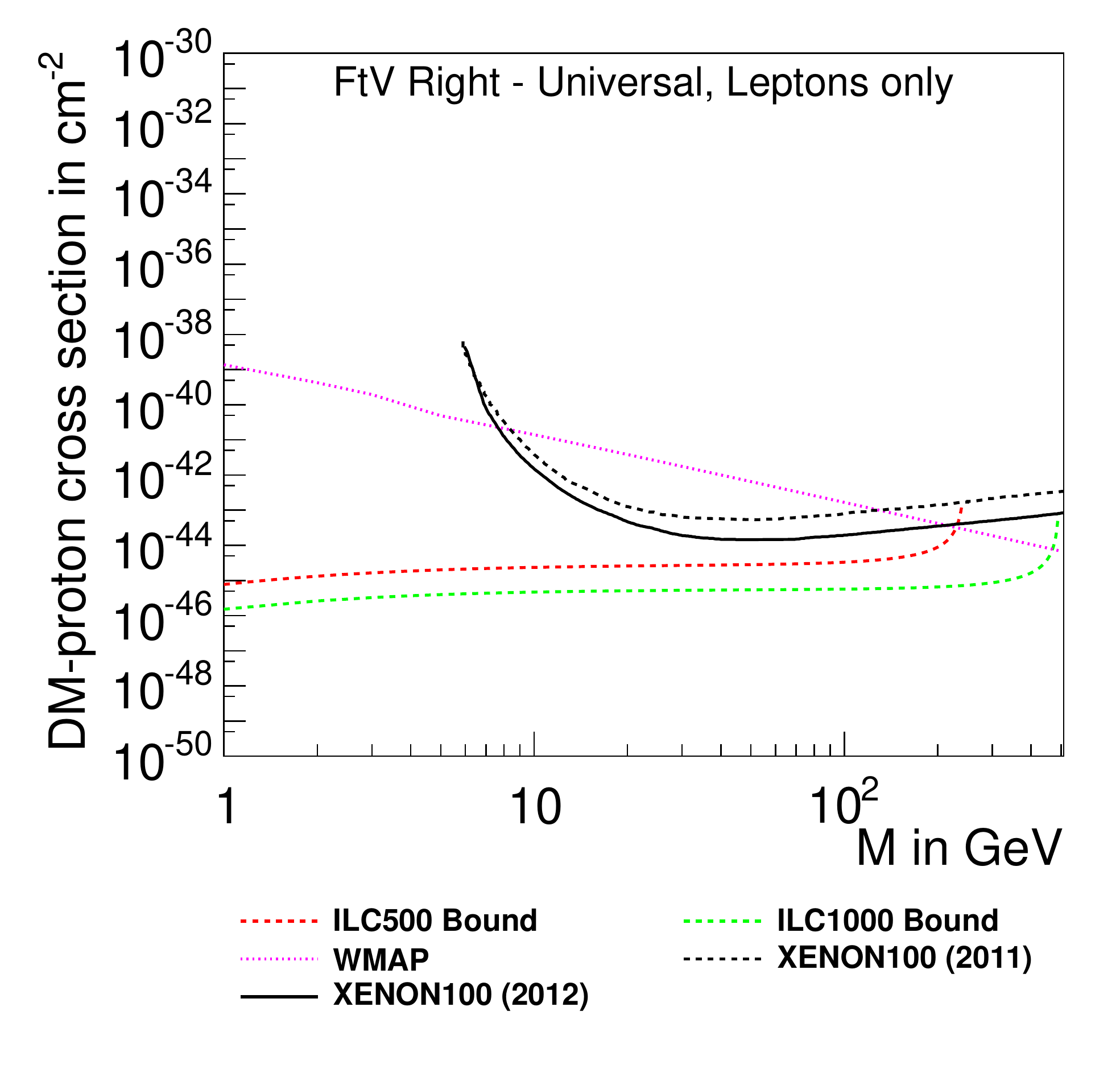} \hfill
 \includegraphics[width=0.475\columnwidth]{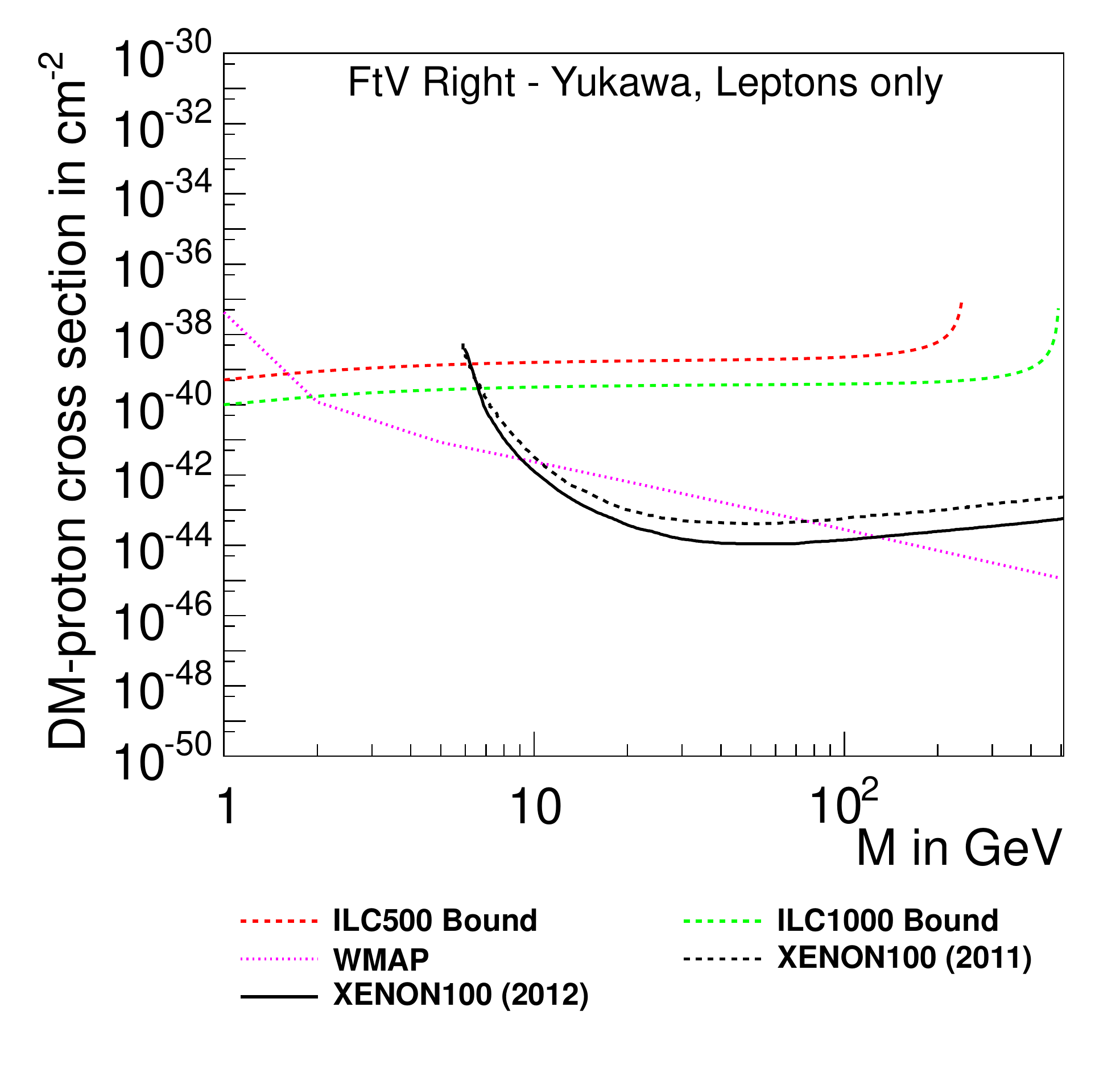} \\
 \caption{Combined \unit{90}{\%} exclusion limits on the \textbf{spin independent} dark matter proton cross
   section from \textsc{Ilc}, \textsc{Wmap} and \textsc{Xenon} for some \textbf{fermion dark matter} models  with \textbf{t--channel vector coupling} to \textbf{leptons only}.}
 \label{img:totalbounds13}
 \end{figure}

\begin{figure}[H]
\centering
 \includegraphics[width=0.475\columnwidth]{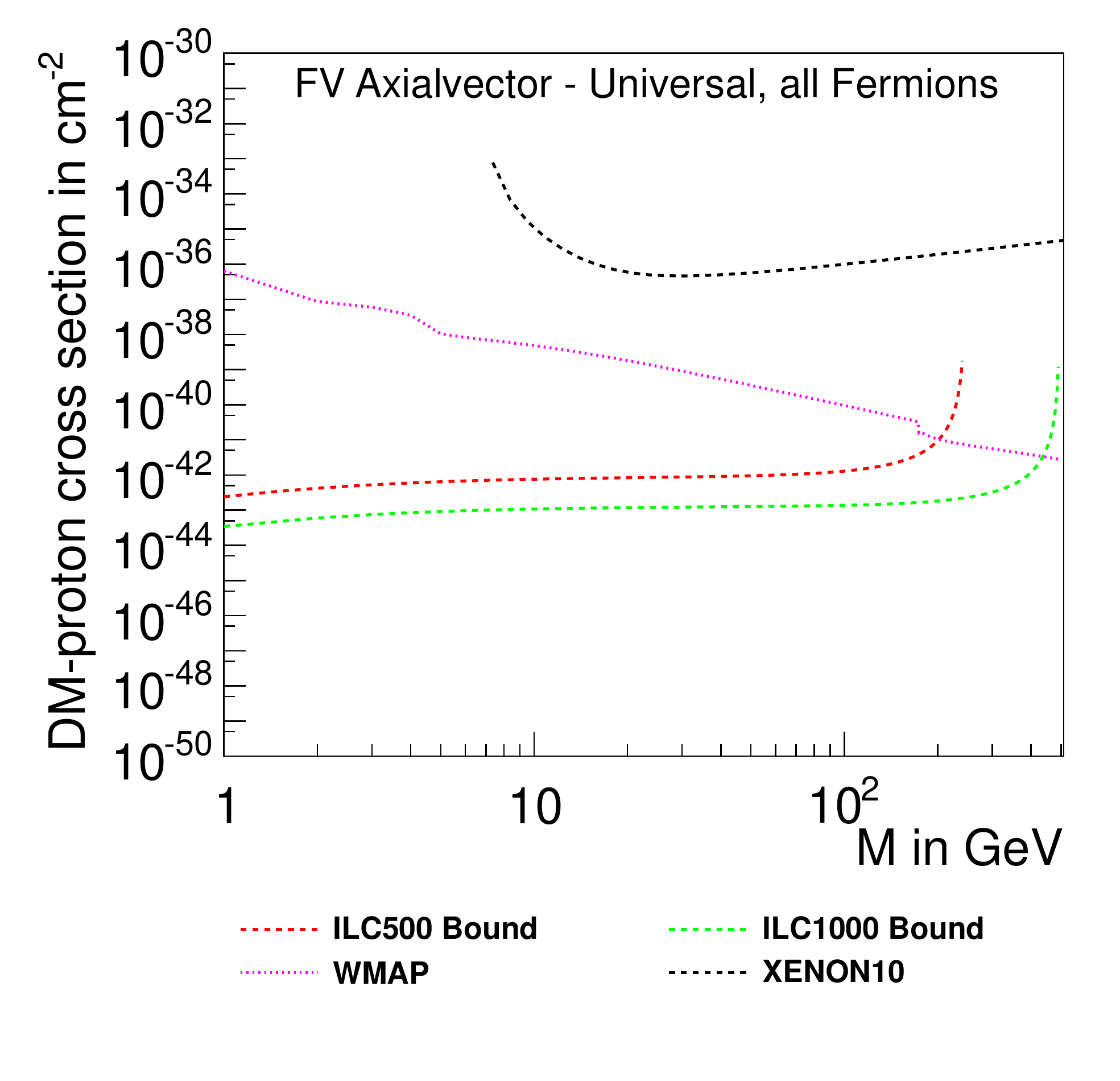} \hfill
 \includegraphics[width=0.475\columnwidth]{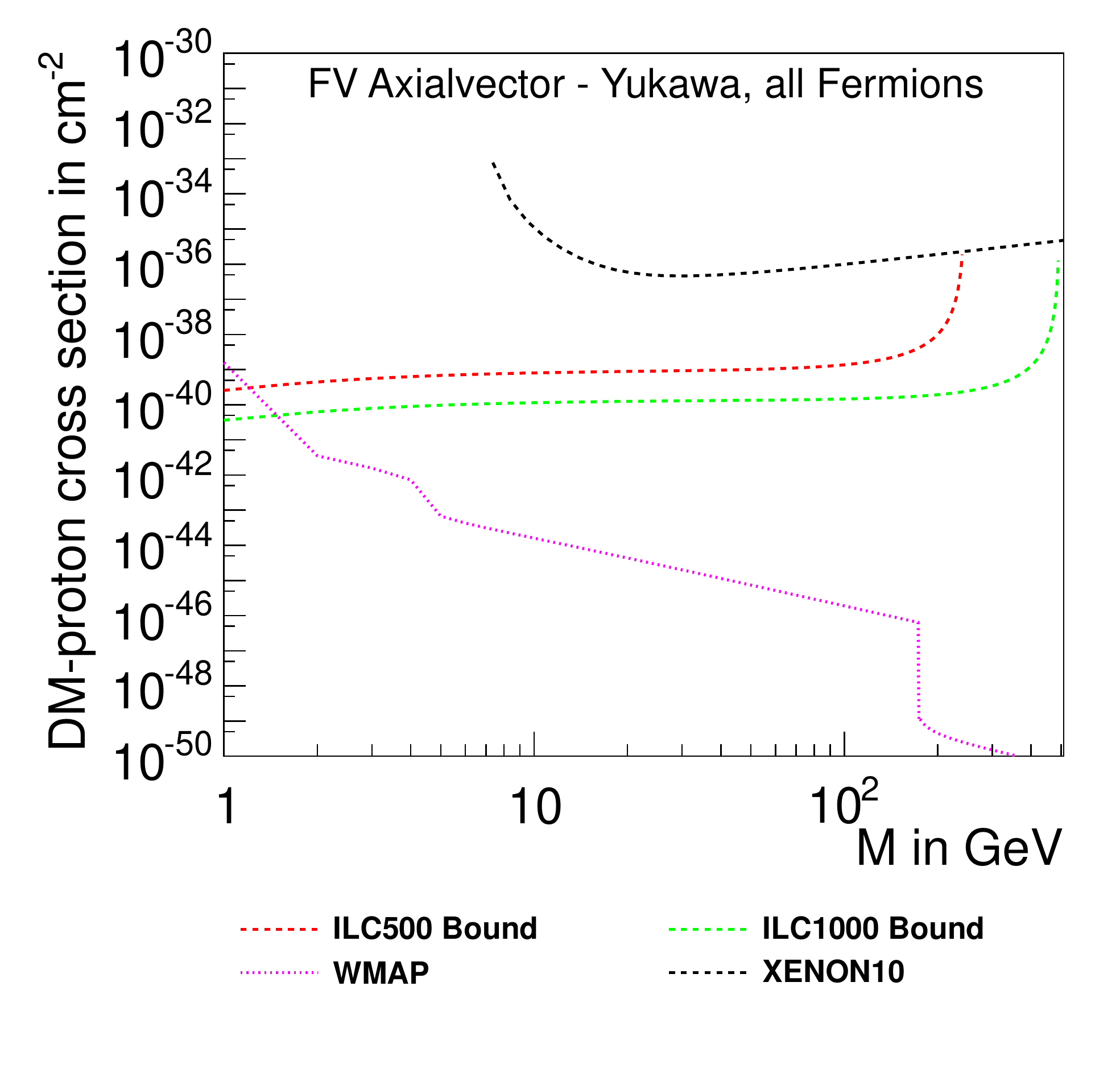} \\
 \caption{Combined \unit{90}{\%} exclusion limits on the \textbf{spin dependent} dark matter proton cross
   section from \textsc{Ilc}, \textsc{Wmap} and \textsc{Xenon} for some \textbf{scalar dark matter} models  with \textbf{s--channel vector coupling} to \textbf{all Standard Model fermions}.}
 \label{img:totalbounds14}
 \end{figure}

\backmatter

\printbibliography
\listoffigures
\listoftables

\pagebreak
\thispagestyle{empty}
\vspace{10cm}
\chapter*{Declaration Of Authorship}
\centering
\includegraphics[width=0.95\textwidth]{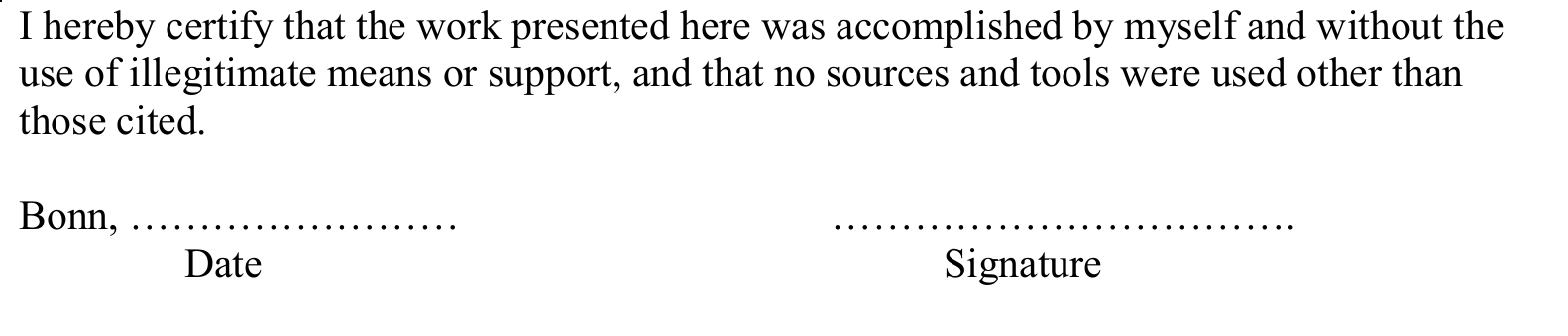}

\end{document}